\documentclass[11pt,a4paper]{article}
\pdfoutput=1
\usepackage{jheppub}
\usepackage{bbm}
\usepackage{youngtab}
\usepackage{graphicx, wrapfig}
\usepackage{rotfloat}
\usepackage{stmaryrd}
\usepackage{amsfonts,amssymb,amsmath,mathtools}
\usepackage{mathrsfs}
\usepackage{tikz}
\usepackage{shadow}
\usepackage{fancybox}
\usepackage{xcolor}
\usepackage{float}
\usepackage[all,cmtip]{xy}

 \usetikzlibrary{decorations.markings}
\newcommand{\maplr}[2]{\begin{picture}(40,20)(0,20)
    \put(18,32){{\small $#1$}}
    \put(3,25){\vector(1,0){34}}
    \put(37,20){\vector(-1,0){34}}
     \put(18,10){{\small $#2$}}
  \end{picture}
}

\def\a{\alpha}
\def\b{\beta}
\def\c{\gamma} \def\g{\gamma}

\def\e{\epsilon}

\def\t{\tau}

\def\R{\Rho}

\DeclareMathOperator{\id}{id}

\DeclareMathOperator{\Mat}{Mat}

\newcommand{\bC}{\ensuremath{\mathbb{C}}}

\newcommand{\bN}{\ensuremath{\mathbb{N}}}

\newcommand{\bP}{\ensuremath{\mathbb{P}}}

\newcommand{\bR}{\ensuremath{\mathbb{R}}}

\newcommand{\bZ}{\ensuremath{\mathbb{Z}}}

\newcommand{\scF}{\ensuremath{\mathscr{F}}}

\newcommand{\scH}{\ensuremath{\mathscr{H}}}

\newcommand{\scM}{\ensuremath{\mathscr{M}}}

\newcommand{\scP}{\ensuremath{\mathscr{P}}}

\newcommand{\frakg}{\ensuremath{\mathfrak{g}}}

\newcommand{\fraksl}{\ensuremath{\mathfrak{sl}}}
\newcommand{\frakso}{\ensuremath{\mathfrak{so}}}

\newcommand{\cA}{\mathcal{A}}

\newcommand{\cC}{\mathcal{C}}
\newcommand{\cD}{\mathcal{D}}

\newcommand{\cF}{\mathcal{F}}

\newcommand{\cH}{\mathcal{H}}
\newcommand{\cI}{\mathcal{I}}

\newcommand{\cL}{\mathcal{L}}

\newcommand{\cN}{\mathcal{N}}
\newcommand{\cO}{\mathcal{O}}
\newcommand{\cQ}{\mathcal{Q}}

\newcommand{\cW}{\mathcal{W}}

\newcommand{\cT}{\mathcal{T}}

\newcommand{\SU}{\mathrm{SU}}
\newcommand{\SO}{\mathrm{SO}}
\newcommand{\SL}{\mathrm{SL}}

\newcommand{\U}{\mathrm{U}}

\newcommand{\Tr}{\mbox{Tr}}

\newcommand{\Li}{{\rm Li}}

\newcommand{\HFK}{\widehat{\mathit{HFK}}}

\newcommand{\hfk}{\mathit{HFK}}
\newcommand{\WL}{{\rm  \textbf{WL}}}
\newcommand{\BR}{{\rm  \textbf{BR}}}
\newcommand{\fin}{{\:\rm  fin}}

\newcommand{\half}{\frac{1}{2}}

\newcommand{\Kauffman}{{\rm Kauff}}
\newcommand{\univ}{{\rm univ}}

\def\e{\epsilon}

\def\bea{\begin{eqnarray}}
\def\eea{\end{eqnarray}}
\def\be{\begin{equation}}
\def\ee{\end{equation}}
\def\ba{\begin{align}}
\def\ea{\end{align}}

\newcommand{\bem}{\begin{pmatrix}}
\newcommand{\eem}{\end{pmatrix}}

\def\={\;  = \;}
\def\+{\, + \,}

\def\wt{\widetilde}
\def\wh{\widehat}
\def\bar{\overline}

\def\rt2{\sqrt{2}}

\newcommand{\unknot}{{\raisebox{-.11cm}{\includegraphics[width=.37cm]{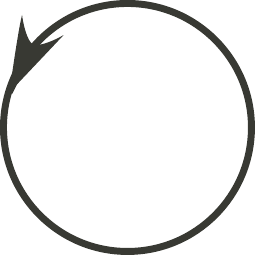}}}\,}

\def\bb#1#2{\left\llbracket\begin{array}{c}{#1}\\{#2}\end{array}\right\rrbracket}

\preprint{CALT-2015-063}

\title{Sequencing BPS Spectra}

\author[1,2]{Sergei Gukov}
\author[1,3]{Satoshi Nawata}
\author[1]{Ingmar Saberi}
\author[4,5]{Marko Sto$\check{\text{s}}$i$\acute{\text{c}}$}
\author[1,6]{Piotr Su{\l}kowski}

\affiliation[1]{Walter Burke Institute for Theoretical Physics, California Institute of Technology,\\ Pasadena, CA 91125, USA}
\affiliation[2]{Max-Planck-Institut f\"ur Mathematik, Vivatsgasse 7, D-53111 Bonn, Germany}
\affiliation[3]{Centre for Quantum Geometry of Moduli Spaces, University of Aarhus, DK-8000, Denmark}
\affiliation[4]{CAMGSD, Departamento de Matem\'atica, Instituto Superior T\'ecnico,\\
Av. Rovisco Pais, 1049-001 Lisbon, Portugal}
\affiliation[5]{Mathematical Institute SANU, Knez Mihajlova 36, 11000 Belgrade, Serbia}
\affiliation[6]{Faculty of Physics, University of Warsaw, ul. Pasteura 5, 02-093 Warsaw, Poland}

\emailAdd{gukov@theory.caltech.edu}
\emailAdd{snawata@gmail.com}
\emailAdd{isaberi@caltech.edu}
\emailAdd{mstosic@isr.ist.utl.pt}
\emailAdd{psulkows@fuw.edu.pl}

\abstract{This paper provides both a detailed study of color-dependence of link homologies, as realized in physics as certain spaces of BPS states, and a broad study of the behavior of BPS states in general. We consider how the spectrum of BPS states varies as continuous parameters of a theory are perturbed. This question can be posed in a wide variety of physical contexts, and we answer it by proposing that the relationship between unperturbed and perturbed BPS spectra is described by a spectral sequence. These general considerations unify previous applications of spectral sequence techniques to physics, and explain from a physical standpoint the appearance of many spectral sequences relating various link homology theories to one another. We also study structural properties of colored HOMFLY homology for links and evaluate Poincar\'e polynomials in numerous examples.
Among these structural properties is a novel ``sliding" property, which can be explained by using (refined) modular $S$-matrix. This leads to the identification of modular transformations in Chern-Simons theory and 3d $\cN=2$ theory via the 3d/3d correspondence.
Lastly, we introduce the notion of associated varieties as classical limits of recursion relations of colored superpolynomials of links, and study their properties.}

\keywords{\begin{figure}[H]\centering \includegraphics[scale=1]{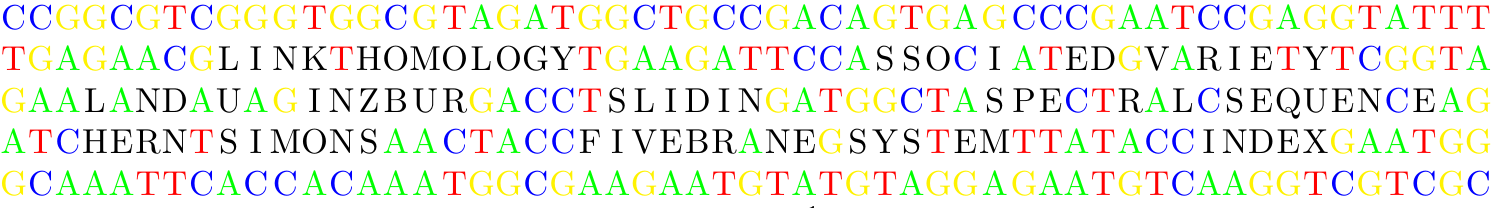} \end{figure}}

\begin{document}
\Yboxdim4pt

\maketitle

\section{BPS spectral sequences}
\label{sec:deformations}

The bulk of this paper deals with the structure of link homologies. As we explain in great details in the next section, link homologies are realized as spaces of BPS states in several M-theory configurations, in which the geometry of certain branes is prescribed by a choice of link $L\subset S^3$ \cite{Ooguri:1999bv,Gukov:2004hz,Gukov:2007ck,Diaconescu:2011xr,Witten:2011zz}.
Schematically,
$$
\scH(L)\cong\scH_{\textrm{BPS}}.
$$
Studying descriptions of this space of BPS states from various viewpoints and in various duality frames then leads to interesting predictions about the structure of link homologies.

This interpretation has led to deep insights regarding the structure and behavior of homological link invariants. In particular, the rich structural properties of link homologies include differentials of various kinds relating different homology theories, which have been formulated in terms of \emph{spectral sequences} in mathematics \cite{Lee:2005,Gornik:2004,Dunfield:2005si,Rasmussen:2006}.
One of the main goals of this paper is to obtain a physical understanding of the spectral sequences between various link homologies. These spectral sequences imply relations between the BPS spectra of different configurations of branes in M-theory, or more generally between different physical theories. Our goal is to understand these relations.

While the notion of a spectral sequence may be unfamiliar or daunting to many physicists, we hope to demonstrate that it can be understood in simple terms, which are clearly connected to physical problems. While certain applications of spectral sequences to physics have been made before (for example, see \cite{Dixon:1991wi,Bouwknegt:1991yg,deBoer:1992sy,Bertolini:2013xga,Wong:2015qnf}), we feel that our general understanding of the context in which a physicist should expect spectral sequences to arise is novel. Put simply, spectral sequences describe how the cohomology of a supercharge $Q$  changes under deformations of continuous parameters.
Generally, $Q$-cohomology does not remain invariant under deformations of the theory, even in contexts where the deformation leaves the Witten index and other supersymmetric indices invariant. For this reason, many of our considerations apply very generally to supersymmetric theories, and should not be construed as limited to the supersymmetric systems we use to describe links.

The first sections of this paper, therefore, discuss spectral sequences and deformation problems in supersymmetric theories in very general terms, and give simple examples of how well-known results (ranging from twists of 2d $\cN=(2,2)$ theories to specializations of 4d indices) fit into our framework. They can be read independently of the rest of the paper. Once we have developed the necessary ideas, we review the physical approach to link homologies in \S \ref{sec:setup}, and then apply them to understand the deformation spectral sequences constructed in \cite{Lee:2005,Gornik:2004}.
The remainder of the paper continues further with our study on ``color'' dependence of link homologies, and its detailed summary is presented in \S \ref{sec:organization}.

Explanations of spectral sequences can be found in many places in the mathematics literature (for instance, see \cite{Bott:1982,Kato:2006}).
However, we will offer a pedestrian exposition below, in language which is hopefully both familiar to physicists and tailored to our purposes. In addition, we will present several examples of spectral sequences in the BPS spectra of physical theories. One of these examples (in the context of Landau-Ginzburg models) will be the relevant case in the context of link homology.

\subsection{Generalities: deformations and spectral sequences}
\label{sec:Qchoice}

We are interested in the following very general question: Suppose we have a supersymmetric quantum field theory, in which a single nilpotent supercharge~$Q$ has been chosen. The supercharge could be a scalar (as in a topological twist of the theory), or more generally any element of the fermionic part of the supersymmetry algebra (for instance, in superconformal indices). We make no assumptions about the action of the Lorentz group on this supercharge.

We would then like to know what happens to $Q$-cohomology as we continuously vary some choice or parameter of the theory. The parameters which can be varied are naturally of two types:
\begin{enumerate}\setlength{\parskip}{-0.2cm}
\item we could change our choice of~$Q$, while keeping the theory the same \cite{Dixon:1991wi,Bouwknegt:1991yg,deBoer:1992sy,Bertolini:2013xga};
\item we could vary some modulus of the theory, such as the superpotential or other coupling constants \cite{Wong:2015qnf}.
\end{enumerate}
\vspace{-.1cm}
While the two types of perturbation are distinct, they both allow us to pose the same broad question: What happens to $Q$-cohomology as continuous parameters are adjusted?

In fact, the spectrum of BPS states (i.e., $Q$-cohomology) changes under the deformation;
this  is part of the reason for introducing genuinely protected quantities, such as the Witten index \cite{Witten:1981nf,Witten:1982im}, in the first place. The Witten index counts the number of vacua in supersymmetric quantum mechanics, with bosonic vacua contributing $+1$ and fermionic vacua $-1$:
\[ \Tr (-1)^F e^{-\beta H} = n_B-n_F ~. \]
It is well-known that the numbers $n_B$ and~$n_F$ are not themselves invariant. However, a long representation of supersymmetry that breaks apart into short representations will always contribute equally to $n_B$ and~$n_F$, so that their difference is honestly invariant under perturbations.

Moreover, while both $n_B$ and $n_F$ can jump, they do so in a particular fashion that can be understood on general grounds. Since the numbers $n_B$ and $n_F$ are degeneracies of a quantum system, one expects enhanced degeneracy to correspond to enhanced symmetry. Moreover, the degeneracies should be enhanced only on a locus of positive codimension in the parameter space, where the parameters are tuned so that symmetry-breaking effects are absent. One can argue for the same conclusion abstractly by considering a one-parameter family of differentials~$d_t$ acting in a graded vector space, and noticing that $d_t$-cohomology jumps only at values of~$t$ for which certain linear subspaces have non-generic intersection, i.e., when certain equations on the parameter are satisfied. In some cases, we will be able to explicitly identify the symmetry that is enhanced when $Q$-cohomology jumps in rank as an $R$-symmetry of the theory.

Let us note that this jumping behavior we are considering is to be contrasted with wall-crossing. Wall-crossing occurs at loci of real codimension one in the moduli space, known as walls of marginal stability. On these walls, the one-particle BPS spectrum is no longer separated from the multiparticle continuum, so that BPS states can decay into BPS constituents. When this occurs,  even the index will jump.


We have identified two types of possible deformations to study. Let us begin with the first kind, namely, which particular supercharge is chosen. The odd part of a supersymmetry algebra is usually a vector space over~$\bC$. Since we will be interested in $Q$-cohomology, scaling a supercharge by a constant does not affect the result, so that the moduli space of possible choices for~$Q$ can be thought of as $\bC \bP^N$ ($N$ being the number of real supercharges of the theory).
Within this moduli space, \emph{nilpotent} supercharges will lie on a certain locus, which need not (in fact, will almost never) be the whole of $\bC \bP^N$. Rather, it is a certain subvariety, described by the equations resulting from the condition that $Q^2=0$. Moreover, the nilpotent locus will not parameterize interesting choices uniquely. There are bosonic symmetries of the theory  that act nontrivially on $\bC \bP^N$ and on the nilpotent locus inside it. To parameterize the possible choices of supercharge without redundancy, one must quotient by these symmetries as well. Let us see how this looks in some examples.
\\

\noindent\textbf{Example 1:}
We begin by considering the $\cN=(2,2)$ supersymmetry algebra in 2d. That algebra has four supercharges, and is defined by the commutation relations
$$
\{Q_+,\overline Q_+\}=\frac{1}{2}(H+P):=H_L~,\quad \{Q_-,\overline Q_-\}=\frac{1}{2}(H-P):=H_R~, \\
$$
with all other anticommutators equal to zero.

\begin{wrapfigure}{L}{0.5\textwidth}
\centering
\begin{tikzpicture}[line join = round, line cap = round, x={(1cm,0cm)}, y={(0cm,1cm)}, z={(0.15cm,0.15cm)}]]
\pgfmathsetmacro{\factor}{1/sqrt(2)};

\newcommand{\tetra}{
\coordinate [label=right:{$Q_+$}] (A) at (2,0,-2*\factor);
\coordinate [label=left:{$\overline Q_+$}] (B) at (-2,0,-2*\factor);
\coordinate [label=above:{$Q_-$}] (C) at (0,2,2*\factor);
\coordinate [label=below:{$\overline Q_-$}] (D) at (0,-2,2*\factor);

  \draw[dashed] (B)--(A);
\draw[-, fill=gray!30, opacity=.5] (A)--(D)--(B)--cycle;
\draw[-, fill=white!30, opacity=.5] (A) --(D)--(C)--cycle;
\draw[-, fill=gray!30, opacity=.5] (B)--(D)--(C)--cycle;

\draw[very thick] (B)--(C) node[midway, above left]{$Q_A$};
\draw[very thick] (A)--(D) node[midway, below right]{$Q_A$};
\draw[dashed, ultra thick] (A)--(C) node[midway, above right]{$Q_B$};
\draw[dashed, ultra thick] (B)--(D) node[midway, below left]{$Q_B$};
}
\tetra
\end{tikzpicture}
\caption{The reduced moduli space of scalar supercharges in the 2d $\cN=(2,2)$ supersymmetry algebra (a tetrahedron in~$\bR \bP^3$). The nilpotent locus (four of the six one-simplices) is indicated with thick lines.\vspace{.8cm}}
\label{fig:tetrahedron}
\end{wrapfigure}
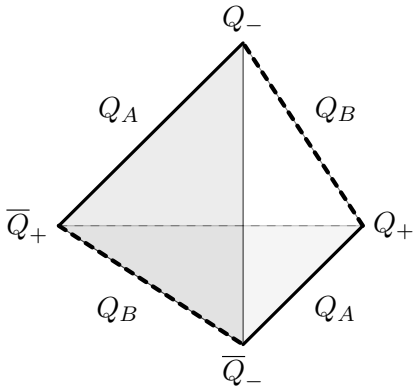

The odd part of the superalgebra is isomorphic (as a vector space) to~$\bC^4$, and so the moduli space of supercharges is~$\bC \bP^3$, corresponding to a general supercharge
\[ Q = a_+ Q_+ + a_- Q_- + b_+ \overline Q_+ + b_- \overline Q_-. \]
Using the supersymmetry algebra, we find that
\[Q^2 = a_+ b_+ H_+ + a_- b_- H_-. \]
It follows that $Q$ is nilpotent only when the equations $a_+ b_+ = a_- b_- = 0$ are satisfied. The nilpotent locus  therefore consists of four distinct complex lines in~$\bC \bP^3$. Each $\bC \bP^1$ intersects two of the others (but not the third) in a point; one should think of two horizontal and two vertical lines in the plane, intersecting to form a square. Concretely, the variety has four components:
\[ a_+ = a_- = 0, \quad b_+ = b_- = 0; \quad a_+ = b_- = 0,\quad b_+ = a_- = 0. \]
The four intersection points correspond to the four individual supercharges $Q_\pm$ and~$\overline Q_\pm$.

Now we can ask how the bosonic Lorentz and $R$-symmetries act on our picture. In this case, the action is simple to describe: it can be used to set all of the phases of the coefficients to zero. The supercharges are $\U(1)$ eigenstates for all three symmetries---Lorentz, $J_L$, and $J_R$. The three symmetries can be used to set the phases of $a_\pm$ and~$b_\pm$ all equal; the remaining overall phase is eliminated upon passing to projective space. Hence, after taking the quotient by these symmetries, the reduced moduli space can be identified with the set of points in~$\bR \bP^3$ for which all coordinates are positive.
 Figure~\ref{fig:tetrahedron} depicts a three-dimensional polyhedral region; in fact, it is a tetrahedron, whose four vertices are the four supercharges $Q_\pm$ and~$\overline Q_\pm$. (Eight such tetrahedra would comprise the whole of~$\bR \bP^3$.) The nilpotent locus lies along four lines which are edges of the tetrahedron, and fit together in a square: each line intersects two of the other three in a single point.

For our purposes, we will often find it more convenient to describe the unreduced nilpotent locus and remember its symmetries separately.
\\

\noindent\textbf{Example 2:}
As another example, let us think about the 4d $\cN=1$ supersymmetry algebra. Again, there are four real supercharges, denoted $Q_\alpha$ and~$\overline Q_{\dot\alpha}$ in usual 4d notations.
The supersymmetry algebra is
\[ \{ Q_\alpha, \overline Q_{\dot \alpha}\} = 2 P_{\alpha \dot\alpha}~, \]
 with all other anticommutators zero.
If we consider a generic fermionic charge,
\[ Q = a_\alpha Q_\alpha + b_{\dot\alpha} \overline Q_{\dot\alpha}~,\]
it is clear that the nilpotent locus is described by the four equations
\[ a_\alpha b_{\dot\alpha} = 0~, \quad \alpha, \dot\alpha = 1,2~.\]
These equations require that either $a=0$ or $b=0$, resulting in two nonintersecting complex lines in~$\bC \bP^3$.
Moreover, the Lorentz group $\SU(2)\times\SU(2)$ acts transitively on each line, so that there are only two points (corresponding to a single supercharge of either chirality) in the reduced moduli space. Thus, the BPS states that contribute to the $\cN=1$ superconformal index~\cite{Romelsberger:2005eg} are the only possible $Q$-cohomology in $\cN=1$ theories. (For a related remark on the 4d chiral ring, see \S\ref{sec:4dchiral}.)
\\

Now, any choice of $Q$ from the nilpotent locus allows one to define the $Q$-cohomology of the theory. But how does the resulting cohomology vary as $Q$ is chosen from different points along the locus?
 To answer this question, let us first consider
a $\bC \bP^1$ family of nilpotent supercharges,
\[ Q_t = t_1 Q_1 + t_2 Q_2~,\]
associated to two anticommuting nilpotent supercharges $Q_1$ and $Q_2$. (This is in fact the situation that applies to each component of the nilpotent locus in Example~1.) A graded Hilbert space equipped with two anticommuting differentials $Q_1$ and~$Q_2$ is called a \emph{bicomplex} or \emph{double complex} by mathematicians, and the relationship between the cohomology of a single differential and the cohomology of a linear combination of the two has been studied in great detail. It is described by a \emph{spectral sequence}, which provides the answer to our familiar question: How does $Q_t$-cohomology depend on~$t$?

In order to give intuition about what the spectral sequence is, we will consider this example in great detail. $Q_t$-cohomology will indeed jump as $Q_t$ varies, and there are essentially three different cases, depending on the anticommutation relations between the supercharges and their adjoints. We treat these from the least to the most general. For more detail, the reader is referred to~\cite{Kim:2015vem}.
\\

\noindent\textbf{Enhanced supersymmetry:}
 We can ask that the two supercharges, together with their adjoints, close into the $\cN=2$ algebra of supersymmetric quantum mechanics:
\[ \{ Q_i, Q_j^\dagger\} = 2 \delta_{ij} H, \quad \{ Q_i, Q_j\} = 0. \]
In this case, the kernel of the positive self-adjoint operator~$H$ can be canonically identified with $Q_i$-cohomology for either supercharge, and therefore with $Q_t$-cohomology for all $t\in\bC \bP^1$. There is no jumping behavior.
\\

\noindent\textbf{Simple (but nontrivial) jumping:} To see jumping behavior, we can imagine relaxing the condition that the two supercharges square to the same Hamiltonian, while preserving the requirement that $\{ Q_1, Q_2^\dagger\} = 0$:
\begin{equation}
 \{ Q_i, Q_j^\dagger \} = 2 \delta_{ij} H_j, \quad \{ Q_i, Q_j\} = 0.
\label{1^2}
\end{equation}
It follows from these relations that $[H_1, H_2]=0$. Note that this algebra is precisely the 2d super-Poincar\'e algebra we studied above in Example~1. In this case, the positive operator whose kernel corresponds to $Q_t$-cohomology is
\[ H_t = \{ Q_t, Q_t^\dagger \} = 2 \left( |t_1|^2 H_1 + |t_2^2| H_2 \right). \]
For generic~$t$, only states with $H_1 = H_2 = 0$ appear in $Q_t$-cohomology. However, there are two special points: the poles $[0:1]$ and~$[1:0]$ of~$\bC \bP^1$.
At such points, a state needs to saturate only \emph{one} of the two BPS bounds $H_i\geq 0$ to contribute to the cohomology.

Let us recall that points of enhanced degeneracy, where $Q_t$-cohomology jumps, would be characterized by enhanced symmetry. This is indeed the case here: there is a $\U(1)_R$ symmetry of the algebra~\eqref{1^2} under which $Q_i$ has charge $(-1)^{i+1}$. This symmetry does not commute with the operator $H_t$ defining the BPS condition, unless~$t$ is at one of the poles.

For this argument to hold up, we must assume that we started with a bigraded vector space, and that the $Q_i$ have bidegrees $(1,0)$ and~$(0,1)$ respectively. These gradings correspond to left- and right-moving $R$-charge in Example~1, and it is usual to assume their existence in the mathematical definition of a bicomplex as well.
\\

\noindent\textbf{A generic bicomplex:} This is the most complicated example, in which it becomes apparent that the notion of a spectral sequence  can be nontrivial. We relax all conditions on the commutation relation $\{ Q_i, Q_j^\dagger\}$; in particular, the commutation relations no longer \emph{a priori} define a closed algebra. A new bosonic operator $S = \{ Q_1, Q_2^\dagger\}$ appears, and there is no way to understand the action of~$S$ or express it in terms of other bosonic operators on general grounds.

\begin{wrapfigure}{l}{0.4\textwidth}
\centering
\begin{tikzpicture}[line join = round, line cap = round, x={(1cm,0cm)}, y={(0cm,1cm)}, z={(0.15cm,0.15cm)}]]
\node(1) at (0,0) {$\bullet$};
\node(2) at (1.5,0) {$\bullet$};
\node(3) at (1.5,-1.5) {$\bullet$};
\node(4) at (3,-1.5) {$\bullet$};
\node(5) at (3,-3) {$\bullet$};
\node(6) at (4.5,-3) {$\bullet$};
\draw[->] (1) to (2);
\draw[->] (2) to (3);
\draw[->] (3) to (4);
\draw[->] (4) to (5);
\draw[->] (5) to (6);
\draw[dashed,->] (3) to (5);
\node at (0.7,0.3) {$Q_1$};
\node at (1.8,-.75) {$Q_2$};
\node at (2,-2.5) {$S$};
\end{tikzpicture}
\caption{}
\label{fig:staircase}
\end{wrapfigure}
In addition to the standard long representations and ``shortened'' representations (in which certain supercharges act by zero), there are now representations of arbitrary dimension that take the form of ``staircases'' in which $S$ acts nontrivially (Figure \ref{fig:staircase}):
The action of~$S$ is indicated by the dashed line in the above picture; $Q_1$ is represented by the horizontal arrows, and~$Q_2$ by the vertical arrows. It is easy to see that the representation drawn in the picture contributes no generators to $Q_1$-cohomology (or in fact to $Q_t$-cohomology for $t\neq [0:1]$), but two generators to $Q_2$-cohomology. Moreover, the two generators are not canceled by the differential induced by $Q_1$ on $Q_2$-cohomology, since they do not have the appropriate bidegrees.

The spectral sequence deals with this issue. It consists of a book with infinitely many pages; each page is a bigraded complex equipped with a certain differential, and the $(k+1)$-st page is the cohomology of the $k$-th page with respect to its differential. One often denotes the pages by $(E_k, d_k)$, where
\[ E_k = \bigoplus E_k^{(p,q)}. \]

The $E_0$ page is the original bicomplex; for the spectral sequence we are considering, $d_0$ is~$Q_2$, and $d_1$ is the differential induced by~$Q_1$. The differential $d_k$ has bidegree $(1-k,k)$; it cancels pairs of generators that lie at opposite ends of a staircase of length~$2k$. The cohomology of a generic supercharge $Q_t$ (where $t$ is away from either pole of~$\bC \bP^1$) is the ``$E_\infty$ page;'' in general, one should be careful about what it means for the spectral sequence to converge, if differentials can occur on infinitely many pages.
\\

In physical language, the bidegree in the original complex corresponds to two commuting $\U(1)$ symmetries of the system. Let us call their generators $U$ and~$V$. The original $Q_i$ are the only choices of supercharge that have well-defined quantum numbers with respect to both of these; a generic $Q_t$ has $U+V$-degree $+1$, but breaks the $U-V$ symmetry. This means that, upon deformation, states cancel in adjacent $U+V$-degree, but can differ by \emph{any} value of the $U-V$ quantum number (corresponding to page number in the spectral sequence).   However, by studying the deformation in detail, physical understanding may allow us to gain information about which pages' differentials may be nontrivial.

To conclude our discussion of generalities, we should say a couple of words about deformation problems of the other type. So far, we have concentrated in examples on the possibility of deforming the choice of~$Q$ in a fixed theory. We could also ask about deforming moduli of the theory: for instance, adding terms to the superpotential in a 2d $\cN=(2,2)$ theory. This will also give rise to a spectral sequence, describing the changing spectrum of BPS states; indeed, it is this type of deformation that will give rise to Lee's and Gornik's spectral sequences  \cite{Lee:2005,Gornik:2004} between link homologies, which we will explore in \S\ref{sec:Gornik}.

\subsection{From elliptic genus to (twisted) chiral ring}\label{sec:EG-CR}

Let us see how the rather abstract considerations of the previous section play out concretely in some simple and down-to-earth examples. We will begin by considering a 2d  theory of an $\cN=(2,2)$ chiral superfield $\Phi=\phi+\theta \psi +\theta^2 F$. The supersymmetry transformations of the chiral multiplet are as follows:
\bea\label{2dsusy-trans}
\hspace{2cm}[Q_\pm,\phi]=\psi_\pm~, \quad [\overline Q_\pm,\overline \phi]=\overline \psi_\pm~, &\quad&  [Q_\pm,\overline\phi]=0~, \quad  [\overline Q_\pm,\phi]=0~,  \\
\{Q_\pm, \psi_\pm\}=\pm\frac{\partial \overline W}{\partial \overline\phi}~,\quad \{\overline Q_\pm, \overline \psi_\pm\}=\pm\frac{\partial W}{\partial \phi}~,&\quad& \{\overline Q_\pm, \psi_\pm\}=-2i \partial_\pm \phi~,\quad \{ Q_\pm, \overline \psi_\pm\}=2i \partial_\pm \overline\phi~.\nonumber
\eea
In~\eqref{2dsusy-trans}, we have included a nonzero superpotential $W(\Phi)$ for later convenience; our first example, though, will be a free chiral, for which $W=0$.

\begin{wraptable}{L}{0.5\textwidth}
\centering
\begin{tabular}{|r|r|r|r|r|}
\hline
 & $ H_L $ & $J_L $ & $K$ & index\tabularnewline
\hline
\hline
$\phi$ & $0$ & $0$ & $1$ & $x$\tabularnewline
$\partial_+{\overline \phi} \strut $ & $1$ & $0$ & $-1$ & $qx^{-1}$\tabularnewline
$\psi_{+}$ & $\frac{1}{2}$ & $-1$ & $1$ & $-q^{\frac{1}{2}}y^{-1}x$\tabularnewline
${\overline \psi}_{+}$ & $\frac{1}{2}$ & $1$ & $-1$ & $-q^{\frac{1}{2}}yx^{-1}$\tabularnewline
\hline
\hline
$\partial_+$ & $1$ & $0$ & $0$ & $q$\tabularnewline
\hline
\end{tabular}
\caption{Single letters in a chiral multiplet annihilated by $\overline Q_-$.}
\label{tab:chiral}
\end{wraptable}

It is well-known that the elliptic genus counts $Q$-cohomology in a 2d $\cN=(2,2)$ theory with signs. More precisely, the elliptic genus in the RR sector counts right-moving ground states:
\begin{equation}
I(x;q,y)=\Tr_{RR} (-1)^F q^{H_L} \overline q^{H_R}y^{J_L} x^K,
\nonumber
\end{equation}
where $J_L$ is the left-moving $R$-charge and $K$ is the generator of a flavor symmetry. Writing the fermion number $F= F_L + F_R$, one can see that only states for which $H_R=0$ can contribute to the elliptic genus, and therefore that~$I$ is a holomorphic function of~$q$. (The same argument does not apply to the left-moving quantum numbers because of the presence of the fugacity $y^{J_L}$.) These states can be identified with either $Q_-$-cohomology or $\overline Q_-$-cohomology. We will choose to consider the cohomology of~$\overline Q_-$ in what follows.

To see spectral sequence explicitly (and for calculational convenience), we shall calculate the elliptic genus in the NS-NS sector instead:
$$
I(x;q,y)=\Tr_{\textrm{NS-NS}} (-1)^F q^{H_L} y^{J_L} x^K.
$$
 In superconformal theories, the two indices contain identical information, thanks to the spectral flow. However, in the NS-NS sector, the vacuum state is unique, so that one can straightforwardly count operators without worrying about subtleties due to fermion zero modes. The Ramond-sector calculation is given, for example, in~\cite{Witten:1993jg,Benini:2013nda}.

From the transformations \eqref{2dsusy-trans}, one can read off the letters annihilated by $\overline Q_-$ that contribute to the elliptic genus \cite{Gadde:2013ftv}. In Table \ref{tab:chiral},  we list them with their charges under all relevant symmetries.
 Each such operator contributes together with all of its left-moving derivatives (which are certain conformal descendants).
As a result, all modes (derivatives) from each letter contribute a factor
\begin{equation}
 (f;q)_\infty = \prod_{k\geq 0} (1-fq^k) ~,
\nonumber
\end{equation}
to the elliptic genus, where $f$ is the product of all fugacities for the field. This factor appears in the numerator for fermions, and in the denominator for bosons.

It is thus easy to see that the elliptic genus of a free chiral is given  by
$$
I_{\chi}(q,y;x)=\frac{(q^{1/2}y^{-1}x;q)_\infty(q^{1/2}yx^{-1};q)_\infty}{(x;q)_\infty (qx^{-1};q)_\infty}~.
$$
After a shift of $y\to q^{1/2}y$ (which accounts, up to an overall normalization factor, for spectral flow), we could also write it in the form
\begin{equation}\label{free-index}
I_{\chi}(q,y;x)= y^{-1/2}\frac{\theta_1(q,y x^{-1})}{\theta_1(q,x^{-1})}.
\end{equation}
The Jacobi theta function which appears here is defined in terms of the variables $q=e^{2\pi i\tau}$ and $y=e^{2\pi iz}$ as
$$
\theta_1(q,y):= -i q^{1/8} y^{1/2} \prod_{k=1}^\infty (1-q^k) (1-y q^k) (1-y^{-1}q^{k-1})  \;.\\$$

Now, let us now consider deforming the supercharge away from $\overline Q_-$, either in the direction of $Q_A = Q_+ + \overline Q_-$ or in the direction of $Q_B = \overline Q_+ + \overline Q_-$ (see Figure \ref{fig:tetrahedron}). Either of these deformations gives rise to a spectral sequence from the elliptic genus to the A- or B-type chiral ring, which moreover collapses at the $E_2$ page. Thus, we can think of it as simply the differential induced by $Q_+$ or $\overline Q_+$ on the $\overline Q_-$-cohomology.

By the 2d $\cN=(2,2)$ supersymmetry algebra relations and the Jacobi identity, one can easily see that all derivative operators are $Q$-exact with respect to either $Q_A$ or~$Q_B$. Hence, only the ``single letters'' can contribute to the chiral rings. More explicitly, for  the twisted chiral ring ($Q_A$-cohomology) of a free chiral, one can read off from \eqref{2dsusy-trans} that the supercharge (differential) $Q_+$ acts on the $\overline Q_-$-cohomology as
$$
Q_+:\partial_+^k\phi\to \partial_+^k \psi_+~, \quad Q_+: \partial_+^k\bar\psi_+\to \partial_+^{k+1}\bar \phi~, \qquad  \textrm{for} \quad k\in\bZ_{\ge0}~.
$$
Therefore, the twisted chiral ring is trivial, consisting only of the identity operator. On the other hand,
 for the chiral ring ($Q_B$-cohomology), the supercharge $\overline Q_+$ acts on the $\overline Q_-$-cohomology as
$$
\overline Q_+: \partial_+^k \psi_+\to \partial_+^{k+1}\phi~, \quad \overline Q_+: \partial_+^{k+1} \overline  \phi\to \partial_+^{k+1}\overline \psi_+~, \qquad  \textrm{for} \quad k\in\bZ_{\ge0}~.
$$
Hence, the chiral ring is generated by the bosonic operator $\phi$ and the fermionic operator $\xi=\overline \psi_+ - \overline \psi_-$. It is isomorphic to the graded polynomial ring $\bC[\phi]\otimes\Lambda^*[\xi]$ (see Figure \ref{fig:LG-dN}), and also to the unknot HOMFLY homology, as we will see in \S\ref{sec:Gornik}. All in all, non-trivial jumps of $Q$-cohomology can be seen as we move away from the vertices in~Figure~\ref{fig:tetrahedron} in either direction along the nilpotent locus.

It follows from this that the index \eqref{free-index} can be written in the form
\begin{equation}\label{split}
I_{\chi}(x,y;q)=\frac{1-y^{-1}x}{1-x}+q(1-qy)B(x,y,q)~,
\end{equation}
where the first term corresponds to the states contributing to the chiral ring and the second term contains the contributions of operators paired by $\overline Q_+$. Note that the charges of $\overline Q_+$ are $H_L=\frac12$ and $J_L=1$, so that the action of $Q_+$ corresponds to the factor $(1-qy)$.
By an identical argument, we could also write
\begin{equation}
I_{\chi}(x;y,q) = 1 + (1 - q y^{-1}) A(x,y,q)~,
\label{A-SS}
\end{equation}
using the spectral sequence from the elliptic genus to the (trivial) twisted chiral ring, which just consists of the differential induced from~$Q_+$.

It is worth remarking that the fugacity parameter $y$ in the elliptic genus records the charges of states under~$J_L$. When we consider the spectral sequence from the elliptic genus to the chiral ring, the interpretation of $y$ changes slightly. States in the A- or B-type chiral rings are graded by only one surviving $\U(1)$ $R$-symmetry: either axial or vector. The remainder is lost to the A- or B-type topological twist. Powers of $y$ in~\eqref{split} and~\eqref{A-SS} above record the appropriate remaining $R$-symmetry properties; however, the identification of $y$ with a vector or axial $R$-symmetry may require a charge conjugation, $y \leftrightarrow y^{-1}$.

\subsection{Deformation by superpotential}\label{sec:def-super}

One of the attractive features of the elliptic genus is that it is invariant along renormalization group flow trajectories, and therefore independent of the superpotential $W$ of a Landau-Ginzburg model, with one minor exception: $W$ determines the allowable $R$-charges of the fields in the superconformal IR theory, and therefore prescribes the graded structure of the index. With this subtlety in mind, the computation for a theory with a Landau-Ginzburg description is identical to that for a free theory.

The reader may be tempted to conclude that no interesting spectral sequence can therefore appear, relating elliptic genera as the superpotential is deformed. But this is not true! While the elliptic genus is always described by the same rational function, it depends on the $R$-charge assignment as well as on the fugacities, and cancellations occur (or fail to occur) between the numerator and denominator, depending on the precise values of the $R$-charges. These cancellations reflect the disappearance of states from the $\overline Q_-$-cohomology.

To convince oneself, one can simply recall that the elliptic genus in the Ramond sector contains the graded dimension of the chiral ring: the latter corresponds, by spectral flow, to the set of Ramond vacua, which are counted by the elliptic genus after setting $q=0$.
However, the chiral ring depends strongly on~$W$, and spectral sequences of type~2 are therefore easy to see. The calculation is easy to do directly in a theory with one chiral superfield.
 If one turns on the superpotential $W=\phi^{N+1}$, corresponding to a renormalization group flow from a free theory in the UV to the $A_N$ superconformal minimal model in the IR, the supersymmetry transformations are perturbed. In particular, the fermionic operator $\xi = \overline \psi_+ + \overline \psi_-$ that contributes to the elliptic genus transforms according to the rule
\begin{equation}\label{EGdiff}
\{Q_B, \overline \psi_+ + \overline \psi_- \} =2\frac{\partial W}{\partial \phi}=2\phi^N~.
\end{equation}
The new differential induced on $\overline Q_B$-cohomology is completely characterized by~\eqref{EGdiff}. Therefore, in the presence of the superpotential  $W=\phi^{N+1}$, the chiral ring of the theory is isomorphic to $\bC[\phi]/(\phi^N)$. In the context of link homology, turning on such a superpotential perturbs the spectrum of BPS states, as described by the $d_N$ differential.

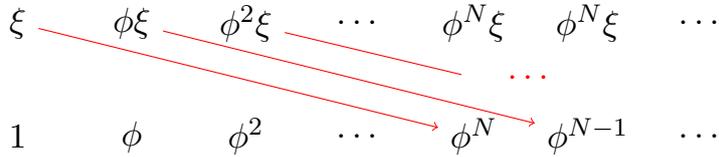
\begin{figure}[H]\centering
\begin{tikzpicture}[scale = 1.5]

\node at (-5,0) [scale =1.2]{$1$};
\node at (-4,0) [scale =1.2]{$\phi$};
\node at (-3,0) [scale =1.2]{$\phi^2$};
\node at (-2,0) [scale =1.2]{$\cdots$};
\node(D) at (-1,0) [scale =1.2]{$\phi^N$};
\node(E) at (0,0) [scale =1.2]{$\phi^{N-1}$};
\node at (1,0) [scale =1.2]{$\cdots$};
\node(F) at (-1,0.52) [scale =1.2]{$$};
\node at (-.5,0.52) [scale =1.2,red]{$\cdots$};
\node(A) at (-5,1) [scale =1.2]{$\xi$};
\node(B) at (-4,1) [scale =1.2]{$\phi\xi$};
\node(C) at (-3,1) [scale =1.2]{$\phi^2\xi$};
\node at (-2,1) [scale =1.2]{$\cdots$};
\node at (-1,1) [scale =1.2]{$\phi^N\xi$};
\node at (0,1) [scale =1.2]{$\phi^N\xi$};
\node at (1,1) [scale =1.2]{$\cdots$};

\draw[red,->] (A) -- (D);
\draw[red,->] (B) -- (E);
\draw[red,] (C) -- (F);
\end{tikzpicture}
\caption{The action of the differential induced by turning on the superpotential $W=\phi^{N+1}$ on the B-type chiral ring of a single free chiral superfield.}
\label{fig:LG-dN}
 \end{figure}

We can again write the elliptic genus in a form corresponding to~\eqref{split}, as the graded dimension of the chiral ring plus an additional term corresponding to operators that pair up after turning on the superpotential $W=\phi^{N+1}$:
\begin{equation}\label{W-SS}
\frac{1-y^{-1}x}{1-x}\quad \xrightarrow{W=\phi^{N+1}} \quad (1+x+\cdots+x^{N-1})+\frac{x^N-y^{-1}x}{1-x}~.
\end{equation}
The different fermionic numerator in~\eqref{W-SS}---as contrasted with~\eqref{split}---corresponds to the different degree assignments for the new differential. A helpful picture of the spectral sequence from the unperturbed to the perturbed chiral ring is shown in Figure~\ref{fig:LG-dN}. This is exactly the $d_N$ differential \cite{Dunfield:2005si,Rasmussen:2006} on the HOMFLY homology of the unknot where the perturbed  chiral ring is isomorphic to the $\fraksl(N)$ homology of the unknot.

 It is interesting to notice that, after the superpotential interactions have been turned on, only bosonic states remain in the BPS spectrum (or $B$-model chiral ring). This leads to an important difference between perturbations corresponding to the $d_N$ differentials and perturbations $W=\phi^{N+1} \to\phi^{N+1} +\phi^{M+1}$ that carry the $\fraksl(N)$ theory to the $\fraksl(M)$ theory for $M<N$. The spectral sequences induced by the latter deformations can only be trivial, since any differential must cancel bosonic and fermionic states in pairs. This means that, under this perturbation, we obtain an $A_M$-type Landau-Ginzburg theory together with some isolated massive vacua, which decouple in the IR limit (see Figure \ref{fig:unknot-def}). As we will see in \S\ref{sec:Gornik}, these factors appear in deformation spectral sequences, such as Gornik's, as trivial $\fraksl(1)$ factors. For this reason, Gornik's spectral sequence never has any interesting differentials at all in the unknot homology.
 \\

For an example of a different type, let us look at one more interesting BPS spectral sequence, related to the 2d $\cN=(2,2)$ superconformal gauged linear sigma model \cite{Witten:1993yc}. (We include this example, as well as the discussion in~\S \ref{sec:4dSCFT}, mostly to convince the reader that our discussion here is not limited to the world of knots.) To that end, we write the on-shell supersymmetric transformation of a vector multiplet $$V=\theta^-\bar \theta^-v_-+\theta^+\bar\theta^+v_+-\theta^-\bar\theta^+\sigma-\theta^+\bar\theta^-\bar\sigma+i\theta^2\bar\theta\bar\lambda+i\bar \theta^2\theta\lambda+\theta^2\bar\theta^2D~$$
 as
\bea\label{2dsusy-trans2}
[\overline Q_{\mp},v_{\pm}]=\pm i\lambda_{\pm} ~,&& [ Q_{\mp},v_{\pm}]=\pm i\overline\lambda_{\pm} ~,\cr
[\overline Q_-,\sigma]=-i \lambda_-~,\quad [ Q_+,\sigma]=i \overline \lambda_+~,&& [ Q_-,\overline\sigma]=-i \overline\lambda_-~, [\overline Q_+,\overline\sigma]=i \lambda_+~, \cr
[\overline Q_+,\sigma]=0~,\quad [ Q_-,\sigma]=0~,&& [ Q_+,\overline\sigma]=0~,\quad [\overline Q_-,\overline\sigma]= 0~,\quad \cr
\{ Q_+,\lambda_+\}=-2\partial_+\overline \sigma~, \quad\{ Q_-,\lambda_-\}=2\partial_- \sigma~, && \{\overline Q_\pm,\lambda_\pm\}=\partial_{\sigma}\wt \cW~,\cr
\{ \overline Q_+,\overline \lambda_+\}=-2\partial_+ \sigma~, \quad\{ \overline Q_-,\overline  \lambda_-\}=2\partial_- \overline \sigma~, && \{ Q_\pm,\overline \lambda_\pm\}=\partial_{\bar\sigma}\wt \cW~.
\eea
\begin{wraptable}{l}{0.4\textwidth}
\centering
\begin{tabular}{|r|r|r|r|}
\hline
 & $ H_L $ & $J_L $ & index\tabularnewline
\hline
\hline
$\partial_+\sigma$ & $\frac32$ & $1$ & $q^{3/2}y$\tabularnewline
\hline
${\overline \sigma}$ & $\frac12$ & $-1$ &  $q^{1/2}y^{-1}$\tabularnewline
\hline
$\lambda_{+}$ & $1$ & $0$ & $-q$\tabularnewline
\hline
${\overline \lambda}_{+}$ & $1$ & $0$ & $-q$\tabularnewline
\hline
\hline
$\partial_+$ & $1$ & $0$ &  $q$\tabularnewline
\hline
\end{tabular}
\caption{Single letters in a vector multiplet annihilated by $\overline Q_-$.\vspace{.3cm}}\label{tab:vector}
\end{wraptable}
The single letters that contribute to the elliptic genus are summarized in Table \ref{tab:vector}. Up to a shift of $y$ with restoring the bosonic zero mode,
the elliptic genus of a vector multiplet can be written as
\bea
I_V(q,y)=\frac{(q;q)_\infty(q;q)_\infty}{(y^{-1};q)_\infty (qy;q)_\infty} =\frac{i\eta(q)^3}{\theta_1(q,y)}~,\nonumber
\eea
where we use the Dedekind eta function
$$
\eta(\tau)=q^{1/24}\prod_{k=1}^\infty (1-q^k)~.
$$

Now, let us study the simplest example of a superconformal gauged linear sigma model:  the $T^*\bC\bP^1$ model, which is
a $\U(1)$ gauge theory with three chirals $\Phi_1,\Phi_2,\Phi_3$, with gauge charges $(1,1-2)$ respectively. The Higgs branch of the theory is the total space of the~$\cO(-2)$ line bundle (i.e.\ cotangent bundle) over $\bC\bP^1$.  Introducing  a flavor fugacity $a_i$ for each chiral, the elliptic genus of the theory~\cite{Benini:2013nda,Gadde:2013ftv} is
\bea\nonumber
I_{T^*\bC\bP^1}(q,y;a_i)&=&\frac{i\eta(q)^3}{\theta_1(q,y)}\int \frac{dz}{2\pi i z}\frac{\theta_1(q,z^{-1}a_1y)}{\theta_1(q,z^{-1}a_1)}\frac{\theta_1(q,z^{-1}a_2y)}{\theta_1(q,z^{-1}a_2)}\frac{\theta_1(q,z^{2}a_3y)}{\theta_1(q,z^{2}a_3)}\cr
&=&\frac{1}{2}\sum_{k,\ell=0}^1y^{-\ell}\frac{\theta_1(q,e^{2\pi i(k+\ell\tau)/2}a_1a_3^{1/2}y)}{\theta_1(q,e^{2\pi i(k+\ell\tau)/2}a_1a_3^{1/2})}\frac{\theta_1(q,e^{2\pi i(k+\ell\tau)/2}a_2a_3^{1/2}y)}{\theta_1(q,e^{2\pi i(k+\ell\tau)/2}a_2a_3^{1/2})}~.
\eea
After integrating out the gauge fugacity $z$, the elliptic genus counts all gauge-invariant operators comprised of letters from the three chirals and the $\U(1)$ vector multiplet.

Next, we deform the supercharge $\overline Q_-$ to either~$Q_A$ or~$Q_B$. It is clear from \eqref{2dsusy-trans} and \eqref{2dsusy-trans2} that there is no non-trivial $Q_B$-cohomology in the vector multiplet, and the $Q_B$-cohomology consists of the gauge invariant operators that can be written in terms of the following letters:
$$\phi_1~,\quad  \phi_2~,\quad \phi_3~,\quad \overline \psi_{1+}-\overline \psi_{1-}~,\quad \overline \psi_{2+}-\overline \psi_{2-}~,\quad
\overline \psi_{3+}-\overline \psi_{3-}~.$$ 
Although it is tedious, one can convince oneself that the other $\overline Q_-$-cohomology states are paired by $\overline Q_+$. As we have seen above, the generating function of the chiral ring can be obtained by taking the $q=0$ limit in the part of the elliptic genus that  the chiral multiplets contribute.  More explicitly, we can write the generating function as
$$
I_{T^*\bC\bP^1\chi\;\textrm{ring}}=y^{-1}+\frac12\sum_{i=0,1}\frac{(1+(-1)^ia_1a_3^{1/2}y)(1+(-1)^ia_2a_3^{1/2}y)(1-y)}{y^{3}(1+(-1)^ia_1a_3^{1/2})(1+(-1)^ia_2a_3^{1/2})}~.
$$

Similarly, the supercharge $Q_+$ pairs all the $Q_-$-cohomology states except the twisted chiral operator $\bar \sigma$. Because of the commutation relation $\{Q_+,\bar\lambda_+\}=\partial_{\bar\sigma}\wt \cW$, the twisted chiral ring depends on the twisted superpotential $\wt \cW$, which is known to receive quantum corrections.
Actually, a one-loop computation yields the effective twisted superpotential of this model~\cite{Witten:1993yc}:
$$
\widetilde \cW_{\textrm{eff}}(\bar\sigma )=-t\bar\sigma-\sum_i q_i(\bar\sigma-a_i) (\log (q_i(\bar\sigma-a_1)) -1 )~,
$$
where $t$ is the complexified FI parameter and $q_i$ are gauge charges. By solving the vacuum equation
$$
\exp\left(\frac{\partial \widetilde \cW_{\textrm{eff}}(\bar\sigma )}{\partial\bar \sigma}\right) =1~,
$$
we obtain the twisted chiral ring of the $T^*\bC\bP^1$ model
$$ \frac{(\bar \sigma-a_1)(\bar \sigma-a_2)}{4(\bar \sigma-a_3)^2}=e^{-t}~.$$
Since the effective twisted superpotential of a gauged linear sigma model is dynamically generated, one can consider that the $Q_A$-cohomology  can be obtained from the $\overline Q_-$-cohomology via a \emph{dynamical BPS spectral sequence}.

\subsection{Remark on 4d chiral ring}\label{sec:4dchiral}

We would like to briefly take note of an important, but subtle, distinction between various constructions used in building invariants ($Q$-cohomology or chiral rings) of quantum field theories. Specifically, every construction we have mentioned so far---the (twisted) chiral rings in 2d and, BPS states that contribute to the elliptic genus  \cite{Witten:1993jg,Benini:2013nda,Gadde:2013ftv} and 4d $\cN=1$ index \cite{Romelsberger:2005eg}---takes the form of a $Q$-cohomology.

A 4d analogue of the chiral ring has also been considered in the literature (for example \cite{Cachazo:2002ry} and references therein). It is defined as the collection of operators annihilated by \emph{all} supercharges of one chirality, modulo any operator that is exact with respect to \emph{any} such supercharge. That is,
\[ \chi = \left( \cap_\alpha \textrm{Ker} \;Q_\alpha \right) / \left( \oplus_\alpha \textrm{Im}\; Q_\alpha \right) ~. \]
%
However, the 4d chiral ring \emph{cannot} be thought of as the $Q$-cohomology of \emph{any} supercharge, be it any of the $Q_\alpha$'s or any combination.

This is easy to see in an example. Consider a chain complex with three generators, one bosonic in degree zero and two fermionic in degree one. Let the action of two supercharges be defined by
\[ Q_1 \phi = \psi_1~, \quad Q_2 \phi = \psi_2~, \quad Q_i \psi_j = 0~. \]
Then both $Q_i$-cohomologies, as well as $(Q_1+Q_2)$-cohomology, are one-dimensional, with support in degree one. Moreover, the Euler characteristic of any possible $Q$-cohomology must equal that of the complex itself, which is $-1$.

However, the chiral ring is empty, and has Euler characteristic zero! This means that (despite its superficially similar construction) the chiral ring is not any kind of $Q$-cohomology, and its behavior cannot be expected to follow the same pattern.

Note that upon dimensional reduction to 2d, the 4d $\cN=1$ algebra becomes the $\cN=(2,2)$ algebra, and the two supercharges $Q_\alpha$ of the same chirality descend to the two supercharges that constitute $Q_B$ in the reduced theory. So it is tempting to suspect that the cohomology of $\sum_\alpha Q_\alpha$ would give the 4d chiral ring. But we have seen in two ways that this cannot be the case: by a direct argument in the preceding paragraph, and in \S\ref{sec:Qchoice} by arguing that any two linear combinations of the $Q_\alpha$ are related by a Lorentz transformation in 4d, and so do not give different choices of~$Q$ in the reduced nilpotent moduli space. When we dimensionally reduce to 2d, we break the Lorentz symmetry from $\SO(4) = \SU(2) \times \SU(2)$ to $\SO(2)\times\SO(2)$, and the quotient by this smaller symmetry no longer identifies all choices of supercharge. This is why it is possible to describe the chiral ring by a $Q$-cohomology in the dimensionally reduced theory.

\subsection{BPS spectral sequences in 4d $\cN=2$ SCFTs}\label{sec:4dSCFT}

In this subsection, we give an example of how the generalities we have been discussing apply in a setting that is quite far removed from the world of Landau-Ginzburg models and link homologies, and therefore from the mainstream of this paper. 
We do so both because we find the example interesting in its own right, and in order to highlight (and convince the reader of) the breadth and generality of the framework we have been discussing. 
We will explain how various limits of the 4d $\cN=2$ superconformal index~\cite{Gadde:2011uv}  can be understood as arising from BPS spectral sequences. To be self-contained, we will briefly review the analysis in~\cite{Gadde:2011uv}; for more details, we refer the reader to the original paper. The indices and relationships between them that we discuss here are not novel; rather, the novelty consists in our interpretation of these results, and the way in which many facts can be organized---and easily and systematically understood---in our picture.

The fermionic part of the 4d $\cN=2$ superconformal algebra $\SU(2,2|2)$ can be expressed as  
\bea\label{sca}
\{Q^i_\alpha,\overline Q_{j\dot{\alpha}} \}=2\delta^i_{\;j} P_{\alpha\dot{\alpha}}~, &\qquad& \{Q^i_\alpha,Q^j_\beta \}= \{\overline Q_{i\dot{\alpha}},\overline Q_{j\dot{\beta}} \}=0~,\cr
\{\overline S^{i\dot{\alpha}},S^\alpha_j \}=2\delta^i_{\;j} K^{\dot{\alpha}\alpha}~, &\qquad& \{S_i^\alpha,S_j^\beta \}= \{\overline S^{i\dot{\alpha}},\overline S^{j\dot{\beta}} \}=0~,\cr
\{Q^i_\alpha,\overline S^{j\dot{\alpha}}\}&=& \{S_i^\alpha,\overline Q_{j\dot{\alpha}} \}=0~,
\eea
as well as
\bea\label{sca2}
\{Q^i_\alpha,S_j^\beta\}&=&4(\delta^i_{\;j} (M_\alpha^{\;\beta}-\tfrac{i}{2}\delta_\a^{\;\beta} D)-\delta_\a^{\;\beta} R^i_{\;j})~,\cr
\{\overline S^{i\dot{\alpha}},\overline Q_{j\dot{\beta}} \}&=&4(\delta^i_{\;j} (\overline M^{\dot{\alpha}}_{\;\dot{\beta}}+\tfrac{i}{2}\delta^{\dot{\alpha}}_{\;\dot{\beta}} D)-\delta^{\dot{\alpha}}_{\;\dot{\beta}} R^i_{\;j})~.
\eea
The 4d $\cN=2$  superconformal index counts the 1/8-BPS states annihilated by one supercharge and its superconformal partner (adjoint), say $\overline Q_{1\dot{-}}$ and $\overline S^{1\dot{-}}$. In other words, it counts $\overline Q_{1\dot{-}}$-cohomology, and these states saturate the bound
$$
\bar\delta_{1\dot{-}}:=\{\overline S^{1\dot{-}},\overline Q_{1\dot{-}} \}=E-2j_2-2R+r=0~.
$$
Recalling that $R$-symmetry indices are raised and lowered with the antisymmetric invariant tensor $\epsilon^{ij}$, it is easy to see from \eqref{sca}  that the three supercharges 
$$
Q_{1-}~,\qquad Q_{1+}~, \qquad \overline Q_{2\dot{+}}~,
$$
commute with $\overline Q_{1\dot{-}}$ and $\overline S^{1\dot{-}}$,
so that the $\cN=2$ superconformal index can be defined by introducing fugacities $\rho,\sigma,\t$ for the corresponding bosonic generators of the commutant subalgebra $\SU(1,1|2)$:
\be\label{indA}
{\mathcal I}(\rho, \sigma, \t)=\Tr(-1)^F\,
\rho^{\half\delta_{{1}{-}}}\,
\sigma^{\half\delta_{{1}{+}}}\,
\t^{\half\tilde \delta_{{2}\dot{+}}}\,
e^{-\beta\,\tilde \delta_{{1}\dot{-}}}\, .
\ee
Here the $\delta$'s can be read off from \eqref{sca2}:
\bea\nonumber
\delta_{1-}&:=& \{Q_{1-},S^{1-}\}=E-2j_1-2R-r ~,\cr
\delta_{1+}&:=&\{Q_{1+},S^{1+}\}= E+2j_1-2R-r~,\cr
\overline \delta_{2\dot{+}}&:=&\{\overline S^{2\dot{+}},\overline  Q_{2\dot{+}}\}= E+2j_2+2R+r~.
\eea
The usual parametrization in terms of fugacities $(p, q, t)$ is related to~\eqref{indA} via the change of variables
 \be
 p = \t \sigma\, , \quad  q = \t \rho\, , \quad  t = \t^2 \,.
 \ee
These parameters gives the $(p,q)$ labels of the elliptic Gamma function as well as the
 $(q,t)$ variables for Macdonald indices. In terms of these fugacities, the index can be expressed as
 \be\label{indqpt}
{\mathcal I}(p, q, t) = \Tr(-1)^F\,p^{\half\delta_{{1}{+}}} \,
q^{\half\delta_{{1}{-}}}\,
t^{R+r}\,
e^{-\beta\,\tilde \delta_{{1}\dot{-}}}~.
\ee

As we did for the elliptic genus in \S \ref{sec:EG-CR} and \S\ref{sec:def-super}, one can list single letters for the $\cN=2$ vector multiplet and half-hypermultiplet that contribute to the index:
\begin{table}[H]
\begin{centering}
\begin{tabular}{|c|r|r|r|r|r|c|c|}
\hline
Letters & $  E$ & $j_1$ & $  j_2$ & $R$ & $r$ & $\mathcal{I}(\sigma ,\rho, \t)$   & $\mathcal{I}(p, q, t)$ \tabularnewline
  \hline
   \hline
$  \phi$ & $1$ & $0$ & $0$ & $0$ & $-1$ & $\sigma \rho  $  & $pq/t$   \tabularnewline
  \hline
$  \lambda_{1\pm}$ & $  \frac{3}{2}$ & $  \pm  \frac{1}{2}$ & $0$ & $  \frac{1}{2}$ & $-  \frac{1}{2}$ & $-\sigma\t,\;-\rho\t$  &  $-p$, $-q$ \tabularnewline
  \hline
$  \bar{\lambda}_{1\dot{+}}$  & $  \frac{3}{2}$ & $0$ & $  \frac{1}{2}$ & $  \frac{1}{2}$ & $  \frac{1}{2}$ & $-\t^2$ &  $-t$ \tabularnewline
  \hline
$  \bar{F}_{\dot{+}\dot{+}}$ & $2$ & $0$ & $1$ & $0$ & $0$ & $\sigma\rho\t^2$  &  $pq$ \tabularnewline
  \hline
  $  \partial_{-\dot{+}}  \lambda_{1+}+  \partial_{+\dot{+}}  \lambda_{1-}=0$ & $  \frac{5}{2}$ & $0$ & $  \frac{1}{2}$ & $  \frac{1}{2}$ &
 $  -\frac{1}{2}$ & $\sigma\rho\t^2$  & $pq$  \tabularnewline
  \hline
\hline
$q$ & $1$ & $0$ & $0$ & $  \frac{1}{2}$ & $0$ & $\t$  &  $\sqrt{t}$ \tabularnewline
  \hline
$  \bar{\psi}_{\dot{+}}$ & $  \frac{3}{2}$ & $0$ & $  \frac{1}{2}$ & $0$ & $-  \frac{1}{2}$ & $-\sigma\rho\t$  & $-pq/\sqrt{t}$ \tabularnewline
  \hline
    \hline
$  \partial_{\pm\dot{+}}$ & $1$ & $  \pm  \frac{1}{2}$ & $  \frac{1}{2}$ & $0$ & $0$ & $\sigma\t,\;\rho\t$   & $p$, $q$ \tabularnewline
\hline
\end{tabular}
\par  \end{centering}
  \caption{Contributions to the index from  ``single letters,'' with the equation of motion $ \partial_{-\dot{+}}  \lambda_{1+}+  \partial_{+\dot{+}}  \lambda_{1-}=0$.
  We denote  the components of the adjoint ${\cal N} = 2$ vector multiplet by $(\phi, \bar \phi,  \lambda_{I,\alpha}, \bar\lambda_{I\,\dot \alpha},  F_{\alpha \beta}, \bar F_{\dot \alpha \dot \beta})$ and the
 components  of the half-hypermultiplet by $(q, \bar q, \psi_\alpha, \bar \psi_{\dot \alpha})$. The letters $\partial_{\alpha \dot \alpha}$ represent the spacetime derivatives.
}
\label{letters}
\end{table}

The authors of \cite{Gadde:2011uv} have considered the four limits of the index shown in Table~\ref{letters}, that count states subject to enhanced BPS shortening conditions. In these limits, the index becomes simpler and is often expressed in terms of certain special functions.

\begin{table}[h]
\begin{centering}
\begin{tabular}{|c|c|c|}
\hline
&Fugacities&states annihilated by\\
\hline
Macdonald index &$\sigma\to0$&$Q_{1+}, \ \overline Q_{1\dot{-}}$\\
\hline
Hall-Littlewood index &$\sigma\to0, \ \rho\to 0$ &$Q_{1\pm}, \ \overline Q_{1\dot{-}}$\\
\hline 
Schur index &$\rho=\tau$& $Q_{1+}, \ \overline Q_{1\dot{-}}$\\
\hline
Coulomb-branch index &$\tau\to 0$& $\overline Q_{2\dot{+}}, \ \overline Q_{1\dot{-}}$\\
\hline
\end{tabular}
\par  \end{centering}
  \caption{Various limits of the $\cN=2$ index. The middle column lists the specializations of fugacities. The right column indicates the supercharges that annihilate states contributing to the corresponding index.}
\label{letters}
\end{table}
In fact, several (but not all) of these limits are related by BPS spectral sequences to the original index. To begin, there is a BPS spectral sequence from $\overline Q_{1\dot{-}}$-cohomology to the states annihilated by $Q_{1+}$ and $\overline Q_{1\dot{-}}$ (that contribute to the Macdonald index). We can identify these states with the cohomology of the perturbed supercharge $Q_{1+} + \overline Q_{1\dot{-}}$.

To see this spectral sequence explicitly, note that the  supersymmetry transformations of the  $\cN=2$ vector multiplet can be written as follows:
\bea\nonumber
\delta_\e A_\mu&=&i\e_i\sigma_\mu\overline\lambda^i-i\lambda_i\sigma_\mu\overline \e^i~,\cr
\delta_\e \lambda_{i}&=&F_{\mu\nu}\sigma^{\mu\nu}\e_i+\sqrt{2}iD_\mu \phi\, \sigma^\mu\overline\e_i+D\e_i~,\cr
\delta_\e \phi&=&\sqrt{2}\e_i\lambda^i~.
\eea
We also recall the transformations of the half-hypermultiplet, which are
\bea\nonumber
\delta_\e q_i&=&-\sqrt{2}\e_i\psi+\sqrt{2}\overline \e_i\overline \psi~,\cr
\delta_\e \psi&=&-\sqrt{2}iD_\mu q_i\sigma^\mu \bar \e^i-2F^i \e_i~.
\eea
From these supersymmetry transformations, one can see that the supercharge $Q_{1+}$ acts on the $\overline Q_{1\dot{-}}$-cohomology by
\bea\nonumber
&&Q_{1+}:\partial_{\pm\dot{+}}^k \phi \to \partial_{\pm\dot{+}}^k \lambda_{1+}~,\qquad Q_{1+}:\partial_{\pm\dot{+}}^k \overline F_{\dot{+}\dot{+}} \to \partial_{\pm\dot{+}}^k\partial_{+\dot{+}}\overline\lambda_{1\dot{+}}~,\cr
&&Q_{1+}:\partial_{\pm\dot{+}}^k\overline \psi_{\dot{+}}\to \partial_{\pm\dot{+}}^k\partial_{+\dot{+}}q~,
\eea
for $k\in\bZ_{k\ge0}$. Thus, the single letters that contribute to the Macdonald index consist of $\lambda_-$, $\overline \lambda_{\dot{+}}$, $q$ and $\partial_{-\dot{+}}$ which satisfy $\delta_{1+}=0$.
At the level of the $\cN=2$ index, one can write
$$
\cI_{\textrm{full}}(\rho, \sigma, \t)=\cI_{\textrm{Mac}}(\rho,\tau)+\sigma(1-\rho^{-1}\tau) A(\rho, \sigma, \t)~.
$$
Since the supercharge $Q_{1+}$ has $\delta_{{1}{-}}=-2$ and $ \delta_{{2}\dot{+}}=2$, the spectral sequence due to $Q_{1+}$ corresponds to the factor  $(1-\rho^{-1}\tau)$. This is why the Schur limit $\rho=\tau$ also leads to the same 1/4-BPS condition as the Macdonald limit $\sigma\to0$ does.

There is also a BPS spectral sequence from $\overline Q_{1\dot{-}}$-cohomology to the states annihilated by $\overline Q_{2\dot{+}}$ and $\overline Q_{1\dot{-}}$ (that contribute to the Coulomb-branch index). For the two kinds of multiplet we are considering, the action of $\overline Q_{2\dot{+}}$ on $\overline Q_{1\dot{-}}$-cohomology is
\bea\nonumber
&&\overline Q_{2\dot{+}}: \partial_{\pm\dot{+}}^k \lambda_{1\pm}\to \partial_{\pm\dot{+}}^{k+1} \phi ~,\qquad \overline Q_{2\dot{+}}:  \partial_{\pm\dot{+}}^k\overline\lambda_{1\dot{+}}\to \partial_{\pm\dot{+}}^k \overline F_{\dot{+}\dot{+}}~,\cr
&&\overline Q_{2\dot{+}}:\partial_{\pm\dot{+}}^kq\to \partial_{\pm\dot{+}}^k\overline\psi_{\dot{+}}~,
\eea
for $k\in\bZ_{k\ge0}$.
Thus, $\phi$ is the only single letter that contributes to the Coulomb-branch index.
At the level of the index, the relation between the full $\cN=2$ index and the Coulomb-branch index can be written as
$$
\cI_{\textrm{full}}(\rho, \sigma, \t)=\cI_{\textrm{Coulomb}}(\sigma,\rho)+\tau(1-\sigma\rho) B(\rho, \sigma, \t)~,
$$
where the action of $\overline Q_{2\dot{+}}$ amounts  to the factor $(1-\sigma\rho)$, since the supercharge $\overline Q_{2\dot{+}}$ has $\delta_{{1}{-}}=2$ and $ \delta_{{2}\dot{+}}=2$. This implies that the Coulomb-branch index can be also obtained by taking the specialization $\sigma=\rho^{-1}$ in the full $\cN=2$ index.

Moreover, by checking the commutation relations~\eqref{sca} and~\eqref{sca2} above, the reader may check using the general methodology we outlined in \S \ref{sec:Qchoice} that both of these spectral sequences must collapse at the $E_2$ page; that is, there are no interesting higher differentials.

In contrast with this, the states satisfying the 3/8-BPS condition and contributing to the Hall-Littlewood index cannot be thought of as the $Q$-cohomology of any choice of supercharge. 
Analogous to what we have seen for the 4d chiral ring in~\S \ref{sec:4dchiral}, the action of the Lorentz group relates the supercharges $Q_{1\pm}$, so that the cohomology of $Q_{1+} + Q_{1-} + \overline Q_{1 \dot{-}}$ must be identical to that of just $Q_{1+} + \overline Q_{1 \dot{-}}$. 
Thus, the Hall-Littlewood index is not the graded dimension of any $Q$-cohomology, and one cannot see a spectral sequence from $\overline Q_{1\dot{-}}$-cohomology (or from the Macdonald index) to the 3/8-BPS states. Indeed, it is easy to see that the supercharge $Q_{1-}$ must act trivially on states contributing to the Macdonald index in another way: it carries charge $\delta_{1+}=-2$ for the fugacity~$\sigma$---but this has already been set to zero.

\subsection{LG models on a strip}\label{sec:strip}

Let us now consider yet another example of jumping behavior of BPS spectra under deformations of a theory, which will bring us back towards our goal of understanding spectral sequences between knot homologies. In contrast with those in~\S \ref{sec:4dSCFT}, this example is a spectral sequence of the second type (i.e. associated to moduli of the theory, rather than to a choice of supercharge). Its novelty consists in two features: (1) we are considering an RG flow trajectory between two interacting theories; and (2) we include certain boundary conditions or defects. The context will still be that of Landau-Ginzburg theories, but we will consider open-string rather than closed-string BPS states.

\begin{wrapfigure}{l}{0.4\textwidth}
\centering
\begin{tikzpicture}
\node (A) at (0,0) {};
\node (B) at (0,4) {};
\node (C) at (3,0) {};
\node (D) at (3,4) {};
\fill [gray!30] (0,0) rectangle (3,4);
\draw[very thick] (-0.01,0)--(-0.01,4);
\draw[very thick] (3.01,0)--(3.01,4);
\node at (-0.4,2) {$\cD_1$};
\node at (3.4,2) {$\cD_2$};
\node at (1.5,2) {$W(x)$};
\end{tikzpicture}
\caption{}
\label{fig:strip}
\end{wrapfigure}

Let us consider the three-parameter family of Landau-Ginzburg superpotentials
\be\label{cubic-potential}
W(x) = (x-u_1)(x-u_2)(x-u_3)~,
\ee
corresponding to the relevant perturbation of the $A_2$ minimal model $W=x^3$.
We are interested in the spectrum of open-string states when the LG model described by~$W(x)$ is placed on a strip (Figure \ref{fig:strip}). Each side has a boundary condition, which may be chosen independently. 
Possible choices of boundary condition that preserve $B$-type supersymmetry in Landau-Ginzburg theories are described by matrix factorizations of the superpotential.

A matrix factorization $\cD$ is a $\bZ_2$-graded free module over a polynomial ring~$R$, equipped with an odd endomorphism $Q_{\textrm{bd}}$ that squares to a polynomial potential $W\in R$:
\[ Q_{\textrm{bd}} : \cD \rightarrow \cD, \quad Q_{\textrm{bd}} ^2 = W \cdot \id_\cD. \]

The variables of~$R$ correspond to the chiral superfields of the model, and $Q_\text{bd}$ specifies the action of the unbroken supercharge~$Q_B$ on the boundary degrees of freedom.
If the matrix factorization has rank one, so that $\cD = R_0 \oplus R_1$, we can write $ Q_{\textrm{bd}} $ in components as
\[
Q_{\textrm{bd}} = \begin{pmatrix} 0 & f \\ g & 0 \end{pmatrix}~.
\]
One could also write the matrix factorization as
$$
R_0\xrightarrow{\,\,f\,\,} R_1\xrightarrow{\,\,  g\,\,} R_0~.
$$
Every choice of superpotential admits a trivial rank-one matrix factorization, for which the boundary supercharge is simply
\be\label{trivial-MF}
R_0\xrightarrow{\,\,1\,\,} R_1\xrightarrow{\,\,W\,\,} R_0~.
\ee
We will often regard two matrix factorizations as equivalent if one can be obtained from the other by taking the direct sum with some number of trivial matrix factorizations.

We will denote the space of maps from~$\cD_1$ to~$\cD_2$ (considered as graded $R$-modules) by  $\Mat(\cD_1, \cD_2)\cong \cD_1\otimes \overline \cD_2$; it can be thought of as a space of matrices.
Furthermore, the boundary supercharge $Q_{\textrm{bd}}$ for the tensor product $\cD_1\otimes \overline \cD_2$ makes $\Mat(\cD_1,\cD_2)$ into a $\bZ/2\bZ$-graded complex.
 The cohomology $\cH^*(\cD_1,\cD_2)$ of this complex is the space of open-string BPS states between these two boundary conditions. As we will see in \S \ref{sec:LG-defect}, the generators in $\cH^*(\cD_1,\cD_2)$  are also called \emph{defect-changing operators}.

Let us see how the spectrum of open-string BPS states jumps in the parameter space of the potential \eqref{cubic-potential}.  We explicitly specify the two boundary conditions for the potential \eqref{cubic-potential} as
\bea
\cD_1:&& \quad R_0\xrightarrow{\,\,x-u_1\,\,} R_1\xrightarrow{\,\,(x-u_2)(x-u_3)\,\,} R_0\cr
\cD_2:&& \quad  R_0\xrightarrow{\,\,x-u_2 \,\,} R_1\xrightarrow{\,\,(x-u_1)(x-u_3)\,\,} R_0~,\nonumber
\eea
where $R=\bC[x]$. Note that these boundary conditions are unobstructed: they make sense over the whole parameter space of the superpotential.

We then calculate their tensor product
\be\nonumber
\cD_1\otimes \overline \cD_2: \quad\left[\begin{array}{c}R\\ \bigoplus \\R\end{array}\right]_0
\xrightarrow{\, d^0\,}
\left[\begin{array}{c}R\\ \bigoplus \\R\end{array}\right]_1
\xrightarrow{\, d^1 \,}
\left[\begin{array}{c}R\\ \bigoplus \\R\end{array}\right]_0~,
\qquad
\ee
where the maps are defined
\be\nonumber
\arraycolsep 6pt
d^0=\left[\!\begin{array}{cc}x-u_1 & u_2-x\\ -(x-u_1)(x-u_3) &(x-u_2)(x-u_3)\end{array}\!\right], \quad d^1={\left[\!\begin{array}{cc}x-u_2 & x-u_2\\ x-u_1 & x-u_1\end{array}\!\right]}~.
\ee
The cohomology of this system is
\bea\nonumber
\cH^0(\cD_1, \cD_2)&=&\textrm{Ker}\; d^0/\textrm{Im} \; d^1\cong \left\{ \begin{array}{ll}\bC[x]/(x-u) \quad&\textrm{for} \quad u_1=u_2=u \\ 1 \quad&\textrm{for} \quad u_1\neq u_2  \end{array} \right. ~,\cr
\cH^1(\cD_1, \cD_2)&=&\textrm{Ker}\; d^1/\textrm{Im} \; d^0\cong \left\{ \begin{array}{ll}\bC[x]/(x-u) \quad&\textrm{for} \quad u_1=u_2=u \\ 1 \quad&\textrm{for} \quad u_1\neq u_2  \end{array} \right. ~.
\eea
Therefore, at the locus $u_1=u_2$ (of complex codimension one) in the parameter space $(u_1,u_2,u_3)$, we can see a jump of the BPS spectrum. Away from the locus, the BPS states disappear in pairs, one bosonic and one fermionic, just as we expected would happen.

It is straightforward to generalize this result to the perturbation of $A_N$ minimal models. Suppose that the potential is
\be\nonumber
W(x)=\prod_{u_i\in S} (x-u_i)~ \quad  \textrm{for} \ |S|=N,
\ee
and we have the two boundary conditions
\be\nonumber
Q_{\textrm{bd}}^{(1)}=\left[\!\begin{array}{cc}0 & \prod_{u_i\in U_2} (x-u_i)\\ \prod_{u_i\in U_1} (x-u_i)&0\end{array}\!\right]~,\quad Q_{\textrm{bd}}^{(2)}=\left[\!\begin{array}{cc}0 & \prod_{u_i\in V_2} (x-u_i)\\ \prod_{u_i\in V_1} (x-u_i)&0\end{array}\!\right]~,
\ee
where $U_1\cup U_2=V_1\cup V_2=S$. (These unions are not necessarily disjoint, since roots will occur in the set $S$ with some multiplicity.) Then, the BPS spectrum turns out to be
\be
\cH^0(\cD_1, \cD_2)\cong\cH^1(\cD_1, \cD_2)\cong \frac{\bC[x]}{\prod_{U_1\cap U_2\cap V_1\cap V_2}(x-u_i)}~.
\label{eq:An-deformations}
\ee
Hence, it is easy to see that the jumping phenomena of this type are ubiquitous in LG model on a strip when one varies parameters of the potential, corresponding to relevant deformations of $A_N$-series minimal models.

There is one other important point to note. We stated that the space $\cH(\cD_1,\cD_2)$ of open-string BPS states should be taken as the space of morphisms in the category of $B$-type boundary conditions. However, our calculation~\eqref{eq:An-deformations} shows that this space can in fact empty, even when $\cD_1$ and~$\cD_2$ are the same defect! For instance, consider the following boundary condition:
\be
\cD : R_0 \xrightarrow{~(x-u)^{p+1}~} R_1 \xrightarrow{~(x-v)^{q+1}~} R_0~.
\label{eq:strange-defect}
\ee
The relevant superpotential is clearly $W = (x-u)^{p+1}(x-v)^{q+1}$. When $u=v$, $\cD$ is a nontrivial rank-one boundary condition in the $A_{p+q+1}$ minimal model, and
\[
\dim \cH^0(\cD,\cD) = \min(p+1,q+1) > 0~.
\]
In fact, we would expect that it must be, since any category must admit at least the identity morphism from an object to itself. Notice, however, that if either boundary condition were the trivial matrix factorization ($p=-1$), the space of morphisms would be zero-dimensional. This is in keeping with our claim above that the trivial defect ``decouples'' from the actual category of boundary conditions: it has zero morphisms with any object, even itself.

Now let us consider deforming $u$ and~$v$ away from one another. Our calculation~\eqref{eq:An-deformations} shows that as soon as this is done, $\dim \cH^0(\cD,\cD)=0$! Not even an identity morphism is present. Nevertheless, this is not a problem, and in fact should have been expected. The theory described by the perturbed superpotential no longer has a unique vacuum state. Rather, it has two vacua; one corresponding to an $A_p$ conformal theory and the other to an $A_q$ conformal theory. The theory splits into two superselection sectors, which do not interact. Therefore, one ought to think of the boundary condition~\eqref{eq:strange-defect} as representing a kind of composite of the trivial defect in the $A_p$ theory and the trivial defect in the $A_q$ theory. The idea that boundary conditions form a category applies to the collection of boundary conditions in a fixed theory; the idea breaks down in the presence of superselection sectors.

Moreover, models of this type actually include LG models on a cylinder, with any number of defects extended in the time direction! This will be exactly the setup that we find in the context of knot homology. For example, consider a LG model with two defects on a cylinder. If the two domains separated by the defects have different potentials in the same variable (chiral field) $x$, \textit{i.e.} $W_1(x)$ and $W_2(x)$, this is equivalent to LG model on a strip with the potential $W_1(x)-W_2(x)$ by using the folding trick as in Figure \ref{fig:folding-trick}. Thus, interesting spectral sequences occur in this model under deformation of the potentials, in contrast to the unknot case where the two domains have the same potential of different variables, $W(x)$ and $W(y)$, as in Figure \ref{fig:LGdefects2b}. Even with more than two defects (as in Figure \ref{fig:LGdefects5}), the LG model on a cylinder undergoes the spectral sequence under the deformations of superpotentials if all the potentials are of the same variable $x$. As we will explain later, one can fuse the defects together one by one, until only two remain.
\\
\\
\\

\section{Fivebranes and links}
\label{sec:setup}

In this section, we study homological invariants of knots and links using the bird's-eye view of fivebrane systems \cite{Ooguri:1999bv,Gukov:2004hz,Gukov:2007ck,Diaconescu:2011xr,Witten:2011zz}. The physical realizations of link homologies predict their rich structural properties, which are accomplished by action of differentials and algebras \cite{Dunfield:2005si,Rasmussen:2006,Gukov:2011ry,Gorsky:2012mk,Gorsky:2013jxa}.
In particular, we build a new vantage point in terms of Landau-Ginzburg model in the fivebrane systems for link homologies \S\ref{sec:LG-defect}. In this viewpoint, the considerations in \S \ref{sec:deformations} will give
 a physical understanding of Lee-Gornik spectral sequences \cite{Lee:2005,Gornik:2004} of $\fraksl(N)$ link homology in \S\ref{sec:Gornik}. Since our focus will change and in fact grow much more specific in what follows, the reader can think of this section as an introduction to the second part of the paper. The following sections of the paper will investigate the color-dependence of link homology, and therefore we will provide the summary and structure of the rest of the paper in \S \ref{sec:organization}.
\\

There are several good reasons why the focus of knot theory  in the twenty-first century has shifted from polynomial invariants to their categorified ({\it i.e.} homological) versions \cite{Crane:1994}. By definition, a categorification of a (Laurent) polynomial with $r$ variables is a $(r+1)$-graded homology theory whose graded Euler characteristic is the polynomial in question. As such, it is usually a stronger invariant of knots and links, since some information is lost in passing to the Euler characteristic. More importantly, there are many operations one can do with vector spaces that are simply impossible at the polynomial level (maps, for instance), which opens a door into the beautiful world of homological algebra.

The quantum group invariants $P^G_{\lambda_1,\ldots,\lambda_n}(L;q)$ computed by Chern-Simons TQFT \cite{Witten:1988hf,Reshetikhin:1990} provide an infinite set of polynomial link invariants that depend on the choice of gauge group $G$, representations $\lambda_i$ of~$G$ (one for each link component), a variable $q$ (related to the coupling constant of Chern-Simons theory), and, of course, the link $L$ itself. All these polynomial invariants have a remarkable property: they are (Laurent) polynomials in variable $q$ with integer coefficients. Therefore, they should be categorified by doubly graded homology theories. Luckily, by now these homology theories have been constructed \cite{Khovanov:2004,Yonezawa,Webster:2010,Wu,Cooper:2010,Frenkel:2010} for any choice of $G$ and $\vec \lambda = (\lambda_1, \ldots, \lambda_n)$, of which the most familiar example is probably Khovanov homology \cite{Khovanov:2000} corresponding to $G = \SU(2)$ and $\lambda_i = \square$. However, how exactly do these invariants depend on the ``representation theory data'' $G$ and $\vec \lambda = (\lambda_1, \ldots, \lambda_n)$?

The dependence of $P^G_{\vec \lambda}(L;q)$ on the group $G$ turns out to be very simple for classical groups of Cartan type $A$, $B$, $C$, or $D$. For each of these root systems, the dependence on the rank is beautifully packaged by a 2-variable polynomial invariant: the colored HOMFLY polynomial $P_{\vec \lambda}(L;a,q)$ for Cartan type $A$ and the so-called colored Kauffman polynomial $F_{\vec \lambda}(L;a,q)$ for groups of type $B$, $C$, or $D$ ({\it cf.} Appendix~\ref{sec:notation}). Thus, for $G = \SU(N)$ one has
\be
P^{\SU(N)}_{\lambda_1,\ldots,\lambda_n}(L;q) \; = \; P_{\lambda_1,\ldots,\lambda_n}(L;a = q^N,q) ~.
\label{PHomSUN}
\ee
This means that quantum group invariants for different groups are not that different after all; there is a lot of regularity captured by the extra variable $a$. Since for each choice of $\lambda_i$ the colored HOMFLY polynomial $P_{\lambda_1,\ldots,\lambda_n}(L;a,q)$ depends on two variables, $a$ and $q$, its categorification must be a triply-graded homology theory. The existence \cite{Gukov:2004hz} and properties \cite{Dunfield:2005si} of this homology theory came from physics as a surprise, contrary to the expectations in math literature; see {\it e.g.} \cite{Khovanov:2003s}. In the ``uncolored'' case, {\it i.e.} when all $\lambda_i = \square$, the triply-graded HOMFLY homology has been given a rigorous mathematical definition \cite{Khovanov:2005,Webster:2009}. Moreover, the Poincar{\'e} polynomial of the colored HOMFLY homology (often called colored \emph{superpolynomial})
\be
\scP_{\vec \lambda} (L;a,q,t) \; := \; \sum_{i,j,k} a^i q^j t^k \textrm{dim}\, \scH_{\vec \lambda} (L)_{i,j,k} 
\label{superPdef}
\ee
is, roughly speaking, an intermediate object: much like colored HOMFLY polynomial, it is still a polynomial (with positive integer coefficients), but it captures more information and provides a window into a homological world.

However, the construction of triply-graded homology to general groups with arbitrary ``color'' label $\vec \lambda = (\lambda_1, \ldots, \lambda_n)$ as well as  the computations of their Poincar\'e polynomials (even for simple knots) remain a major challenge. This makes any predictions especially valuable. In fact, using bird's-eye view of fivebrane systems, many predictions and conjectures on structure of colored knot homologies have been made \cite{Dunfield:2005si,Gukov:2005qp,Gukov:2011ry,Gorsky:2013jxa,Nawata:2013mzx}, which enables us to determine colored superpolynomials of many knots \cite{Fuji:2012pm,Fuji:2012nx,Fuji:2012pi,Nawata:2012pg,Nawata:2013mzx}. The structure of knot homology appears as a large set of differentials, called $d_N$ and colored differentials, which should be formulated in terms of spectral sequences \cite{Rasmussen:2006}. Hence, we first provide a physical meaning of spectral sequences in link homology based on the understanding in \S\ref{sec:deformations}.

From the next section, we will study the structure of link homology as well as the ``color''-dependence $\vec \lambda = (\lambda_1, \ldots, \lambda_n)$. Actually, there is an important difference between HOMFLY invariants of knots and links. Unlike the reduced colored HOMFLY polynomial for a knot which is a genuine Laurent polynomial, in the case of links it has a nontrivial denominator. Since  the corresponding homology theory that categorifies it is consequently infinite-dimensional,  colored HOMFLY homology of links loses some of properties for that of knots, which makes the analysis more difficult. However, the link homology turns out to enjoy surprising regularities on color dependence $\vec \lambda = (\lambda_1, \ldots, \lambda_n)$ that can be captured by \emph{sliding property}, \emph{homological blocks} and \emph{associated varieties}. In \S\ref{sec:organization}, we will summarize these results and present the organization of the remainder of the paper.

\subsection{Link homology as $\cQ$-cohomology}

In every extant physical approach to homological link invariants, they are realized as spaces of BPS states
annihilated by a supercharge $\cQ$ (modulo $\cQ$-exact states), which are also known
as BPS states \cite{Gukov:2004hz,Gukov:2007ck,Dimofte:2010tz,Witten:2011zz} (see also \cite{Aganagic:2011sg} for torus knots):
\be
\scH_{\vec \lambda} (L) \; \cong \; \{ \cQ\text{-cohomology} \} \; \equiv \; \scH_{\text{BPS}} \,.
\label{HHBPS}
\ee
Moreover, in every physical realization of link homologies
all spatial dimensions are effectively compact, so that the system reduces
to supersymmetric quantum mechanics in one non-compact ``time'' direction $\bR_t$.

All physical approaches to link homologies are essentially different ways to look at the same
physical system, which has two phases---``deformed conifold'' and ``resolved conifold''---related
by the {\it geometric transition} \cite{Gopakumar:1998ki}.
The former describes doubly-graded $\fraksl(N)$ link homologies for fixed $N$,
whereas the latter reproduces triply-graded HOMFLY homologies:
 \begin{center}
 \shadowbox{\begin{minipage}{.85\linewidth}
\be\nonumber
\begin{array}{rcl}
\text{deformed conifold} & \quad \phantom{\int} \Leftrightarrow \phantom{\int} \quad & \textrm{doubly-graded $\fraksl(N)$ homology}  \\
\text{resolved conifold} & \quad \phantom{\int} \Leftrightarrow \phantom{\int} \quad & \textrm{triply-graded HOMFLY homology}
\end{array}
\ee
\end{minipage}}
\end{center}

On the ``deformed conifold'' side the physical setup is:
\be\label{surfeng2}\tag{def-M5}
\begin{matrix}
{\mbox{\rm space-time:}} && \qquad  \bR_t \times &TN_4 & \times & T^* S^3 \\
{\mbox{\rm $N$ M5-branes:}} && \qquad  \bR_t \times &  D&  \times &S^3 \\
{\mbox{\rm M5'-branes:}} && \qquad  \bR_t \times &D&  \times & M_L
\end{matrix}
\ee
where $TN_4 \cong \bR^4_{\epsilon_1,\epsilon_2}$ is the Taub-NUT 4-manifold,
$D \cong \bR^2_{\epsilon_1}$ is the two-dimensional ``cigar'' (holomorphic Lagrangian submanifold) in the Taub-NUT space,
and the Lagrangian 3-manifold $M_L \subset T^* S^3$ is the conormal bundle to the link $L \subset S^3$,
such that
\be\nonumber
L = S^3 \cap M_L~.
\ee
After the geometric transition ({\it i.e.} on the ``resolved conifold'' side) the corresponding system is
\be\tag{res-M5}
\begin{matrix}
{\mbox{\rm space-time:}} & \qquad & \bR_t & \times & TN_4 & \times & X \\
{\mbox{\rm M5'-branes:}} & \qquad & \bR_t & \times & D & \times & M_L
\end{matrix}
\label{surfeng}
\ee
The effective 3d $\cN=2$ theory on $\bR_t  \times  D$ is what is often called $T[M_L]$;
it is a 3d $\cN=2$ theory labeled by a 3-manifold $M_L$, or equivalently by the link $L$ \cite{Dimofte:2010tz,Terashima:2011qi,Dimofte:2011ju,Dimofte:2011py,Fuji:2012pm,Fuji:2012nx,Fuji:2012pi,DGG-Kdec,Yagi:2013fda,Lee:2013ida,Cordova:2013cea,3d3drevisited}.

To be more precise, the BPS states in question are the so-called {\it open} BPS states $\scH(L)\cong\scH_{\text{BPS}}^{\text{open}}$, meaning that they are represented by bordered Riemann surfaces in $X$ with boundary on $M_L$. In contrast, the {\it closed} BPS states are represented by membranes supported on closed holomorphic curves in $X$, with no boundary. In other words, the space of open BPS states is determined by the pair $(X,M_L)$, whereas the space of closed BPS states depends only on $X$. It has been conjectured \cite{Harvey:1995fq} that closed BPS states form an algebra ${\mathcal A}$. Similarly, in \cite{Gukov:2011ry} it was argued that open BPS states furnish a module $M$ for this algebra:
 \begin{center}
 \shadowbox{\begin{minipage}{.8\linewidth}
 \be\label{HHrep}
\begin{array}{rcl}
\text{(refined) open BPS states}: &  & M := \scH_{\text{BPS}}^{\text{open}} \qquad\\
 & & \circlearrowleft \\
\text{(refined) closed BPS states}: &  & {\mathcal A} := \scH_{\text{BPS}}^{\text{closed}}\qquad\qquad
\end{array}
\ee
\end{minipage}}
\end{center}
In fact, this relation between closed and open BPS states is supposed to hold in general, regardless of the specific nature of the Calabi-Yau 3-fold $X$ and the Lagrangian subvariety $M_L$. (In this general context, the notion of {\it refined} BPS states exists only for rigid $X$.) However, in application to knots, the action of ${\mathcal A}$ on triply-graded homology \eqref{HHBPS} is especially useful and accounts for certain differentials that act on colored HOMFLY homology \cite{Gukov:2011ry}. Although at present one can only identify certain elements of ${\mathcal A}$ but not the entire algebra, it is expected to be related to the \emph{double affine Hecke algebra} \cite{Cherednik:2011nr} and the \emph{rational Cherednik algebra} \cite{Gorsky:2012mk} for torus knots. (See also \cite{Nawata:2015wya} for a review.)

Both systems \eqref{surfeng2} and \eqref{surfeng} enjoy a $\U(1)_P \times \U(1)_F$ symmetry
that acts on $TN_4$ and gives rise to two gradings (= conserved charges),
$q$-grading and $t$-grading\footnote{Up to exchange of $\epsilon_1$ and $\epsilon_2$,
this identification and its relation with the refinement of topological string \cite{Iqbal:2007ii},
the Nekrasov-Shatashvili limit \cite{Nekrasov:2009rc}, {\it etc.} is explained in \cite[\S 4]{Fuji:2012pm}.
Note that the unrefined limit corresponds to $\epsilon_1 = - \epsilon_2$, {\it i.e.} $t=-1$.}:
\be
TN_4 \cong \bR^4_{\epsilon_1,\epsilon_2}
~ , \qquad
D \cong \bR^2_{\epsilon_1}
~ , \qquad
q = e^{\epsilon_1}
~ , \qquad
t = - e^{-(\epsilon_1 + \epsilon_2)}~.
\label{Omegabkgr}
\ee
In addition, the system \eqref{surfeng} has $\U(1)_a$ gauge symmetry in five dimensions $\bR_t \times TN_4$
that gives rise to the $a$-grading of HOMFLY homology.
Indeed, compactification of eleven-dimensional M-theory on the resolved conifold $X$
is a simple example of {\it geometric engineering} of SUSY gauge theories \cite{Katz:1996fh},
that in the present case engineers a super-Maxwell theory on $\bR_t \times TN_4$ with gauge group $\U(1)_a$.

The Omega-background \cite{Nekrasov:2002qd} associated with the isometry $\U(1)_P \times \U(1)_F$ implements equivariant localization
to the fixed point, the origin $\{ 0 \} \in TN_4$. As a result, the effective theory on the branes in
\eqref{surfeng2} and \eqref{surfeng} is $(0+1)$-dimensional quantum mechanics with only one supercharge $\cQ$
and its hermitian conjugate $\cQ^{\dagger}$:
\be\nonumber
\{ \cQ , \cQ^{\dagger} \} \; = \; 2H.
\ee

There are two standard ways to describe knots and links in this setup:
either via Wilson / 't Hooft line operators
(that are naturally labeled by representations of $G = \SU(N)$)
or via monodromy defects (that are labeled by $G$-valued holonomies).
The system \eqref{surfeng2},
where knots and links are introduced via M5'-branes supported on $\bR_t \times D  \times  M_L$,
provides a realization of the monodromy defects.

In order to obtain a realization based on Wilson and 't Hooft operators one needs to replace the
set of M5'-branes in \eqref{surfeng2} by M2'-branes (a.k.a. membranes):
\be \label{surfeng3}\tag{def-M2}
\begin{matrix}
{\mbox{\rm space-time:}} && \qquad  \bR_t \times &TN_4&  \times  &T^* S^3 \\
{\mbox{\rm $N$ M5-branes:}} && \qquad  \bR_t \times& D &\times& S^3 \\
{\mbox{\rm M2'-branes:}} && \qquad  \bR_t \times& \{ 0 \} & \times & \Sigma_L
\end{matrix}
\ee
where $\Sigma_L$ is the total space of $T^* L \subset T^* S^3$ that meets the zero section, $S^3$, along the link $L$. Interestingly, these two systems are connected by the Hannay-Witten effect \cite{Diaconescu:2011xr,Frenkel:2015rda}; when we push the M5'-branes to the cotangent fiber of $T^*S^3$, the M2'-branes are dynamically generated.

Now, if we perform the geometric transition starting with the system \eqref{surfeng3},
then on the ``resolved conifold'' side we get
\be\tag{res-M2}
\begin{matrix}
{\mbox{\rm space-time:}} & \qquad & \bR_t & \times & TN_4 & \times & X \\
{\mbox{\rm M2'-branes:}} & \qquad & \bR_t & \times & \{ 0 \} & \times & \Sigma_L
\end{matrix}
\label{surfeng4}
\ee
which seems to be a new duality frame, little studied in the literature.\footnote{Part of the reason this system is more subtle than its
counterpart \eqref{surfeng} involving M5-branes is that it is more difficult to ``lift'' M2-branes off the M5-branes in \eqref{surfeng3}.
The resulting surface $\Sigma_L$ in \eqref{surfeng4} has asymptotic boundary $\partial \Sigma_L = L \subset S^3 \subset \partial X$
and presumably can be constructed along the lines of \cite{Taubes}.}

Our goal is to use the physical systems \eqref{surfeng2} and \eqref{surfeng3}
(resp. \eqref{surfeng} and \eqref{surfeng4}) to explain the structural properties and, ideally, compute
the doubly-graded $\fraksl(N)$ homologies (resp. the triply-graded HOMFLY homologies) denoted by $\scH^{\fraksl(N)}_{\vec \lambda} (L)$ (resp. $\scH_{\vec \lambda} (L)$) in \eqref{HHBPS}.

\subsection{Effective quantum mechanics}

In all of the four duality frames, equivariant localization ({\it i.e.} the Omega-background in the directions of $TN_4$)
effectively reduces the theory to a quantum mechanics with ``time'' direction $\bR_t$
and two real supercharges, $\cQ$ and $\cQ^{\dagger}$.
The space of supersymmetric states in this effective quantum mechanics is the desired space $\scH (L)$,
which is invariant under isotopies of $L \subset S^3$
due to the topological twist along the $S^3$ directions. Because of this, in \eqref{surfeng2} and \eqref{surfeng3}
we could, in fact, replace $S^3$ by {\it any} 3-manifold $M_3$ without breaking any of the symmetries or supersymmetries.
In particular, one could still associate a Hilbert space $\scH (M_3,L)$ of supersymmetric states
to a link $L \subset M_3$ defined as $\cQ$-cohomology, which should be an exciting direction for future research.
In this paper, we focus on the simplest case $M_3 = S^3$ merely for simplicity and in order to keep contact with
current mathematical developments, which at this stage are limited to colored link homologies in $S^3$.

As explained {\it e.g.} in \cite{Gukov:2007ck}, closing up the time direction $\bR_t$ into a circle
corresponds to ``decategorification.''
In other words, replacing $\bR_t$ by $S^1$ in \eqref{surfeng2}, \eqref{surfeng}, \eqref{surfeng3}, and \eqref{surfeng4}
gives physical systems effectively compact in all directions. Therefore, instead of producing the Hilbert space $\scH (L)$
as in $0+1$ dimensional quantum mechanics, this $0$-dimensional system computes the partition function,
which is the graded Poincar\'e polynomial of $\scH (L)$.
In duality frames \eqref{surfeng} and \eqref{surfeng4} it yields the colored superpolynomial \eqref{superPdef},
whereas in duality frames \eqref{surfeng2} and \eqref{surfeng3}
it gives the Poincar\'e polynomial of the doubly-graded $\fraksl(N)$ link homology,
\be\nonumber
\bar \scP^{\fraksl(N)}_{\vec \lambda} (L;q,t) \; := \; \sum_{i,j} q^i t^j \textrm{dim}\, \bar \scH^{\fraksl(N)}_{\vec \lambda} (L)_{i,j} \,.
\ee

The richness of physics does not stop with the four duality frames \eqref{surfeng2}, \eqref{surfeng}, \eqref{surfeng3}, and \eqref{surfeng4}.
In each of the duality frames, one can look at the $\cQ$-cohomology \eqref{HHBPS} in a number of different ways. (See  \cite{Gukov:2011ry,Nawata:2015wya} for overviews of different vantage points.)
For example, looking at it from the vantage point of the Calabi-Yau 3-fold gives a description of $\scH (L)$ in terms of enumerative invariants. A look from the vantage point of $\bR_t \times TN_4$ gives a description of the space $\scH (L)$ in terms of gauge theory with a surface operator. Focusing on the physics of the surface operator leads to yet another equivalent description of the same system (and hence the space $\scH (L)$) as the space of BPS states in the 3d $\cN=2$ theory $T[M_L]$ on $\bR_t \times D$.
In particular, this provides an identification of the effective quantum mechanics in question with a 3d-5d coupled system in the Omega-background. Schematically,
$$
\begin{array}{ccc}
\text{Quantum mechanics} \quad & \cong & \quad \text{3d } \cN=2 \text{ theory } T[M_L] \\
\text{on } \bR_t && \text{ on } \bR_t \times \bR^2_{\epsilon_1}
\end{array}
$$
where we used the identification \eqref{Omegabkgr}.

\subsection{Link homology and fusion of defect lines in LG models}\label{sec:LG-defect}

Note that in \eqref{surfeng2}, \eqref{surfeng}, and \eqref{surfeng3} the most ``interesting'' part of the setup
(where the physical degrees of freedom that contribute to $\scH (L)$ ``live'') is actually the four-dimensional factor
$$
\bR_t \times \bR^2_{\epsilon_1} \times L~,
$$
where we used \eqref{Omegabkgr} to emphasize that $D \cong \bR^2_{\epsilon_1}$ is effectively compact due to Omega-background.
(For instance, the equivariant volume $\text{Vol} (D) = \frac{1}{\epsilon_1}$ is finite for finite values of $\epsilon_1$.)
In the setup \eqref{surfeng4} the physical degrees of freedom that contribute to $\scH (L)$ live on $\bR_t \times L$
(so that $D$ does not even appear as part of their support).
In either of these cases, what is the effective theory on $\bR_t \times L \cong \bR_t \times S^1$?

In the deformed conifold side with either the probe M5'-branes \eqref{surfeng2} or the probe M2'-branes \eqref{surfeng3},
it was argued \cite{Gukov:2005qp,Gukov:2011ry} that the answer to this question involves Landau-Ginzburg (LG) models
that appear \cite{Khovanov:2004,Khovanov:2005,Khovanov:2007} in Khovanov-Rozansky formulation of link homologies.
Indeed, in order to understand the effective theory on $\bR_t \times L$ away from crossings, one can choose $L$
to be a piece of strand or an unknot that runs, say, along the great circle (the ``equator'')
\be\label{equator}
S^1_{\sigma} \subset S^3 \,,
\ee
parametrized by a (periodic) coordinate $\sigma$ (Figure \ref{fig:LGdefects1}).
Then, the effective theory on $\bR_t \times L$ describes a surface operator coupled
to a 4d topological field theory on $\bR_t \times S^3$ \cite{Chun:2015gda}. And, if $L$ covers $S^1_{\sigma}$ $k$ times,
the physics on $\bR_t \times S^1_{\sigma}$ away from crossings can be described by a LG model with the superpotential
$$
W = W_{\lambda_1} + \ldots + W_{\lambda_k}
$$
where $\lambda_i$'s are representations (``colors'') carried by the strands, {\it cf.} Figure \ref{fig:LGdefects1}.
For instance, when $\lambda = \square$ is the fundamental representation one recovers the potential
\be
\label{slN-fund}
W = x^{N+1}
\ee
used by Khovanov-Rozansky in construction of $\fraksl(N)$ link homology \cite{Khovanov:2004}.
We will use the physical realization of Khovanov-Rozansky homology \cite{Khovanov:2004,Khovanov:2005}
to understand the physical origin/meaning of the differentials and spectral sequences connecting different link homology to one another.

\noindent
\begin{minipage}{\linewidth}
      \centering
      \begin{minipage}{0.32\linewidth}
          \begin{figure}[H]\centering
              \includegraphics[scale=1]{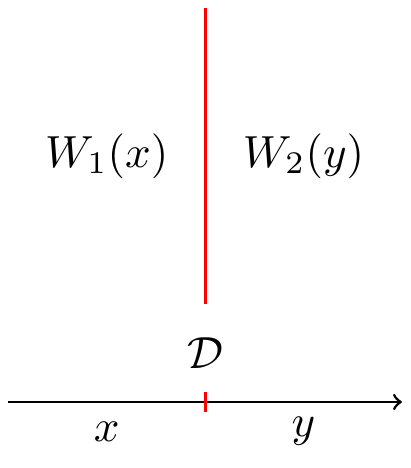}
              \caption{A defect between two LG models}\label{fig:LGdefects3}
          \end{figure}
      \end{minipage}
      \hspace{0.05\linewidth}
      \begin{minipage}{0.58\linewidth}
  \begin{figure}[H]
 \centering
    \includegraphics[width=9cm]{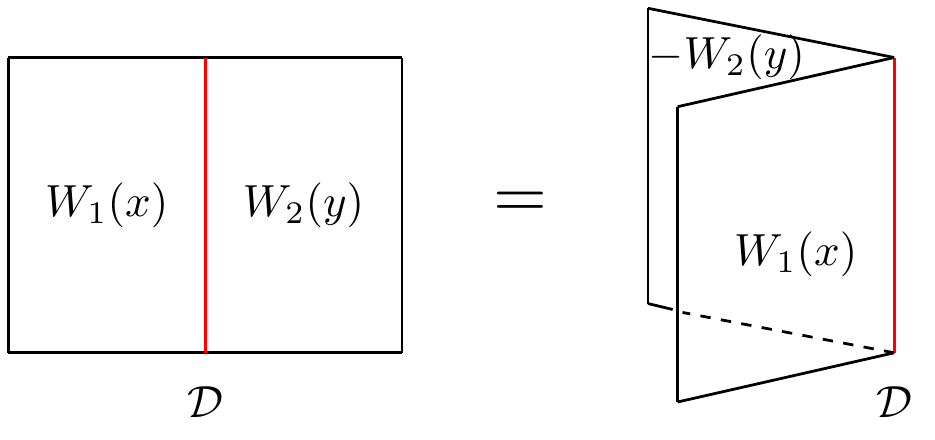}
    \caption{Folding trick}\label{fig:folding-trick}
\end{figure}
      \end{minipage}
  \end{minipage}
\vspace{.3cm}

In the construction of link homology given by Khovanov-Rozansky, a cone of two matrix factorizations is assigned to each crossing of a knot, as we shall review in greater detail in what follows.
Recall from \S\ref{sec:strip} that matrix factorizations represent B-type boundary conditions in Landau-Ginzburg models. Moreover, defects (interfaces) can also be represented by matrix factorizations \cite{Brunner:2007qu}, via the \emph{folding trick} (Figure \ref{fig:folding-trick}). A B-type defect  $\cD$ joints together a LG theory with chiral superfields $x_i$ and a superpotential $W_1(x_i)$, and a different LG theory with superfields $y_i$ and a superpotential $W_2(y_i)$ (Figure \ref{fig:LGdefects3}). The B-type defect
is described in the language of  a matrix factorization with the potential $W(x_i,y_i) = W_1(x_i)-W_2(y_i)$:
$$
R_0\xrightarrow{\,\,d_+\,\,} R_1\xrightarrow{\,\,  d_-\,\,} R_0
~,\qquad d_+\circ d_-=d_-\circ d_+=W_1(x_i)-W_2(y_i)~,
$$
where $R_0$ and $R_1$ are the polynomial ring $\bC[x_i,y_i]$.
\begin{figure}[H]
 \centering
    \includegraphics[width=11cm]{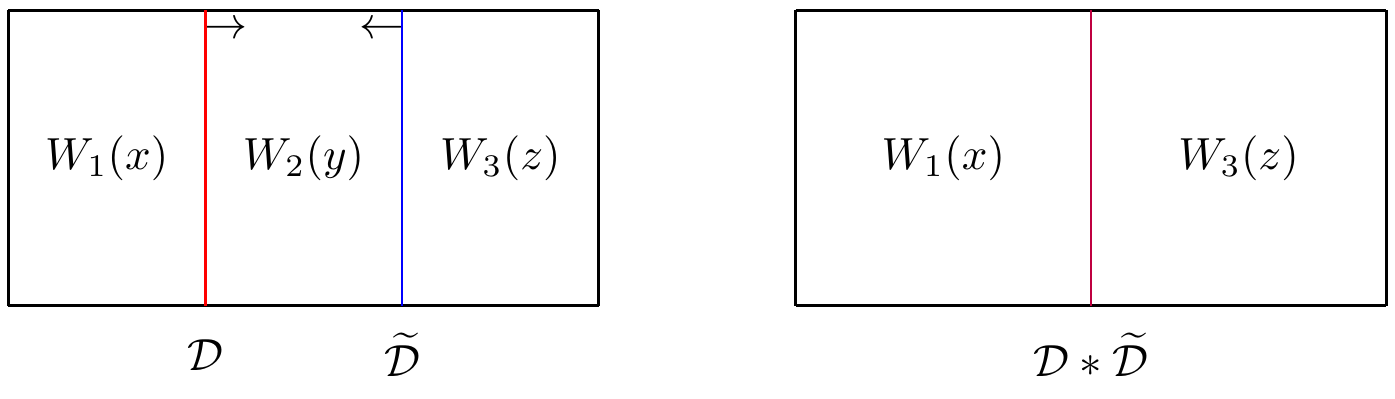}
    \caption{Fusion of topological defects that is equivalent to a tensor product of the corresponding matrix factorizations.}\label{fig:LGdefects4}
\end{figure}

Since these defects are topological, one can move and distort them in general. In particular, two defects $\cD$ and $\wt \cD$ can be put  on top of each other creating a new defect $\cD\ast\wt\cD$, which is usually called a \emph{fusion} of the defects (Figure \ref{fig:LGdefects4}). The fusion amounts to taking the tensor product of the matrix factorizations $\cD$ and $\wt \cD$  defined by
\be
\cD \otimes \wt \cD: \quad
\left[\begin{array}{c}R_0 \otimes  \wt R_0 \\ \bigoplus \\ R_1 \otimes  \wt R_1 \end{array}\right]_0
\xrightarrow{\quad d^0\quad}
\left[\begin{array}{c} R_1 \otimes \wt R_0 \\ \bigoplus \\ R_0 \otimes \wt R_1 \end{array}\right]_1
\xrightarrow{\quad d^1 \quad}
\left[\begin{array}{c} R_0 \otimes \wt R_0 \\ \bigoplus \\ R_1 \otimes \wt R_1 \end{array}\right]_0~,
\label{DDbar}
\ee
where the differentials are defined by
$$\arraycolsep 6pt
d^0=\left[\!\begin{array}{cc} d_+ \otimes \text{id} & \text{id} \otimes \wt d_+ \\
-\text{id} \otimes \wt d_- & d_- \otimes \text{id} \end{array}\!\right]
\ ,\quad
d^1={\left[\!\begin{array}{cc} d_- \otimes \text{id} & - \text{id} \otimes \wt d_+ \\
 \text{id} \otimes \wt d_- & d_+ \otimes \text{id} \end{array}\!\right]}~,
$$
which satisfies $d^0\circ d^1=d^1\circ d^0=W_1(x)-W_3(z)$. In fact, the resulting matrix factorization is generally of infinite rank.
Physically speaking, the chiral fields $y$ squeezed in between the two defects are promoted to new degrees of freedom on the new defects when the two defects coincide. However, if both $\cD$ and $\wt \cD$ are of finite rank, the matrix factorization $(\cD \otimes \wt \cD)_{\bC[x,z]}$ can be reduced to finite rank by throwing away an infinite number of trivial matrix factorizations \eqref{trivial-MF}. After this reduction, we denote the resulting matrix factorization by
$$
\cD\ast \wt \cD:=(\cD \otimes \wt \cD)^{\textrm{red}}_{\bC[x,z]}~.
$$

\begin{figure}
 \centering
    \includegraphics[width=13cm]{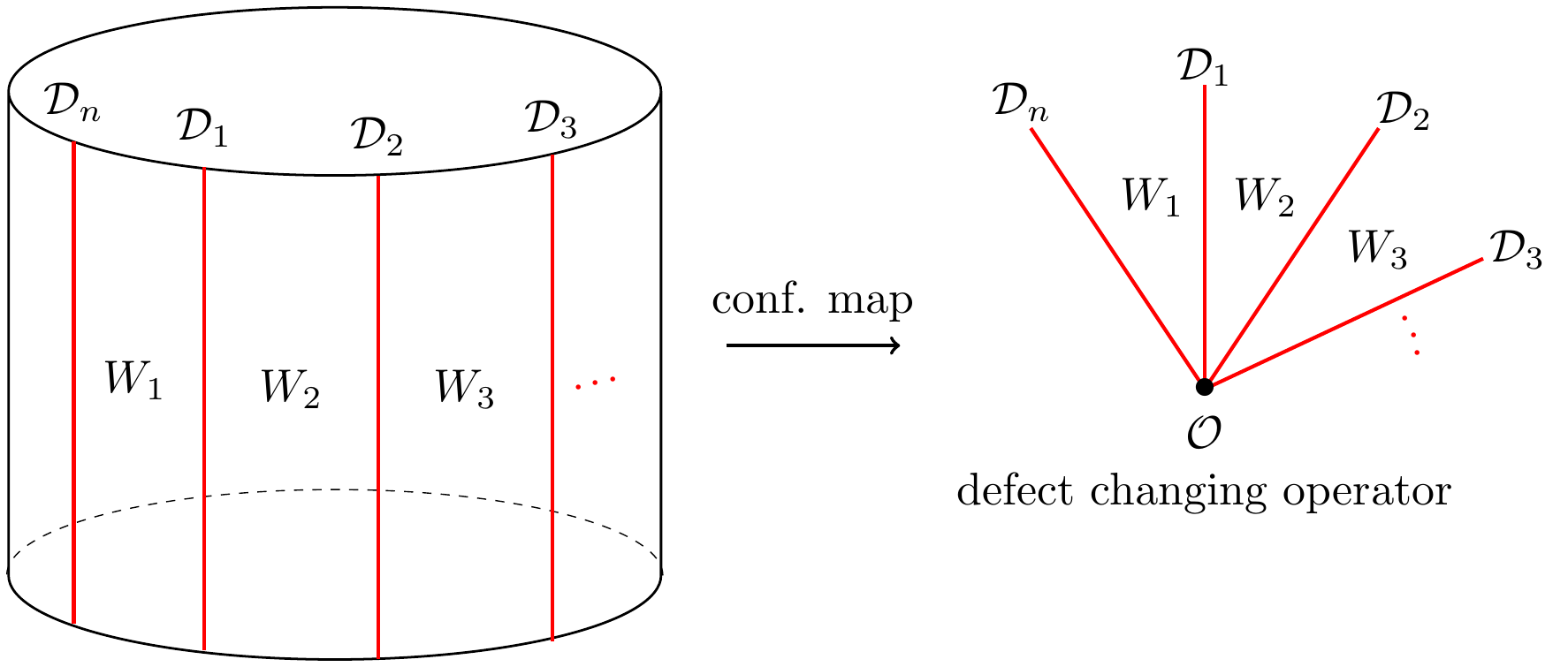}
    \caption{Landau-Ginzburg theory with defects on a cylinder. Via the state-operator correspondence, the BPS states of the LG model can be identified with the defect-changing operators.}\label{fig:LGdefects5}
\end{figure}

If there are $n$ defects with the superpotentials $W_1, \ldots, W_n$ and $W_{n+1}=W_1$, one can think that the LG model with the defects is placed on a cylinder. The tensor product $\cC = \otimes_i \cD_i$ is a tensor product of the matrix factorizations $\cD_i$ of the potentials $W_i-W_{i+1}$; since $d^2=\sum_{i=1}^{n}W_i-W_{i+1}=0$, it is a complex. The homology $\cH^*(\cD_1,\cD_2,\cdots,\cD_n)$ of this complex is the BPS spectrum of the theory. By the state-operator correspondence, the BPS spectrum can be identified with the spectrum of operators at the junctions of the defects  \cite{Brunner:2007qu} (Figure \ref{fig:LGdefects5}).

\begin{figure}
 \centering
    \includegraphics[width=11cm]{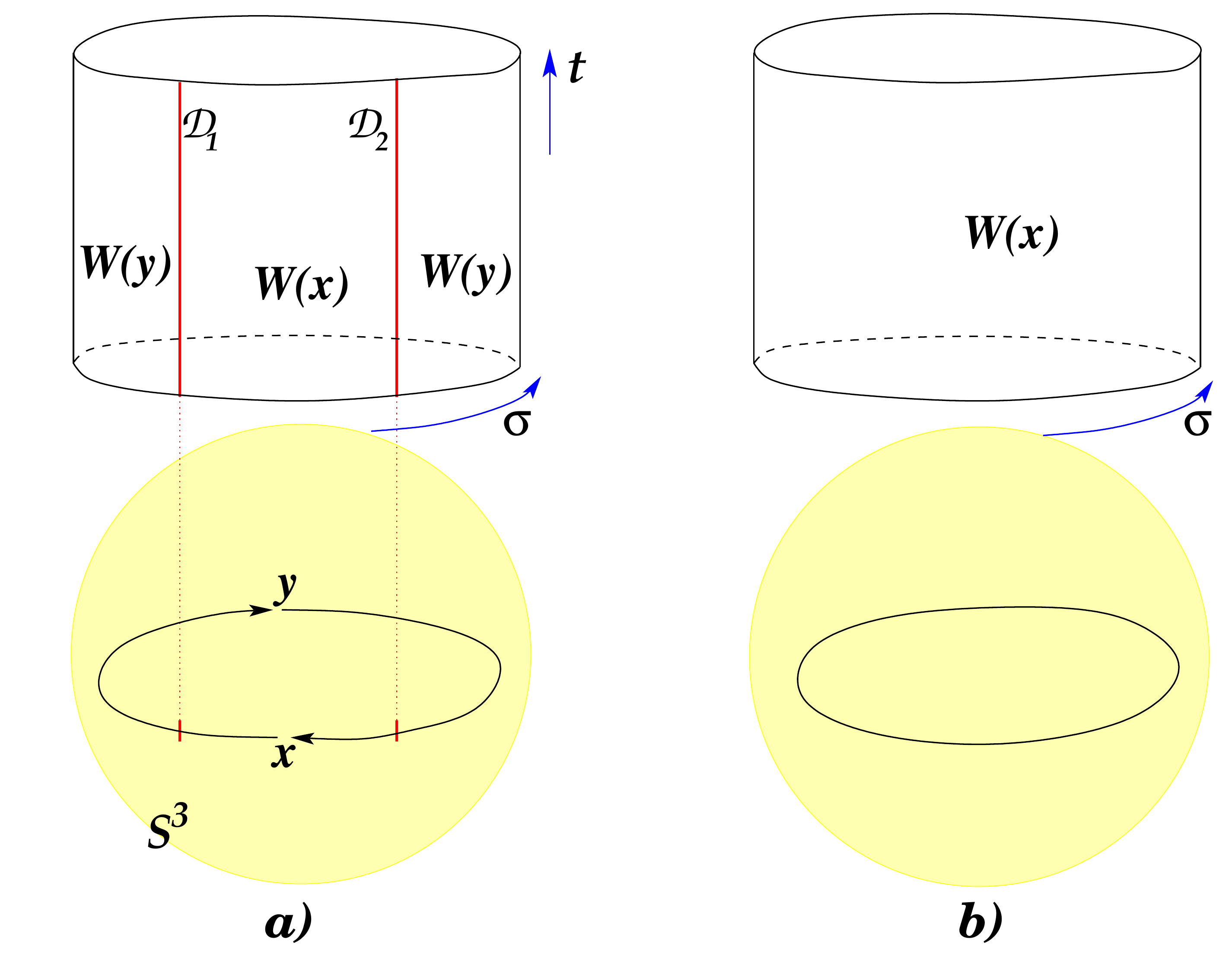}
    \caption{The Landau-Ginzburg model for the unknot. Since $\cD_2$ is the inverse of $\cD_1$, they can annihilate each other
leaving the Landau-Ginzburg model without any defect lines.}\label{fig:LGdefects2b}
\end{figure}

\begin{figure}
\begin{center}
\includegraphics[scale=0.4]{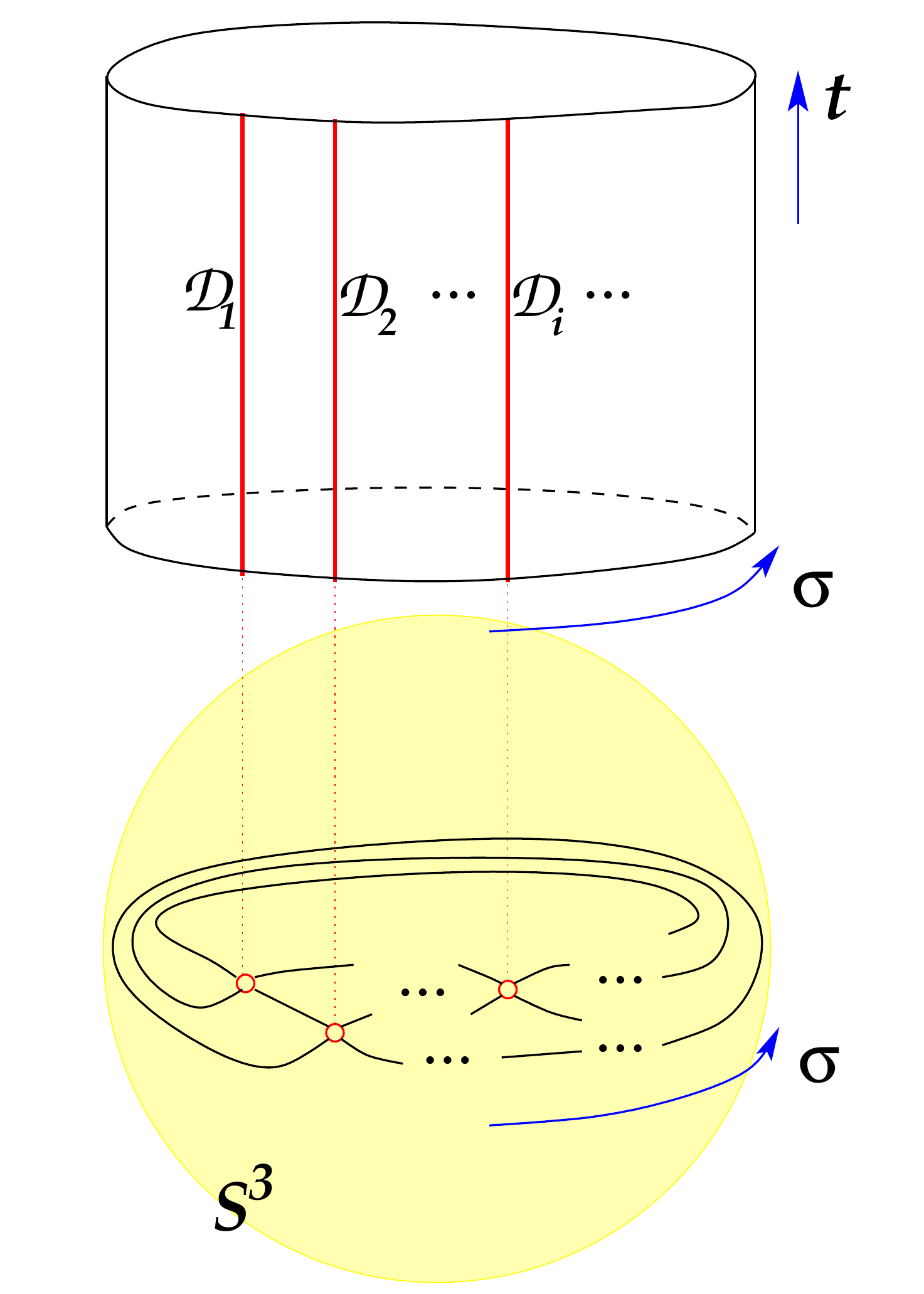}
\end{center}
\caption{LG model with defects on $\bR_t\times S^1_\sigma$ in the fivebrane system.\vspace{.5cm}}
\label{fig:LGdefects1}
\end{figure}

The LG model with defects on a cylinder is very akin to the Khovanov-Rozansky construction.
For example, the unknot can be represented as a closure of braid on one strand, {\it i.e.} as two semi-circles glued together (Figure \ref{fig:LGdefects2b}.a).
The first semi-circle is represented by a matrix factorization $\cD_1$ of $W(x)-W(y)$
\be\label{mf1}
R_0\xrightarrow{\,\,x-y\,\,} R_1\xrightarrow{\quad  p\quad} R_0~,
\ee
and the second semi-circle corresponds to a matrix factorization $\cD_2$ of $W(y)-W(x)$
\be\label{mf2}
R_0\xrightarrow{\,\,y-x\,\,} R_1\xrightarrow{\quad  p\quad} R_0~,
\ee
where $R=\bC[x,y]$ and
$$
p=p(x,y)=\frac{W(x)-W(y)}{x-y}~.
$$
The homology of the unknot is given as the homology of the tensor product \eqref{DDbar} of the matrix factorizations (\ref{mf1}) and (\ref{mf2}):
\be\nonumber
\left[\begin{array}{c}R\\ \bigoplus \\R\end{array}\right]_0
\xrightarrow{\, d^0\,}
\left[\begin{array}{c}R\\ \bigoplus \\R\end{array}\right]_1
\xrightarrow{\, d^1 \,}
\left[\begin{array}{c}R\\ \bigoplus \\R\end{array}\right]_0~,
\qquad\arraycolsep 6pt
d^0=\left[\!\begin{array}{cc}x-y & y-x\\ -p &p\end{array}\!\right], \quad d^1={\left[\!\begin{array}{cc}p & x-y\\ p & x-y\end{array}\!\right]}~.
\ee
It is easy to see that $d^0\circ d^1=d^1\circ d^0=0$.
Consequently, a simple computation yields
\be\nonumber
H^0=\textrm{Ker} ~d^0/\textrm{Im} ~d^1\cong \bC[x]/W'(x) ~, \quad H^1=\textrm{Ker} ~d^1/\textrm{Im} ~d^0\cong \left\{ \begin{array}{cr} 0& \quad  \textrm{for}\ p\ne 0\\
\bC[x]& \quad \textrm{for}\ p=0 \end{array}\right.~.
\ee

With the potential $W(x)=x^{N+1}$ as in \eqref{slN-fund}, we obtain the $\fraksl(N)$ knot homology of the unknot
\be
\overline\scH_{\yng(1)}^{\fraksl(N)}(\unknot) \cong \mathbb{C}[x] / ( x^N ).
\ee
The $q$-grading of the chiral field $x$ is two, so that the Poincar\'e polynomial is $$\overline\scP_{\yng(1)}^{\fraksl(N)}(\unknot)=1+q^2+\cdots+q^{2N-2} ~.$$
When  $W=0$, we have the HOMFLY homology of the unknot. In this case, both $H^0$ and $H^1$ are nontrivial and isomorphic to $\bC[x]$, so the homology is supported in two different homological degrees. In particular, the Poincar\'e polynomial is
\be\nonumber
\overline\scP_{\yng(1)}(\unknot)=\frac {1} {1-q^2}+\frac{a^2t}{1-q^2}~,
\ee
where the first term corresponds to homological degree zero whereas the second term corresponds to homological degree one as well as the $a$-grading is shifted by two. On the resolved side \eqref{surfeng}, the free $\cN =2$ chiral superfield $x$ comes from M2-branes ending on M5-branes.

In fact, in the above discussion $\cD_2 = \bar \cD_1$ is the inverse of the defect $\cD_1$, so that $\cD_1$ and $\cD_2$ can annihilate each other
leaving behind a LG theory with superpotential $W$ on a cylinder  without any defect lines. (Figure \ref{fig:LGdefects2b}) Hence, we have $\cH(\cD_1,\cD_2)\cong \textrm{Jac}(W)$, the Jacobi ring of~$W$ (or space of closed-string BPS states).
The effective theory on the most ``interesting'' part of the unknot theory (which is the product $\bR_t\times S^1_\sigma$ of the time and the equator \eqref{equator}) in the deformed conifold side, \eqref{surfeng2} or \eqref{surfeng3}, is described by a LG model.

\vspace{.3cm}

Having identified B-type defects between LG models as matrix factorizations, one can make use of the defects for physical realization of the Khovanov-Rozansky construction of link homology on $ \bR_t \times S^1_\sigma$.

Recall, that in the Khovanov-Rozansky formulation \cite{Khovanov:2004,Khovanov:2005,Rasmussen:2006}
one first constructs a cube of resolutions by replacing each crossing either with ${\raisebox{-.1cm}{\includegraphics[width=.5cm]{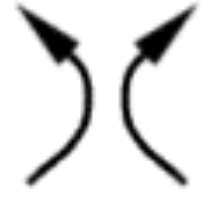}}}$ or with ${\raisebox{-.1cm}{\includegraphics[width=.35cm]{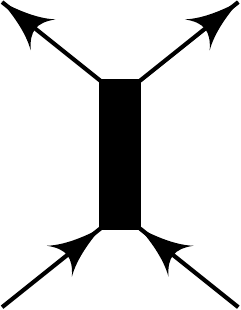}}}$. Each resulting resolution is a planar Murakami-Ohtsuki-Yamada graph \cite{Murakami:1998}. Then, one associates a matrix factorization to each resolved crossing, together with a map from one $\cD_{\includegraphics[width=.4cm]{smoothing}}$ to the other $\cD_{\includegraphics[width=.3cm]{resolution}}$ at a positive crossing ${\raisebox{-.1cm}{\includegraphics[width=.4cm]{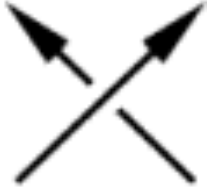}}}$, corresponding to an edge of the cube. The part of the complex associated to one such crossing is called \emph{a mapping cone}:
\be\label{cone}
\begin{tikzpicture}[scale=0.8]
\node at (0,0) {$R_0$};
\node at (3,0) {$R_1$};
\node at (6,0) {$R_0$};
\node at (0,2) {$\wt R_0$};
\node at (3,2) {$\wt R_1$};
\node at (6,2) {$\wt R_0$};
\draw[->] (0,0.5)--(0,1.5) ;
\node at (-0.3,1) {$d_v^{0}$};
\draw[->] (3,0.5)--(3,1.5) ;
\node at (2.7,1) {$d_v^{1}$};
\draw[->] (6,0.5)--(6,1.5) ;
\node at (5.7,1) {$d_v^{0}$};
\draw[->] (0.5,0)--(2.5,0) ;
\node at (1.5,.3) {$d_+$};
\draw[->] (3.5,0)--(5.5,0) ;
\node at (4.5,.3) {$d_-$};
\draw[->] (0.5,2)--(2.5,2) ;
\node at (1.5,2.3) {$\wt d_+$};
\draw[->] (3.5,2)--(5.5,2) ;
\node at (4.5,2.3) {$\wt d_-$};
\node at (-2,0) {$\includegraphics[width=.6cm]{smoothing}$};
\draw[->] (-2,0.5)--(-2,1.5) ;
\node at (-2,2) {$\includegraphics[width=.6cm]{resolution}$};
\node at (-8.5,1.1) {$\includegraphics[width=.6cm]{overcrossing}$};
\node at (-7.8,1.1) {$:$};
\node at (-5,1) {$\textrm{Cone}(d_v:\cD_{\includegraphics[width=.4cm]{smoothing}}\to \cD_{\includegraphics[width=.3cm]{resolution}}):$};
\node at (-8.9,.6) {$x_1$};
\node at (-8.1,.6) {$x_2$};
\node at (-8.9,1.6) {$x_3$};
\node at (-8.1,1.6) {$x_4$};
\end{tikzpicture}
\ee
where all rings are isomorphic to $R=\bC[x_1,x_2,x_3,x_4]/(x_3+x_4-x_1-x_2)$ and the maps are defined by
\bea\label{map}
\wt d_+=(x_3-x_1)(x_2-x_3)~,&\quad& \wt d_-=\frac{W(x_3)+W(x_4)-W(x_1)-W(x_2)}{(x_3-x_1)(x_2-x_3)}~,\cr
d_+=(x_3-x_1)~,&\quad& d_-=\frac{W(x_3)+W(x_4)-W(x_1)-W(x_2)}{x_3-x_1}~,\cr
d_v^{0}=x_2-x_3~,&\quad&d_v^{1}=1~.
\eea
Note that the map $d_v$ is induced in such a way that the diagram commutes. At a negative crossing ${\raisebox{-.1cm}{\includegraphics[width=.4cm]{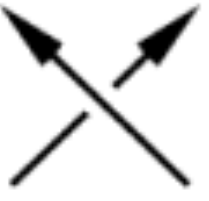}}}$, we associate the inverse of \eqref{cone}. (See  \cite{Khovanov:2004,Khovanov:2005,Rasmussen:2006} for more detail.) One then takes the tensor product of all these complexes to produce the total complex
\bea\label{complex}
\begin{tikzpicture}
\node at (-1.1,0) {$\raisebox{.6cm}{\maplr{d_+}{d_-}}$};
\node at (0.2,0) {$C_{j,i}$};
\node at (1.5,0) {$\raisebox{.6cm}{\maplr{d_+}{d_-}}$};
\node at (3,0) {$C_{j,i+1}$};
\node at (4.5,0) {$\raisebox{.6cm}{\maplr{d_+}{d_-}}$};
\node at (5.9,0) {$C_{j,i+2}$};
\node at (7.4,0) {$\raisebox{.6cm}{\maplr{d_+}{d_-}}$};
\node at (-1.1,2) {$\raisebox{.6cm}{\maplr{d_+}{d_-}}$};
\node at (0.2,2) {$C_{j+1,i}$};
\node at (1.5,2) {$\raisebox{.6cm}{\maplr{d_+}{d_-}}$};
\node at (3,2) {$C_{j+1,i+1}$};
\node at (4.5,2) {$\raisebox{.6cm}{\maplr{d_+}{d_-}}$};
\node at (5.9,2) {$C_{j+1,i+2}$};
\node at (7.4,2) {$\raisebox{.6cm}{\maplr{d_+}{d_-}}$};
\draw[->] (0.2,0.5)--(0.2,1.5) ;
\node at (-0.1,1) {$d_v$};
\draw[->] (3,0.5)--(3,1.5) ;
\node at (2.7,1) {$d_v$};
\draw[->] (5.9,0.5)--(5.9,1.5) ;
\node at (5.6,1) {$d_v$};
\draw[->] (0.2,2.5)--(0.2,3.5) ;
\node at (-0.1,3) {$d_v$};
\draw[->] (3,2.5)--(3,3.5) ;
\node at (2.7,3) {$d_v$};
\draw[->] (5.9,2.5)--(5.9,3.5) ;
\node at (5.6,3) {$d_v$};
\draw[->] (0.2,-1.5)--(0.2,-0.5) ;
\node at (-0.1,-1) {$d_v$};
\draw[->] (3,-1.5)--(3,-0.5) ;
\node at (2.7,-1) {$d_v$};
\draw[->] (5.9,-1.5)--(5.9,-0.5) ;
\node at (5.6,-1) {$d_v$};
\end{tikzpicture}~.
\eea
The homology of this complex turns out to be a knot invariant. When $W=0$, it is easy to see from \eqref{map} that $d_-$ is the zero map. Thus, the complex for HOMFLY homology has $\bZ\times\bZ\times \bZ$-gradings corresponding to the $(q,t,a)$-grading where the $q$-grading is the internal grading $(q(x_i)=2)$ and $(t,a)$-grading is inherited from the bi-grading $(j,i)$ of \eqref{complex}. Note that $d_v$ is of degree $(0,1,0)$ and $d_+$ is of degree $(2,0,2)$. When $W\neq0$,  the complex is well-defined only after collapsing the $\bZ$-grading to $\bZ_2$-grading because of the presence of $d_-$.
Therefore, the complex for $\fraksl(N)$ homology loses the $a$-grading so that it is $\bZ\times\bZ\times \bZ_2$-graded where the $\bZ_2$-grading comes from a matrix factorization. It is worth mentioning that this link homology is supported in a range of $t$-degrees that is at most the crossing number plus one, because a map $\raisebox{-.1cm}{\includegraphics[width=.45cm]{overcrossing}}: \raisebox{-.1cm}{\includegraphics[width=.5cm]{smoothing}} \xrightarrow{d_v}  \raisebox{-.1cm}{\includegraphics[width=.35cm]{resolution}}$, which increases $t$-degree by one, is assigned at each crossing.

To provide a physical realization of this formulation, using topological invariance along $S^3$
we represent a link $L$ as a thin braid of $k$ strands that runs along the equator $S^1_{\sigma}\subset S^3$:
\be\nonumber
k\text{-tuple cover}: \quad
L \to S^1_{\sigma},
\ee
as shown in Figure \ref{fig:LGdefects1}. In Figure \ref{fig:LGdefects1}, ``$\circ$'' stands for any of the basic crossings ${\raisebox{-.1cm}{\includegraphics[width=.4cm]{overcrossing}}}$ and ${\raisebox{-.1cm}{\includegraphics[width=.4cm]{undercrossing}}}$. Since we associate matrix factorizations \eqref{cone} at each crossing, this means that the corresponding defect is placed at the position of the crossing in LG model on $\bR_t\times S^1_\sigma$.

More precisely, a mapping cone~\eqref{cone} of two matrix factorizations admits a physical interpretation as a bound state of the two defects $\cD_{\includegraphics[width=.4cm]{smoothing}}$ and $\cD_{\includegraphics[width=.3cm]{resolution}}$, formed due to a fermionic \emph{defect changing operator}
$d_v\in\cH^1(\cD_{\includegraphics[width=.4cm]{smoothing}},\cD_{\includegraphics[width=.3cm]{resolution}})$ \cite{Walcher:2004tx,Brunner:2009zt}. Hence, we assign the bound state of the two defects at each defect line in Figure \ref{fig:LGdefects1}. In fact, the bound state formation can be understood as a perturbation of the supercharge. Since the mapping cone \eqref{cone} can be expressed as
\be\nonumber
\cD_{\includegraphics[width=.3cm]{overcrossing}}=\textrm{Cone}(d_v:\cD_{\includegraphics[width=.4cm]{smoothing}}\to \cD_{\includegraphics[width=.3cm]{resolution}}):\; R_1\oplus \wt R_0\maplr{c_1}{c_0}R_0\oplus\wt R_1
\ee
with
\bea
c_1=\left(\begin{array}{cc} d_- & 0 \\ d_v & \wt d_+\end{array}\right)\,,\quad
c_0=\left(\begin{array}{cc} d_+ & 0 \\ d_v & \wt d_-\end{array}\right)\,,\nonumber
\eea
we see that the two non-interacting (direct sum) defects $\cD_{\includegraphics[width=.4cm]{smoothing}}$ and $\cD_{\includegraphics[width=.3cm]{resolution}}$ are bound together by perturbing the boundary supercharge with  the defect changing operator $d_v$
\be\label{def-bd}
Q_{\text{bd}}=\left(\begin{array}{cc} d_{\mp} & 0 \\ 0 & \wt d_{\pm}\end{array}\right) \quad\longrightarrow\quad  Q_{\text{bd}}+\delta Q_{\text{bd}} =\left(\begin{array}{cc} d_{\mp} & 0 \\ d_v & \wt d_{\pm}\end{array}\right) ~.
\ee
In summary, in LG theory with the potential $W$ on  $\bR_t\times S^1_\sigma$, a bound state, either $\cD_{\includegraphics[width=.3cm]{overcrossing}}$ or $\cD_{\includegraphics[width=.3cm]{undercrossing}}$, is assigned at each crossing (Figure \ref{fig:LGdefects1}). Each strand carries a $\cN =2$ chiral superfield $x_i$ on a strip between the defect lines. As a result, the tensor product $\cD_1\ast\cdots\ast \cD_n$ of the defects provides the complex \eqref{complex}, and therefore the BPS spectrum in this setup can be identified with the link homology
\be\nonumber
\cH(\cD_1,\cdots,\cD_n)=\overline\scH(L)~.
\ee

The topological invariance of link homology tells us the fusion structure of the defects. The invariance under Reidemeister move II implies that the product of an over and under-crossing is the identity
$$
{\raisebox{-.6cm}{\includegraphics[width=2cm]{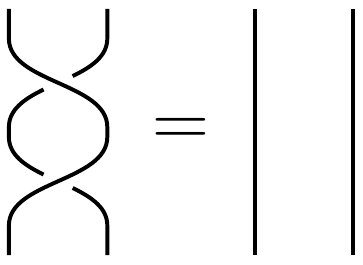}}}\quad \longrightarrow \quad
\cD_{\includegraphics[width=.3cm]{overcrossing}} \ast \cD_{\includegraphics[width=.3cm]{undercrossing}} = \textrm{Id} =\cD_{\includegraphics[width=.3cm]{undercrossing}} \ast \cD_{\includegraphics[width=.3cm]{overcrossing}} ~.
$$
In addition, the Reidemeister move III means it obeys the braid group relation
\be\label{braid}
{\raisebox{-1cm}{\includegraphics[width=4cm]{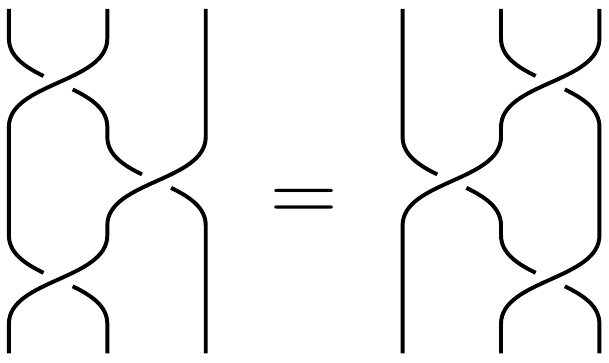}}}\quad \longrightarrow \quad
\cD^{(i)}\ast \cD^{(i+1)} \ast \cD^{(i)} =\cD^{(i+1)} \ast \cD^{(i)} \ast \cD^{(i+1)}~.
\ee
with the obvious relation $ \cD^{(i)} \ast \cD^{(j)} =\cD^{(j)} \ast \cD^{(i)} $ for $|i-j|>1$
where we denote the defect corresponding to a braid of $i$ and $(i+1)$-strands  by $\cD^{(i)}$.
In other words, $(\cD,\ast)$ is an object of a braided monoidal category.

The structure of defects is similar  to that of B-branes considered in Seidel-Thomas \cite{Seidel:2000ia}. Therefore, the construction above strongly suggests that it should satisfy a spherical condition
\bea\label{conddim}
\cH^i(\cD,\cD)=\left\{\begin{array}{ll}\bC\,,& i=0,n\\ 0\,,&{\rm otherwise}\end{array}\right.
\eea
for some $n$. In addition, the defects should form a sequence of spherical matrix factorizations
\bea
\dim\;\cH(\cD^{(i)},\cD^{(j)})=\left\{\begin{array}{ll}1\,,& |i-j|=1\\ 0\,& |i-j|>1\end{array}\right.\,.
\eea
These conditions indeed give rise to the braid group relation \eqref{braid} \cite{Brunner:2009zt}. See also \cite[\S 2.2]{Gukov:2007ck}.

\subsection{Gornik's spectral sequence}\label{sec:Gornik}

Having realized the Khovanov-Rozansky formulation of link homology in the fivebrane systems, let us now investigate $\fraksl(N)$ homology of the trefoil knot $\bf 3_1$.
It is known  for two-bridge knots with a small number of crossings \cite{Lewark:2015} that the homology $H_*(C_{j,i},d_{\pm})$ with respect to $d_\pm$ in \eqref{complex} at each $j$ is isomorphic to a direct sum of several copies of $\bC[x]/(W^\prime)$, and the induced differential $d_v$ on $H_*(C_{j,i},d_{\pm})$ is multiplication by $W^{\prime\prime}$. In particular, in the case of the trefoil, at $\deg t=j=0,2,3$, the homology  $H_*(C_{j,i},d_{\pm})$ with respect to $d_\pm$ in \eqref{complex} is  isomorphic to $\bC[x]/(W^\prime)$, and the induced differential $d_v$ is non-trivial between $t$-degrees 2 and 3.
For instance,  let's consider $\fraksl(2)$ homology, for which $W=x^3$. Then, the complex can be expressed as follows:
\bea\label{ss-trefoil}
\begin{tikzpicture}
\node at (.5,3.8) {$t$-degree};
\node at (2.5,3.8) {$0$};
\node at (5,3.8) {$1$};
\node at (7.5,3.8) {$2$};
\node at (10,3.8) {$3$};
\draw[thick] (-.5,3.5) -- (11,3.5);
\draw[thick] (1.5,4) -- (1.5,1.2);
\node at (0.5,3) [scale=1]{complex};
\node at (0.5,1.8) [scale=1]{generators};
\node(A) at (2.5,3) {$\bC[x]/(x^2)$};
\node(B) at (5,3) {$0$};
\node(C) at (7.5,3) {$\bC[x]/(x^2)$};
\node(D) at (10,3) {$\bC[x]/(x^2)$};
\node at (2.5,2.5) {\rotatebox{90}{$\in$}};
\node at (7.5,2.5) {\rotatebox{90}{$\in$}};
\node at (10,2.5) {\rotatebox{90}{$\in$}};
\node at (2.5,2) {$x$};
\draw (2.5,2)[red] circle (.2cm);
\node at (7.5,2) {$x$};
\draw (7.5,2)[red] circle (.2cm);
\node(F) at (10,2) {$x$};
\node at (2.5,1.5) {$1$};
\draw (2.5,1.5)[red] circle (.2cm);
\node(E) at (7.5,1.5) {$1$};
\node at (10,1.5) {$1$};
\draw (10,1.5)[red] circle (.2cm);
\draw[->,red] (E) -- (F) node [pos=.5,above,sloped,scale=.8] {${d_v= x}$};
\draw[->] (A) -- (B) node [pos=.5,above,scale=.8] {${d_v}$};
\draw[->] (B) -- (C) node [pos=.5,above,scale=.8] {${d_v}$};
\draw[->] (C) -- (D) node [pos=.5,above,scale=.8] {${d_v}$};
\end{tikzpicture}
\eea
This complex is essentially the same as the complex $\llbracket {\bf 3_1} \rrbracket$ coming from the cube of resolutions in \cite[Figure (3)]{BarNatan:2002}. Here, the induced differential $d_v$ is a multiplication by $W^{\prime\prime}=2x$, and its non-trivial action is depicted by the red arrow above. Thus, the unreduced $\fraksl(2)$ homology $\overline \scH^{\fraksl(2)}_{\yng(1)}( {\bf 3_1})\cong H_*(H_*(C_{j,i},d_{\pm}),d_v)$ is four-dimensional (the red circles above), and with appropriate shifts of $q$-degrees, its Poincar\'e polynomial can be written as
$$
\overline\scP^{\fraksl(2)}_{\yng(1)}({\bf 3_1})=1+q^2+q^4t^2+q^8t^3~.
$$

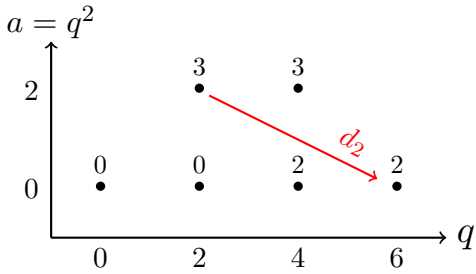
\begin{wrapfigure}{l}{0.5\textwidth}\centering
\begin{tikzpicture}[scale=1.3]
  \draw[thick,->] (-5.5,-.5) -- (-1.5,-.5);
\draw[thick,->] (-5.5,-.5) -- (-5.5,1.5);
\node at (-5,-.7) {0};
\node at (-4,-.7) {2};
\node at (-3,-.7) {4};
\node at (-2,-.7) {6};
\node at (-1.3,-.5) [scale =1.3]{$q$};
\node at (-5.7,0) {0};
\node at (-5.7,1) {2};
\node at (-5.5,1.7) [scale =1.1] {$a=q^2$};
\node at (-5,0) [scale =3]{$\cdot$};
\node  at (-5,0.25) [scale =.9]{$0$};
\node at (-4,0) [scale =3]{$\cdot$};
\node  at (-4,0.25) [scale =.9]{$0$};
\node at (-3,0) [scale =3]{$\cdot$};
\node  at (-3,0.25) [scale =.9]{$2$};
\node at (-2,0) [scale =3]{$\cdot$};
\node  at (-2,0.25) [scale =.9]{$2$};
\node at (-4,1) [scale =3]{$\cdot$};
\node  at (-4,1.25) [scale =.9]{$3$};
\node at (-3,1) [scale =3]{$\cdot$};
\node  at (-3,1.25) [scale =.9]{$3$};
\draw[thick,->,red] (-3.9,0.95) -- (-2.2,.1) node [pos=.8,above,sloped] {$d_2$};
\end{tikzpicture}
\caption{unreduced $\fraksl(2)$ homology of the trefoil} \label{fig:sl2-trefoil}
\end{wrapfigure}

The induced differential $d_v$ can be understood in the following way. It was proposed in \cite{Dunfield:2005si} that the HOMFLY homology is endowed  with a set of differentials $\{d_N\}_{N\in\bZ}$ where the homology with respect to $d_{N}$ is isomorphic to the $\fraksl(|N|)$ homology. To obtain the unreduced $\fraksl(2)$ homology $\overline \scH^{\fraksl(N)}_{\yng(1)}(K)$ from the reduced HOMFLY homology $\scH_{\yng(1)}(K)$, first one takes the tensor product $\scH_{\yng(1)}(K)\otimes \bC[x]/(x^N)$ with the $\fraksl(N)$ unknot homology, and then computes the homology with respect to $d_N$ whose $(a,q,t)$-degree is $(-2,2N,-1)$. For instance, the reduced HOMFLY homology of the trefoil is three-dimensional and its Poincar\'e polynomial (up to factor $a^2 q^{-2}$) is
$$
\scP_{\yng(1)}({\bf 3_1})=1+q^4t^2+a^2q^2t^3~.
$$
To get the unreduced $\fraksl(2)$ homology, one takes the $d_2$ differential after multiplying the $\fraksl(2)$ unknot factor $(1+q^2)$ as illustrated in Figure \ref{fig:sl2-trefoil}. In this example, it is easy to see that $\scH_{\yng(1)}({\bf  3_1})\otimes \bC[x]/(x^2)$ is isomorphic to $H_*(C_{j,i}({\bf  3_1}),d_{\pm})$, and the action of $d_2$ amounts to that of the induced differential $d_v$ while the direction of the arrow is opposite. Hence, in this simple example, the induced differential $d_v$ for $\fraksl(N)$ homology is essentially equivalent to the $d_N$ differential.

In fact, the spectral sequences corresponding to differentials $d_{N>0}$ have been constructed explicitly \cite{Rasmussen:2006}. In physics,
one should think of these spectral sequences as triggered by the deformation of the boundary supercharge \eqref{def-bd}.
Let us note that since the superpotential is homogeneous, the LG model has $\U(1)_V \equiv \U(1)_P $ vector $R$-symmetry and the  $\U(1)_V$ $R$-charge is intrinsically the $q$-degree. Indeed, one can see from Figure \ref{fig:sl2-trefoil} that the differential $d_v$ preserves the $q$-degree.

\vspace{.3cm}
Deformation spectral sequences in knot homology have been investigated  by Lee~\cite{Lee:2005} for $\fraksl(2)$ and by Gornik~\cite{Gornik:2004} for $\fraksl(N)$ homology. There, changing the superpotential $W=x^{N+1}$ to $W=x^{N+1} +(N+1)\beta^N x$ leads to the deformation of the boundary supercharge $d_{\pm}$ in \eqref{map}. Furthermore, because the potential is no longer homogeneous, the $\U(1)_V$ $R$-symmetry is broken, so that the deformed complex $C^{\textrm{def}} (L)$ is filtered instead of bi-graded. Then, one can summarize the results of~\cite{Gornik:2004} as follows (see \S \ref{sec:deformations} for notations.):
\begin{itemize}\setlength{\parskip}{-0.1cm}
\item the $E_1$ page of the spectral sequence associated to $C^{\textrm{def}} (L)$ is isomorphic to the undeformed $\fraksl(N)$ homology $\overline\scH_{\yng(1)}(L)$.
\item the homology of $C^{\textrm{def}} (L)$ in the $E_\infty$ page is isomorphic to the tensor product $[\overline \scH^{\fraksl(N)}_{\yng(1)}(\unknot)]^{\otimes n}$ of $n$-copies of the $\fraksl(N)$ homology of the unknot where $n$ is the number of the component of $L$.
\end{itemize}

Actually, the Chinese remainder theorem tells us that the Jacobi ring of the deformed potential is isomorphic to
$$
\bC[x]/(x^N-\beta^N)\cong \bigoplus_{i=0}^{N-1}  \bC[x]/(x-\beta\omega^i)~,
$$
where $\omega$ is an $N$-th root of  unity. Its physical interpretation is as follows.
When the superpotential is perturbed by the relevant operator $x^{N+1}\to x^{N+1}+(N+1)\beta x$, the Landau-Ginzburg model $\textrm{LG}[\fraksl(N)]$ undergoes RG flow and decomposes into $N$ decoupled theories  $\textrm{LG}[\fraksl(1)] \oplus \cdots \oplus \textrm{LG}[\fraksl(1)]$, each with the superpotential $x-\beta \omega^i$ ($i=0,\cdots, N-1$).
\begin{wrapfigure}{l}{0.5\textwidth}\centering
\begin{tikzpicture}
\node at (3,3.1) {$\bullet$};
\node at (3.1,3) {$\bullet$};
\node at (2.9,3) {$\bullet$};
\node at (3,2.9) {$\bullet$};
\node at (4.2,4.2) {$x$};
\draw  (1.5,1.5) -- (4.5,1.5) -- (4.5,4.5) -- (1.5,4.5) -- (1.5,1.5);
\draw  (4,4.3) -- (4,4) -- (4.3,4);

\node at (7,2) {$\bullet$};
\node at (8,3) {$\bullet$};
\node at (7,4) {$\bullet$};
\node at (6,3) {$\bullet$};
\node at (8.2,4.2) {$x$};
\draw  (5.5,1.5) -- (8.5,1.5) -- (8.5,4.5) -- (5.5,4.5) -- (5.5,1.5);
\draw[very thick,->] (4.7,3)--(5.3,3);
\draw  (8,4.3) -- (8,4) -- (8.3,4);
\end{tikzpicture}
\caption{Deformation of the potential $x^5 \to  x^5 +x$ breaks a degenerate vacuum into massive vacua.} \label{fig:unknot-def}
\end{wrapfigure}
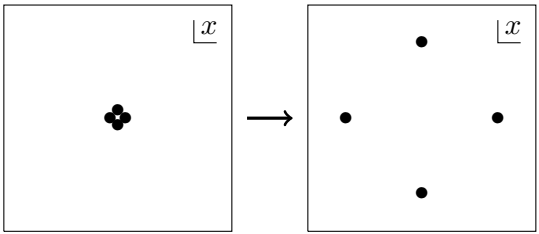

However, in the LG model $\textrm{LG}[\fraksl(1)]$, there is only the trivial defect (matrix factorization)! It is therefore immediate that the defects in the unperturbed theory (which encode the choice of link~$L$) become trivial, \emph{i.e.} invisible, from the standpoint of the $\fraksl(1)$ theories which remain after the perturbation.
The homology of $\textrm{LG}[\fraksl(1)]$ is therefore just the Jacobi ring $\bC[x]/(x-\beta\omega^i)$, which is one-dimensional. In sum, the spectrum of BPS states in the IR is $N^n$-dimensional for an $n$-component link $L$, just as Gornik proved.

Let us now see examples.
In the case of the unknot, there is no non-trivial spectral sequence, \textit{i.e.} the $E_1$ page is the same as the $E_\infty$ page. Under the deformation of the superpotentials $x^{N+1}\to x^{N+1}+x$, the dimension of the Jacobi ring is invariant while critical points of the potential become non-degenerate (Figure \ref{fig:unknot-def}).

To see the non-trivial spectral sequence, we again look at the example of the $\fraksl(2)$ homology of the trefoil. When we perturb the potential
\be\label{sl3-def}
W=x^3 \quad \longrightarrow \quad W=x^3 -3x~,
\ee
the complex \eqref{ss-trefoil} is deformed as shown in the following table.
\begin{figure}[h]
\bea\nonumber
\begin{tikzpicture}
\node at (0.2,3.8) {$t$-degree};
\node at (2.5,3.8) {$0$};
\node at (5.2,3.8) {$1$};
\node at (7.9,3.8) {$2$};
\node at (10.6,3.8) {$3$};
\draw[thick] (-.7,3.5) -- (12,3.5);
\draw[thick] (1.2,4) -- (1.2,1.2);
\node at (0.2,3) [scale=1]{complex};
\node at (0.2,1.8) [scale=1]{generators};
\node(A) at (2.5,3) [scale=.9]{$\bC[x]/(x^2-1)$};
\node(B) at (5.2,3) {$0$};
\node(C) at (7.9,3) [scale=.9]{$\bC[x]/(x^2-1)$};
\node(D) at (10.6,3) [scale=.9]{$\bC[x]/(x^2-1)$};
\node at (2.5,2.5) {\rotatebox{90}{$\in$}};
\node at (7.9,2.5) {\rotatebox{90}{$\in$}};
\node at (10.6,2.5) {\rotatebox{90}{$\in$}};
\node at (2.5,2) {$x$};
\draw (2.5,2)[red] circle (.2cm);
\node(G) at (7.9,2) {$x$};
\node(F) at (10.6,2) {$x$};
\node at (2.5,1.5) {$1$};
\draw (2.5,1.5)[red] circle (.2cm);
\node(E) at (7.9,1.5) {$1$};
\node(H) at (10.6,1.5) {$1$};
\draw[->,red] (E) -- (F) ;
\draw[->,red,dashed] (G) -- (H) ;
\draw[->] (A) -- (B) node [pos=.5,above,scale=.8] {${d_v}$};
\draw[->] (B) -- (C) node [pos=.5,above,scale=.8] {${d_v}$};
\draw[->] (C) -- (D) node [pos=.5,above,scale=.8] {${d_v}$};
\end{tikzpicture}
\eea
\end{figure}

Since the induced differential $d_v=x$ is a multiplication by $x$, the generator $x$ at $t=2$ is mapped to the generator $x^2 \sim 1$ at $t=3$ in the deformed ring $\bC[x]/(x^2-1)$ (the red dotted arrow). This is exactly the Gornik's spectral sequence, and the homology is isomorphic to the $\fraksl(2)$ unknot homology. Translating this into physics language, the perturbation \eqref{sl3-def} of the potential triggers RG flow from LG$[\fraksl(2)]$ to LG$[\fraksl(1)]\oplus $LG$[\fraksl(1)]$. Hence, in the infrared limit, the complex can be expressed as follows. In the ring $\bC[x]/(x\pm1)$, there is only one generator $1$ and the induced differential is a multiplication by $1$. Therefore, the homology of the complex is localized at $t=0$ and it is two-dimensional.
\bea\nonumber
\begin{tikzpicture}
\node at (0.2,3.8) {$t$-degree};
\node at (2.5,3.8) {$0$};
\node at (5.2,3.8) {$1$};
\node at (7.9,3.8) {$2$};
\node at (10.6,3.8) {$3$};
\draw[thick] (-.5,3.5) -- (12,3.5);
\draw[thick] (1.2,4) -- (1.2,.2);
\node at (0.2,2.5) [scale=1]{complex};
\node at (0.2,.8) [scale=1]{generators};
\node at (2.5,3) [scale=.9]{$\bC[x]/(x+1)$};
\node at (7.9,3) [scale=.9]{$\bC[x]/(x+1)$};
\node at (10.6,3) [scale=.9]{$\bC[x]/(x+1)$};
\node(A) at (2.5,2.5) {\rotatebox{90}{$\oplus$}};
\node at (7.9,2.5) {\rotatebox{90}{$\oplus$}};
\node at (10.6,2.5) {\rotatebox{90}{$\oplus$}};
\node at (2.5,2) [scale=.9]{$\bC[x]/(x-1)$};
\node(B) at (5.2,2.5) {$0$};
\node(C) at (7.9,2) [scale=.9]{$\bC[x]/(x-1)$};
\node(D) at (10.6,2) [scale=.9]{$\bC[x]/(x-1)$};
\node at (2.5,1.5) {\rotatebox{90}{$\in$}};
\node at (7.9,1.5) {\rotatebox{90}{$\in$}};
\node at (10.6,1.5) {\rotatebox{90}{$\in$}};
\node at (2.5,1) {$1$};
\draw (2.5,1)[red] circle (.2cm);
\node(G) at (7.9,1) {$1$};
\node(F) at (10.6,1) {$1$};
\node at (2.5,.5) {$1$};
\draw (2.5,.5)[red] circle (.2cm);
\node(E) at (7.9,.5) {$1$};
\node(H) at (10.6,.5) {$1$};
\draw[->,red] (E) -- (H) ;
\draw[->,red] (G) -- (F) ;
\draw[->] (3.8,2.5) -- (B) node [pos=.5,above,scale=.8] {${d_v}$};
\draw[->] (B) -- (6.6,2.5) node [pos=.5,above,scale=.8] {${d_v}$};
\draw[->] (9,2.5) -- (9.5,2.5) node [pos=.5,above,scale=.8] {${d_v}$};
\end{tikzpicture}\nonumber
\eea

It is easy to generalize Gornik's spectral sequence. One can perturb the potential $W=x^{N+1}$ in such a way that the deformed potential $W^{\textrm{def}}$ has the form
$$
\partial_x W^{\textrm{def}}=\prod_{i=1}^k (x-u_i)^{n_i} ~, \qquad \textrm{with}\quad N=\sum_{i=1}^{k}n_i~.
$$
Then, the same argument tells us that LG[$\fraksl(N)$] flows to $\textrm{LG}[\fraksl(n_1)] \oplus \cdots \oplus \textrm{LG}[\fraksl(n_k)]$ under this relevant perturbation. Thus, there exists the spectral sequence associated to the deformed complex where the complex in the $E_2$ page is isomorphic to the $\fraksl(N)$ homology, and the homology of the deformed complex in the $E_\infty$ page is isomorphic to $\overline\scH^{\fraksl(n_1)}_{\yng(1)}(L)\oplus \cdots \oplus \overline\scH^{\fraksl(n_k)}_{\yng(1)}(L)$. For instance, one can schematically express this spectral sequence in the trefoil homology when one perturbs the potential $x^{6} \to x^6  - 12 x^5/5 +3 x^4/2$ so that $\partial_x W^{\textrm{def}}=x^3(x-1)^2$:
\bea\nonumber
\begin{tikzpicture}
\node at (.5,3.9) {$t$-degree};
\node at (2,3.9) {$0$};
\node at (4,3.9) {$1$};
\node at (6,3.9) {$2$};
\node at (8,3.9) {$3$};
\draw[] (-1,3.6) -- (9,3.6);
\draw[] (1.5,4) -- (1.5,1);
\node at (0.2,3) [scale=1]{UV: LG[$\fraksl(5)$]};
\node at (-.6,1.75) [scale=1]{IR: };
\node at (2,3) [scale =2]{$\cdot$};
\node at (2.1,2.9) [scale =2]{$\cdot$};
\node at (1.9,2.9) [scale =2]{$\cdot$};
\node at (2.1,3.1) [scale =2]{$\cdot$};
\node at (1.9,3.1) [scale =2]{$\cdot$};
\node at (6,3) [scale =2]{$\cdot$};
\node at (6.1,2.9) [scale =2]{$\cdot$};
\node at (5.9,2.9) [scale =2]{$\cdot$};
\node(A) at (6.1,3.1) [scale =2]{$\cdot$};
\node at (5.9,3.1) [scale =2]{$\cdot$};
\node at (8,3) [scale =2]{$\cdot$};
\node at (8.1,2.9) [scale =2]{$\cdot$};
\node at (7.9,2.9) [scale =2]{$\cdot$};
\node at (8.1,3.1) [scale =2]{$\cdot$};
\node(B) at (7.9,3.1) [scale =2]{$\cdot$};
\draw[dashed] (-1,2.5) -- (9,2.5);
\node at (.6,2) [scale=1]{LG[$\fraksl(3)$]};
\node at (.6,1.5) [scale=1]{LG[$\fraksl(2)$]};
\node at (2,2) [scale =2]{$\cdot$};
\node at (2.1,2.1) [scale =2]{$\cdot$};
\node at (1.9,2.1) [scale =2]{$\cdot$};
\node at (2.1,1.5) [scale =2]{$\cdot$};
\node at (1.9,1.5) [scale =2]{$\cdot$};
\node at (6,2) [scale =2]{$\cdot$};
\node(C) at (6.1,2.1) [scale =2]{$\cdot$};
\node at (5.9,2.1) [scale =2]{$\cdot$};
\node(E) at (6.1,1.5) [scale =2]{$\cdot$};
\node at (5.9,1.5) [scale =2]{$\cdot$};
\node at (8,2) [scale =2]{$\cdot$};
\node at (8.1,2.1) [scale =2]{$\cdot$};
\node(D) at (7.9,2.1) [scale =2]{$\cdot$};
\node at (8.1,1.5) [scale =2]{$\cdot$};
\node(F) at (7.9,1.5) [scale =2]{$\cdot$};
\draw[->,red] (6.3,3.1) -- (7.7,3.1);
\draw[->,red] (6.3,2.1) -- (7.7,2.1);
\draw[->,red] (6.3,1.5) -- (7.7,1.5);
\draw[thick,->] (-.6,2.7) to  (-.6,2);
\node at (-1.5,2.5) [scale =1]{RG};
\end{tikzpicture}
\eea

\subsection{Summary and outline of the rest of the paper}
\label{sec:organization}

Having clarified the physical realizations of link homologies and spectral sequences,  the rest of the paper will study simple rules that control ``color-dependence'' of homological link invariants. First, we will introduce the so-called \emph{colored differentials} that allow us to extract colored HOMFLY homology for smaller representations from that for larger representations. Second, we shall see a peculiar \emph{sliding property} of colored link homology that will lead us to a notion of \emph{homological blocks}. These properties will enable us to compute Poincar\'e polynomials of HOMFLY homology for links. In addition, we will obtain recursion relations for these link invariants, which will give rise to \emph{associated varieties}. All these structural properties that control ``color-dependence'' are deeply interrelated and fit together into a consistent framework, which ultimately encodes the representation theory data $\vec \lambda = (\lambda_1, \ldots, \lambda_n)$ via the classical geometry of the associated variety.
While most of the results in the following sections are of a mathematical nature, we stress that much of the structure we reveal has been inspired by physics and follows from the M-theory configurations of knots or links explained in this section. Here we provide a summary of our results and outline the organization of the paper.
\\

\noindent $\bullet$ {\bf Structural properties of link homology}

As we have seen in \eqref{HHrep}, link homologies are endowed with a rich algebraic structure that can be seen as actions of differentials. The differential structure of knot homology was first revealed in \cite{Dunfield:2005si}, which conjectures that $\fraksl(N)$ homology can be obtained from HOMFLY homology by using the action of the $d_N$ differential. Then, the existence of colored differentials $d_{\lambda\to\mu}$ in $\lambda$-colored HOMFLY homology of a knot was postulated in \cite{Gukov:2011ry} so that the homology with respect to $d_{\lambda\to\mu}$ is isomorphic to its $\mu$-colored homology for $|\mu| \le |\lambda|$. Generalizations of these results to quadruply-graded knot homologies and rectangular diagrams are given in \cite{Gorsky:2013jxa}, which makes remarkably rich properties manifest:
\begin{enumerate}\setlength{\parskip}{-0.15cm}
\item colored differentials
\item mirror symmetry
\item self-symmetry
\item refined exponential growth property
\end{enumerate}
\vspace{-.15cm}
The use of these properties to determine colored superpolynomials has been presented in \cite{Gukov:2011ry,Gorsky:2013jxa}, and $q$-holonomic expressions for a number of knots have been obtained in \cite{Fuji:2012pm,Fuji:2012nx,Fuji:2012pi,Nawata:2012pg,Nawata:2013mzx} in the case of rectangular Young tableaux. While the forms of superpolynomials determined in this way are conjectural,  they are reproduced by calculations in the refined Chern-Simons theory \cite{Aganagic:2011sg,Fuji:2012pm} in the case of torus knots.
In this work, we extend these investigations to the case of links.

The first main result is that we postulate, in \S\ref{sec:quadruply}, a consistent structure of quadruply-graded homological invariants of links  whose components are all colored by the same rectangular representation.
To this end, we consider two related classes of homology theories for links: one infinite and one finite. Unlike the reduced colored HOMFLY polynomial of a knot, which is a genuine Laurent polynomial, the invariant of a link has a nontrivial denominator. Consequently, the homology theory that categorifies it is infinite-dimensional. This infinite homology theory satisfies some (but not all) of the properties existing in the case of knots. As such, we also define a finite-dimensional colored homology of links, which categorifies an appropriate, link-independent multiple of the reduced colored HOMFLY invariant. These two versions of the colored link homology have quite similar properties, but are still different and independent. In particular, the finite-dimensional homology possesses the full set of differentials and structural properties that exist in the case of the colored homology of knots. We focus in this paper on thin links and on torus links, but we also conjecture that these structural properties hold for arbitrary links.

In fact, there is a physical perspective on quadruple-gradings and colored differentials, which comes from yet another variation of the brane systems (\ref{surfeng2}), in which the $N$ M5-branes are replaced by two sets spanning two orthogonal two-dimensional subspaces, $D\cap D_{\perp}=\{0\}$, inside the Taub-NUT geometry:
\be\label{surfeng5}
\begin{matrix}
{\mbox{\rm space-time:}} && \qquad  \bR_t~ \times &TN_4 & \times & T^*S^3 \\
{\mbox{\rm $n$ M5-branes:}} && \qquad  \bR_t~ \times &D& \times & S^3  \\
{\mbox{\rm $m$ M5-branes:}} && \qquad \bR_t~ \times &D{_{\perp}}&  \times  &S^3 \\
{\mbox{\rm M5'-branes:}} && \qquad \bR_t~ \times &D&  \times  &M_L
\end{matrix}
\ee
In this system the numbers of the M5-branes are adjusted so that
$
N=n-m.
$
Now, reducing the Taub-NUT space along its circle direction produces a system which consists of a set of $n$ semi-infinite D4-branes ending on one side of a D6-brane, and another set of $m$ D4-branes ending on the other side of this D6-brane. This system is naturally related to the representation theory of the Lie superalgebras  $\mathfrak{gl}(n|m)$, which also turn out to be deeply related to quadruply graded homologies. In particular, the structure of $\mathfrak{gl}(n|m)$ is encoded in properties of positive and negative colored differentials $d^{\pm}_{\lambda\to \mu}$ in knot homologies \cite{Gorsky:2013jxa}. In this paper we explain that these two sets of differentials also arise---albeit in a more intricate way---in colored link homologies. Furthermore, the brane system (\ref{surfeng5}) has  been considered in \cite{Witten:2011zz}, where it was shown that $\mathfrak{gl}(n|m)$ homological knot invariants can be reconstructed from the counting of the number of solutions of a certain system of partial differential equations, the so-called Kapustin-Witten equations, defined in the above background. The boundary conditions for these differential equations are specified by a knot or a link embedded in the neighborhood of the interface between the two sets of D4-branes. From this perspective, both knots and links can be considered on an equal footing, which indicates that the structural properties of knot homologies should indeed generalize to the case of links. While we do not analyze the framework of \cite{Witten:2011zz} in this paper, our results are consistent with these expectations.

In \S \ref{sec:example}, we  see these properties concretely for both finite- and infinite-dimensional homology in the example of the Hopf link.
Moreover, we derive explicit forms of the Poincar{\'e} polynomials of both finite- and infinite-dimensional HOMFLY homology  for the Hopf link with both components colored by the same (anti-)symmetric representation (``diagonal'' superpolynomials), by using the structural properties. In addition, we obtain the ``diagonal'' superpolynomials for various links in Appendix \ref{sec:app-ExamplesDiag}. Through these derivations, we have found remarkably simple expressions for colored superpolynomials as well as colored HOMFLY polynomials for hyperbolic links and knots obtained by surgeries of the Borromean rings. The expressions are  generalizations of Habiro's cyclotomic expansions \cite{Habiro:2008} for colored Jones polynomials, which will be presented in \S \ref{sec:cyclotomic}.

Having determined homologies for links colored by the same representation, we are in fact able to deduce the form of homologies colored by an arbitrary set of symmetric  representations in some cases. This follows from the remarkable observation, presented in \S \ref{sec:sliding}, that a certain part of the homological structure changes in a universal way when one changes the color of a single unknot component of a link. We generalize this observation to link components colored by arbitrary symmetric representations, and call the above mentioned part of the homological structure a ``homological block''. The ``color-shifting" property then states that as we change the color of a link component, the only variation in the homology is that these homological blocks shift in $(q,t_c)$-degree in a universal way. Taking advantage of the behavior of homological blocks, we find explicit forms of superpolynomials labeled by arbitrary symmetric representations for the $(2,2p)$ torus links in \S \ref{sec:sliding}. Furthermore, we give an explanation of the sliding property of link invariants based on surgeries of 3-manifolds. From this perspective, it turns out that the modular $S$-matrix plays a crucial role for the sliding property. The properties of homological blocks are very interesting in themselves;  they undoubtedly deserve further mathematical study, which will be a part of future work.

Even though the approach presented in this paper is very powerful and leads to explicit and intricate results, it still relies on several conjectural properties of link homologies. To convince ourselves and the reader that the results we find are correct, we test them and compare to other computations whenever possible. In particular, in \S\ref{sec:refined} we compute refined Chern-Simons invariants of the $(2,2p)$ torus links based on braiding operations \cite{Aganagic:2011sg,Fuji:2012pm}, and we prove in Appendix \ref{sec:RR-id} that they are equal to the colored superpolynomials obtained in \S\ref{sec:sliding}. Although the two complicated expressions look completely different, the proof using multiple-series Rogers-Ramanujan type identities \cite{Andrews:1984} turns out to be surprisingly simple and elegant. These results provide the strongest tests of the axiomatic structure of colored link homologies proposed in this paper.
\\

\noindent $\bullet$ {\bf Associated varieties}

There is another way to understand  the ``color-dependence'' of homological link invariants in terms of classical geometry that plays a role analogous to \emph{associated varieties} in geometric representation theory \cite{Beilinson:1981,Vogan:1998}. As we have seen in \eqref{HHrep}, the link homology can be regarded as an $\cA$-module $M$ so that the associated variety can be schematically understood as  the support of generators  in $\cA$ that annihilate any element in $M$. However,  until a mathematically rigorous formulation is found, we use
 a ``working'' definition of the associated variety  as the ``limit shape'' for colored HOMFLY homology or its Poincar{\'e} polynomial \eqref{superPdef}. Indeed, whenever one has a set of objects labeled by an integer or by an element of a (weight) lattice or by a partition, it is always a good idea to consider the asymptotic behavior of that set and ask how fast the size of its objects grows. In our case, this question means understanding how the size of $\scH_{\vec \lambda} (L)_{i,j,k}$ grows as $\vec \lambda \to \infty$. Although both the colored HOMFLY homology and its Poincar\'e polynomial \eqref{superPdef} contain a lot of intricate information, it nicely ``averages out'' into a classical geometrical shape as $\vec \lambda \to \infty$. This shape is precisely what we call the associated variety of $L$.

A prototypical example of an associated variety is the planar algebraic curve defined by the zero locus of the $A$-polynomial~\cite{Cooper:1994}. It is defined as the $\SL(2,\mathbb{C})$ character variety of the knot group and, via various versions of the volume conjecture \cite{Kashaev:1996kc,Murakami:1999,Gukov:2003na,Garoufalidis:2003,Garoufalidis:2004}, it also determines the color-dependence of colored Jones polynomials.
Recently, this conjecture has been generalized to various deformations of the $A$-polynomial, in particular to the augmentation polynomial \cite{Ng,NgFramed}, which (conjecturally) equals the $Q$-deformed $A$-polynomial defined as the classical limit of recursion relations satisfied by the colored HOMFLY polynomials \cite{Aganagic:2012jb,Aganagic:2013jpa,Arthamonov:2013rfa}. A more general version of the volume conjecture and a corresponding class of algebraic curves arises from analysis of colored HOMFLY homologies, whose color-dependence is captured by the so-called super-$A$-polynomial \cite{Fuji:2012pm,Fuji:2012nx,Fuji:2012pi,Fuji:2013rra,Nawata:2012pg,Nawata:2013mzx}.
As shown in \cite{Fuji:2012pm,Fuji:2012nx,Fuji:2012pi,Fuji:2013rra}, Poincar\'e polynomials of colored HOMFLY homologies satisfy recursion relations (with respect to color $\vec \lambda$) that can be expressed in a form of ``quantum algebraic curve'' obtained by $(a,q,t)$-deformation of the $A$-polynomial. The $q$-deformation is non-commutative (or ``quantum''), whereas the other two deformations are commutative, with the same variables $a$ and $t$ as in \eqref{PHomSUN} and \eqref{superPdef}.  Besides combining these results in a single unified framework, we propose a generalization to links (with more than one component) that involves a number of surprising new features.


For this purpose, let us first state the volume conjecture for links in the most general context.
For an $n$-component link $L$, the large color asymptotic behavior of colored quantum invariants $P^{\SU(N)}_{\lambda_1,\cdots,\lambda_n}(L;q)$ will determine  an algebraic variety of complex dimension $d = \sum_{i=1}^n r_i$, where $r_i$ is the number of rows (or, columns, depending on which is kept fixed in the $\vec \lambda \to \infty$ limit) in the Young diagram $\lambda_i$. In fact, it is a holomorphic Lagrangian subvariety in the moduli space of flat connections $\prod_{i=1}^n \frac{(\mathbb{C}^*\times\mathbb{C}^*)^{r_i}}{S_{r_i}}$ where $S_{r_i}$ is the Weyl group symmetry of $\SL(r_i,\mathbb{C})$. If all the components are colored by symmetric representations, the ambient moduli space is  the product of $n$-copies of the Hitchin moduli space over a torus:
\be
\scM_H ( \Sigma,\SU(2)) \; = \;
\scM_{\text{flat}} ( \Sigma,\SL(2,\mathbb{C})) \; = \;
\left( \frac{\mathbb{C}^*\times\mathbb{C}^*}{\mathbb{Z}_2} \right)^n
\ee
where  $\Sigma := \partial (S^3 \setminus L)$ consists of $n$ disjoint tori.

In a similar way, we determine associated varieties in \S \ref{sec:associated} and Appendix \ref{sec:app-varieties} by using the closed-form expressions for colored superpolynomials and HOMFLY polynomials obtained in this paper.
 First, we can determine them from the asymptotics of colored superpolynomials of links for large colors. More generally, we can determine a system of recursion relations satisfied by colored superpolynomials, and present it as a ``quantum associated variety''. In the limit $q\to 1$, this quantum variety reproduces ``classical associated variety''. We also introduce a related concept of diagonal super-$A$-polynomials; we define them as complex curves encoding asymptotics of diagonal superpolynomials of finite-dimensional HOMFLY homology. In Appendix \ref{sec:diagonal}, we compute diagonal super-$A$-polynomials in several examples and show that they satisfy so-called quantizability conditions \cite{Gukov:2011qp}.

Furthermore, in \S \ref{sec:associated}, we demonstrate various properties of associated varieties, which make them very interesting objects to study. First, as already indicated above, they satisfy a certain generalized version of the volume conjecture. However, for torus links, the $\mathbb{Z}_2$ Weyl group symmetry is not manifest when the parameters $a$ and $t$ are turned on, while the associated variety for a hyperbolic link enjoys the $\mathbb{Z}_2$ symmetry. This occurs because the $\SL(2,\bC)$ character variety for a torus link has more than one component, and each component is deformed by the complex structure $a$ (or $t$) in a different way. Therefore, the associated varieties are holomorphic Lagrangian subvarieties in $(\bC^*\times\bC^*)^n$ in general. With homological grading turned off, i.e. for $t=-1$, the associated varieties can be interpreted as augmentation varieties of knot contact homology \cite{Aganagic:2012jb,Aganagic:2013jpa,Arthamonov:2013rfa}.
\\

\noindent $\bullet$  {\bf Mutants and thick knots}

In \S \ref{sec:mutants} we discuss how homological invariants can be applied to detect mutant pairs of links. Although it is known that quantum invariants colored by the fundamental representation \emph{cannot} distinguish mutant pairs \cite{Morton:1996}, unreduced $\fraksl(2)$ homological invariants \emph{can} detect some mutant pairs of links \cite{Wehrli:2003}. We present an explicit example for this statement in \S\ref{sec:mutants}. In \S \ref{sec:HFK}, we discuss properties of HFK-like differentials and invariants of some thick knots. It was conjectured in \cite{Dunfield:2005si} that the $d_0$ differential acts non-trivially on uncolored HOMFLY homology of thick knots, and its homology is isomorphic to the knot Floer homology. We study this property in the context of quadruply-graded colored HOMFLY homology by using thick torus knots in \S\ref{sec:HFK}.
\\

\noindent $\bullet$  {\bf 3d/3d correspondence}

 As we have discussed, the brane systems, \eqref{surfeng} and \eqref{surfeng2}, suggest that the data associated to a knot or a link should be encoded in the data of the effective theory that lives on the $\mathbb{R}_t\times D$ part of the Lagrangian brane. This is indeed the case, and this correspondence between knots and 3d $\mathcal{N}=2$ theories is referred to as the 3d/3d correspondence. For knots this duality has been analyzed from various viewpoints in \cite{Dimofte:2010tz,Terashima:2011qi,Dimofte:2011ju,Dimofte:2011py,Yagi:2013fda,Lee:2013ida,Cordova:2013cea}. In particular, it was generalized to the case of homological knot invariants in \cite{Fuji:2012pm,Fuji:2012nx,Fuji:2012pi,3d3drevisited}. In \S\ref{sec:phys}, we extend this duality to the case of link invariants. Surprisingly, it turns out that the 3d $\cN=2$ theory for the Hopf link can be obtained by acting with the modular $S$-transformation \cite{Witten:2003ya} on the 3d $\cN=2$  theory for the unknot. Moreover, we find the following remarkable equivalence that is one of the novel physical insights in this paper:
\begin{center}
 \shabox{\parbox{.85\hsize}{The 3d/3d correspondence identifies the $\SL(2,\bZ)$ transformations in Chern-Simons theory \cite{Witten:1988hf} with the $\SL(2,\bZ)$ transformations in 3d $\cN=2$ gauge theory \cite{Witten:2003ya}.}}
\end{center}
This identification tells us that the surgery of two solid tori with links inside by the $S$-transformation corresponds to joining two 3d $\cN=2$ theories by a 2d $S$-transformation wall \cite{Gadde:2013wq}. This actually explains the sliding property of the link invariants.
In the last section \S \ref{sec:phys}, we further discuss relations of our results to refined BPS invariants and several related open problems.
\\

\noindent $\bullet$ {\bf Open problems}

The analysis presented in this paper undoubtedly deserves continuation, and many problems still remain open. Among these, it is important to determine superpolynomials colored by more general representations (in particular not only rectangular), and for other knots and  links (in particular thick links and links with more than two components)---and to reveal properties of the corresponding associated varieties. An outstanding mathematical problem is to provide a combinatorial formulation of the corresponding colored link homologies, and to understand the role of homological blocks. It would also be interesting to understand associated varieties in terms of geometric representation theory, and to interpret  the role of higher-dimensional associated varieties in physics.\\
\\
\\


\section{Colored HOMFLY homology of links}
\label{sec:quadruply}

In this section, we outline the properties of the colored HOMFLY homology
that categorifies the colored HOMFLY invariant $P_{\lambda_1, \ldots, \lambda_n} (L;a,q)$ of a link $L$
whose components are colored by the same rectangular Young diagram $\lambda_1 = \ldots = \lambda_n = [r^{\rho}]$ with $\rho$ rows and $r$ columns.
These properties extend the structure of colored HOMFLY homology of knots motivated by the relationship of the link homologies and the homologies of BPS states \eqref{HHBPS} introduced in \cite{Gukov:2011ry} and later expanded in \cite{Gorsky:2013jxa}.
In particular, we argue that the link homology $\scH_{\lambda,\ldots,\lambda}(L)$ can be endowed with two
homological gradings, which are generically independent but happen to coincide when $\lambda = \square$.
Following \cite{Gorsky:2013jxa}, we call the two homological gradings $t_r$- and $t_c$-gradings (for {\it rows} and {\it columns}).

We focus first on the reduced homology $\scH_{\lambda,\ldots,\lambda}(L)$ (and return to the unreduced one $\bar{\scH}_{\lambda,\ldots,\lambda}(L)$ in \S\ref{sec:unknot}). One crucial difference between knots and links is that for links with two or more components, the colored HOMFLY homology is \emph{infinite-dimensional} even in the reduced case. The reason is that the reduced HOMFLY ``polynomial'' of a link is not a polynomial, but rather a power series with respect to the variable $q$. This will have a large impact on the properties of link homology. In particular, the Poincar\'e polynomial  of the infinite-dimensional homology is no longer a Laurent polynomial, but rather a \emph{rational function}. Moreover, since a colored HOMFLY invariant can be expanded in positive or negative powers of $q$, depending on whether $|q|<1$ or $|q|>1$, we propose two ways to categorify the colored HOMFLY invariant, resulting in ``positive'' HOMFLY homology $\scH^{(+)}_{\lambda,\ldots,\lambda}(L)$ for $|q|<1$ and ``negative'' HOMFLY homology $\scH^{(-)}_{\lambda,\ldots,\lambda}(L)$ for $|q|>1$. Remarkably, both $\scH^{(\pm)}_{\lambda,\ldots,\lambda}(L)$ can be endowed with four independent gradings ($a$, $q$, $t_r$, and $t_c$-gradings).

One advantage of endowing the HOMFLY homology with four gradings is that it will make many structural properties very explicit and manifest \cite{Gorsky:2013jxa}. Due to the infinite support, though, some of the properties of colored HOMFLY homology of knots ({\it e.g.} the so-called ``self-symmetry'') do not extend to the case of links. Moreover, while the colored HOMFLY homology of knots comes equipped with both positive and negative colored differentials $d^\pm_{\lambda \to \mu}$, there exist only positive colored differentials $d^+_{\lambda \to \mu}$ in the positive colored HOMFLY homology of links $\scH^{(+)}_{\lambda,\ldots,\lambda}(L)$ and only negative colored differentials $d^-_{\lambda \to \mu}$ in the negative colored link HOMFLY homology $\scH^{(-)}_{\lambda,\ldots,\lambda}(L)$.
These differentials have the property that the homology of $\scH^{(\pm)}_{\lambda,\ldots,\lambda}(L)$, with respect to $d^\pm_{\lambda \to \mu}$ is isomorphic to $\scH^{(\pm)}_{\mu,\ldots,\mu}(L)$
where $\mu$ is obtained from $\lambda$ by erasing some of its rows or columns. As explained in \cite{Gorsky:2013jxa}, the properties of these differentials correspond to the representation theory of the Lie superalgebras  $\mathfrak{gl}(n|m)$ and were denoted by $d_{n|m}^{col}$.
Here we simply denote these differentials by $d_N$, where $N=n-m$.

In addition, as in the case of knots, all these properties become particularly simple and elegant when expressed in terms of the $Q$-grading
\be
\label{Q}
Q :=\frac{q+t_r-t_c}{\rho} \,.
\ee
Although it is just a regraded version of the original homology $\scH_{\lambda,\ldots,\lambda}(L)$,
due to its prominent role we denote the $Q$-graded colored HOMFLY homology by $\wt{\scH}_{\lambda,\ldots,\lambda}(L)$:
\be\nonumber
(\wt\scH_{\lambda,\ldots,\lambda} (L) )_{i,j,k,l} \; := \; (\scH_{\lambda,\ldots,\lambda} (L))_{i,\rho j-k+l,k,l} \,.
\ee
In particular, our strategy in deriving the explicit formulae for the ``diagonal'' colored HOMFLY homology will be first to produce the results for $\wt{\scH}_{\lambda,\ldots,\lambda}(L)$ and then, via re-grading, to carry them over to other versions of the colored HOMFLY homology.

Furthermore, we introduce another variant of the reduced colored HOMFLY homology of links that is finite-dimensional.
This finite-dimensional homology, denoted by $\scH^{\rm fin}_{[r^\rho]}(L)$, categorifies the following multiple of the reduced colored HOMFLY invariant of an $n$-component link $L$:
\be
\label{finpol}
P^{\fin}_{[r^\rho]}(L;a,q):= \Big[(aq^{-\rho})^{r\rho}\prod_{i=1}^\rho (q^{2i};q^2)_r\Big]^{n-1}P_{\begin{scriptsize}\underbrace{\underline{[r^{\rho}]},[r^\rho],\cdots,[r^\rho]}_n\end{scriptsize}}(L;a,q) \ .
\ee
The overall factor is the denominator of the unreduced $[r^{\rho}]$-colored HOMFLY invariant of $n-1$ unlinked copies of the unknots, and is independent of the link $L$. This polynomial $P^{\fin}_\lambda(L)$ is a genuine Laurent polynomial and its categorification is a homology $\scH^{\rm fin}_\lambda$ with finite support. We  conjecture that the finite-dimensional homology $\scH^{\rm fin}_\lambda(L)$ of links enjoys all of the properties of the colored HOMFLY homology of knots, summarized in the following subsection below. We also note that in the case of a knot $K$ ({\it i.e.} one component link), $\scH^{\rm fin}_\lambda(K)$ and $\scH_\lambda(K)$ coincide, as expected.

In the case of the symmetric representations, $\scH^\fin_{[r]}(L)$ has two additional features. First of all, one can introduce another grading, the so-called $\delta$-grading, which on a generator $x$ of $\scH^\fin_{[r]}(L)$ is given by:
\bea
\label{delta-grading-HOMFLY}
\delta(x):= a(x)+\frac{q(x)}{2}-\frac{t_r(x)+t_c(x)}{2} \,.
\eea
If all generators of the $[r]$-colored homology of a link $L$ have the same $\delta$-grading, we say that the link $L$ is $[r]$-thin. We conjecture that a link is $[r]$-thin for every $r$ whenever it is thin in the uncolored homology. In particular, we conjecture that all two-bridge links are $[r]$-thin for all $r$.
The second feature is the existence of a differential $d_{1-r}$ on the $[r]$-colored finite-dimensional homology.
In the uncolored case, the homology  $H_*(\scH_{\yng(1)},d_{0})$ with respect to $d_0$ is conjectured to be isomorphic to the link Floer homology
that categorifies the Alexander polynomial \cite{Dunfield:2005si}.
This differential can be extended to the colored theory and the Poincar\'e polynomial of the homology
$H_*(\scH_{[r]},d_{1-r})$ exhibits the so-called refined exponential growth property.
These properties will be explained in \S\ref{sec:HFK} in more detail.

In the following subsections we list the structural properties of both $\scH^{\rm fin}_\lambda(L)$ and $\scH^{(\pm)}_{\lambda,...,\lambda}(L)$ for an arbitrary link $L$, as well as uncolored Kauffman homology of thin links.

\subsection{Finite-dimensional homology}
\label{sec:finite-dimension}

The finite-dimensional HOMFLY homology $\wt\scH^{\fin}_{[r^\rho]}(L)$ of a link $L$ with $n$ components categorifies
the Laurent polynomial \eqref{finpol} obtained by multiplying the reduced colored HOMFLY invariant by the appropriate factor.
We require the finite-dimensional HOMFLY homology $\wt\scH^{\fin}_{[r^\rho]}(L)$
to have all the structural properties of the colored HOMFLY homology of knots \cite{Gorsky:2013jxa}.
In particular, we build $\wt\scH^{\fin}_{[r^\rho]}(L)$ as a quadruply-graded homology theory, with $(a,Q,t_r,t_c)$-gradings
related to the $(a,q,t_r,t_c)$-gradings of $\scH^{\fin}_{[r^\rho]}(L)$ via
\be\nonumber
(\wt\scH^{\fin}_{[r^\rho]}(L) )_{i,j,k,l} \; := \; (\scH^{\rm fin}_{[r^\rho]}(L))_{i,\rho j-k+l,k,l} \,.
\ee
Although  the four gradings are independent in general, a link $L$ is called \emph{homologically thin} if all generators of $\scH^{\rm fin}_{[r]}(L)$ have the same $\delta$-grading \eqref{delta-grading-HOMFLY}. For a thin link $L_{\rm thin}$ with $n$ components, the $\delta$-grading of any generator $x$ of the $[r]$-colored quadruply-graded HOMFLY homology can be written in terms of the $S$-invariant \cite{Rasmussen:2004}:
\bea
\label{delta-grading}
\delta(x)=\frac{r}{2}\left[S(L_{\rm thin})+n-1\right] \,.
\eea

The main advantage of the quadruply-graded theory is that it makes
all of the structural properties and isomorphisms completely explicit.
In order to see them, let us define the Poincar\'e polynomials of the two variants of the quadruply-graded homology
\bea\nonumber
\scP^{\rm fin}_{[r^\rho]}(L;a,q, t_r,t_c)&:=& \sum_{i,j,k,\ell} a^i q^j t_r^k t_c^\ell ~\dim\;(\scH^{\rm fin}_{[r^\rho]}(L))_{i,j,k,\ell}~,\cr
\wt\scP^{\fin}_{[r^\rho]}(L;a,Q, t_r,t_c)&:=& \sum_{i,j,k,\ell} a^i Q^j t_r^k t_c^\ell ~\dim\;(\wt\scH^{\fin}_{[r^\rho]}(L))_{i,j,k,\ell}~,
\eea
related by
\bea\nonumber
\wt\scP^{\fin}_{[r^\rho]}(L;a,q^\rho,t_rq^{-1},t_cq)=\scP^{\rm fin}_{[r^\rho]}(L;a,q,t_r,t_c)~.
\eea
In the uncolored case (but not more generally), the two homological gradings ($t_r$ and $t_c$) coincide,  so that the resulting homology is triply-graded in agreement with \cite{Dunfield:2005si}.
It also follows from \eqref{Q} that in this case the $q$- and $Q$-gradings of the uncolored homology are the same.
Hence, we can use the following notation for the Poincar\'e polynomial  of the uncolored homology
\bea\nonumber
\scP^{\rm fin}_{\yng(1)}(L;a,q,t)=\scP^{\rm fin}_{\yng(1)}(L;a,q,t_r=1,t_c=t)=\wt \scP^{\fin}_{\yng(1)}(L;a,Q=q,t_r=1,t_c=t)~.
\eea

Now, let us summarize the structural properties of the quadruply-graded colored HOMFLY homology
(see \cite{Gorsky:2013jxa} for more details):

\begin{itemize}
\item {\bf Self-symmetry} \\
The finite-dimensional quadruply-graded HOMFLY homology enjoys the self-symmetry
\bea\nonumber
(\wt\scH^{\fin}_{[r^\rho]}(L))_{i,j,k,\ell}&\cong&(\wt\scH^{\fin}_{[r^\rho]}(L))_{i,-j,k-\rho j,\ell-rj}~,
\eea
which can be stated at the level of the Poincar\'e polynomials
\bea\label{self-symmetry}
\wt\scP^{\fin}_{[r^\rho]}(L;a,Q,t_r,t_c)&=&\wt\scP^{\fin}_{[r^\rho]}(L;a,Q^{-1}t_r^{-\rho}t_c^{-r},t_r,t_c)~.
\eea

\item {\bf Mirror symmetry}\\
The $[\rho^r]$-colored and $[r^\rho]$-colored HOMFLY homology theories are related via the ``mirror'' symmetry
\bea\nonumber
(\wt\scH^{\fin}_{[\rho^r]}(L))_{i,j,k,\ell}\cong(\wt\scH^{\fin}_{[r^\rho]}(L))_{i,j,\ell,k}\cong(\wt\scH^{\fin}_{[r^\rho]}(L))_{i,-j,\ell- \rho j,k-r j}~,
\eea
which implies the following property of the Poincar\'e polynomials
\bea
\label{mirror-finite}
\wt\scP^{\fin}_{[\rho^r]}(L;a,Q,t_r,t_c)&=&\wt\scP^{\fin}_{[r^\rho]}(L;a,Q,t_c,t_r)=\wt\scP^{\fin}_{[r^\rho]}(L;a,Q^{-1}t_c^{-\rho}t_r^{-r},t_c,t_r)~.~~~~
\eea
\item {\bf Refined exponential growth}  \\
Let $L^*$ be either a rational link or a torus link.
The finite-dimensional quadruply-graded HOMFLY homology of the link $L^*$ obeys the refined exponential growth property
\bea
\wt\scP^{\fin}_{[r^\rho]}(L^*;a,Q,t_r,t_c=1)&=&\left[\wt\scP^{\fin}_{[1^{\rho}]}(L^*;a,Q,t_r,t_c=1)\right]^r\label{exp-growth-fin-1}~,\\
\wt\scP^{\fin}_{[r^\rho]}(L^*;a,Q,t_r=1,t_c)&=&\left[\wt\scP^{\fin}_{[r]}(L^*;a,Q,t_r=1,t_c)\right]^{\rho}\nonumber~.
\eea
It follows immediately that
\bea\nonumber
\dim \scH^{\fin}_{[r^\rho]}(L^*)=\left[\dim \scH^{\fin}_{\yng(1)}(L^*)\right]^{r\rho}<\infty~.
\eea
This property has been discovered at the level of the colored HOMFLY polynomial invariants in \cite{Nawata:2013qpa}.
All the examples considered in this paper enjoy this property.

\item {\bf Colored differentials} \\
We postulate that the finite-dimensional HOMFLY homology $\wt\scH^\fin_{[r^\rho]}(L)$
colored by a rectangular Young diagram comes equipped with ``column-removing'' and ``row-removing'' colored differentials.
Specifically, for every $k$ such that $r>k\ge 0$,
there are two different column-removing differentials $d^\pm_{[r^\rho]\to [k^\rho]}$
on $\wt\scH^\fin_{[r^\rho]}(L)$, with the $(a,Q,t_r,t_c)$-grading:
\be\label{diff-degree}
\wt\deg ~d^+_{[r^\rho]\to [k^\rho]}=(-2,2,-1,-2k-1)\,, \quad \wt\deg ~d^-_{[r^\rho]\to [k^\rho]}=(-2,-2,-2\rho-1,-2r-2k-1)\,.
\ee
Due to the presence of an extra factor in the definition (\ref{finpol}) of $P^{\fin}_{\lambda}$, the homology $H_*(\wt\scH^{\fin}_{[r^\rho]}(L),d^\pm_{[r^\rho]\to [k^\rho]})$ with respect to the colored differential $d^\pm_{[r^\rho]\to [k^\rho]}$ is not  isomorphic to $\wt\scH^{\fin}_{[k^\rho]}(L)$, but the relationship between the two is rather simple. 
It is easiest to see this relationship at the level of the Poincar\'e polynomials. Namely, the Poincar\'e polynomial of the homology $H_*(\wt\scH^{\fin}_{[r^\rho]}(L),d^\pm_{[r^\rho]\to [k^\rho]})$ is just a multiple of the Poincar\'e polynomial of $\wt\scH^{\fin}_{[k^\rho]}(L)$:
\bea
\label{regrading-finite-column}
&&\wt \scP(\wt\scH^{\fin}_{[r^\rho]}(L),d^+_{[r^\rho]\to [k^\rho]})(a,Q,t_r,t_c)\\
&&=(aQ^{-1}t_c^{ k})^{\rho (r - k)(S(L)+n-1)} \Big[\prod_{i=1}^\rho (-Q^2t_r^{2i-1}t_c;t_c^2)_{r-k}\Big]^{n-1} \wt \scP^{\fin}_{[k^\rho]} (L; a,Qt_c^{r-k},t_r,t_c)~,\cr
&&\wt \scP(\wt\scH^{\fin}_{[r^\rho]}(L),d^-_{[r^\rho]\to [k^\rho]})(a,Q,t_r,t_c)\cr
&&=(aQt_r^\rho t_c^{r+k}) ^{\rho(r-k)(S(L)+n-1)} \Big[\prod_{i=1}^\rho(-Q^{-2}t_r^{1-2i}t_c^{-1};t_c^{-2})_{r-k}\Big]^{n-1} \wt \scP^{\fin}_{[k^\rho]} (L; a,Q,t_r,t_c)~.\nonumber
\eea
In a similar way, for every $\sigma$ with $\rho>\sigma\ge 0$, there are two different row-removing differentials  $d^\pm_{[r^\rho]\to [r^\sigma]}$ on $\wt\scH^\fin_{[r^\rho]}(L)$ whose $(a,Q,t_r,t_c)$-gradings are given by
$$
\wt\deg ~d^+_{[r^\rho]\to [r^\sigma]}=(-2,2,-2\sigma-1,-1)\,, \quad \wt\deg ~d^-_{[r^\rho]\to [r^\sigma]}=(-2,-2,-2\rho-2\sigma-1,-2r-1)\,.
$$
The Poincar\'e polynomial of the homology $H_*(\wt\scH^{\fin}_{[r^\rho]}(L),d^\pm_{[r^\rho]\to [r^\sigma]})$ can be written in terms of the Poincar\'e polynomial of $\wt\scH^{\fin}_{[r^\sigma]}(L)$:
\bea\nonumber
&&\wt \scP(\wt\scH^{\fin}_{[r^\rho]}(L),d^+_{[r^\rho]\to [r^\sigma]})(a,Q,t_r,t_c)\cr
&&=(aQ^{-1}t_r^{ \sigma})^{ r(\rho - \sigma)(S(L)+n-1)} \Big[\prod_{i=1}^r (-Q^2t_rt_c^{2i-1};t_r^2)_{\rho-\sigma}\Big]^{n-1} \wt \scP^{\fin}_{[r^{\sigma}]} (L; a,Qt_r^{\rho-\sigma},t_r,t_c)~,\cr
&&\wt \scP(\wt\scH^{\fin}_{[r^\rho]}(L),d^-_{[r^\rho]\to [r^\sigma]})(a,Q,t_r,t_c)\cr
&&=(aQt_c^r t_r^{\rho+\sigma}) ^{r(\rho-\sigma)(S(L)+n-1)} \Big[\prod_{i=1}^r (-Q^{-2}t_r^{-1}t_c^{1-2i};t_r^{-2})_{\rho-\sigma}\Big]^{n-1} \wt \scP^{\fin}_{[r^{\sigma}]} (L; a,Q,t_r,t_c)~.
\eea

\item  {\bf Universal colored differentials}   \\
When $\lambda$ is a rectangular Young tableau with either two rows or two columns,
there exists yet another set of colored differentials, $d^\uparrow$ and $d^\leftarrow$, called the universal colored differentials.
The universal colored differential $d^\uparrow$ (resp. $d^\leftarrow$) ``removes'' a row (resp. a column):
\bea
\label{universal-color-diff}
H_*(\wt\scH^\fin_{[r^2]}(L),d^\uparrow)\cong \wt\scH^\fin_{[r]}(L)~,\quad H_*(\wt\scH^\fin_{[2^\rho]}(L),d^\leftarrow)\cong \wt\scH^\fin_{[1^\rho]}(L)~.
\eea
The $(a, Q, t_r, t_c)$-degrees  of the differentials $d^\leftarrow$ and $d^\uparrow$ are given by
\bea\nonumber
\wt\deg ~d^\uparrow=(0,0,-2,0)~, \quad \wt\deg ~d^\leftarrow=(0,0,0,2)~,
\eea
and the re-gradings in the isomorphisms \eqref{universal-color-diff} can be expressed at the level of the Poincar\'e polynomials:
\bea\nonumber
\wt \scP(\wt\scH^{\fin}_{[r^2]}(L),d^\uparrow)(a, Q, t_r, t_c)&=&\wt\scP^{\fin}_{[r]}(L;a^2, Q^2, t_r^4, t_c^2)~,\cr
\wt \scP(\wt\scH^{\fin}_{[2^\rho]}(L),d^\leftarrow)(a, Q, t_r, t_c)&=&\wt\scP^{\fin}_{[1^\rho]}(L;a^2, Q^2, t_r^2, t_c^4)~.
\eea
We call these colored differentials ``universal'' because their $a$-degree is equal to $0$.

\end{itemize}

\subsection{Infinite-dimensional homology}
\label{infdim}

By definition, a reduced colored HOMFLY invariant of an $n$-component link $L$
is obtained from the unreduced colored HOMFLY invariant via dividing by the unknot factor
\be\label{simprel}
\bar{P}_{\lambda_1,\lambda_2,\ldots,\lambda_n}(L)={P}_{
\underline{\lambda_1},\lambda_2,\ldots,\lambda_n}(L)\bar{P}_{\lambda_1}(\unknot)~,
\ee
where we underline the component on which the invariant is ``reduced''.
Since the reduced colored HOMFLY invariant $P_{\underline{\lambda_1},\lambda_2,\ldots,\lambda_n}(L;a,q)$ is a rational function,
it has infinitely many terms (monomials) when expanded in powers of $q$.
As a result, its categorification necessarily must be infinite-dimensional,
which makes a rather dramatic distinction from reduced HOMFLY homology of knots.

Furthermore, for $|q|<1$, factors appearing in the denominator of the colored HOMFLY invariant $P_{\underline{\lambda_1},\lambda_2,\ldots,\lambda_n}(L;a,q)$
should be expanded in positive powers of $q$, while for $|q|>1$ the invariant should be expanded in negative powers of $q$.
As a result, in each case, there exists the corresponding colored HOMFLY homology which categorifies $P_{\underline{\lambda_1},\lambda_2,\ldots,\lambda_n}(L;a,q)$.
In this paper, we denote these two variants of the reduced colored HOMFLY homology
by $\scH^{(+)}_{\underline{\lambda_1},\lambda_2,\ldots,\lambda_n}(L)$ and $\scH^{(-)}_{\underline{\lambda_1},\lambda_2,\ldots,\lambda_n}(L)$,
respectively.
When the colors $(\lambda_1,\ldots,\lambda_n)$ are all given by the same rectangular Young diagram $\lambda$, then the colored HOMFLY homology of links shares some of the properties
of the knot homology \cite{Gorsky:2013jxa}. In what follows, we shall describe all the structural properties of $\scH^{(\pm)}_{\lambda,\ldots,\lambda}(L)$. We no longer identify the component on which the reduction \eqref{simprel} is performed when all colors are identical, since there is no ambiguity in this case.
Since the infinite-dimensionality makes the concrete study of arbitrary rectangular Young tableaux rather difficult,
we further restrict our attention to the case of symmetric representations $[r]$ and anti-symmetric representations $[1^r]$.

As in the case of knots, we postulate that the colored HOMFLY homology $\scH^{(\pm)}_{\lambda,\ldots,\lambda}(L)$ of a link $L$ admits four independent gradings. Introducing the $Q$-grading, we can also see one of the mirror symmetries and the refined exponential growth property in $\scH^{(\pm)}_{\lambda,\ldots,\lambda}(L)$.
However, the positive colored HOMFLY homology $\scH^{(+)}_{\lambda,\ldots,\lambda}(L)$ is equipped only with positive colored differentials $d^+_{\lambda\to\mu}$ while  the negative colored HOMFLY homology $\scH^{(-)}_{\lambda,\ldots,\lambda}(L)$ is equipped only with negative colored differentials $d^-_{\lambda\to\mu}$. Also, the colored HOMFLY homology $\scH^{(\pm)}_{\lambda,\ldots,\lambda}(L)$ of links does not enjoy the self-symmetry, since it has an infinite ``tail'' of generators for which the powers of $q$ grow either to $+\infty$ or to $-\infty$. The universal colored differentials do not exist in this case either.\\

In order to see these properties more explicitly, let us introduce the Poincar\'e polynomials of the two variants of the quadruply-graded homology:
\bea\nonumber
\scP^{(\pm)}_{\lambda,\cdots,\lambda}(L;a,q, t_r,t_c)&:=& \sum_{i,j,k,\ell} a^i q^j t_r^k t_c^\ell ~\dim\;(\scH^{(\pm)}_{\lambda,\ldots,\lambda}(L))_{i,j,k,\ell}~,\cr
\wt\scP^{(\pm)}_{\lambda,\cdots,\lambda}(L;a,Q, t_r,t_c)&:=& \sum_{i,j,k,\ell} a^i Q^j t_r^k t_c^\ell ~\dim\;(\wt\scH^{(\pm)}_{\lambda,\ldots,\lambda}(L))_{i,j,k,\ell}~,
\eea
related by
\bea\nonumber
\wt\scP^{(\pm)}_{\lambda,\ldots,\lambda}(L;a,q^\rho,t_rq^{-1},t_cq)=\scP^{(\pm)}_{\lambda,\ldots,\lambda}(L;a,q,t_r,t_c)~.
\eea
As in the homology theory with finite support considered earlier,
the $t_r$-grading and $t_c$-grading coincide on every generator of the uncolored homology $\scH^{(\pm)}_{\yng(1)}(L)$.
We therefore use the following notation:
$$
\scP^{(\pm)}_{\yng(1), \ldots,\yng(1)}(L;a,q,t)=\scP^{(\pm)}_{\yng(1), \ldots,\yng(1)}(L;a,q,t_r=1,t_c=t)=\wt \scP^{(\pm)}_{\yng(1), \ldots,\yng(1)}(L;a,Q=q,t_r=1,t_c=t)~.
$$
In general, there is no relation between ``positive'' and ``negative'' homologies; however, in the uncolored case,
\be\nonumber
\scP^{(-)}_{\yng(1), \ldots,\yng(1)}(L;a,q,t)=t^\#\scP^{(+)}_{\yng(1),  \ldots,\yng(1)}(L;at^2,q^{-1},t^{-1})~.
\ee

\begin{itemize}
\item {\bf Mirror symmetry}\\
The $[1^r]$-colored and $[r]$-colored HOMFLY homology theories are related by the mirror symmetry
\bea\nonumber
(\wt\scH^{(\pm)}_{[1^r],\ldots, [1^r]}(L))_{i,j,k,\ell}\cong(\wt\scH^{(\pm)}_{[r],\ldots,[r]}(L))_{i,j,\ell,k}~.
\eea
At the level of the Poincar\'e polynomials, the mirror symmetry reads
\bea\label{mirror-infinite}
\wt\scP^{(\pm)}_{[1^r],\ldots,[1^r]}(L;a,Q,t_r,t_c)&=&t_r^{(n-1)r(r-1)}\wt\scP^{(\pm)}_{[r],\ldots,[r]}(L;a,Q,t_c,t_r)~.
\eea
\item {\bf Refined exponential growth}\\
Let $L^*$ be either a rational link or a torus link.
The infinite-dimensional quadruply-graded HOMFLY homology of such a link $L^*$ obeys the refined exponential growth property
\bea\label{REGP-infinite}
\wt\scP^{(\pm)}_{[r],\ldots,[r]}(L^*;a,Q,t_r,t_c=1)&=&\left[\wt\scP^{(\pm)}_{\yng(1),\ldots,\yng(1)}(L^*;a,Q ,t_r,t_c=1)\right]^r~,\cr
\wt\scP^{(\pm)}_{[1^r],\ldots,[1^r]}(L^*;a,Q,t_r=1,t_c)&=&\left[\wt\scP^{(\pm)}_{\yng(1),\ldots,\yng(1)}(L^*;a,Q ,t_r=1,t_c)\right]^r~.
\eea
\item {\bf Colored differentials}\\
For every $k$ such that $r>k\ge 0$, there is a colored  differential $d^\pm_{[r]\to [k]}$ on $\wt\scH^{(\pm)}_{[r]}(L)$, and  there is a colored differential $d^\pm_{[1^r]\to [1^k]}$ on $\wt\scH^{(\pm)}_{[1^r]}(L)$ with exactly the same degrees as in the case of the finite-dimensional homology.
We stress that $\wt\scH^{(+)}(L)$ comes equipped only with  differentials $d^+$, and $\wt\scH^{(-)}(L)$ comes equipped only with the differentials $d^-$; there are no negative colored differentials $d^{-}$ on $\wt\scH^{(+)}(L)$ and there are no positive colored differentials $d^{+}$ on $\wt\scH^{(-)}(L)$. The $(a,Q,t_r,t_c)$-degree of these differentials are the same as \eqref{diff-degree}.
Then, these differentials satisfy
\bea\nonumber
H_*(\scH^{(\pm)}_{[r],\ldots,[r]}(L),d^\pm_{[r]\to [k]})\cong\scH^{(\pm)}_{[k],\cdots,[k]}(L), \cr
H_*(\scH^{(\pm)}_{[1^r],\ldots,[1^r]}(L),d^\pm_{[1^r]\to [1^k]})\cong\scH^{(\pm)}_{[1^k],\ldots,[1^k]}(L).
\eea
The isomorphisms involve some regradings, which can be conveniently described at the level of Poincar\'e polynomials:
\bea\label{color-diff-infinite}
&&\wt\scP(\scH^{(+)}_{[r],\ldots,[r]}(L),d^+_{[r] \to [k]})(a,Q,t_r,t_c)\\
&&\hspace{3cm}=[aQ^{-1}t_c^{k}]^{(r-k)S(L)}\wt\scP^{(+)}_{[k],\ldots,[k]}(L;a,Qt_c^{r-k},t_r,t_c)~,\cr
&&\wt\scP(\scH^{(+)}_{[1^r],\ldots,[1^r]}(L),d^+_{[1^r] \to [1^k]})(a,Q,t_r,t_c)\cr
&&\hspace{2cm}=t_r^{(n-1)[r(r-1)-k(k-1)]}[aQ^{-1}t_r^{k}]^{(r-k)S(L)}\wt\scP^{(+)}_{[1^k],\ldots,[1^k]}(L;a,Q t_r^{r-k},t_r,t_c)~,\cr
&&\wt\scP(\scH^{(-)}_{[r],\ldots,[r]}(L),d^-_{[r] \to [k]})(a,Q,t_r,t_c)\cr
&&\hspace{3cm}=[t_rt_c]^{(n-1)(r-k)}[aQt_rt_c^{r+k}]^{(r-k)S(L)}\wt\scP^{(-)}_{[k],\ldots,[k]}(L;a,Q,t_r,t_c)~,\cr
&&\wt \scP(\wt\scH^{(-)}_{[1^r],\ldots,[1^r]}(L),d^-_{[1^r]\to [1^k]})(a,Q,t_r,t_c)\cr
&&\hspace{2cm}=[t_r^{r+k}t_c]^{(n-1)(r-k)}[aQ t_r^{r+k}t_c] ^{(r-k)S(L)} \wt \scP^{(-)}_{[1^{k}],\ldots,[1^{k}]} (L; a,Q,t_r,t_c)~.\nonumber
\eea
\end{itemize}

\subsection{Uncolored Kauffman homology}

Even though our main interest is in colored HOMFLY homology of links,
the uncolored Kauffman homology often serves as a useful intermediate step.
Therefore, in this subsection, we briefly describe the basic properties of the uncolored Kauffman homology of links.
Like HOMFLY homology, the \emph{reduced} Kauffman homology $\scH^\Kauffman(L)$ of links is also infinite-dimensional,
since the reduced Kauffman invariant $F(L;a,q)$ of a link $L$
is not polynomial in~$q$.
For this reason, the Kauffman homology of links inherits only some of the properties \cite{Gukov:2005qp,Nawata:2013mzx}
of the Kauffman homology of knots.

In order to see these properties, we label the three gradings by $a$, $q$, and~$t$,
and introduce the Poincar\'e polynomial of the Kauffman homology $\scH^\Kauffman(L)$:
\bea\nonumber
\scF(L;a,q,t):=\sum_{i,j,k} a^i q^j t^k \dim \;(\scH^\Kauffman(L))_{i,j,k}.
\eea
At $t=-1$, it specializes to the Kauffman invariant: $\scF(L;a,q,-1)=F(L;a,q)$.
While there are many differentials in the Kauffman homology of knots,
only a few are present in the case of links:

\begin{itemize}
\item {\bf Canceling differential}\\
There are three canceling differentials in the uncolored Kauffman homology of knots, {\it cf.} (6.16) in \cite{Gukov:2005qp}.
Due to the infinite-dimensional nature, the uncolored Kauffman homology of links has only one canceling differential $d_{\rm cancel}$,
whose $(a,q,t)$-degree is $(-1,1,-1)$.
This differential corresponds to $d_2$ in \cite{Gukov:2005qp} or $d^+_{[1]\to[0]}$ in \cite{Nawata:2013mzx}.
The homology with respect to the canceling differential $d_{\rm cancel}$ is one-dimensional and its Poincar\'e polynomial
is equal to $(aq^{-1})^{S(L)}$.

\item {\bf Universal differential}\\
The most prominent feature of the uncolored Kauffman homology of knots is the universal differential.
There exists only one universal differential in the uncolored Kauffman homology of links (unlike the case of knots, where there are two).
The $(a,q,t)$-degree of the universal differential is given by $\deg d^{\rm univ}_{\to}=(0,2,1)$,
and the homology with respect to the universal differential is isomorphic to the uncolored HOMFLY homology:
\bea\nonumber
H_*(\scH^{\Kauffman}_{\yng(1)}(L),d^{\rm univ}_{\to})\cong \scH^{(+)}_{\yng(1)}(L)~,
\eea
up to the regrading
\bea
\label{univ-diff-regrading}
 \scP(\scH^{\Kauffman}(L),d^{\rm univ}_{\to})(a, q,t)&=&q^{S(L)}\scP^{(+)}_{\yng(1)}(L;a q^{-1}, q, t)~.
\eea
\end{itemize}

For the examples of two-component thin links in \S\ref{sec:example} and Appendix \ref{sec:app-ExamplesDiag},
we observe the following two relations between the Kauffman and  HOMFLY homology.
First, there is a relation between the uncolored Kauffman homology and the $(\yng(2),\yng(2))$-colored HOMFLY homology
\be
\label{Kauffman-thin}
\scF(L_{\rm thin};a,q,t)=\wt\scP^{(+)}_{\yng(2),\yng(2)}(L_{\rm thin};a q^{-1},a^{1/2} q^{-1/2}, a^{1/2} q^{-3/2}t, a^{-1/2}q^{3/2}),
\ee
which is reminiscent of the isomorphism of representations $(\frakso(3),\yng(1))\cong (\fraksl(2),\yng(2))$.
In fact, the specialization $a=q^2$ yields (4.48) in \cite{Nawata:2013mzx}.
Furthermore, we find the following relation:
\bea\nonumber
\scF(L_{\rm thin};q^3,q,t)=[\wt\scP^{(+)}_{\yng(1),\yng(1)}(L_{\rm thin};q^2,q,t)]^2~,
\eea
which can be regarded as a consequence of the isomorphism of algebras, $\frakso(4)\cong \fraksl(2)\oplus\fraksl(2)$.
It would be interesting to explore these relations further, and  to understand whether they hold for all thin links.

\subsection{Unknot and unreduced homology}
\label{sec:unknot}

The unreduced homology $\bar{\scH}^{(\pm)}_{\lambda_1,\lambda_2,\ldots,\lambda_n}(L)$ of a link is a tensor product of the reduced homology ${\scH}^{(\pm)}_{\underline{\lambda_1},\lambda_2,\ldots,\lambda_n}(L)$ (where we underline the component labeled by $\lambda_1$ on which the ``reduction'' is performed) and the unreduced homology of the unknot $\bar{\scH}^{(\pm)}_{\lambda_1}(\unknot)$.
In terms of the corresponding Poincar\'e polynomials we have:
\be\label{simprel2}
\bar{\scP}^{(\pm)}_{\lambda_1,\lambda_2,\ldots,\lambda_n}(L)={\scP}^{(\pm)}_{
\underline{\lambda_1},\lambda_2,\ldots,\lambda_n}(L)\bar{\scP}^{(\pm)}_{\lambda_1}(\unknot).
\ee
The unreduced $[r^{\rho}]$-colored HOMFLY polynomial of the unknot is given by:
\bea\nonumber
\bar{P}_{[r^{\rho}]}(\unknot;a,q)&=&(a^{-1}q^{\rho})^{r\rho}
\prod_{i=1}^{\rho}\prod_{j=1}^r \frac{1-a^2q^{2(j-i)}}{1-q^{2\rho}q^{2(j-i)}}\\
&=&(a^{-1}q^{\rho})^{r\rho}\prod_{i=1}^{\rho} \frac{(a^2q^{2(1-i)};q^2)_r}{(q^{2(\rho-i+1)};q^2)_r}=(a^{-1}q^{\rho})^{r\rho}\prod_{j=1}^{r} \frac{(a^2q^{2(j-1)};q^{-2})_r}{(q^{2(\rho+j-1)};q^{-2})_r}~.\nonumber
\eea
The quadruply-graded homology $\bar{\scH}^{(+)}_{[r^{\rho}]}(\unknot)$ that categorifies $\bar{P}_{[r^{\rho}]}(\unknot)$ with $|q|<1$ and $\bar{\scH}^{(-)}_{[r^{\rho}]}(\unknot)$ that categorifies $\bar{P}_{[r^{\rho}]}(\unknot)$ with $|q|>1$ have the following Poincar\'e polynomials:
\bea\nonumber
\overline{\scP}^{(+)}_{[r^{\rho}]}(\unknot;a,q,t_r,t_c)&=&(a^{-1}q^{\rho})^{r\rho}
\prod_{i=1}^{\rho}\prod_{j=1}^r \frac{1+a^2q^{2(j-i)}t_r^{2i-1}t_c^{2j-1}}{1-q^{2\rho}q^{2(j-i)}t_r^{2i-2}t_c^{2j-2}}\\
 \overline \scP^{(-)}_{[r^\rho]}(\unknot;a,q,t_r,t_c)&=&(aq^\rho)^{-r\rho}\prod_{i=1}^{\rho}\prod_{j=1}^r \frac{1+a^2q^{2(j-i)}t_r^{2i+1}t_c^{2j+1}}{1-q^{-2\rho}q^{2(i-j)}t_r^{2-2i}t_c^{2-2j}}
\eea
In the case $|q|<1$, the first expression is related to the Macdonald dimension \eqref{Mac_sp}
\bea\nonumber
\overline{\scP}^{(+)}_{[r^{\rho}]}(\unknot;a,q,t_r=1,t_c=t)=M^*_{[r^\rho]}(\unknot;A=-a^2t,q_1=q^2t^2,q_2=q^2)~,
\eea
where the Macdonald dimension of a rectangular representation $[r^\rho]$ is given by
\bea\nonumber
M^*_{[r^\rho]}(\unknot;A,q_1,q_2)=(A^{-1/2}q_2^{\rho/2})^{r\rho}\prod_{i=1}^{\rho}\prod_{j=1}^r\frac{1-Aq_1^{j-1}q_2^{1-i}}{1-q_1^{j-1}q_2^{\rho-i+1}}
\eea
In the version with $(a,Q,t_r,t_c)$-gradings, we have
\bea\nonumber
\widetilde{\bar{\scP}}^{(+)}_{[r^{\rho}]}(\unknot;a,Q,t_r,t_c)&=&(a^{-1}Q)^{r\rho}\prod_{i=1}^{\rho}\prod_{j=1}^r \frac{1+a^2t_r^{2i-1}t_c^{2j-1}}{1-Q^2t_r^{2i-2}t_c^{2j-2}}\\
&=&(a^{-1}Q)^{r\rho}\prod_{i=1}^{\rho}\frac{ (-a^2t_r^{2i-1}t_c;t_c^2)_r}{(Q^2t_r^{2i-2};t_c^2)_r}=(a^{-1}Q)^{r\rho}\prod_{j=1}^{r} \frac{(-a^2t_c^{2j-1}t_r;t_r^2)_\rho}{(Q^2t_c^{2j-2};t_r^2)_\rho}~,\cr
\wt {\overline \scP}^{(-)}_{[r^\rho]}(\unknot;a,Q,t_r,t_c)&=&(aQ)^{-r\rho}\prod_{i=1}^{\rho}\prod_{j=1}^r \frac{1+a^2t_r^{2i+1}t_c^{2j+1}}{1-Q^{-2}t_r^{2-2i}t_c^{2-2j}}\cr
 &=&(aQ)^{-r\rho}\prod_{i=1}^{\rho} \frac{ (-a^2t_r^{2i+1}t_c^{3};t_c^{2})_r}{ (Q^{-2}t_r^{2(1-i)};t_c^{-2})_r}=(aQ)^{-r\rho}\prod_{j=1}^{r} \frac{(-a^2t_c^{2j+1}t_r^3;t_r^{2})_\rho}{ (Q^{-2}t_c^{2(1-j)};t_r^{-2})_\rho}~.\nonumber
\eea
Many properties of reduced knot homology extend to the unreduced homology of the unknot.
Thus, one finds all positive row- and column-removing colored differentials with exactly the same properties
as described for the reduced homology in \S \ref{infdim}.
Furthermore, the refined exponential growth property holds:
\bea\nonumber
\widetilde{\bar{\scP}}^{(\pm)}_{[r^{\rho}]}(\unknot;a,Q,t_r=t,t_c=1)&=&
\left[\widetilde{\bar{\scP}}^{(\pm)}_{[1^{\rho}]}(\unknot;a,Q,t_r=t,t_c=1)\right]^r\cr
\widetilde{\bar{\scP}}^{(\pm)}_{[r^{\rho}]}(\unknot;a,Q,t_r=1,t_c=t)&=&
\left[\widetilde{\bar{\scP}}^{(\pm)}_{[r]}(\unknot;a,Q,t_r=1,t_c=t)\right]^\rho
\eea
along with the mirror symmetry:
\be\nonumber
\widetilde{\bar{\scP}}^{(\pm)}_{[r^{\rho}]}(\unknot;a,Q,t_r,t_c)=\widetilde{\bar{\scP}}^{(\pm)}_{[{\rho}^r]}(\unknot;a,Q,t_c,t_r).
\ee
By~(\ref{simprel2}), these properties therefore hold for the unreduced colored homology of any link.



\section{Detailed example: the Hopf link $T_{2,2}$}
\label{sec:example}

In this section we explicitly construct the various homology theories and corresponding Poincar\'e polynomials introduced in \S \ref{sec:quadruply}, in the simplest possible case: the Hopf link. By analyzing this example we illustrate that various properties of link homologies postulated in \S \ref{sec:quadruply} are satisfied. We also show that certain expressions for Poincar\'e polynomials for the Hopf link agree with results of computations in refined Chern-Simons theory, which we will present in detail in \S \ref{sec:refined}. In fact, in \S \ref{sec:refined} we will show that refined Chern-Simons theory reproduces Poincar\'e polynomials for torus links in a much more general setting. This gives a strong test of the formalism introduced in \S \ref{sec:quadruply}.

We have also constructed various homologies and Poincar\'e polynomials of many other links: $(2,2p)$ torus links, twist links $L_p$, the Borromean rings, and the $(3,3)$ torus link. While these computations are more involved, they follow the same strategy as is used for the Hopf link. We therefore relegate the details of these computations to appendix~\ref{sec:app-ExamplesDiag}.

\noindent
\begin{minipage}{\linewidth}
      \centering
      \begin{minipage}{0.3\linewidth}
          \begin{figure}[H]
              \includegraphics[scale=0.4]{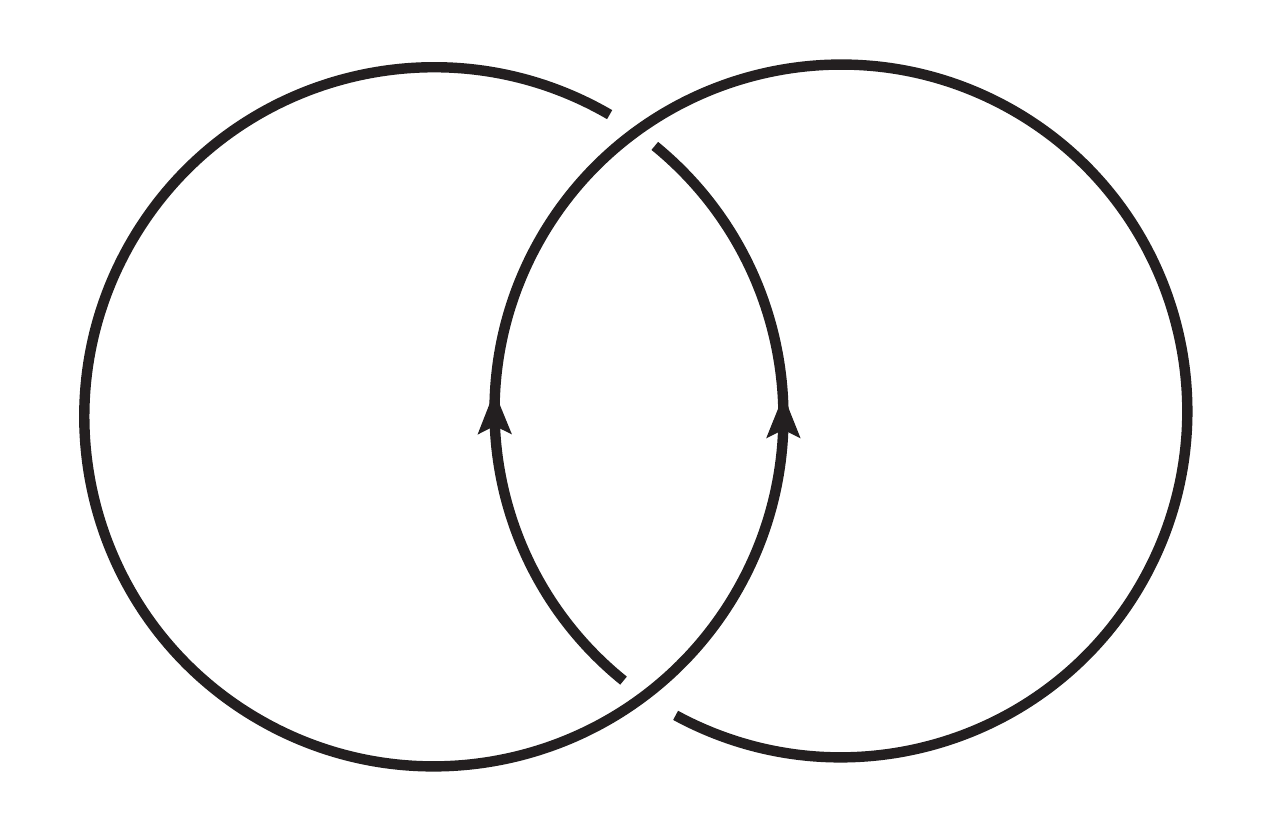}
              \caption{The Hopf link $T_{2,2}$.}\label{fig:Hopf}
          \end{figure}
      \end{minipage}
      \hspace{0.05\linewidth}
      \begin{minipage}{0.58\linewidth}
          \begin{figure}[H]
          \centering
              \includegraphics[scale=1.2]{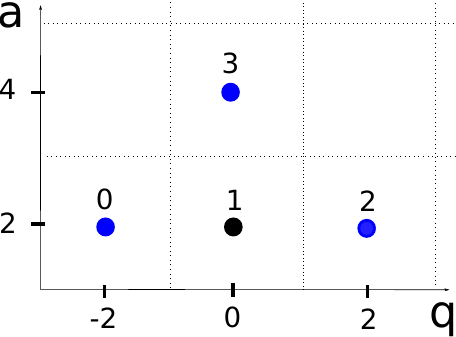}
              \caption{The uncolored finite-dimensional HOMFLY homology of the Hopf link. The blue dots correspond to the generators of HOMFLY homology of the trefoil.}\label{fig:uncolor-Hopf-finite}
          \end{figure}
      \end{minipage}
  \end{minipage}

\subsection{Finite-dimensional HOMFLY homology }

\subsubsection*{Fundamental representation}

The reduced uncolored HOMFLY invariant of the Hopf link is
\be\nonumber
P(T_{2,2};a,q)=\frac{a(1 - q^2 + q^4- a^2 q^2 )}{q(1-q^2)}~.
\ee
The uncolored finite-dimensional homology $\scH^{\rm fin}_{\yng(1)}(T_{2,2})$ of the Hopf link categorifies
\bea\nonumber
P^\fin_{\yng(1)}({T_{2,2}};a,q)=aq^{-1}(1-q^2)P_{\yng(1),\yng(1)}({T_{2,2}};a,q)=a^2 (q^{-2}-1+q^2)-a^4~.
\eea
Since the Hopf link is homologically thin, it is straightforward to read off the homological degree of each generator using \eqref{delta-grading}:
\bea\nonumber
\scP^{\rm fin}_{\yng(1)}({T_{2,2}};a,q,t)&=&a^2 (q^{-2}+t+q^2 t^2)+a^4 t^3~.
\eea
As shown in Figure \ref{fig:uncolor-Hopf-finite}, the finite-dimensional HOMFLY homology of the Hopf link contains the  HOMFLY homology of the trefoil knot.

\subsubsection*{Second symmetric representation}
\begin{figure}[h]
 \centering
    \includegraphics[width=13cm]{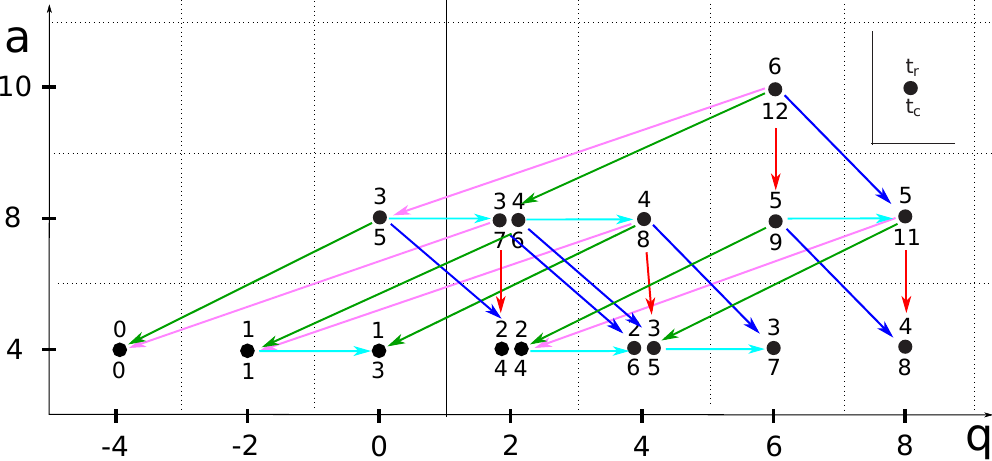}
    \caption{The finite-dimensional $[2]$-colored quadruply-graded HOMFLY homology of the Hopf link in the $(a,q,t_r,t_c)$-gradings. The red and pink arrows represent the action of the colored differentials $d^{\pm}_{[2]\to [1]}$, respectively. The blue and green arrows represent the action of the canceling differentials $d^\pm_{[2]\to [0]}$, respectively. The finite-dimensional colored HOMFLY homology enjoys the action of both positive and negative colored differentials. The light blue arrows show the action of the universal colored differential $d^\leftarrow$.}\label{fig:HOMFLY-prime-Hopf-2-2}
\end{figure}

To obtain the finite-dimensional $[2]$-colored HOMFLY homology, one needs to categorify the following polynomial
\bea\nonumber
P^{\fin}_{\yng(2)}({T_{2,2}};a,q)&=&a^2q^{-2}(1-q^2)(1-q^4)P_{\yng(2),\yng(2)}({T_{2,2}};a,q)\\
&=&   a^4 (q^{-4} - q^{-2}-1 + 2 q^2 - q^6 + q^8) + a^6 (-1 + q^4 - q^6 - q^8)+a^8 q^6~.\nonumber
\eea
Using the differentials, one can obtain the finite-dimensional $[2]$-colored HOMFLY homology of the Hopf link with its $(a,q,t_r,t_c)$-gradings:
\bea\nonumber
\scP^{\rm fin}_{\yng(2)}({T_{2,2}};a,q,t_r,t_c)&=&
a^4 q^{-4}(1 +   q^2 t_r t_c) (1 + q^4  t_r t_c^3)\cr
&&+ a^4   q^2  t_r^2 t_c^4 (1 + q^2 t_c^2) (1 +   q^2 t_r t_c) (1 + a^2 q^{-2} t_r t_c ) \cr
      && +
  a^4   q^8  t_r^4 t_c^8 (1 + a^2 q^{-2} t_r t_c ) (1 + a^2  t_r t_c^3) ~.
\eea
The ``dot diagram'' representation of the homology and the actions of various differentials on it are illustrated in Figure \ref{fig:HOMFLY-prime-Hopf-2-2}.

\begin{figure}[h]
 \centering
    \includegraphics[width=9cm]{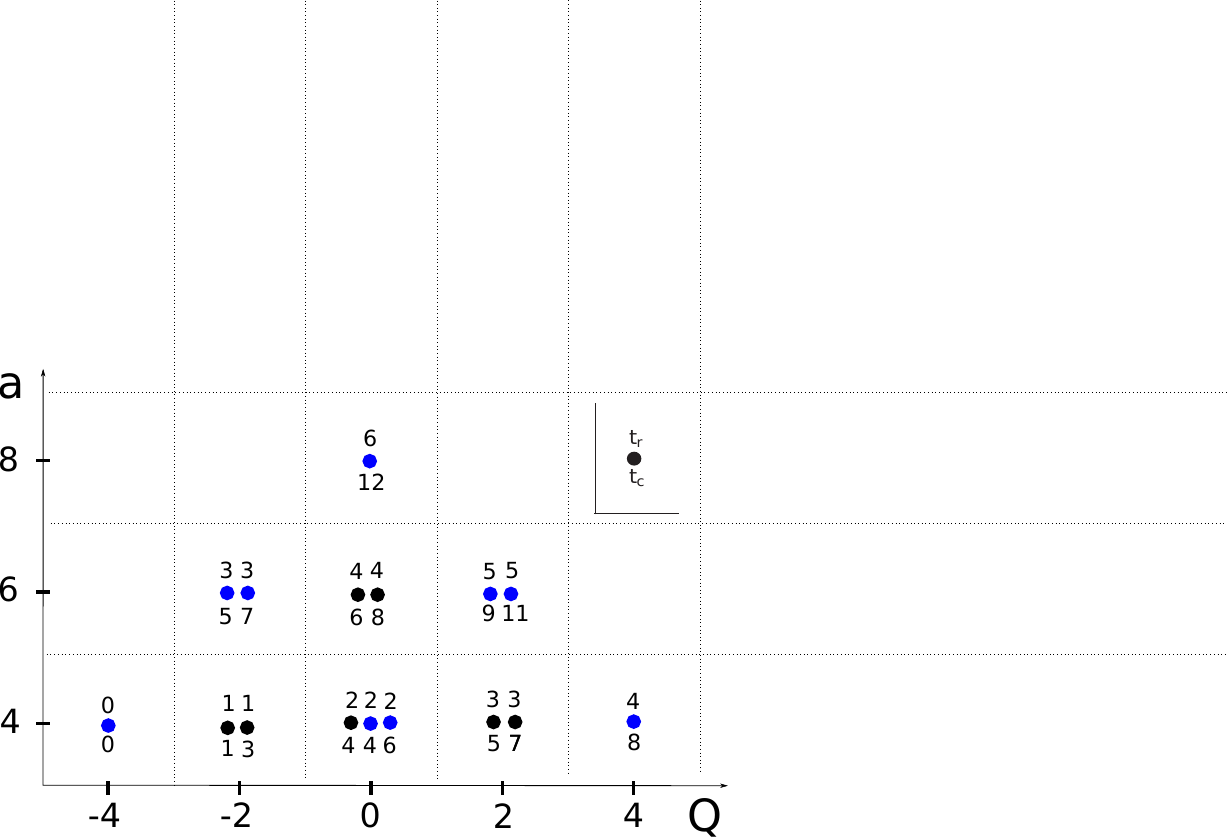}
    \caption{The finite-dimensional $[2]$-colored quadruply-graded HOMFLY homology of the Hopf link in the $(a,Q,t_r,t_c)$-gradings. The self-symmetry property becomes manifest in these gradings. The blue dots correspond to the generators of the $[2]$-colored HOMFLY homology of the trefoil.}\label{fig:HOMFLY-prime-Hopf-2}
\end{figure}

It is straightforward to obtain the finite-dimensional $[2]$-colored HOMFLY homology of the Hopf link in the $(a,Q,t_r,t_c)$-gradings:
\bea\label{Hopf-quad-finite-2}
\wt\scP^{\fin}_{\yng(2)}({T_{2,2}};a,Q,t_r,t_c)&=& \scP^{\rm fin}_{\yng(2)}({T_{2,2}};a,Q,t_rQ,t_cQ^{-1})\cr
&=&
a^4Q^{-4}(1 +  Q^2 t_r t_c) (1 + Q^2 t_r  t_c^3) \cr
    &&  +  a^4 t_r^2 t_c^4 (1 + t_c^2)  (1 + a^2Q^{-2} t_r t_c) (1 +
      Q^2 t_r t_c)\cr
      &&+  a^4Q^4t_r^4 t_c^8  (1 + a^2Q^{-2} t_r t_c) (1 + a^2Q^{-2}  t_rt_c^3)~.
\eea
In the  $(a,Q,t_r,t_c)$-gradings, the self-symmetry \eqref{self-symmetry} becomes manifest:
\bea\nonumber
\wt\scP^{\fin}_{\yng(2)}({T_{2,2}};a,Q,t_r,t_c)=\wt\scP^{\fin}_{\yng(2)}({T_{2,2}};a,Q^{-1}t_r^{-1}t_c^{-2},t_r,t_c)~.
\eea
As shown in Figure~\ref{fig:HOMFLY-prime-Hopf-2}, the finite-dimensional $[2]$-colored HOMFLY homology of the Hopf link contains the $[2]$-colored HOMFLY homology of the trefoil.

Let us look at the actions of colored differentials on  the finite-dimensional $[2]$-colored HOMFLY homology $\wt\scH^\fin_{\yng(2)}(T_{2,2})$ of the Hopf link, in the $(a,Q,t_r,t_c)$-gradings.
For the finite-dimensional version of the theory, the homology with respect to the canceling differential $d^\pm_{[2]\to[0]}$ is \emph{not} one-dimensional, as explained in \eqref{regrading-finite-column}. The Poincar\'e polynomial of the homology with respect to the canceling differential $d^\pm_{[2]\to[0]}$ is given by
\bea\nonumber
\wt\scP(\wt \scH^\fin_{\yng(2)}(T_{2,2}),d^+_{[2]\to[0]})(a,Q,t_r,t_c)&=&a^4Q^{-4} (1+Q^2t_rt_c)(1+Q^2t_rt_c^3)~,\\
\wt\scP(\wt \scH^\fin_{\yng(2)}(T_{2,2}),d^-_{[2]\to[0]})(a,Q,t_r,t_c)&=&a^4Q^{4}t_r^4t_c^8 (1+Q^{-2}t_r^{-1}t_c^{-1})(1+Q^{-2}t_r^{-1}t_c^{-3})~,\nonumber
\eea
where the $(a,Q,t_r,t_c)$-gradings \eqref{diff-degree} of the canceling differentials $d^\pm_{[2]\to[0]}$ are given by
\be\nonumber
{\wt \deg}~d^+_{[2]\to[0]}=(-2,2,-1,-1)~,\qquad {\wt \deg}~d^-_{[2]\to[0]}=(-2,-2,-3,-5)~.
\ee
On the other hand, the $(a,Q,t_r,t_c)$-gradings \eqref{diff-degree} of the colored differentials $d^\pm_{[2]\to[1]}$ are given by
\be\nonumber
{\wt \deg}~d^+_{[2]\to[1]}=(-2,2,-1,-3)~,\qquad {\wt \deg}~d^-_{[2]\to[1]}=(-2,-2,-3,-7).
\ee
One can easily verify the following relations with the uncolored homology:
\bea\nonumber
\wt\scP(\wt \scH^\fin_{\yng(2)}(T_{2,2}),d^+_{[2]\to[1]})(a,Q,t_r,t_c)&=&a^2Q^{-2} (1+Q^2t_rt_c)\wt\scP^\fin_{\yng(1)}(T_{2,2};a,Qt_c,t_r,t_c)~,\\
\wt\scP(\wt \scH^\fin_{\yng(2)}(T_{2,2}),d^-_{[2]\to[1]})(a,Q,t_r,t_c)&=&a^2Q^{2} t_r^2t_c^6(1+Q^{-2}t_r^{-1}t_c^{-1})\wt\scP^\fin_{\yng(1)}(T_{2,2};a,Q,t_r,t_c)~.\nonumber
\eea
In addition, the $[2]$-colored finite-dimensional HOMFLY homology $\wt\scH_{\yng(2)}(T_{2,2})$ of the Hopf link enjoys the action of the universal colored differential $d^\leftarrow$ with ${\wt \deg}~d^\leftarrow=(0,0,0,2)$. The homology with respect to $d^\leftarrow$  is isomorphic to the uncolored homology \eqref{universal-color-diff}
\bea\nonumber
\wt \scP(\wt\scH^{\fin}_{[2]}(L),d^\leftarrow)(a, Q, t_r, t_c)&=&\wt\scP^{\fin}_{[1]}(L;a^2, Q^2, t_r^2, t_c^4)~.
\eea

\subsubsection*{Arbitrary symmetric representations}

The structural properties of the ``diagonal homology'' (where all components are colored by the same representation)
are so rigid that we can obtain an explicit formula for the superpolynomials
$\scP^{\fin}_{[r]}(T_{2,2})$ for any positive integer $r$.
In the $(a,Q,t_r,t_c)$-gradings, the refined growth property \eqref{exp-growth-fin-1} determines the $t_c=1$ specialization of the $[r]$-colored HOMFLY homology:
\bea\label{Hopf-specialization-final}
\wt\scP^{\fin}_{[r]}(T_{2,2};a,Q,t_r,t_c=1)&=&a^2 Q^{-2}\left[ (1+Q^2t_r)+ Q^4 t_r^2(1+a^2Q^{-2} t_r)\right]^r\\
&=&a^{2r}Q^{-2r}  \sum_{i=0}^r Q^{4i}t_r^{2i} (1+a^2Q^{-2}t_r)_{i}(1+Q^2t_r)_{r-i}\binom{r}{i}~.\nonumber
\eea
The $t_c$-grading can then be determined using properties of the colored differentials. To that end, we make the following ansatz: the binomial $\binom{r}{i}$ in \eqref{Hopf-specialization-final} is replaced by quantum binomial ${r \brack i}_{t_c^2}$ with respect to the variable $t_c^2$. The factors $(1+a^2Q^{-2}t_r)^{i}$ and $(1+Q^{2}t_r)^{r-i}$ in \eqref{Hopf-specialization-final} are replaced by the $t_c^2$-Pochhammer symbols $(-a^2Q^{-2}t_rt_c^*;t_c^2)_{i}$ and $(-Q^{2}t_rt_c^{*};t_c^2)_{r-i}$. The differentials determine the $t_c$-grading uniquely, so that the Poincar\'e polynomial of the finite-dimensional $[r]$-colored HOMFLY homology of the Hopf link is
\be\label{Hopf-prime-quad}
\wt\scP^{\fin}_{[r]}(T_{2,2};a,Q,t_r,t_c)=a^{2r}Q^{-2r}  \sum_{i=0}^r Q^{4i}t_r^{2i}t_c^{2r i} (-a^2Q^{-2}t_rt_c;t_c^2)_{i}(-Q^2t_rt_c;t_c^2)_{r-i}{r \brack i}_{t_c^2}~.
\ee
It is straightforward to express the Poincar\'e polynomial of the triply-graded $[r]$-colored HOMFLY homology in the standard $t_r$-grading
\bea
\scP^{\rm fin}_{[r]}({T_{2,2}};a,q,t)&=&\wt\scP^{\fin}_{[r]}(T_{2,2};a,q,tq^{-1},q)\cr
&=&a^{2r} q^{-2r} \sum_{i=0}^r q^{2(r+1)i}t^{2i}(-a^2q^{-2}t;q^2)_{i}(-q^{2}t;q^2)_{r-i}{r \brack i}_{q^2}.~~~   \label{T22-HOMFLYprime}
\eea

\begin{figure}[h]
 \centering
    \includegraphics[width=13cm]{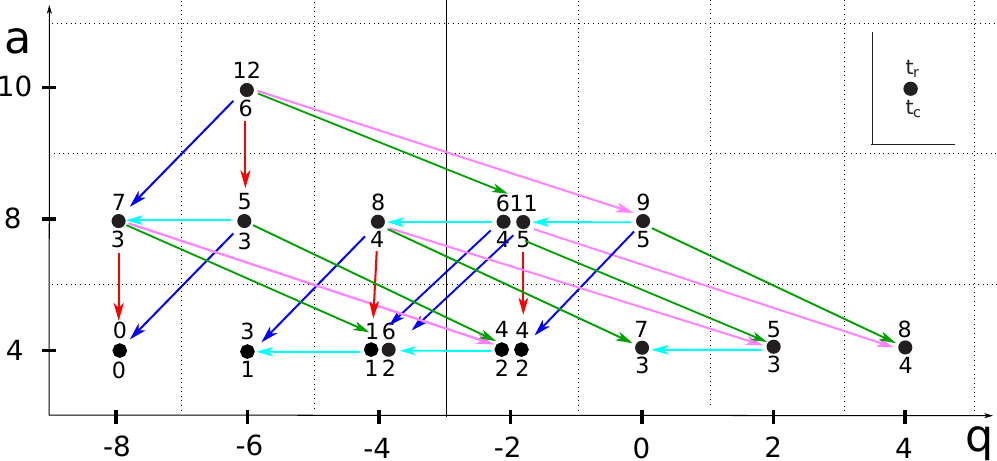}
    \caption{The finite-dimensional $[1,1]$-colored quadruply-graded HOMFLY homology of the Hopf link in the $(a,q,t_r,t_c)$-gradings. The red and pink arrows represent the action of the colored differentials $d^{\pm}_{[1,1]\to [1]}$, respectively. The blue and green arrows represent the action of the canceling differentials $d^\pm_{[1,1]\to [0]}$, respectively. The light blue arrows show the action of the universal colored differential $d^\uparrow$.}\label{fig:HOMFLY-prime-Hopf-11-2}
\end{figure}

\subsubsection*{Anti-symmetric representations}

Next, let us consider the finite-dimensional HOMFLY homology of the Hopf link colored by anti-symmetric representations. The finite-dimensional $[1,1]$-colored HOMFLY homology $\scH^\fin_{\yng(1,1)}(T_{2,2})$ of the Hopf link categorifies
\bea\nonumber
P^\fin_{[1,1]}(T_{2,2};a,q)&=&(-aq)^{r}(q^2;q^2)_rP_{[1,1],[1,1]}(T_{2,2};a,q)\\
&=&a^4 ( q^{-8} - q^{-6} + 2q^{-2} -1 - q^2 + q^4)+a^6 ( - q^{-8} - q^{-6} + q^{-4}-1) + a^8q^{-6} ~.\nonumber
\eea
The Poincar\'e polynomial of this theory in the $(a,q,t_r,t_c)$-grading is given by
\bea\nonumber
&&\scP^{\rm fin}_{\yng(1,1)}({T_{2,2}};a,q,t_r,t_c) = \cr
&=& a^4q^{-8} (1 + q^4 t_r t_c + q^2 t_r^3t_c  + 2 q^6 t_r^4t_c^2  + q^{10}  t_r^5 t_c^3 +
     q^4 t_r^6t_c^2  + q^8 t_r^7t_c^3  + q^{12} t_r^8t_c^4 )\cr
     && + a^6q^{-8} (q^2  t_r^5t_c^3 + q^6t_r^6 t_c^4  + t_r^7t_c^3  + q^4t_r^8 t_c^4  +
    q^8 t_r^9 t_c^5  + q^6 t_r^{11} t_c^5)+a^8q^{-6} t_r^{12} t_c^6  ~.
\eea
Figure \ref{fig:HOMFLY-prime-Hopf-11-2} shows the actions of colored differentials on $\scH^\fin_{\yng(1,1)}(T_{2,2})$.

\begin{figure}[h]
 \centering
    \includegraphics[width=9cm]{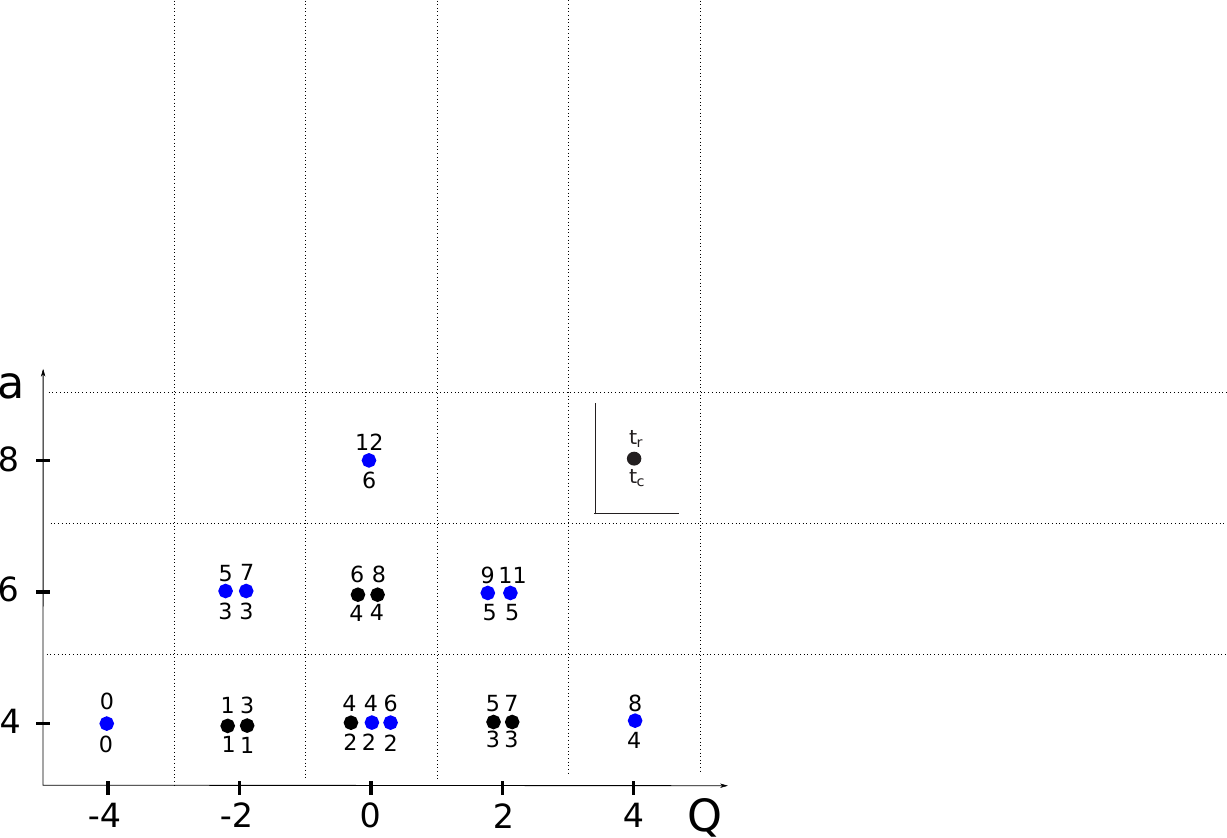}
    \caption{The finite-dimensional $[1,1]$-colored quadruply-graded HOMFLY homology of the Hopf link in the $(a,Q,t_r,t_c)$-gradings. The blue dots correspond to the generators of the $[1,1]$-colored quadruply-graded HOMFLY homology of the trefoil.}\label{fig:HOMFLY-prime-Hopf-11}
\end{figure}

Since the Hopf link is homologically thin, it is straightforward to obtain the finite-dimensional $[1,1]$-colored HOMFLY homology of the Hopf link in the $(a,Q,t_r,t_c)$-gradings
\bea\label{Hopf-quad-finite-11}
\wt\scP^{\rm fin}_{\yng(1,1)}({T_{2,2}};a,Q,t_r,t_c)&=& \scP^{\rm fin}_{\yng(1,1)}({T_{2,2}};a,Q^{1/2},t_rQ^{1/2},t_cQ^{-1/2})\cr
&=&
a^4Q^{-4}(1 +  Q^2 t_c t_r) (1 + Q^2 t_c  t_r^3) \cr
    &&  +  a^4 t_c^2 t_r^4 (1 + t_r^2)  (1 + a^2Q^{-2} t_c t_r) (1 +
      Q^2 t_c t_r)\cr
      &&+  a^4Q^4t_c^4 t_r^8  (1 + a^2Q^{-2} t_c t_r) (1 + a^2Q^{-2}  t_ct_r^3)~.
\eea
As shown in Figure \ref{fig:HOMFLY-prime-Hopf-11}, the finite-dimensional $[1,1]$-colored HOMFLY homology of the Hopf link contains the  $[1,1]$-colored HOMFLY homology  of the trefoil.  It is easy to see from \eqref{Hopf-quad-finite-2} and \eqref{Hopf-quad-finite-11} that the mirror symmetry \eqref{mirror-finite} holds for the finite-dimensional colored HOMFLY homology
\bea\nonumber
\wt\scP^{\rm fin}_{\yng(1,1)}({T_{2,2}};a,Q,t_r,t_c)=\wt\scP^{\rm fin}_{\yng(2)}({T_{2,2}};a,Q,t_c,t_r)~.
\eea
Therefore, the mirror symmetry \eqref{mirror-finite} gives the finite-dimensional $[1^r]$-colored HOMFLY homology of the Hopf link
\bea\nonumber
&&\wt\scP^{\fin}_{[1^r]}(T_{2,2};a,Q,t_r,t_c)=\wt\scP^{\fin}_{[r]}(T_{2,2};a,Q,t_c,t_r) \cr
&=&a^{2r}Q^{-2r}  \sum_{i=0}^r Q^{4i}t_r^{2r i}t_c^{2i} (-a^2Q^{-2}t_rt_c;t_r^2)_{i}(-Q^2t_rt_c;t_r^2)_{r-i}{r \brack i}_{t_r^2}~.
\eea

\subsubsection*{$[r,r]$-colored HOMFLY homology}

The finite-dimensional $[r,r]$-colored HOMFLY homology of the Hopf link contains the $[r,r]$-colored HOMFLY homology of the trefoil, which has been obtained in \cite{Nawata:2013mzx}. Using the expression in \cite{Nawata:2013mzx}, one can deduce
the Poincar\'e polynomial of the finite-dimensional $[r,r]$-colored HOMFLY homology of the Hopf link
\bea\label{quad-HOMFLY-2row-trefoil}
&&\wt\scP^{\fin}_{[r,r]}(T_{2,2};a,Q,t_r,t_c)\\
&=&a^{4 r}\sum_{k=0}^{r}\sum_{j=0}^{k}\sum_{i=0}^{r-k} Q^{4 (i - j)}  t_r^{2 (2 i - 3 j + k + 2 r)} t_c^{2 r (-j + r) + 2 i (j + r)}{r \brack k}_{t_c^2} {k \brack j }_{t_c^2} {r-k\brack i }_{t_c^2}\cr
 && \times (-Q^{2} t_r t_c;t_c^2)_{k}(-Q^{-2} t_r^{-3} t_c^{1-2r};t_c^2)_{k-j}(- a^2 Q^{2}t_r^{5}t_c^{2r+1} ;t_c^2)_j (-a^2Q^{-2} t_r^3 t_c;t_c^2)_{r-k}\cr
 &&\times  (-Q^{2} t_r t_c^{2k+1};t_c^2)_{r-k-i}(-a^2 Q^{-2} t_r t_c;t_c^2)_i\nonumber
\eea
This formula satisfies all the properties summarized in \S \ref{sec:finite-dimension}.

\subsection{Infinite-dimensional HOMFLY homology}

\subsubsection*{Fundamental representation}

\begin{figure}[h]
 \centering
    \includegraphics[width=12cm]{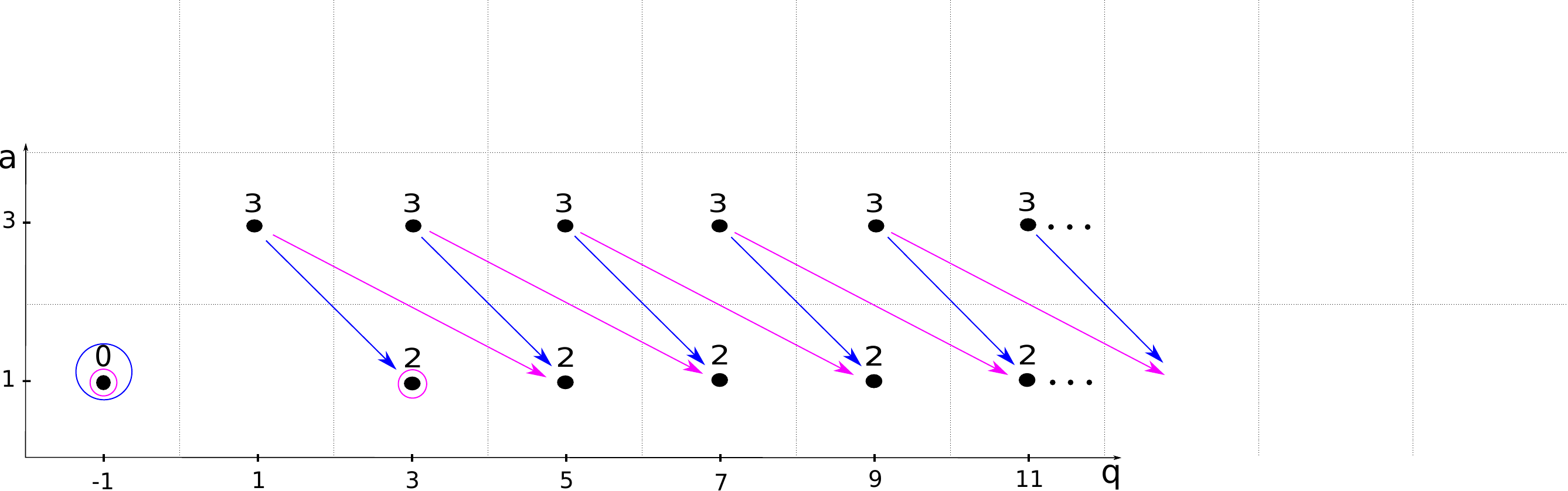}
    \caption{The uncolored positive HOMFLY homology $\scH^{(+)}_{[1],[1]}(T_{2,2})$ of the Hopf link. The ellipses stand for the geometric progressions originating from $1/(1-q^2)$. The blue arrows represent the action of the canceling differential. The dot encircled in blue stands for the generator of $H_*(\scH^{(+)}_{[1],[1]}(T_{2,2}),d^+_{[1]\to[0]})$, which is one-dimensional. The pink arrows represent the action of the differential $d_2$. The dots encircled in pink stand for the generators of the Khovanov homology $H_*(\scH^{(+)}_{[1],[1]}(T_{2,2}),d_{2})$.}\label{fig:HOMFLY-Hopf-1}
\end{figure}

\begin{figure}[h]
\centering
\includegraphics[width=5cm]{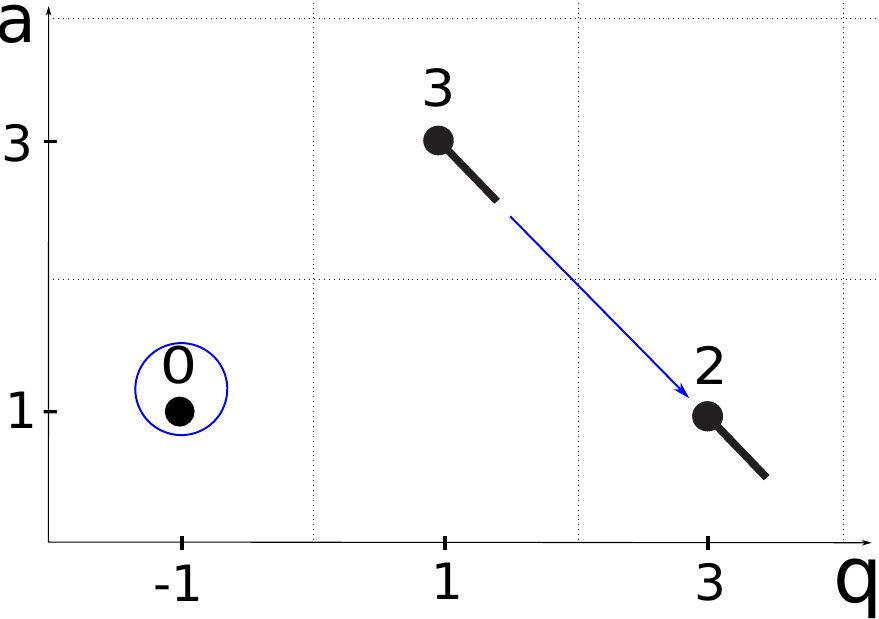}\caption{The uncolored  positive HOMFLY homology $\scH^{(+)}_{[1],[1]}(T_{2,2})$ of the Hopf link in the compact notation, where a straight tail denotes a geometric progression originating from $1/(1-q^2)$. The blue arrow represents the action of the canceling differential. The $S$-invariant of the Hopf link is $S(T_{2,2})=1$.}
\label{fig:uncolor-Hopf+}
\end{figure}

The Poincar\'e polynomial of the uncolored HOMFLY homology $\scH^{(+)}_{[1],[1]}(T_{2,2})$ of the Hopf link (that corresponds to $|q|<1$) is given by
\bea\label{Hopf-uncolor}
\scP^{(+)}_{\yng(1),\yng(1)}(T_{2,2};a,q,t)=aq^{-1}+a q^3 t^2\frac{1+a^2q^{-2}t}{1-q^2}~.
\eea
In the range $|q|<1$, the denominator in the second term of \eqref{Hopf-uncolor} gives a geometric power series $1/(1-q^2)=1+q^2+q^4+\cdots$. Therefore, the reduced HOMFLY homology of the Hopf link is indeed infinite-dimensional, as illustrated in Figure~\ref{fig:HOMFLY-Hopf-1}. However, apart from the geometric progressions, the HOMFLY homology of the Hopf link is very similar to that of the trefoil. (See Figure 12 in \cite{Gorsky:2013jxa} for the HOMFLY homology of the trefoil.) The comparison becomes easier to see if we use a short-hand notation for the geometric power series illustrated in Figure~\ref{fig:uncolor-Hopf+}.


The canceling differential $d^+_{[1]\to[0]}$ acts on the entire infinite-dimensional ``tail'' associated with geometric progressions, so that the homology with respect to $d^+_{[1]\to[0]}$ is one dimensional:
\bea\nonumber
\scP(\scH^{(+)}_{\yng(1),\yng(1)}(T_{2,2}),d^+_{[1]\to[0]})(a,q,t)=\frac{a}{q}~.
\eea
Note that the $S$-invariant of the Hopf link is $S(T_{2,2})=1$. Unlike for knot homology, the differential $d_N$ acts on the link homology non-trivially even in the case of thin links. Therefore, although the HOMFLY homology is infinite-dimensional, the reduced $\frak\fraksl(N)$ homology of a link is finite-dimensional because the differential $d_N$ kills infinitely many generators in $\scH^{(+)}_{\yng(1),\yng(1)}(L)$. For instance, the reduced $\fraksl(2)$ homology of the Hopf link can be obtained by applying the $d_2$ differential to the uncolored HOMFLY homology
\bea\nonumber
\scP^{\fraksl(2)}_{\yng(1),\yng(1)}(T_{2,2};q,t)=\scP(\scH^{(+)}_{\yng(1),\yng(1)}(T_{2,2}),d_2)(a=q^2,q,t) =q+q^5t^2~.
\eea
This result agrees with the reduced Khovanov homology of the Hopf link.
The action of the $d_2$ differential is illustrated in Figure~\ref{fig:HOMFLY-Hopf-1}.
Moreover, for $N>1$, the reduced $\frak\fraksl(N)$ homology of the Hopf link is given by
\be\nonumber
\scP^{\fraksl(N)}_{\yng(1),\yng(1)}(T_{2,2};q,t)=\scP(\scH^{(+)}_{\yng(1),\yng(1)}(T_{2,2}),d_N)(a=q^N,q,t) =q+q^{3+N}t^2\frac{1-q^{2(N-1)}}{1-q^2}~.
\ee

\begin{figure}[h]
 \centering
    \includegraphics[width=11cm]{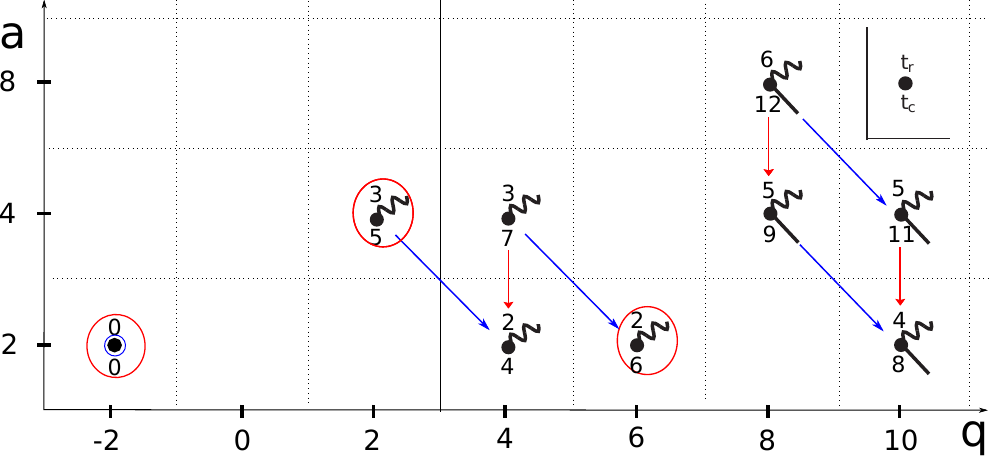}
    \caption{The $([2],[2])$-colored positive HOMFLY homology  $\scH^{(+)}_{[2],[2]}(T_{2,2})$ of the Hopf link in $(a,q,t_r,t_c)$-gradings. The straight and wavy tails represent the factors $1/(1-q^2)$ and $1/(1- q^{4}t_c^{2})$, respectively. The blue and red arrows represent the action of the canceling differential $d^+_{[2]\to [0]}$ and the colored differential $d^+_{[2]\to [1]}$, respectively. The homology $H_*(\scH^{(+)}_{[2],[2]}(T_{2,2}),d^+_{[2]\to [1]})$  is isomorphic to $\scH^{(+)}_{[1],[1]}(T_{2,2})$ depicted in Figure \ref{fig:uncolor-Hopf+}.}\label{fig:HOMFLY-Hopf-2}
\end{figure}

\subsubsection*{Second symmetric representation}

Next, let us consider the colored HOMFLY homology of the Hopf link. The $([2],[2])$-colored HOMFLY polynomial of the Hopf link looks like
\be\label{HOMFLY-poly-Hopf-2}
P_{\yng(2),\yng(2)}({T_{2,2}};a,q)=\frac{a^2(1 - q^2 - q^4 + 2 q^6 - q^{10}  + q^{12} -
 a^2 (q^4 - q^8 + q^{10} + q^{12})+ a^4 q^{10})}{q^2(1-q^2)(1-q^4)}~.
\ee
Starting with this expression, the differentials determine the $([2],[2])$-colored HOMFLY homology of the Hopf link
\bea\label{Hopf-infinite-22}
\scP^{(+)}_{\yng(2),\yng(2)}({T_{2,2}};a,q,t_r,t_c) &=&\frac{a^2}{q^{2}}+\frac{a^2q^4 t_r^2 t_c^4 (1 + q^2 t_c^2)  (1 + a^2 q^{-2}t_r t_c)}{1 - q^4 t_c^2}\cr
&&+\frac{ a^2 q^{10}  t_r^4 t_c^8 (1 + a^2q^{-2} t_c t_r) (1 + a^2 t_r t_c^3 )}{(1 -  q^2) (1 - q^4 t_c^2)}~.
\eea
Even in this case, the $([2],[2])$-colored HOMFLY homology of the Hopf link is also similar to that of the trefoil, apart from the geometric progressions ({\it cf.} Figure 13 in \cite{Gorsky:2013jxa}).

\begin{figure}[h]
 \centering
    \includegraphics[width=8.5cm]{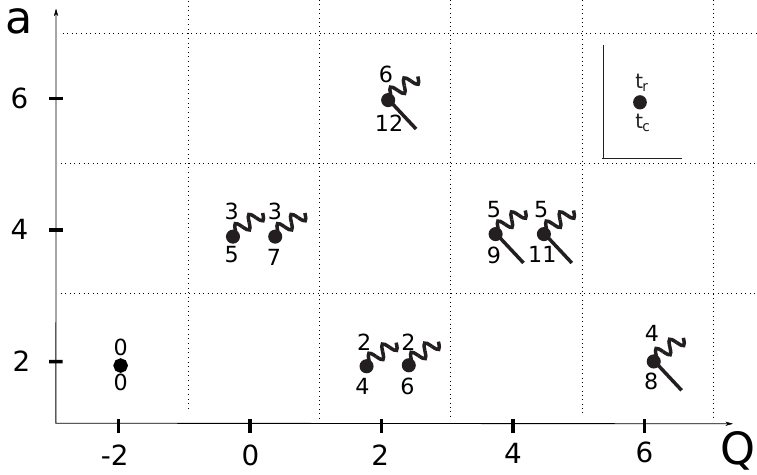}
    \caption{The $([2],[2])$-colored positive HOMFLY homology $\wt\scH^{(+)}_{[2],[2]}(T_{2,2})$ of the Hopf link in $(a,Q,t_r,t_c)$-gradings. The straight and wavy tails represent the factors $1/(1-Q^2)$ and $1/(1- Q^{2}t_c^{2})$, respectively.}\label{fig:quad-Hopf-2}
\end{figure}

Using the definition \eqref{Q} of the auxiliary $Q$-grading, the tilde-version of the $([2],[2])$-colored HOMFLY homology takes the form:
\bea\nonumber
&&\wt\scP^{(+)}_{\yng(2),\yng(2)}(T_{2,2};a,Q,t_r,t_c) =\scP^{(+)}_{\yng(2),\yng(2)}({T_{2,2}};a,Q,t_rQ,t_cQ^{-1})\\
 &=&a^2\left[Q^{-2}+\frac{Q^2 t_r^2 t_c^4 (1 +  t_c^2)  (1 + a^2 Q^{-2}t_r t_c)}{1 - Q^2t_c^2}+\frac{ Q^6  t_r^4 t_c^8 (1 + a^2Q^{-2} t_c t_r) (1 + a^2Q^{-2} t_r t_c^3 )}{(1 -  Q^2) (1 - Q^2 t_c^2)}\right]~.\nonumber
\eea
The ``dot diagram'' is depicted in Figure \ref{fig:quad-Hopf-2}. It is easy to see that the Poincar\'e polynomial satisfies the refined exponential growth property \eqref{REGP-infinite}:
\bea\nonumber
\wt\scP^{(+)}_{\yng(2),\yng(2)}({T_{2,2}};a,Q,t_r=t,t_c=1)=\left[ \wt\scP^{(+)}_{\yng(1),\yng(1)}({T_{2,2}};a,Q,t) \right]^2~.
\eea
In addition, the change of grading \eqref{color-diff-infinite} for the positive colored differential $d^+_{\yng(2)\to \yng(1)}$ acting on the $(\yng(2),\yng(2))$-colored HOMFLY homology becomes manifest in the $(a,Q,t_r,t_c)$-grading conventions:
\be\nonumber
\scP(\wt\scH^{(+)}_{\yng(2),\yng(2)}(T_{2,2}),d^{+}_{\yng(2)\to\yng(1)} )(a,Q,t_r,t_c)=aQ^{-1}t_c\wt\scP^{(+)}_{\yng(1),\yng(1)}(a,Qt_c,t_r,t_c)~.
\ee

\subsubsection*{Arbitrary symmetric representations}
\noindent
As in the case of the finite-dimensional homology, it is convenient to use the $(a,Q,t_r,t_c)$-gradings in order to obtain $([r],[r])$-colored HOMFLY homology.
In fact, the refined growth property \eqref{REGP-infinite} determines the $t_c=1$ specialization of the $([r],[r])$-colored HOMFLY homology
\bea\label{Hopf-specialization}
\wt\scP^{(+)}_{[r],[r]}(T_{2,2};a,Q,t_r,t_c=1)&=&\left[aQ^{-1}\left(1+Q^4 t_r^2\frac{1+a^2Q^{-2}t_r}{1-Q^2}\right)\right]^r\cr
&=&a^{r}Q^{-r} \sum_{i=0}^{r}Q^{4i}t_r^{2i}  \frac{(1+a^2Q^{-2}t_r)^{i}}{(1-Q^{2})^{i}} \binom{r}{i}~.
\eea
The $t_c$-grading then can be determined with the help of differentials. Thus, the binomial $\binom{r}{i}$ in \eqref{Hopf-specialization} is replaced by the quantum binomial ${r \brack i}_{t_c^2}$ with respect to $t_c^2$. In addition, the factors $(1+a^2Q^{-2}t_r)^{i}$ and $(1-Q^{2})^{i}$ in \eqref{Hopf-specialization} are replaced by $t_c^2$-Pochhammer symbols $(-a^2Q^{-2}t_rt_c^*;t_c^2)_{i}$ and $(Q^{2}t_c^{*};t_c^2)_{i}$. The differentials determine the $t_c$-gradings uniquely, so that the Poincar\'e polynomial of the $([r],[r])$-colored positive HOMFLY homology of the Hopf link takes the form
\be\label{Hopf-quad+}
\wt\scP^{(+)}_{[r],[r]}(T_{2,2};a,Q,t_r,t_c)=a^{r}Q^{-r} \sum_{i=0}^{r}Q^{4i}t_r^{2i}t_c^{2{r} i}\frac{(-a^2Q^{-2}t_rt_c;t_c^2)_{i}}{(Q^{2}t_c^{2({r}-i)};t_c^2)_{i}}  {r \brack i}_{t_c^2}~.
\ee
Furthermore, taking into account the unknot factor described in \S \ref{sec:unknot}, the Poincar\'e polynomial of the triply-graded \emph{unreduced} HOMFLY homology of the Hopf link with the $t_c$-grading turns out to agree with the refined Chern-Simons invariant \eqref{refined-CS-sym+} colored by the diagonal representation $([r],[r])$
\bea\label{poincare-rcs}
\overline\scP^{(+)}_{[r],[r]}(T_{2,2};a,q,t_r=1,t_c=t)&=&\overline\scP^{(+)}_{[r]}(\unknot;a,q,t_r=1,t_c=t)\scP^{(+)}_{[r],[r]}(T_{2,2};a,q,t_r=1,t_c=t)\cr
&=&\overline{\rm rCS}^{(+)}_{[r],[r]}(T_{2,2};a,q,t)~.
\eea
In fact, the original motivation behind introducing the four gradings in \cite{Gorsky:2013jxa} was to reconcile different conventions for homological invariants labeled by symmetric representations: $t_r$-grading used in \cite{Gukov:2011ry} and $t_c$-grading used in \cite{Aganagic:2011sg,DuninBarkowski:2011yx,Cherednik:2011nr}.
Note that the relation \eqref{poincare-rcs} holds in the $t_c$-grading because the refined Chern-Simons invariant $\overline{\rm rCS}^{(+)}_{[r]}(T_{2,2p})$ in \eqref{refined-CS-sym+} is obtained by the change of variables \eqref{change-of-variables} used in \cite{Aganagic:2011sg,DuninBarkowski:2011yx,Cherednik:2011nr},

\subsubsection*{Anti-symmetric representations}

\begin{figure}[h]
 \centering
    \includegraphics[width=11cm]{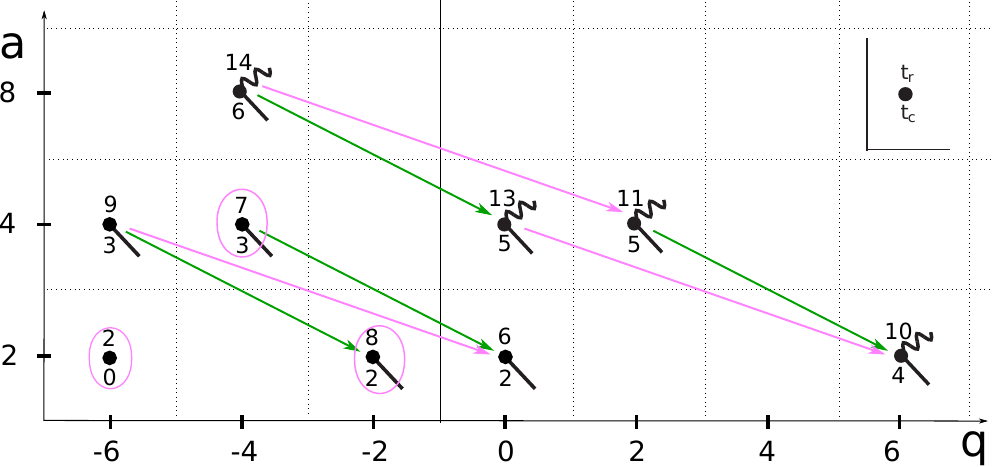}
    \caption{The $([1,1],[1,1])$-colored positive HOMFLY homology $\scH^{(+)}_{[1,1],[1,1]}(T_{2,2})$ of the Hopf link in $(a,q,t_r,t_c)$-gradings where the straight and wavy tails represent the factors $1/(1-q^{2}t_r^2)$ and $1/(1- q^{4})$, respectively. The green and pink arrows represent the actions of the canceling differential $d^+_{[1,1]\to [0]}$ and the colored differential $d^+_{[1,1]\to [1]}$, respectively. The homology $H_*(\scH^{(+)}_{[1,1],[1,1]}(T_{2,2}),d^+_{[1,1]\to [1]})$  is isomorphic to $\scH^{(+)}_{[1],[1]}(T_{2,2})$ shown in Figure \ref{fig:uncolor-Hopf+}.}\label{fig:HOMFLY-Hopf-11+}
\end{figure}

The Poincar\'e polynomial of the $([1^r],[1^r])$-colored positive HOMFLY homology  $\scH^{(+)}_{[1^r],[1^r]}(T_{2,2})$ of the Hopf link follows immediately from the mirror symmetry \eqref{mirror-infinite}:
\bea\label{Hopf-quad-anti+}
\wt\scP^{(+)}_{[1^r],[1^r]}(T_{2,2};a,Q,t_r,t_c)&=&t_r^{r(r-1)}\wt\scP^{(+)}_{[r],[r]}(T_{2,2};a,Q,t_c,t_r)\\
&=&a^{r}Q^{-r}t_r^{r(r-1)} \sum_{i=0}^{r}Q^{4i}t_r^{2ri}t_c^{2 i}\frac{(-a^2Q^{-2}t_rt_c;t_r^2)_{i}}{(Q^{2}t_r^{2({r}-i)};t_r^2)_{i}}  {r \brack i}_{t_r^2}\nonumber.
\eea
As in the case of symmetric representations, the Poincar\'e polynomial of $\overline\scH^{(+)}_{[1^r],[1^r]}(T_{2,2})$  with the $t_c$-degree coincides with the $([1^r],[1^r])$-colored refined Chern-Simons invariant \eqref{refined-CS-anti+}:
\be\nonumber
\overline\scP^{(+)}_{[1^r],[1^r]}(T_{2,2};a,q,t_r=1,t_c=t) \; = \; \overline{\rm rCS}^{(+)}_{[1^r],[1^r]}(T_{2,2};a,q,t)~.
\ee
The $([1^r],[1^r])$-colored positive HOMFLY homology  $\scH^{(+)}_{[1^r],[1^r]}(T_{2,2})$ of the Hopf link enjoys the action of the positive colored differentials $d^{+}_{[1^r]\to [1^k]}$.
For instance, the $(\yng(1,1),\yng(1,1))$-colored HOMFLY homology in the $(a,q,t_r,t_c)$-gradings can be written as
\bea\nonumber
&&\scP^{(+)}_{\yng(1,1),\yng(1,1)}(T_{2,2};a,q,t_r,t_c)=\wt\scP^{(+)}_{\yng(1,1),\yng(1,1)}(T_{2,2};a,q^2,t_r/q,t_cq)\\
&=&a^2\left[q^{-6} t_r^2 + \frac{ t_c^2 t_r^6 (1 + a^2 q^{-4}t_r t_c) (1 +q^{-2} t_r^2 )}{
   1 - q^2 t_r^2} +
   \frac{ q^6t_r^{10} t_c^4  (1 + a^2 q^{-4}t_r t_c) (1 + a^2 q^{-6}t_r^3 t_c)}{(1 -
       q^4) (1 - q^2 t_r^2)}\right]~.\nonumber
\eea
and it is easy to verify that the homology with respect to $d^{+}_{\yng(1,1)\to \yng(1)}$ (pink arrows in Figure \ref{fig:HOMFLY-Hopf-11+}) is isomorphic to the uncolored HOMFLY homology
\be\nonumber
H_*(\scH^{(+)}_{\yng(1,1)}(T_{2,2}),d^{+}_{\yng(1,1)\to \yng(1)})\cong\scH^{(+)}_{\yng(1)}(T_{2,2})~.
\ee

\begin{figure}[h]
 \centering
    \includegraphics[width=12cm]{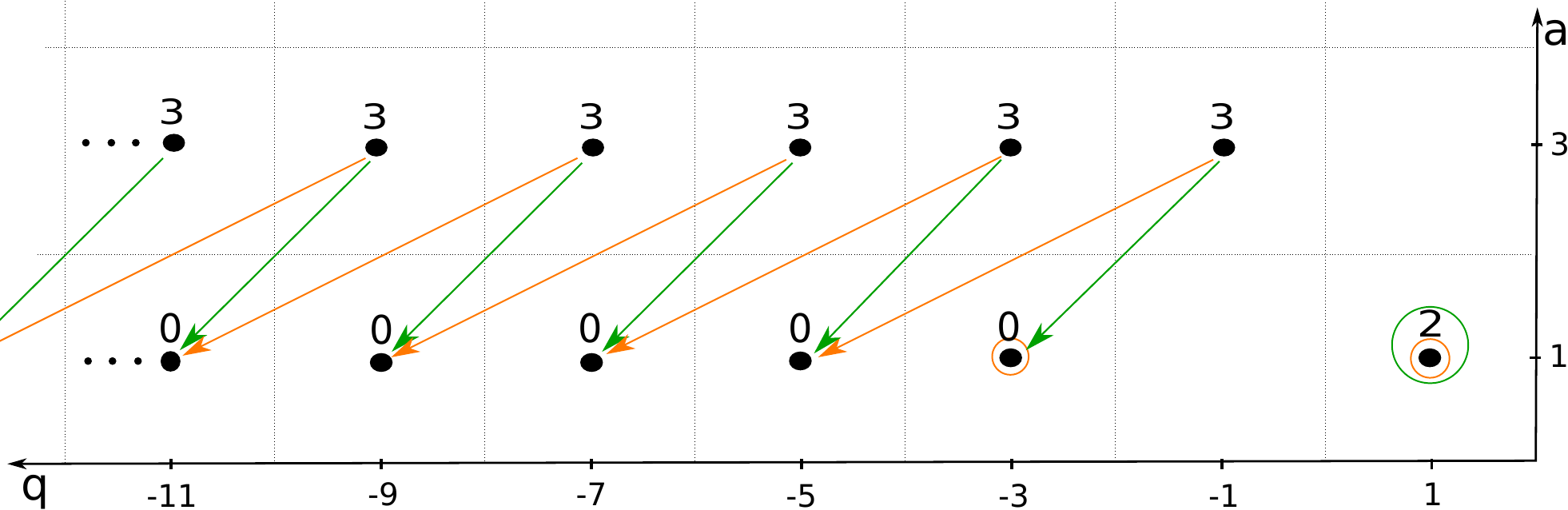}
    \caption{The uncolored negative HOMFLY homology $\scH^{(-)}_{[1],[1]}(T_{2,2})$ of the Hopf link. The ellipses stand for the geometric progressions originating from $1/(1-q^{-2})$. The green and orange arrows represent the actions of the canceling differential $d^-_{[1]\to[0]}$ and the $d_{-2}$ differential, respectively. The dots encircled in orange represent the generators of $H_*(\scH^{(-)}_{[1],[1]}(T_{2,2}),d_{-2})$.}\label{fig:HOMFLY-Hopf-1-}
\end{figure}

\noindent
In the range $|q|>1$ one should expand $1/(1-q^{-2})=1+q^{-2}+q^{-4}+\cdots$.
Therefore, the Poincar\'e polynomial of the corresponding HOMFLY homology can be written as
\bea\label{Hopf-uncolor-}
\scP^{(-)}_{\yng(1),\yng(1)}(T_{2,2};a,q,t)=aqt^2+a q^{-3} \frac{1+a^2q^{2}t^3}{1-q^{-2}}~.
\eea

\begin{figure}[h]
\centering
\includegraphics[width=5cm]{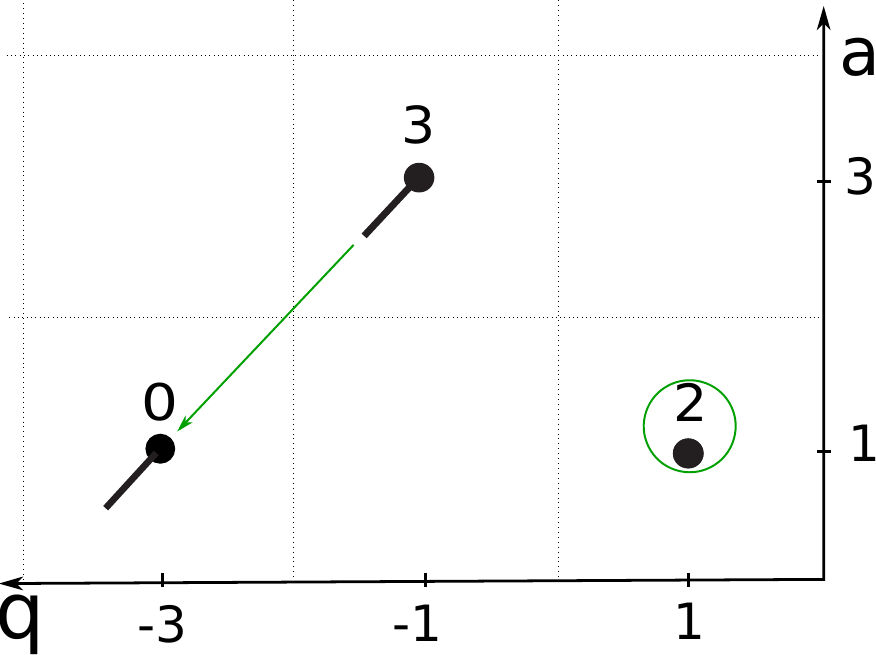}\caption{The uncolored negative HOMFLY homology $\scH^{(-)}_{[1],[1]}(T_{2,2})$ of the Hopf link. To avoid clutter, we use the straight tails to represent the geometric progressions associated with the factor $1/(1-q^{-2})$. }
\label{fig:uncolor-Hopf-}
\end{figure}

\noindent
For the uncolored HOMFLY homology, the dot diagram (in Figure \ref{fig:uncolor-Hopf-}) for $|q|>1$  is a mirror image of that (Figure \ref{fig:uncolor-Hopf+}) for $|q|<1$. Moreover, the two are related via
\bea\nonumber
\scP^{(-)}_{\yng(1),\yng(1)}(T_{2,2};a,q,t)=\scP^{(+)}_{\yng(1),\yng(1)}(T_{2,2};at^2,q^{-1},t^{-1})~.
\eea
The HOMFLY homology $\scH^{(-)}_{\yng(1),\yng(1)}(T_{2,2})$ comes equipped with the action of $d_N$ differentials with $N<0$. For instance, the action of the $d_{-2}$ differential is depicted in Figure~\ref{fig:HOMFLY-Hopf-1-}, where the Poincar\'e polynomial of the homology with respect to $d_{-2}$ can be related to the $\fraksl(2)$ Khovanov homology:
\be\nonumber
\scP(\scH^{(-)}_{\yng(1),\yng(1)}(T_{2,2}),d_{-2})(a=q^{-2},q,t)=q^{-5}+q^{-1}t^2=t^2\scP^{\fraksl(2)}_{\yng(1),\yng(1)}(T_{2,2};q^{-1},t^{-1})~.
\ee

\noindent
Taking the geometric progressions of the $(\yng(2),\yng(2))$-colored HOMFLY polynomial \eqref{HOMFLY-poly-Hopf-2} in the negative powers of $q$, one can obtain its categorification $\scH^{(-)}_{\yng(2),\yng(2)}(T_{2,2})$ in the $(a,q,t_r,t_c)$-gradings by using the differentials. The result is
\bea
\label{Hopf-2-}
&&\scP^{(-)}_{\yng(2),\yng(2)}(T_{2,2};a,q,t_r.t_c) = \\
&=&a^2 \left[q^2 t_r^4t_c^6 + q^{-2}  t_r^2t_c^2\frac{(1 + q^2 t_c^2) (1 + a^2 q^4t_r^3 t_c^5 )}{ (1 - q^{-2}) } + q^{-8}t_c^{-2}\frac{(1 + a^2 q^4 t_r^3 t_c^5 ) (1 +
      a^2 q^6t_r^3 t_c^7 )}{(1  -q^{-2}) (1 -
      q^{-4} t_c^{-2} )}\right]~.\nonumber
\eea
\begin{figure}[h]
 \centering
    \includegraphics[width=11cm]{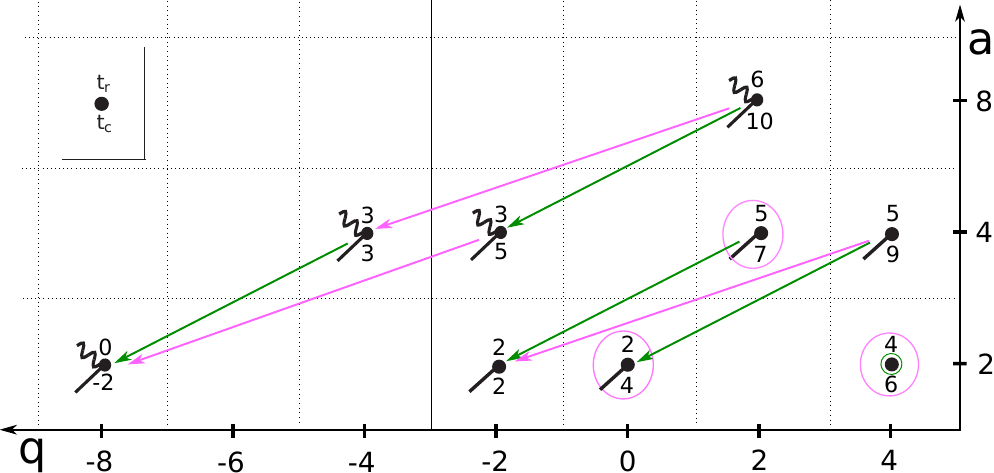}
    \caption{The $([2],[2])$-colored negative HOMFLY homology $\scH^{(-)}_{[2],[2]}(T_{2,2})$ of the Hopf link in $(a,q,t_r,t_c)$-gradings. The straight and wavy tails represent the factors $1/(1-q^{-2})$ and $1/(1- q^{-4}t_c^{-2})$, respectively. The green and pink arrows represent the actions of the canceling differential $d^-_{[2]\to[0]}$ and the colored differential $d^-_{[2]\to [1]}$, respectively. The homology $H_*(\scH^{(-)}_{[2],[2]}(T_{2,2}),d^-_{[2]\to [1]})$  is isomorphic to $\scH^{(-)}_{[1],[1]}(T_{2,2})$ depicted in Figure \ref{fig:uncolor-Hopf-}.}\label{fig:HOMFLY-Hopf-2-}
\end{figure}

\begin{figure}[h]
 \centering
    \includegraphics[width=8.5cm]{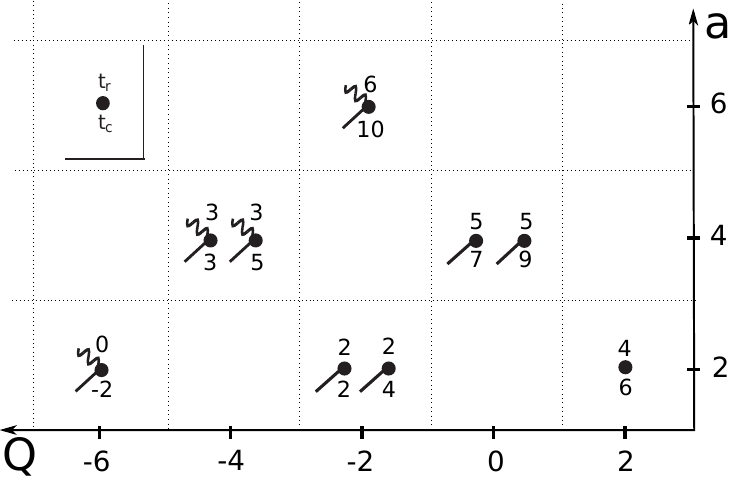}
    \caption{The $([2],[2])$-colored negative HOMFLY homology $\wt\scH^{(-)}_{[2],[2]}(T_{2,2})$ of the Hopf link in $(a,Q,t_r,t_c)$-gradings where the straight and wavy tails represent the factors $1/(1-Q^{-2})$ and  $1/(1- Q^{-2}t_c^{-2})$, respectively.}\label{fig:quad-Hopf-2-}
\end{figure}

\noindent
Using the definition \eqref{Q} of the $Q$-grading, the tilde version of the $(\yng(2),\yng(2))$-colored negative HOMFLY homology of the Hopf link is
\bea\nonumber
&&\wt\scP^{(-)}_{\yng(2),\yng(2)}(T_{2,2};a,Q,t_r,t_c) =\scP^{(-)}_{\yng(2),\yng(2)}({T_{2,2}};a,Q,t_rQ,t_cQ^{-1})\\
&=&a^2 \left[Q^2 t_r^4 t_c^6  +Q^{-2} t_r^2 t_c^2 \frac{(1 + t_c^2) (1 + a^2 Q^2 t_r^3 t_c^5 )}{ (1 - Q^{-2})}+Q^{-6}  t_c^{-2}  \frac{(1 + a^2 Q^2t_r^3 t_c^5 ) (1 +
      a^2 Q^2 t_r^3t_c^7 )}{(1 - Q^{-2})   (1 - Q^{-2} t_c^{-2}) }\right]~.\nonumber
\eea
The dot diagram is shown in Figure~\ref{fig:quad-Hopf-2-}. It is easy to see that the Poincar\'e polynomial satisfies the refined exponential growth property \eqref{REGP-infinite}:
\bea\nonumber
\wt\scP^{(-)}_{\yng(2),\yng(2)}({T_{2,2}};a,Q,t_r=t,t_c=1)=\left[ \wt\scP^{(-)}_{\yng(1),\yng(1)}({T_{2,2}};a,Q,t) \right]^2~.
\eea
The action of the negative colored differentials is illustrated in Figure~\ref{fig:HOMFLY-Hopf-2-}.
The regrading \eqref{color-diff-infinite} associated to this action
becomes manifest in the $(a,Q,t_r,t_c)$-gradings:
\be\nonumber
\scP(\wt\scH^{(-)}_{\yng(2),\yng(2)}(T_{2,2}),d^{-}_{\yng(2)\to\yng(1)} )(a,Q,t_r,t_c)=aQt_r^2t_c^4\wt\scP^{(-)}_{\yng(1),\yng(1)}(T_{2,2};a,Q,t_r,t_c)~.
\ee

The refined exponential growth \eqref{REGP-infinite} and the differentials \eqref{color-diff-infinite} also determine the Poincar\'e polynomial of the $([r],[r])$-colored negative HOMFLY homology $\scH^{(-)}_{[r],[r]}(T_{2,2})$ of the Hopf link:
\bea\label{Hopf-quad-}
\wt\scP^{(-)}_{[r],[r]}(T_{2,2};a,Q,t_r.t_c)&=&a^r Q^r t_r^{2 r} t_c^{r(r+1)}\sum_{i=0}^{r} Q^{-4 i} t_r^{-2 i} t_c^{-2  ri} \frac{(-a^2 Q^2 t_r^3 t_c^{1 + 2 r} ;t_c^2)_i}{(Q^{-2};t_c^{-2})_i} {r \brack i}_{t_c^2}.\qquad\quad
\eea
To compare with the refined Chern-Simons invariant \eqref{refined-CS-sym-} in the range of $|q|>1$,
one should use the $t_r$-grading due to the change of variables \eqref{cv-FGS} used in \cite{Gukov:2011ry}.
In fact, the Poincar\'e polynomial of the unreduced HOMFLY homology $\overline\scH^{(-)}_{[r],[r]}(T_{2,2})$
with the $t_r$-grading is equal to the $([r],[r])$-colored refined Chern-Simons invariant \eqref{refined-CS-sym-}:
\be\nonumber
\overline\scP^{(-)}_{[r],[r]}(T_{2,2};a,q,t_r=t,t_c=1)=\overline{\rm rCS}^{(-)}_{[r],[r]}(T_{2,2};a,q,t)~.
\ee

\noindent
The Poincar\'e polynomial of the $([1^r],[1^r])$-colored negative HOMFLY homology of the Hopf link  simply follows from the mirror symmetry \eqref{mirror-infinite}:
\bea\label{Hopf-quad-anti-}
\wt\scP^{(-)}_{[1^r],[1^r]}(T_{2,2};a,Q,t_r.t_c)&=&t_r^{r(r-1)}\wt\scP^{(-)}_{[r],[r]}(T_{2,2};a,Q,t_c.t_r)\\
&=&a^r Q^r t_r^{2 r^2} t_c^{2r} \sum_{i=0}^{r} Q^{-4 i}  t_r^{-2 ri}t_c^{-2 i}\frac{(-a^2 Q^2 t_r^{1 + 2 r} t_c^3 ;t_r^2)_i}{(Q^{-2};t_r^{-2})_i} {r \brack i}_{t_r^2}.\nonumber
\eea
Unlike the above three cases of $\overline \scH^{(\pm)}_{[r],[r]}(T_{2,2})$ and $\overline \scH^{(+)}_{[1^r],[1^r]}(T_{2,2})$,
the Poincar\'e polynomial of $\overline \scH^{(-)}_{[1^r],[1^r]}(T_{2,2})$  does \emph{not} match
the refined Chern-Simons invariant \eqref{refined-CS-anti-}, even in the $t_r$-grading.

\begin{figure}[h]
 \centering
    \includegraphics[width=11cm]{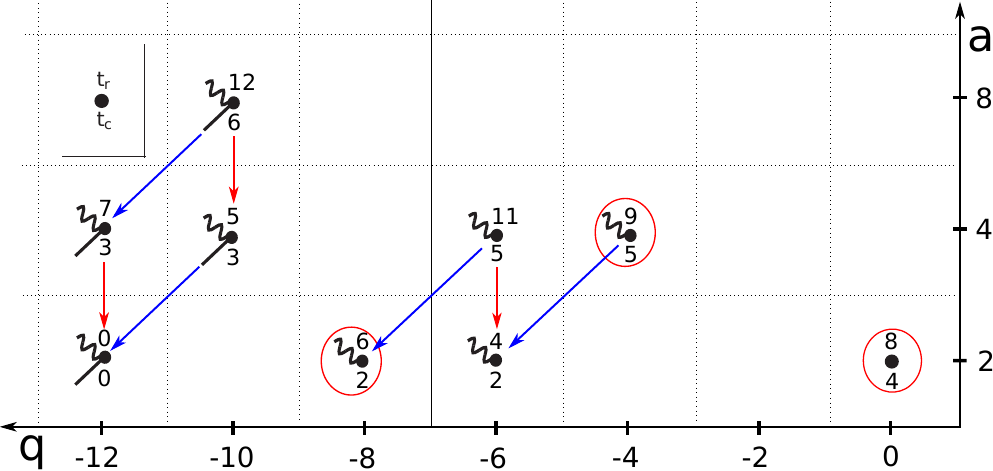}
    \caption{The $([1,1],[1,1])$-colored negative HOMFLY homology $\scH^{(-)}_{[1,1],[1,1]}(T_{2,2})$ of the Hopf link in $(a,q,t_r,t_c)$-gradings where the straight and wavy tails represent the factors $1/(1-q^{-2}t_r^{-2})$ and $1/(1- q^{-4})$, respectively. The blue and red arrows represent the actions of the canceling differential $d^-_{[1,1]\to [0]}$ and the colored differential $d^-_{[1,1]\to [1]}$, respectively. The homology $H_*(\scH^{(-)}_{[1,1],[1,1]}(T_{2,2}),d^-_{[1,1]\to [1]})$  is isomorphic to $\scH^{(-)}_{[1],[1]}(T_{2,2})$ depicted in Figure \ref{fig:uncolor-Hopf-}.}\label{fig:HOMFLY-Hopf-11-}
\end{figure}

Recall that the negative colored HOMFLY homology of the Hopf link is equipped only with negative colored differentials $d^-_{[1^r]\to[1^k]}$.
For instance, the action of the negative colored differentials on the $(\yng(1,1),\yng(1,1))$-colored HOMFLY homology $\scH^{(-)}_{\yng(1,1),\yng(1,1)}(T_{2,2})$ is shown in Figure~\ref{fig:HOMFLY-Hopf-11-}.
The Poincar\'e polynomial in the $(a,q,t_r,t_c)$-gradings has the form
\bea
&&\scP^{(-)}_{\yng(1,1),\yng(1,1)}(T_{2,2};a,q,t_r,t_c)=\wt\scP^{(-)}_{\yng(1,1),\yng(1,1)}(T_{2,2};a,q^2,t_r/q,t_cq)\cr
&=&a^2\left[  t_r^8t_c^4+ q^{-6}t_r^4t_c^2\frac{(1 + q^{-2}t_r^2) (1 + a^2 q^2 t_c^3 t_r^5)}{  (1 -q^{-4})}+q^{-12} \frac{(1 +
      a^2 q^2 t_r^5 t_c^3) (1 + a^2t_r^7 t_c^3 )}{(1 -q^{-4})  (1 - q^{-2} t_r^{-2})}\right]~.\qquad
\eea
In particular, the homology with respect to $d^-_{\yng(1,1)\to\yng(1)}$  is isomorphic to the uncolored homology with the following regrading:
\be
\scP(\wt\scH^{(-)}_{\yng(1,1),\yng(1,1)}(T_{2,2}),d^{-}_{\yng(1,1)\to\yng(1)} )(a,Q,t_r,t_c)=aQt_r^6t_c^2\wt\scP^{(-)}_{\yng(1),\yng(1)}(T_{2,2};a,Q,t_r,t_c)~.
\ee

\begin{figure}[h]
 \centering
    \includegraphics[width=8cm]{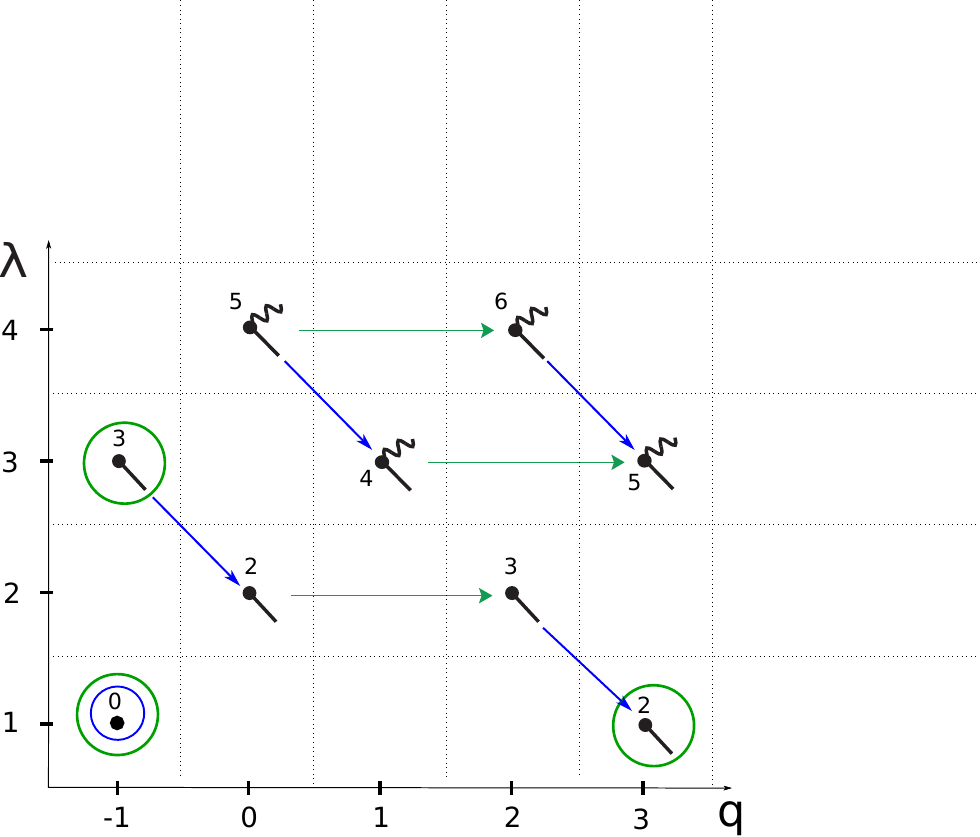}
    \caption{The uncolored Kauffman homology of the Hopf link where the straight and wavy tails represent the power series expansion of the factors $1/(1-q^2)$ and $1/(1-a q^{-1})$, respectively. The blue and green arrows show the action of the canceling differential and the universal differential, respectively. It is easy to see that the homology $H_*(\scH^{\Kauffman}(T_{2,2}),d^\univ_{\to})$ is isomorphic to the HOMFLY homology $\scH^{(+)}_{[1],[1]}(T_{2,2})$ shown in Figure \ref{fig:uncolor-Hopf+}.}\label{fig:Kauffman-Hopf}
\end{figure}

\subsection{Uncolored Kauffman homology}
The Kauffman polynomial of the Hopf link looks like
\bea
F({T_{2,2}};a,q) =\frac{a (1 - q^2 + q^4 + q a -
   q^3 a - a^2 + q^2 a^2-
   q^4 a^2)}{q (1 - q^2) }~.
\eea
Using the structural properties and its relation to the HOMFLY homology found earlier, it is easy to
reconstruct the categorification of this Kauffman invariant, which is a triply-graded homology theory with the Poincar\'e polynomial
\be
\scF({T_{2,2}};a,q,t) = a q^{-1} + \frac{a^2t^2  (1+a^{-1} q^3  ) (1+ a q^{-1} t)}{1-q^2}  +  \frac{a^3 q t^4 (1+ a q^{-1} t)(1+  q^{2} t) }{(1-q^2)(1-a q^{-1})} \,.
\ee
As we will see in other examples below,
the factors $(1-q^2)$ and $(1-a q^{-1})$ in the denominator are ubiquitous in the Poincar\'e polynomial of the Kauffman homology of links.
Note that the ``dot diagram'' shown in Figure \ref{fig:Kauffman-Hopf} is similar to that of the trefoil \cite{Gukov:2005qp},
apart from the geometric progressions.
The homology $H_*(\scH^{\Kauffman}(T_{2,2}),d^\univ_{\to})$ with respect to the universal differential $d^\univ_{\to}$ is isomorphic to the HOMFLY homology \eqref{Hopf-uncolor} of the Hopf link
\bea
\scF(\scH^{\Kauffman}(T_{2,2}),d^\univ_{\to})(a,q,t)&=& a q^{-1} + \frac{a q^3 t^2+a^3q^{-1}t^3}{1-q^2}\cr
&=&q^{S(T_{2,2})}\scP^{(+)}_{\yng(1),\yng(1)}(T_{2,2};a/q,q,t)~.
\eea


\begin{figure}[h]
 \centering
 \includegraphics[width=14cm]{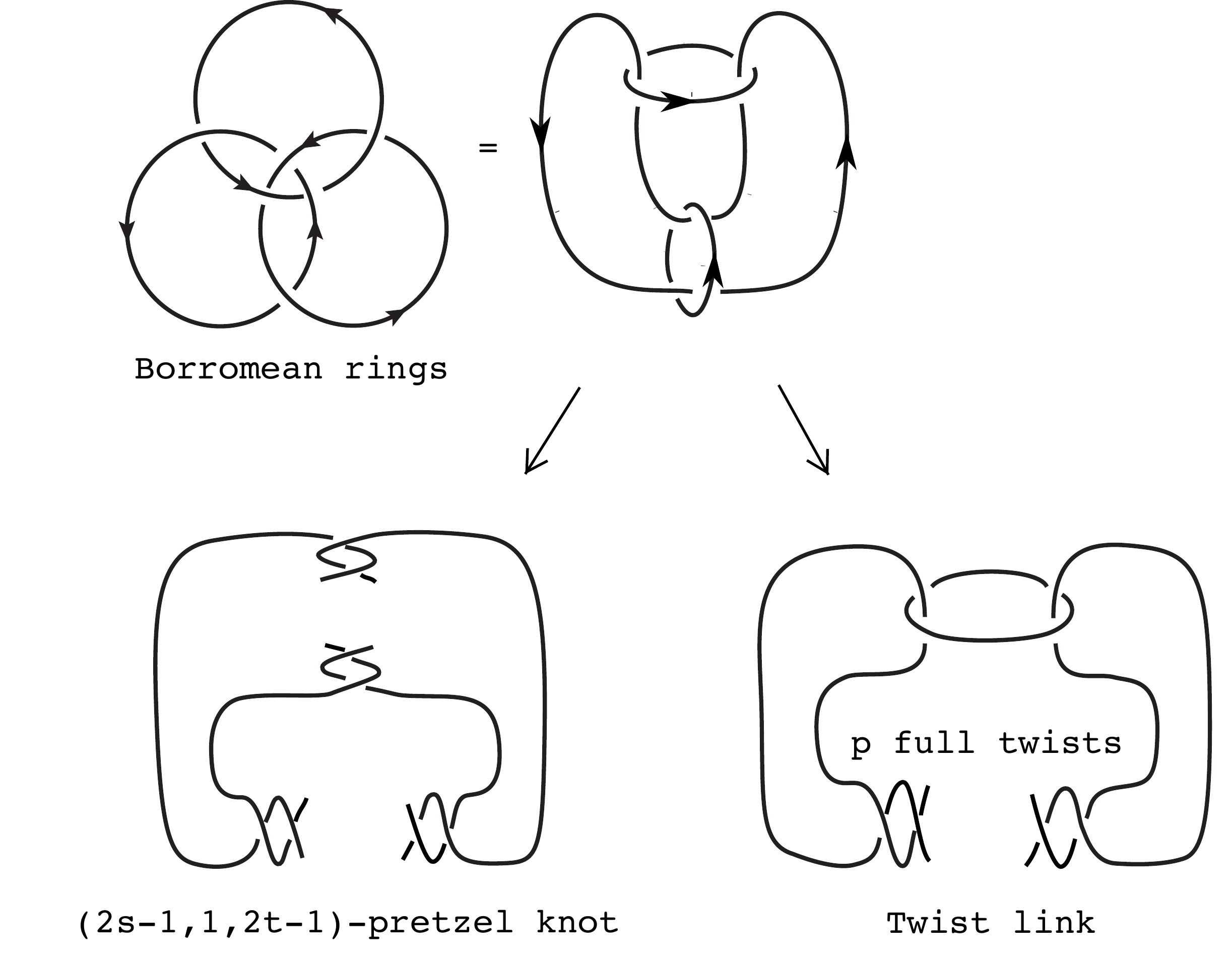}
    \caption{Surgeries along the Borromean rings.}\label{fig:surgery}
\end{figure}


\section{Cyclotomic expansions of link invariants}
\label{sec:cyclotomic}

In this section we provide succinct expressions, in the form of cyclotomic expansions, for the colored HOMFLY invariants and Poincar\'e polynomials (which we also refer to as superpolynomials) of hyperbolic knots and links obtained from the Borromean rings by surgeries. Although the Rosso-Jones formulae \cite{Rosso:1993vn,lin2010hecke,Brini:2011wi} provide a closed form for quantum group invariants of torus links colored by arbitrary representations, it is very difficult to compute the colored quantum group invariants for other knots and links. So far, closed form expressions for the HOMFLY invariants colored by arbitrary symmetric representations have been obtained for only a small number of non-torus links \cite{Kawagoe:2012bt,Arthamonov:2013rfa}. As we show below, for a certain family of hyperbolic links, Habiro's cyclotomic expansions of colored Jones polynomials can be generalized not only to HOMFLY invariants colored by arbitrary symmetric representations, but also to Poincar\'e polynomials of putative link homologies.

We will use the results of this section to determine and discuss properties of the corresponding associated varieties in \S \ref{sec:associated}. However, the results presented in this section are interesting in their own right, and deserve further analysis.

\subsection{Cyclotomic expansions of colored Jones polynomials}

In \cite{Habiro:2000,Habiro:2002,Habiro:2008} Habiro introduced the cyclotomic expansion of colored Jones polynomials. In particular, he studied a large class of knots and links constructed from the Borromean rings by surgeries. Let $A_1$, $A_2$, and $A_3$ denote the components of the Borromean rings. Following the standard terminology, we call the two-component link $(A_1 \sqcup A_2)_{A_3; -1/p}$ obtained by the surgery along the component $A_3$ with framing $-1/p$ the twist link $L_p$. Similarly, the knot $(A_1)_{A_2,A_3;-1/s,-1/t}$ obtained by the rational surgery on the components $A_2$ and $A_3$ with framing $-1/s$ and $-1/t$, respectively, is the $(2s-1,1,2t-1)$-pretzel knot $K_{s,t}$ shown in Figure~\ref{fig:surgery}. For special values of the surgery coefficients we have further identifications. Thus, for $p=1$ the twist link $L_{p=1}$ is the famous Whitehead link. The knot $K_{s=2,t=2}$ is the knot $\bf 7_4$ in the Rolfsen table, while the $(1,1,2t-1)$-pretzel knot $K_{s=1,t}$ is better known as the twist knot $K_t$.

The cyclotomic expansions of the unreduced colored Jones polynomials for these knots and links take the following beautiful forms \cite{Habiro:2008}:
\bea\label{Habiro}
\overline J^{\fraksl(2)}_{[r_1]}(K_{s,t};q)  &=&\sum_{i=0}^{\infty}(-1)^i \omega_{s,i}(q)\omega_{t,i}(q)\bb{{r_1}+1+i}{2i+1}\{2i+1\}_{2i}\,, \\
\overline  J^{\fraksl(2)}_{[r_1],[r_2]}(L_p;q)  &=&\sum_{i=0}^{\infty}(-1)^i \omega_{p,i}(q)\bb{{r_1}+1+i}{2i+1}\bb{{r_2}+1+i}{2i+1}
    \{i\}!\{2i+1\}_{2i}\,,\cr
     \overline J^{\fraksl(2)}_{[r_1],[r_2],[r_3]}(\BR;q)
    &=&\sum_{i=0}^{\infty}(-1)^i
    \bb{{r_1}+1+i}{2i+1} \bb{{r_2}+1+i}{2i+1}\bb{{r_3}+1+i}{2i+1}
    (\{i\}!)^2\{2i+1\}_{2i}\,.\nonumber
    \eea
where we use the following notations
\bea\nonumber
& & \{i\}={q}^{i}-{q}^{-i}\ ,  \quad \{i\}_{n}=\{i\}\{i-1\}\cdots\{i-n+1\}  ~, \cr
&& \{n\}!=\{n\}\{n-1\}\cdots \{1\} \ ,  \quad
\bb{n}{k}= \frac{\{n\}!}{\{k\}!\{n-k\}!}\,~.
\eea
It is easy to see that the surgery with framing $-1/p$ along the $j$-th component results in replacing the factor $\bb{r_j+i+1}{2i+1}\{i\}! $ by the twist element $\omega_{p,i}$:
\bea\label{twist-element-jones}
\omega_{p,i}(q)= q^{i(i+3)/2} \sum_{j=0}^i  (-1)^j  q^{j(2p(j+1)+j-1)} \frac{1-q^{2(2j+1)}}{(q^{2(j+1)};q^2)_{i+1}} { i \brack j }_{q^2}~.
\eea
In fact, the twist elements are cyclotomic functions. As shown in \cite[Appendix A]{Nawata:2012pg}, the expression \eqref{twist-element-jones} is equivalent to the twist element defined in  \cite[Prop. 14.4]{Habiro:2008}.
To obtain the formulae \eqref{Habiro}, it is convenient to introduce a basis  $\{p'_{k}\}$ of $U_q(\fraksl(2))$ representations
\bea\label{basechange}
[r]=\sum_{i=0}^r\bb{r+i+1}{2i+1}\{i\}! \ p'_{i}~.
\eea
It has been proven that the $\fraksl(2)$ quantum invariants of the Borromean rings colored by these representations are particularly simple
\bea\label{BR-simple}
\overline J^{\fraksl(2)}_{p'_{i_1},p'_{i_2},p'_{i_3}}(\BR;q)=\left\{ \begin{array}{ll}(-1)^{i}\frac{\{2i+1\}_{i+1}}{\{1\}}& {\rm if}  \ i_1=i_2=i_3=i\\ 0 & {\rm otherwise} \end{array} \right.~.
\eea
Therefore, using \eqref{basechange} one can obtain \eqref{Habiro}.

\subsection{Cyclotomic expansions of colored HOMFLY invariants}

It was pointed out in \cite{Nawata:2012pg} that the colored HOMFLY invariants of the twist knots also have cyclotomic expansions similar to those of the colored Jones polynomials \eqref{Habiro}. Therefore, it is natural to expect that the colored HOMFLY invariants of the Borromean rings have a similar structure. In order to see that this is indeed the case, let us introduce a basis  $\{p'_{k}\}$ of representations of $U_q(\frak\fraksl(N))$:
\bea\nonumber
[r]=\sum_{i=0}^r F_{r,i}p'_{i}~,
\eea
where, replacing $q^N$ with $a$, we get
\bea\label{basechange2}
F_{r,i}(a,q) &=&(-1)^i a^{-(r - i )} q^{(i (-3 + 3 i - 4 r) + 2 r)/2}\frac{(q^2;q^2)_i(a^2q^{4i};q^2)_{r-i}}{(q^2;q^2)_{r-i}} \, .
\eea
Then, as in the $\fraksl(2)$ case \eqref{BR-simple}, the quantum $\frak\fraksl(N)$ invariants of the Borromean rings colored by these representations all vanish except in the special case
\bea\label{BR-simple-2}
\overline P_{p'_{i},p'_{i},p'_{i}}(\BR;a,q)=a^{-3i}q^{(7-3i)i/2} \frac{(a^2;q^2)_{2i}(a^2q^{-2};q^2)_{i}}{[(q^2;q^2)_i]^2} ~.
\eea
It is straightforward to check that \eqref{basechange2} and \eqref{BR-simple-2} reduce to \eqref{basechange} and \eqref{BR-simple}, respectively, when $a=q^2$. Hence, if we define $G_i(a,q)\equiv \overline P_{p'_{i},p'_{i},p'_{i}}(\BR;a,q)$, then the unreduced  HOMFLY invariants for this class of knots and links obtained from the Borromean rings and colored by arbitrary symmetric representations can be written in the succinct form
\bea\label{cyclotomic-HOMFLY}
\overline P_{[r_1]}(K_{s,t};a,q)  &=&\sum_{i=0}^{\infty}\wt \omega_{s,i}(a,q)\wt\omega_{t,i}(a,q)F_{r_1,i}(a,q)G_{i}(a,q)\,, \cr
\overline  P_{[r_1],[r_2]}(L_p;a,q)  &=&\sum_{i=0}^{\infty}\wt \omega_{p,i}(a,q)F_{r_1,i}(a,q)F_{r_2,i}(a,q)G_{i}(a,q)\,,\cr
     \overline P_{[r_1],[r_2],[r_3]}(\BR;a,q)
    &=&\sum_{i=0}^{\infty}
   F_{r_1,i}(a,q)F_{r_2,i}(a,q)F_{r_3,i}(a,q)G_{i}(a,q)\,,
\eea
where the twist element $\tilde \omega_{p,i}$ is given by
\bea
\tilde \omega_{p,i}(a,q)&=&a^i q^{i(i-1)/2} \sum_{j=0}^i  (-1)^j a^{2pj} q^{(2p+1)j(j-1)} \frac{1-a^2q^{2(2j-1)}}{(a^2q^{2(j-1)};q^2)_{i+1}} { i\brack j }_{q^2}.
\eea
Here the summations in \eqref{cyclotomic-HOMFLY} are indeed finite because $F_{r,i}(a,q)=0$ for $i>r$.
We verified that the expressions \eqref{cyclotomic-HOMFLY} are consistent with the colored HOMFLY invariants of the pretzel knots \cite{Mironov:2013qaa}, the Whitehead link \cite{Kawagoe:2012bt,Arthamonov:2013rfa} and the Borromean rings \cite{Arthamonov:2013rfa}.

Furthermore, Habiro has introduced unified quantum invariants of 3-manifolds which become the WRT invariants at a root of unity~\cite{Habiro:2008}. Actually, the formula
\bea\nonumber
\overline P(M_{s,t,u};a,q)  &=&\sum_{i=0}^{\infty}\wt \omega_{s,i}(a,q)\wt\omega_{t,i}(a,q)\wt\omega_{u,i}(a,q)G_{i}(a,q)
\eea
provides an $a$-deformation of his unified WRT invariants for the integral homology sphere $M_{s,t,u}$. It would be interesting to further investigate its mathematical and physical meaning, as well as its relation to modular forms \cite{Bringmann:2011}.

\subsection{Cyclotomic expansions of colored superpolynomials}

In the case of the pretzel knot $K_{s,t}$, the cyclotomic expansion can be extended to the level of  superpolynomials. In particular, the formulae for the superpolynomials of the twist knots $K_{s=1,t}$ are given in \cite{Nawata:2012pg,Fuji:2012pi}.
We postulate that their generalization to the \emph{reduced} quadruply-graded colored HOMFLY homology of the pretzel knot $K_{s,t}$ takes form
\bea\label{superpoly-pretzel}
&&\wt\scP_{[r]}(K_{s,t};a,Q,t_r,t_c)=\cr
&=&(-t_rt_c)^{-r({\rm sgn}(s)+{\rm sgn}(t))/2} \sum_{i=0}^{r}\varpi_{s,i}(a,t_r,t_c)\varpi_{t,i}(a,t_r,t_c)\\
&& \hspace{4cm}\times (-a^2t_c^{2(r-1)})^{-i} (-a ^2Q^{-2}  t_rt_c;t_c^2)_{i} (- a ^2Q^2t_r^3 t_c^{1+2r};t_c^2)_{i } {r \brack i}_{t_c^2} .\nonumber
\eea
where the twist elements are defined as
\bea\label{twisting3}
&&\varpi_{p,i}(a,t_r,t_c)=\cr
&=&a^{i} t_c^{{i}(i-1)/2}\sum_{j=0}^{i}(-1)^j (at_rt_c)^{2pj} t_c^{(2p+1)j(j-1)} \frac{1-a^2t_r^2t_c^{4j}}{(a^2t_r^2 t_c^{2j};t_c^2)_{i+1}} { i \brack j }_{t_c^2} \\
&=&\left\{\begin{array}{ll}a^i t_c^{i(i-1)/2}\displaystyle\sum_{i=s_p\ge\cdots\ge s_1\ge0}  \prod^{p-1}_{j=1}(at_rt_c)^{2s_j} t_c^{2s_j(s_j-1)}{s_{j+1} \brack s_{j}}_{t_c^2}&  {\rm for}\quad p>0\\
(-at_r^2t_c^2)^{-i} t_c^{-i(i-1)/2} \displaystyle\sum_{i=s_{|p|}\ge\cdots\ge s_1\ge0}  \prod^{|p|-1}_{j=1}(at_rt_c)^{-2s_j} t_c^{-2s_j(s_{j+1}-1)}{s_{j+1} \brack s_{j}}_{t_c^2}& {\rm for} \quad p<0~.\end{array} \right.\nonumber
\eea
Note that the Poincar\'e polynomial of the standard triply-graded homology can be easily obtained via the specialization
\bea\nonumber
\scP_{[r]}(K_{s,t}\;;a,q,t) \; = \; \wt\scP_{[r]}(K_{s,t}\;;a,Q=q,t_r=t/q,t_c=q)~.
\eea
The naive application of this method to the colored HOMFLY homology of the twist links and the Borromean rings does not work, since it would violate the properties of link homology explained in \S \ref{sec:quadruply}. Although the diagonal colored homology has been calculated for these links in Appendix \ref{sec:app-ExamplesDiag}, different colors for non-torus links are out of reach in this paper. Since even colored $\fraksl(2)$ homology has not been evaluated yet for non-torus links, we hope that it will be computed soon.


\section{Refined Chern-Simons invariants for torus links}
\label{sec:refined}

Another distinctive family of knots and links that has been extensively studied in the physics and math literature consists of torus links. In this case, the system \eqref{surfeng} has an extra symmetry that is ultimately responsible for further simplifications and additional methods that can only work for this particular class of links. One of these methods, proposed by Aganagic and Shakirov \cite{Aganagic:2011sg}, is based on a generalization of the modular $S$ and $T$ matrices by expressions involving Macdonald polynomials, so that they still satisfy the Verlinde formula. Another approach, based on double affine Hecke algebras (DAHA), was proposed by Cherednik~\cite{Cherednik:2011nr}; it also applies only to torus links and, in all cases that have been studied, it leads to the same results as the refined Chern-Simons theory. For certain colors (in particular, for symmetric representations) these methods produce results that agree with Poincar\'e polynomials of the colored link homology \cite{DuninBarkowski:2011yx,Fuji:2012pm,Gorsky:2013jna}, but more generally they lead to expressions with minus signs, which therefore cannot be interpreted as Poincar\'e polynomials \eqref{superPdef}. Furthermore, for torus links with multiple components colored by small symmetric representations, the refined Chern-Simons invariants were computed in \cite{DuninBarkowski:2011yx} using generalizations of the Rosso-Jones formula.

In this section we compute the refined invariants of $T_{2,2p}$ torus links colored by arbitrary symmetric and anti-symmetric representations in the formalism of refined Chern-Simons theory, using braiding operations described in~\cite[\S3.1]{Fuji:2012pm}. We will show that these results agree with computations based on structural properties postulated in \S \ref{sec:quadruply}. This is a strong consistency check for both approaches. Moreover, even though these results agree, they have different algebraic representations. This gives rise to interesting mathematical identities, and also has an interesting interpretation in the physics of 3d $\mathcal{N}=2$ gauge theories, which will be the subject of section \S \ref{sec:phys}.

The ``refinement'' of the standard modular $S$ and $T$ matrices looks like
\bea
\label{ST-refine}
S^{\rm ref}_{\lambda \mu}=S^{\rm ref}_{00}\frac{M_{\lambda}(q_2^{\varrho})M_{\mu}(q_2^{\varrho}q_1^{\lambda})}{q_1^{\left|\lambda\right|\left|\mu\right|/N}}\,, \qquad T^{\rm ref}_{\lambda \mu}= T^{\rm ref}_{00}\cdot
\frac{q_1^{\frac{1}{2}||\lambda||^2}}{q_2^{\frac{1}{2}||\lambda^{t}||^2}} \frac{q_2^{\frac{N}{2}|\lambda|}}{q_1^{\frac{1}{2N}|\lambda|^2}}\cdot\delta_{\lambda \mu}\,,
\eea
where $M(x;q_1,q_2)$ is the Macdonald polynomial,  $\left\Vert \lambda \right\Vert^2 =\sum \lambda_{i}^{2}$, and  $\lambda^{t}$ denotes the transposition of the Young diagram $\lambda$.
The principal specialization $M_{\lambda}(t^{\varrho})=M_{\lambda}(t^{\varrho_{1}},...,t^{\varrho_{N}})$ of the Macdonald polynomial is the refined version of the quantum dimension of the representation $\lambda$, and $M_{\mu}(q_2^{\varrho}q_1^{\lambda})=M_{\mu}(q_2^{\varrho_{1}}q_1^{\lambda_{1}},...,q_2^{\varrho_{N}}q_1^{\lambda_{n}})$ where $\varrho$ is the Weyl vector, such that $\varrho_{j}=(N+1)/2-j$ for $G=SU(N)$.
The combinatorial expression for the Macdonald polynomial at the principal specialization takes the following form
\be
M_{\lambda} (q_2^{\varrho};q_1,q_2) \; = \; \prod_{(i,j)\in \lambda}
\frac{q_2^{\frac{N-i+1}{2}}q_1^{\frac{j-1}{2}}-q_2^{-\frac{N-i+1}{2}}q_1^{-\frac{j-1}{2}}}{q_2^{\frac{\lambda^{t}_j-i+1}{2}}q_1^{\frac{\lambda_i-j}{2}}-q_2^{-\frac{\lambda^{t}_j-i+1}{2}}q_1^{-\frac{\lambda_i-j}{2}}} \, .
\label{Mac_sp}
\ee
One can verify that Macdonald polynomials satisfy
\be
M_{\lambda_1}(x;q_1,q_2) \, M_{\lambda_2}(x;q_1,q_2) \; = \; \sum_{\mu \in \lambda_1 \otimes \lambda_2} N^\mu_{\lambda_1 \lambda_2}(q_1,q_2) M_{\mu}(x;q_1,q_2) \,,
\ee
where $N^{\mu}_{\lambda_1 \lambda_2}(q_1,q_2)$ is
the refinement of the Littlewood-Richardson coefficient, given by a rational function of $q_1$ and $q_2$.
Identifying $q^{C_{\lambda}}$ with $\frac{T_{\lambda \lambda}}{T_{00}}$, the refined
braiding eigenvalues in the vertical framing are given by
\bea
{B}_{\mu}^{(\pm)}(\lambda_{1},\, \lambda_{2})(q_{1},q_{2})&=&\epsilon_{\lambda_{1}\lambda_{2}}^{\mu}\left(\frac{T_{\lambda_{1}\lambda_{1}}}{T_{00}}\right)^{\pm1/2}\left(\frac{T_{\lambda_{2}\lambda_{2}}}{T_{00}}\right)^{\pm1/2}\left(\frac{T_{\mu \mu}}{T_{00}}\right)^{\mp1/2}\\
&=&	\epsilon_{\lambda_{1}\lambda_{2}}^{\mu}\frac{q_{1}^{\pm\frac{1}{4}(||\lambda_{1}||^{2}+||\lambda_{2}||^{2})}}{q_{2}^{\pm \frac{1}{4}(||\lambda_{1}^{t}||^{2}+||\lambda_{2}^{t}||^{2})}}\frac{q_{2}^{\pm\frac{N}{4}(|\lambda_{1}|+|\lambda_{2}|)}}{q_{1}^{\pm\frac{1}{4N}(|\lambda_{1}|^{2}+|\lambda_{2}|^{2})}}\frac{q_{1}^{\mp\frac{1}{4}||\mu||^{2}}}{q_{2}^{\mp\frac{1}{4}||\mu^{t}||^{2}}}\frac{q_{2}^{\mp\frac{N}{4}|\mu|}}{q_{1}^{\mp\frac{1}{4N}|\mu|^{2}}}\,,\nonumber
 \eea
where the sign $\epsilon_{\lambda_1 \lambda_2}^{\mu}=\pm 1$ is determined by whether $\mu$ appears symmetrically or anti-symmetrically in $\lambda_1 \otimes \lambda_2$.
For the symmetric representations and the anti-symmetric representations,
the explicit expressions for the Littlewood-Richardson coefficients can be obtained from the Pieri formula \cite{Macdonald:1995}:
\begin{eqnarray}
\wh N_{[r_1][r_2]}^{[r_1+r_2-\ell,\ell]}(q_1,q_2)
& = & \frac{(q_2;q_1)_\ell(q_1^{r_1-\ell+1};q_1)_\ell(q_1^{r_2-\ell+1};q_1)_\ell(q_1^{r_1+r_2-2\ell}q_2^2;q_1)_\ell}{(q_1;q_1)_\ell (q_1^{r_1-\ell}q_2;q_1)_\ell(q_1^{r_2-\ell}q_2;q_1)_\ell(q_1^{r_1+r_2-2\ell+1}q_2;q_1)_\ell}  \,, \nonumber\\
\wh N_{[1^{r_1}][1^{r_2}]}^{[2^\ell,1^{r_1+r_2-2\ell}]}(q_1,q_2)
& = &\frac{(q_1;q_2)_{r_2-\ell}(q_2^{r_1-\ell+1};q_2)_{r_2-\ell}}{(q_2;q_2)_{r_2-\ell}(q_1q_2^{r_1-\ell};q_2)_{r_2-\ell}} \ .\nonumber
\end{eqnarray}

Unlike the case of  the $(2,2p+1)$ torus knots,  the gamma factor \cite{DuninBarkowski:2011yx,Fuji:2012pm,Shakirov:2013moa} is not required for the computation of refined link invariants.  Hence, the refined Chern-Simons invariants of the $(2,2p)$ torus link colored by symmetric and anti-symmetric representations are given by
\bea\label{refined-Wilson-sym}
&&\langle W^{\rm ref}_{[r_1],[r_2]}(T_{2,2p})\rangle(A,q_1,q_2)\cr
&=&\sum_{\lambda \in [r_1]\otimes [r_2]}\left[B^{(+)}_{\lambda}([r_1],[r_2])\right]^{2p} \wh N_{[r_1],[r_2]}^{\lambda}(q_1,q_2) M_{\lambda}({q_2}^{\varrho};q_1,q_2) \cr
&=&q_1^{p r_1 r_2/N}(A^{-1}q_2)^{\frac12(r_1+r_2)} \sum_{\ell=0}^{\min(r_1,r_2)}  q_1^{-(\ell - r_1) (\ell - r_2)p}  q_2^{(p+1)\ell} \cr
&&\times\frac{(q_1^{r_1-\ell+1};q_1)_\ell(q_1^{r_2-\ell+1};q_1)_\ell (Aq_2^{-1};q_1)_\ell(A;q_1)_{r_1+r_2-\ell}}{(q_1;q_1)_\ell (q_1^{r_1-\ell}q_2;q_1)_\ell(q_1^{r_2-\ell}q_2;q_1)_\ell(q_1^{r_1+r_2-2\ell}q_2;q_1)_\ell(q_2;q_1)_{r_1+r_2-2\ell} }
\eea
\bea\label{refined-Wilson-anti}
&&\langle W^{\rm ref}_{[1^{r_{1}}],[1^{r_{2}}]}(T_{2,2p})\rangle(A,q_1,q_2)\cr
&=&\sum_{\lambda \in [1^{r_{1}}]\otimes [1^{r_{2}}]}\left[B^{(+)}_{\lambda}([1^{r_{1}}],[1^{r_{2}}])\right]^{2p} \wh N_{[1^{r_{1}}],[1^{r_{2}}]}^{\lambda} (q_1,q_2)M_{\lambda}({q_2}^{\varrho};q_1,q_2) \cr
&=&(-1)^{r1 + r2}  q_1^{p r_1 r_2/N}(Aq_2)^{\frac12(r_1+r_2)} \sum_{\ell=0}^{\min(r_1,r_2)}  q_1^{-p\ell } q_2^{p(\ell - r_1) (\ell - r_2)- \ell ( r_1 + r_2-\ell+1)}\cr
&&\times\frac{(q_1;q_2)_{r_2-\ell}(q_2^{r_1-\ell+1};q_2)_{r_2-\ell} (A^{-1}q_1^{-1};q_2)_\ell(A^{-1};q_2)_{r_1+r_2-\ell}}{(q_2;q_2)_{r_2-\ell}(q_1q_2^{r_1-\ell};q_2)_{r_2-\ell} (q_2;q_2)_{r_1+r_2-2\ell}(q_2^{-\ell};q_2)_{\ell}(q_1^{-1}q_2^{\ell-r_1-r_2};q_2)_{\ell}}
\eea
where $A$ stands for $q_2^N$.
In particular, the refined Chern-Simons invariants of the Hopf link colored by arbitrary representations $(\lambda,\mu)$ can be expressed as
\bea\label{Hopf-refine}
\langle W^{\rm ref}_{\lambda,\mu}(T_{2,2})\rangle(q_1,q_2)&=&\sum_{\nu \in \lambda \otimes \mu}(B^{(+)}_{\nu}(\lambda,\mu))^{2} \wh N_{\lambda,\mu}^{\nu} (q_1,q_2) M_{\nu}({q_2}^{\varrho};q_1,q_2) \cr
&=&q_1^{\frac{|\lambda||\mu|}{N}} q_2^{-\frac{|\lambda|+|\mu|}{2}N} M_{\lambda}(q_2^{\varrho}) M_{\mu}(q_2^{\varrho}q_1^{\lambda}).
\eea
This expression can be identified with the refined modular $S$-matrix \eqref{ST-refine} up to a certain framing factor, as was first demonstrated in \cite{Iqbal:2011kq} in the context of refined topological strings. In fact, the derivation of the second line from the first line in \eqref{Hopf-refine} was given in \cite[Appendix B]{Iqbal:2011kq}.

In the case of torus knots, the following change of variables is used in \cite{Aganagic:2011sg,DuninBarkowski:2011yx,Cherednik:2011nr} to relate refined Chern-Simons invariants to Poincar\'e polynomials of colored HOMFLY homology:
\bea\label{change-of-variables}
q_1\to q^2 t^2 \,,\quad q_2\to q^2\,,\quad A\to -a^2 t.
\eea
Applying the same change of variables in the case of $(2,2p)$ torus links colored by arbitrary symmetric representations, we find
\begin{align}\label{refined-CS-sym+}
&\overline {\rm rCS}^{(+)}_{[r_1],[r_2]}(T_{2,2p};a,q,t):=\langle W^{\rm ref}_{[r_1],[r_2]}(T_{2,2p})\rangle(A=-a^2 t ,q_1=q^2 t^2 ,q_2=q^2)\cr
&=a^{( p-1) (r_1 + r_2)} q^{2 p r_1 r_2 - ( p-1) (r_1 + r_2)} t^{ 2 p r_1 r_2}\sum_{\ell=0}^{\min(r_1,r_2)}  \frac{q^{2 \ell (p+1)} (q t)^{-2 p (\ell -r_1) (\ell - r_2)}}{(q^2t^2;q^2t^2)_\ell (q^2;q^2t^2)_{r_1+r_2-2\ell} }  \tag{rCS+s} \\
&\times\frac{((qt)^{2(r_1-\ell+1)};q^2t^2)_\ell((qt)^{2(r_2-\ell+1)};q^2t^2)_\ell (-a^2 q^{-2}t;q^2t^2)_\ell(-a^2t;q^2t^2)_{r_1+r_2-\ell}}{(q^{2(r_1-\ell+1)}t^{2(r_1-\ell)};q^2t^2)_\ell(q^{2(r_2-\ell+1)}t^{2(r_2-\ell)};q^2t^2)_\ell(q^{2(r_1+r_2-2\ell+2)}t^{2(r_1+r_2-2\ell+1)};q^2t^2)_\ell}~.\quad \nonumber
\end{align}
If we expand this result as a power series in $q$,
the coefficients of these refined Chern-Simons invariants all turn out to be positive. Moreover, as we discuss in  \S \ref{sec:sliding} and Appendix~\ref{sec:RR-id}, the refined Chern-Simons invariants \eqref{refined-CS-sym+} for symmetric representations coincide with the Poincar\'e polynomials \eqref{torus-link-diff-rk+} (see also \eqref{poincare-rcs}) of colored HOMFLY homology obtained by the analysis of structural properties of link homologies.

Similarly, the refined Chern-Simons invariants associated with anti-symmetric representations can be written as
\begin{align}\label{refined-CS-anti+}
&\overline {\rm rCS}^{(+)}_{[1^{r_1}],[1^{r_2}]}(T_{2,2p};a,q,t)=\langle W^{\rm ref}_{[1^{r_1}],[1^{r_2}]}(T_{2,2p})\rangle(A=-a^2 t ,q_1=q^2 t^2 ,q_2=q^2) \cr
&=(-1)^{r_1 + r_2}\sum_{\ell=0}^{\min(r_1,r_2)} q^{2p(\ell - r_1) (\ell - r_2)-2 \ell ( r_1 + r_2-\ell+1)-2p\ell} t^{-2p\ell }\tag{rCS+a}  \\
&\times\frac{(q^2t^2;q^2)_{r_2-\ell}(q^{2(r_1-\ell+1)};q^2)_{r_2-\ell} (-a^{-2}q^{-2}t^{-3};q^2)_\ell(-a^{-2}t^{-1};q^2)_{r_1+r_2-\ell}}{(q^2;q^2)_{r_2-\ell}(q^{2(r_1-\ell+1)}t^2;q^2)_{r_2-\ell} (q^2;q^2)_{r_1+r_2-2\ell}(q^{-2\ell};q^2)_{\ell}(q^{2(\ell-r_1-r_2-1)}t^{-2};q^2)_{\ell}} ~.
 \nonumber
\end{align}
Expanding the denominators gives a power series with positive coefficients.

Curiously, one can use a different change of variables \cite{Fuji:2012pm}
to relate the refined Chern-Simons invariants to colored HOMFLY homology with the different (FGS) grading conventions \cite{Gukov:2011ry}:
\bea\label{cv-FGS}
q_1\to q^{-2}t^{-2} ~,\quad q_2\to q^{-2}~,\quad A\to-a^2 t^3 ,
\eea
Compared to the grading conventions in \eqref{change-of-variables},
one simply needs to exchange the role of symmetric and anti-symmetric representations.
As a result, the refined Chern-Simons invariants colored by symmetric representations with the FGS grading can be obtained from \eqref{refined-Wilson-anti} by the change of variables \eqref{cv-FGS}:
\begin{align}\label{refined-CS-sym-}
&\overline {\rm rCS}^{(-)}_{[r_1],[r_2]}(T_{2,2p};a,q,t):=\langle W^{\rm ref}_{[1^{r_1}],[1^{r_2}]}(T_{2,2p})\rangle(A=-a^2 t^3 ,q_1=\tfrac{1}{q^{2}t^{2}} ,q_2=\tfrac{1}{q^{2}})\cr
&=(-t^3)^{r_1+r_2} a^{2 (r_1 + r_2)} q^{2 (r_1 - 1) (r_2 - 1)-2}  \sum_{\ell=0}^{\min(r_1,r_2)} \frac{q^{2 \ell ( r_1 + r_2-\ell+1)+2p\ell-2p(\ell - r_1) (\ell - r_2)} t^{2p\ell }}{(q^{-2};q^{-2})_{r_2-\ell}(q^{-2};q^{-2})_{r_1+r_2-2\ell}}\qquad\qquad   \tag{rCS-s}\\
&\times\frac{(q^{-2}t^{-2};q^{-2})_{r_2-\ell}(q^{-2(r_1-\ell+1)};q^{-2})_{r_2-\ell} (-a^{-2}q^{2}t^{-1};q^{-2})_\ell(-a^{-2}t^{-3};q^{-2})_{r_1+r_2-\ell}}{(q^{-2(r_1-\ell+1)}t^{-2};q^{-2})_{r_2-\ell} (q^{-2(\ell-r_1-r_2-1)}t^{2};q^{-2})_{\ell}(q^{2\ell};q^{-2})_{\ell}} ~.\nonumber
\end{align}
Likewise, the refined Chern-Simons invariants colored by anti-symmetric representations with the FGS grading can be read off from \eqref{refined-Wilson-sym} by the change of variables \eqref{cv-FGS}
\begin{align}\label{refined-CS-anti-}
&\overline {\rm rCS}^{(-)}_{[1^{r_1}],[1^{r_2}]}(T_{2,2p};a,q,t):=\langle W^{\rm ref}_{[r_1],[r_2]}(T_{2,2p})\rangle(A=-a^2 t^3 ,q_1=\tfrac{1}{q^{2}t^{2}} ,q_2=\tfrac{1}{q^{2}})\cr
&=a^{( p-1) (r_1 + r_2)} q^{( p-1) (r_1 + r_2)-2 p r_1 r_2 } t^{ -2 p r_1 r_2}\sum_{\ell=0}^{\min(r_1,r_2)}  \frac{q^{-2 \ell (p+1)} (q t)^{2 p (\ell -
    r_1) (\ell - r_2)}}{((qt)^{-2};(qt)^{-2})_\ell (q^{-2};(qt)^{-2})_{r_1+r_2-2\ell} }  \cr
&\times\frac{((qt)^{2(\ell-r_1-1)};(qt)^{-2})_\ell((qt)^{2(\ell-r_2-1)};(qt)^{-2})_\ell (-a^2 q^{2}t^3;(qt)^{-2})_\ell}{(q^{2(\ell-r_1-1)}t^{2(\ell-r_1)};(qt)^{-2})_\ell(q^{2(\ell-r_2-1)}t^{2(\ell-r_2)};(qt)^{-2})_\ell}\cr
&\times\frac{(-a^2t^3;(qt)^{-2})_{r_1+r_2-\ell}}{(q^{2(2\ell-r_1-r_2-2)}t^{2(2\ell-r_1-r_2-1)};(qt)^{-2})_\ell}. \tag{rCS-a}
\end{align}
These invariants have only positive coefficients if we expand in negative powers of $q$ (for instance, $1/(1-q^{-1})=1+q^{-1}+q^{-2}\cdots$).

\begin{figure}[H]
\centering
\includegraphics[scale=0.4]{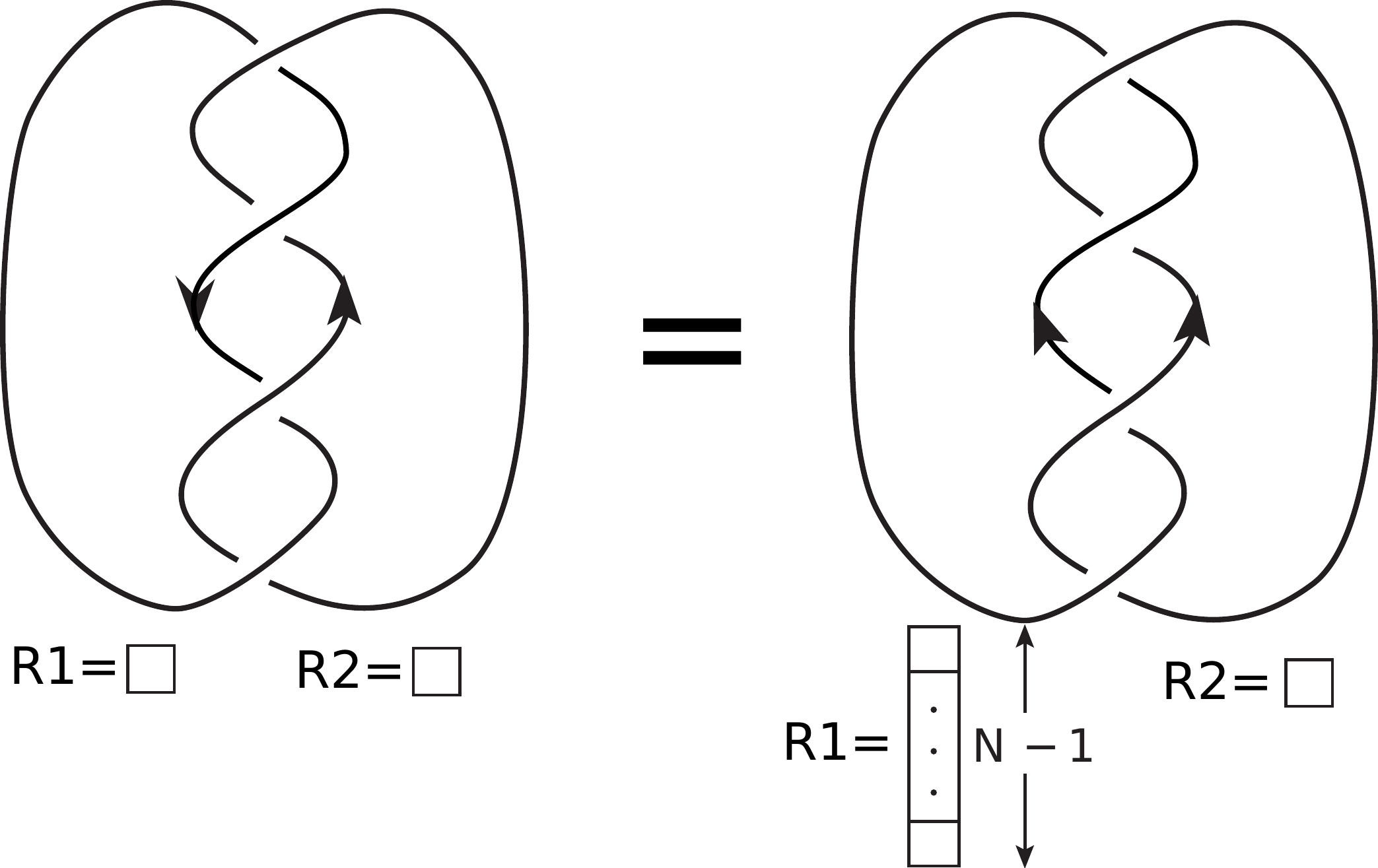}\caption{Changing relative orientation of the $(2,4)$ torus link is equivalent to changing its color.}
\label{fig:4a1-2}
\end{figure}

Finally, we conclude this section by considering torus links with a different relative orientation, which in general affects link invariants in a rather non-trivial way. Since relative orientation reversal acts trivially on the Hopf link, we take the $(2,4)$ torus link as our basic example, denoting its relative orientation-reversed version by $\wt T_{2,4}$. In Chern-Simons theory, changing orientation is equivalent to assigning the conjugate representation, as illustrated in Figure~\ref{fig:4a1-2}. If the same property holds true in the refined Chern-Simons theory, one can evaluate the uncolored invariant of the $\wt T_{2,4}$ link as
\bea
\langle W^{\rm ref}_{\yng(1),\yng(1)}(\wt T_{2,4})\rangle(A,q_1,q_2)&=& \sum_{\lambda \in [1]\otimes [1^{N-1}]}\left[ B^{(-)}_{\lambda}([1],[1^{N-1}])\right]^4 N_{[1],[1^{N-1}]}^{\lambda}(q_1,q_2) M_{\lambda} (q_2^{\varrho};q_1,q_2)\cr
&=&\frac{q_1^{2/N}q_2(1-A)( q_2 - q_1 q_2 - q_2^2 + q_1 q_2^2 + A (q_1 - q_1 q_2 + q_1^2 q_2)-A^2 q_1^2)}{q_1^2A^2(1-q_2)^2}~,\nonumber\eea
where each building block is given by
\bea\nonumber
M_{[2,1^{N-2}]}(q_2^{\varrho};q_1,q_2)&=&\frac{q_2^{1 - N} (1-q_2^N) (q_2 - q_2^N) (1-q_1 q_2^N)}{(1-
   q_2)^2 (q_2 - q_1 q_2^N)}~,\cr
   N_{[1],[1^{N-1}]}^{\emptyset}(q_1,q_2)&=&\frac{(1 - q_2^N) (1 - q_1)}{(1 - q_2) (1 - q_1 q_2^{N - 1})}~,\qquad
   N_{[1],[1^{N-1}]}^{[2,1^{N-2}]}(q_1,q_2)=1~,\cr
   B^{(-)}_{\emptyset}([1],[1^{N-1}])&=&q_1^{( N-1)/(2 N)} q_2^{(N-1)/2 } ~, \qquad B^{(-)}_{[2,1^{N-2}]}([1],[1^{N-1}])=q_1^{1/(2 N)}~.\quad
\eea
Modifying the framing factor and implementing the change of variables \eqref{change-of-variables}, the refined Chern-Simons invariant of the $\wt T_{2,4}$ link can be written as
\bea\label{rCS-4a1}
\overline {\rm rCS}_{\yng(1),\yng(1)}(\wt T_{2,4};a,q,t)=\frac{q(1+a^{2}t)}{a(1-q^{2})}\left[a(q^{-1}-qt^2)-a^3\left(q^{-1}t^3+\frac{q^3t^5}{1-q^2}\right)-\frac{a^5q t^6}{1-q^2}\right]~.\qquad
\eea
Clearly, this expression contains both positive and negative coefficients, implying that refined Chern-Simons invariants are different from the Poincar\'e polynomials of the HOMFLY homology, {\it cf.} \eqref{HOMFLY-4a1+}.


\section{Homological blocks and sliding property}
\label{sec:sliding}

In \S \ref{sec:example} and Appendix \ref{sec:app-ExamplesDiag}, we determined the reduced HOMFLY homology of links colored by symmetric representations using the structural properties of link homology.
In this section, we describe how the homology changes when the color of one of the components changes. Remarkably, we discover that the effect on homology is especially simple and ``universal'' when the link component whose color is changed is the unknot: the homology generators can be divided into certain groups (that we call \emph{homological blocks}), such that the only effect of the color-change is to shift these blocks in $(q,t_c)$-degree.

Let us describe in more detail this ``color-sliding'' behavior of the colored homologies. Let $\bar{\scP}^{(+)}_{[r_1],\ldots,[r_n]}(L;a,q,t_c)$
be the unreduced colored homology of a $n$-component link $L$ whose components are colored by symmetric representations
$[r_1],\ldots,[r_n]$. (By mirror symmetry, the analogous statements hold for anti-symmetric representations $[1^{r_1}],\ldots,[1^{r_n}]$.)
Let $\scP^{(+)}_{\underline{[r_1]},\ldots,[r_n]}(L;a,q,t_c)$ be the Poincar\'e polynomial of the reduced homology of $L$, reduced with respect to the first component.
As explained in \S \ref{sec:unknot}, we have a simple relationship between these two theories:
\be\nonumber
\bar{\scP}^{(+)}_{[r_1],[r_2],\ldots,[r_n]}(L)={\scP}^{(+)}_{
\underline{[r_1]},[r_2],\ldots,[r_n]}(L)\bar{\scP}^{(+)}_{[r_1]}(\unknot).
\ee

We claim that when the first component of $L$ is the unknot, $\scP^{(+)}_{\underline{[r_1]},\ldots,[r_n]}(L;a,q,t_c)$ depends on $r_1$ in a very simple and universal way. Namely, there exists a function $H_{[r_2],\ldots,[r_n]}(L;a,q,t_c,x_1)$, which is a polynomial in $x_1$ and a power series in variables $a,q,t_c$, such that
for every $r_1\ge 0$ we have:
\be\label{colslide}
\scP^{(+)}_{\underline{[r_1]},[r_2],\ldots,[r_n]}(L;a,q,t_c)=\left(\frac{a}{q}\right)^{r_1} H_{[r_2],\ldots,[r_n]}(L;a,q,t_c,x_1=(qt_c)^{2r_1})\,\,.
\ee
An important particular case is obtained by  setting $r_1=0$, {\it i.e.} by coloring the first component by the trivial representation, which is equivalent to erasing that component. In fact, for $r_1=0$ the left hand side becomes the Poincar\'e polynomial $\bar{\scP}^{(+)}_{[r_2],\ldots,[r_n]}(L';a,q,t_c)$ of the unreduced homology of the remaining $(n-1)$-component link $L'$, which therefore coincides with the function $H$ specialized to $x_1=1$.\\

The terms in $H$ with the same power of $x_1$ represent a piece of homology that we call a homological block. From the form of (\ref{colslide}) we see that the difference between various Poincar\'e polynomials $\scP^{(+)}_{\underline{[r_1]},[r_2],\ldots,[r_n]}(L;a,q,t_c)$ when $r_1$ changes, is that these blocks shift away in $(q,t_c)$-degree, with the ``speed'' determined by the power of $x_1$. In addition, when $r_2=\ldots=r_n$, these homological blocks are closed under the action of positive colored differentials from \S\ref{sec:quadruply}: there are no differentials acting between different homological blocks.\\

Below we illustrate this structure in the case of torus links, where it works beautifully --- for both thin and thick links --- in part due to the well-known ``positivity'' property of torus links: their colored HOMFLY polynomials are such that all terms with the fixed $a$-degree have the same sign, and all generators of the colored HOMFLY homology with the same $a$-degree have the $t$-degrees of the same parity. Therefore, while sliding homological blocks, one has no ``cancellations''  between the generators of different blocks.
For more general links, although the sliding-property can be also seen from the result in \eqref{cyclotomic-HOMFLY}, there is always possibility of such cancellations.  Because one can treat them using the appropriate spectral sequence with non-trivial higher differentials, we conjecture that the color-sliding property holds in general. To exploit it in detail, it requires concrete calculations of colored knot homology of non-torus links.

Confirmations of our conjecture already exist. It was shown in \cite{Wedrich:2014} that in the case of links with an unknot component, the polynomial (decategorified) version of color-sliding property (\ref{colslide}) holds. Moreover, our conjecture should lift the Batson-Seed spectral sequence on Khovanov homology of links, which relates the Khovanov homology of a link and the Khovanov homology of the disjoint union of its sublinks \cite{BatsonSeed:2013}. In the particular case when one of the components of a link $L$ is the unknot, the Batson-Seed spectral sequence starts at the Khovanov homology of $L$ and converges to the tensor product of the Khovanov homology of the sublink $L'$ (obtained from $L$ by erasing the unknot component) and the Khovanov homology of the unknot. In the triply-graded homology case this corresponds exactly to an instance of our conjecture when all link components are colored by fundamental representation and when we change the color/representation of the unknot component from the fundamental representation to the trivial one.
\\

In \S\ref{sec:sliding-CS}, the explanation of the sliding property is presented by the modular $S$-matrix at the decategorified level. Because the link $L$ can be obtained by the surgery of the link $L'$ and the unknot $\unknot$, quantum invariants of the link $L$ can be expressed by using the modular $S$-matrix, which accounts for the sliding property. This explanation can be uplifted to torus link homology by means of refined Chern-Simons theory.
In \S\ref{sec:phys}, we discuss an interpretation of homological blocks from the physics perspective, in the context of the dual 3d $\mathcal{N}=2$ theories, as well as in the BPS state counting.

The existence of homological blocks and color-sliding is among our main results in this paper. We believe that a deeper mathematical and physical understanding of the structure of homological blocks, as well as the precise generalization of the color-sliding property to arbitrary links, are very interesting and potentially very fruitful. These investigations will be a part of future work.

\subsection{$(2,2p)$ torus links}

As our first example, let us consider the reduced $([1],[r])$-colored HOMFLY homology of the Hopf link,
\be\nonumber
\scH_{[1],\underline{[r]}}(T_{2,2}) \,,
\ee
{\it i.e.} the homology where one of the components is colored by the fundamental representation, while the other is colored by an arbitrary symmetric representation $[r]$, with $r$ being a non-negative integer. As we shall show, all these homologies are related in a very simple way, as predicted by our conjectures on the color-sliding property and the existence of homological blocks.

First of all, the $([1],[2])$-colored unreduced HOMFLY invariant of the Hopf link is given by
\bea\label{Hopf-12-poly}
\overline P_{[1],[2]}(T_{2,2};a,q)=\frac{q^2(a^2;q^2)_2}{a^2(q^2;q^2)_2}\left[\frac{a^2}{q^2}+\frac{a^2q^2(1-a^2q^{-2})}{(1-q^2)} \right]~.
\eea
As pointed out in \cite{Nawata:2013qpa}, the unreduced colored HOMFLY invariant of a link always contains the unknot factor colored by the largest colors assigned to the components of the link. In addition, we use the $([1],[2])$-colored unreduced $\fraksl(2)$ homology of the Hopf link computed in \cite{Cooperprivate}:
\bea\label{Hopf-sl2-12}
\overline \scP^{\fraksl(2)}_{[1],[2]}(T_{2,2};q,t)=1 + q^2 + q^4 t^2 + q^8 t^3 + q^8 t^4 + q^{10} t^4+(1+q^2)\frac{q^6t^4+q^{10}t^5}{1-q^4t^2}~.
\eea
Since it contains the denominator $1/(1-q^4t^2)$, this expression suggests that one should use the $t_c$-grading for a homological grading of the unreduced HOMFLY homology. The only possible way to categorify the $([1],[2])$-colored HOMFLY invariants \eqref{Hopf-12-poly} of the Hopf link consistent with the structural properties is
\bea\label{Hopf-triple-12}
\overline \scP^{(+)}_{[1],[2]}(T_{2,2};a,q,t_c)=\frac{q^2(-a^2t_c;q^2t_c^2)_2}{a^2(q^2;q^2t_c^2)_2}\left[\frac{a^2}{q^2}+\frac{a^2q^4t_c^4(1+a^2q^{-2}t_c)}{(1-q^2)}\right]~.
\eea
Furthermore, since the $(a,q,t_c)$-degree of the $d_2$ differential is $(-2,2,-1)$, the Poincar\'e polynomial of the homology with respect to this differential has the form
\bea
\scP(\overline \scH^{(+)}_{[1],[2]}(T_{2,2}),d_2)(a,q,t_c)=1 + q^2 + q^4 t^2 + a^2 q^4 t^3 + q^8 t^4 + q^{10} t^4+q^6t^4(1+q^2)\frac{1+a^2t^5}{1-q^4t^2}~,\nonumber\eea
which indeed yields \eqref{Hopf-sl2-12} at $a=q^2$.\\

\begin{figure}[h]
 \centering
    \includegraphics[width=15cm]{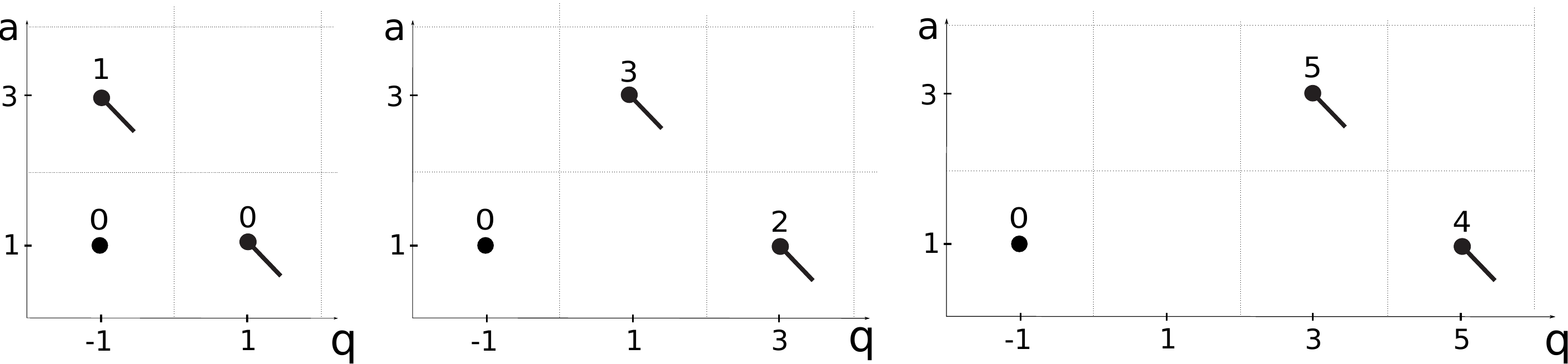}
    \caption{The dot diagrams for $a\overline \scP^{(+)}_{[1],[0]}(T_{2,2})/q\overline \scP^{(+)}_{[0]}(\bigcirc)$ (left), $\overline \scP^{(+)}_{[1],[1]}(T_{2,2})/\overline \scP^{(+)}_
{[1]}(\bigcirc)$ (middle), and $q\overline \scP^{(+)}_{[1],[2]}(T_{2,2})/a \overline \scP^{(+)}_{[2]}(\bigcirc)$ (right) with $t_c$-grading shown next to each generator. The left diagram is the uncolored HOMFLY homology of the unknot.} \label{fig:HOMFLY-Hopf-1-sliding}
\end{figure}

In the case $r=1$, the corresponding homology was computed in \eqref{Hopf-uncolor}:
\bea
\scP^{(+)}_{[1],[1]}(T_{2,2};a,q,t)=aq^{-1}+a q^3 t^2\frac{1+a^2q^{-2}t}{1-q^2}~.
\eea
Assigning the singlet representation $[0]$ to one component of the Hopf link is equivalent to ignoring that component,
effectively reducing the problem to that of the unknot. Therefore, we can write
\bea
\overline \scP^{(+)}_{[1],[0]}(T_{2,2};a,q,t_c)=\frac{a}{q}\overline \scP^{(+)}_{[1]}(\unknot;a,q,t_c)=\frac{1+a^2q^2t}{1-q^2}=1+\frac{q^2(1+a^2q^{-2}t)}{1-q^2}~.
\eea
As the rank of one of the representations is increased, the generators in the tails shift by two units in $(q,t_c)$-degree, as illustrated in Figure~\ref{fig:HOMFLY-Hopf-1-sliding}. Based on this, the $([1],[r])$-colored unreduced homology $\overline \scH_{[1],[r]}(T_{2,2})$ can be extrapolated to be
\bea\nonumber
\overline \scP^{(+)}_{[1],[r]}(T_{2,2};a,q,t_c)=\overline \scP^{(+)}_{[r]}(\unknot;a,q,t_c)\frac{a^{r-1}}{q^{r-1}}\left[\frac{a}{q}+(q t_c)^{2r}\frac{a q(1+a^2q^{-2}t_c)}{(1-q^2)}\right]~.
\eea

\begin{figure}[h]
 \centering
    \includegraphics[width=15cm]{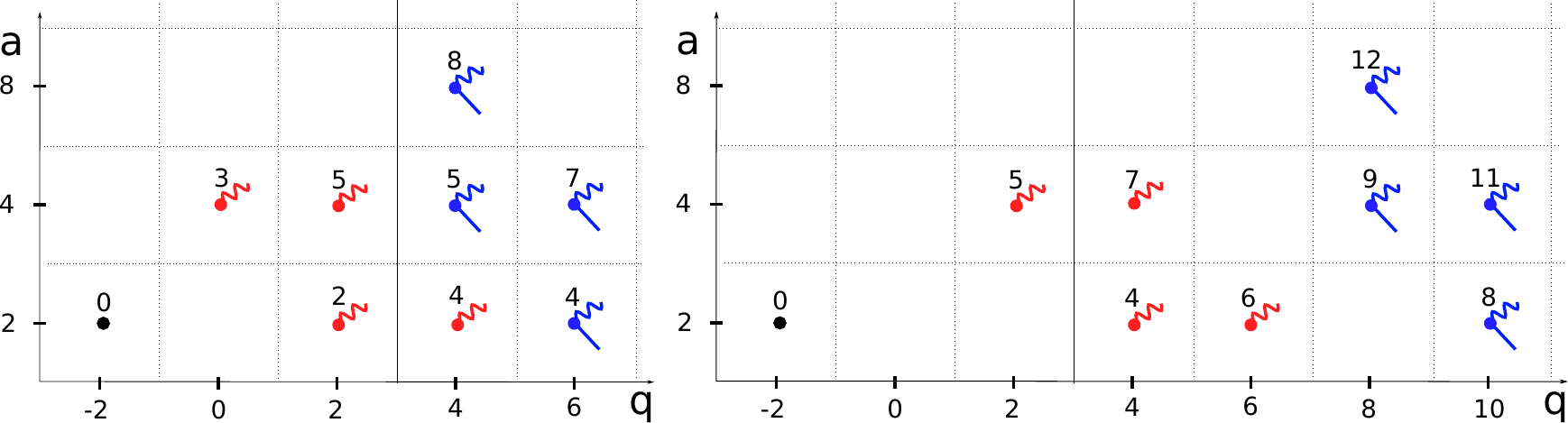}
    \caption{The dot diagrams for $a\overline \scP^{(+)}_{[2],[1]}(T_{2,2})/q\overline \scP^{(+)}_{[1]}(\bigcirc)$ (left) and $\overline \scP^{(+)}_{[2],[2]}(T_{2,2})/ \overline \scP^{(+)}_{[2]}(\bigcirc)$ (right) with $t_c$-grading shown next to each generator.} \label{fig:HOMFLY-Hopf-2-sliding}
\end{figure}

Next, let us manipulate the Poincar\'e polynomial of the $([1],[2])$-colored HOMFLY homology \eqref{Hopf-triple-12} of the Hopf link
\begin{small}
\bea
&&\overline \scP^{(+)}_{[2],[1]}({T_{2,2}};a,q,t_c)=\overline \scP^{(+)}_{[1],[2]}({T_{2,2}};a,q,t_c)=\cr
 &=& \overline \scP^{(+)}_{[2]}(\unknot;a,q,t_c) \left[\frac{a^2}{q^2}+\frac{a^2q^4t_c^4(1+a^2q^{-2}t_c)}{(1-q^2)}\right]\\
  &=&\overline \scP^{(+)}_{[1]}(\unknot;a,q,t_c)\frac{(1+a q t_c^3)}{(1-q^4t_c^2)}\left[\frac{a}{q}+\frac{aq^5t_c^4(1+a^2q^{-2}t_c)}{(1-q^2)}\right]\cr
 &=&\overline \scP^{(+)}_{[1]}(\unknot;a,q,t_c)\frac{q}{a}\Big[\frac{a^2}{q^{2}}+{\color{red}\frac{a^2(q t_c)^{2} (1 + q^2 t_c^2)  (1 + a^2 q^{-2} t_c)}{1 - q^4 t_c^2}}+{\color{blue}\frac{ a^2 q^{2}  (q t_c)^{4} (1 + a^2q^{-2} t_c) (1 + a^2  t_c^3 )}{(1 -  q^2) (1 - q^4 t_c^2)}}\Big]~,\nonumber
\eea
\end{small}
and compare it with the $([2],[2])$-colored HOMFLY homology of the Hopf link, which can be read off from \eqref{Hopf-infinite-22}:
\begin{small}
\bea\nonumber
&&\overline\scP^{(+)}_{[2],[2]}({T_{2,2}};a,q,t_c) = \\
&=&\overline \scP^{(+)}_{[2]}(\unknot;a,q,t_c) \left[\frac{a^2}{q^{2}}+{\color{red}\frac{a^2(q t_c)^4 (1 + q^2 t_c^2)  (1 + a^2 q^{-2} t_c)}{1 - q^4 t_c^2}}+{\color{blue}\frac{ a^2 q^{2} (q  t_c)^8 (1 + a^2q^{-2} t_c) (1 + a^2  t_c^3 )}{(1 -  q^2) (1 - q^4 t_c^2)}}\right]~.\nonumber
\eea
\end{small}
Here, the terms in red correspond to generators that slide in $q$-grading and $t_c$-grading by two units, while those in blue represent generators that slide by four units, {\it cf.} Figure~\ref{fig:HOMFLY-Hopf-2-sliding}.
Therefore, using this sliding property, one can deduce
\begin{footnotesize}
\bea\nonumber
&&\overline \scP^{(+)}_{[2],[r]}({T_{2,2}};a,q,t_c) = \\
 &=&\overline \scP^{(+)}_{[r]}(\unknot;a,q,t_c)\frac{a^{r-2}}{q^{r-2}} \Big[\frac{a^2}{q^{2}}+{\color{red}\frac{a^2(q t_c)^{2r} (1 + q^2 t_c^2)  (1 + a^2 q^{-2} t_c)}{1 - q^4 t_c^2}}+{\color{blue}\frac{ a^2 q^{2}  (q t_c)^{4r} (1 + a^2q^{-2} t_c) (1 + a^2  t_c^3 )}{(1 -  q^2) (1 - q^4 t_c^2)}}\Big]~.\nonumber
\eea
\end{footnotesize}

The simple behavior of the terms in the above equation as the color is changed is the \emph{sliding property;} each term slides at the same ``speed,'' and corresponds to a homological block.   In fact, writing the unreduced $([r],[r])$-colored HOMFLY homology of the Hopf link \eqref{Hopf-quad+} in the $(a,q,t_c)$-grading
\bea
\overline\scP^{(+)}_{[r],[r]}({T_{2,2}};a,q,t_c)&=&\overline \scP^{(+)}_{[r]}(\unknot;a,q,t_c) \sum_{i=0}^{r}q^{2(r+1)i}t_c^{2ri} \frac{(-a^2q^{-2}t_c;q^2t_c^2)_{i}}{(q^{2(r-i+1)}t_c^{2(r-i)};q^2t_c^2)_{i}}{r \brack i}_{q^2t_c^2},\nonumber\eea
we see that the summands for each value of $i$ in fact correspond to separate homological blocks. Taking into account the speed at which a given homological block ``slides'', one can obtain the Poincar\'e polynomial of the unreduced $([r_1],[r_2])$-colored  HOMFLY homology of the Hopf link in the $(a,q,t_c)$-gradings
\bea\label{unreduced-Hopf+}
\overline\scP^{(+)}_{[r_1],[r_2]}({T_{2,2}};a,q,t_c)&=&\frac{(-a^2t_c;q^2 t_c^2)_{r_1}}{(q^2;q^2 t_c^2)_{r_1}} \sum_{i=0}^{r_2}q^{2(r_1+1)i}t_c^{2r_1i} \frac{(-a^2q^{-2}t_c;q^2t_c^2)_{i}}{(q^{2(r_2-i+1)}t_c^{2(r_2-i)};q^2t_c^2)_{i}}{r_2\brack i}_{q^2t_c^2}.\cr
&&
\eea
As expected, this expression is symmetric under the exchange of $r_1$ and $r_2$, although it may not be apparent from its form and the way we derived it here. Moreover, we will prove in Appendix \ref{sec:RR-id} that \eqref{unreduced-Hopf+} is equal to the refined Chern-Simons invariants \eqref{refined-CS-sym+} at $p=1$.

\vskip 0.5 cm

The colored HOMFLY homology of the $(2,4)$ torus link $T_{2,4}$ also enjoys the sliding property.
The unreduced $([r],[r])$-colored HOMFLY homology of the $(2,4)$ torus link $T_{2,4}$ in the $(a,q,t_c)$-gradings is \eqref{T24-infinite+}:
\bea\label{unreduced-T24+}
\overline\scP^{(+)}_{[r],[r]}({T_{2,4}};a,q,t_c)
&=&a^{2r}q^{-2r}\frac{(-a^2t_c;q^2 t_c^2)_{r}}{(q^2;q^2 t_c^2)_{r}}\\
&& \sum_{r \ge j\ge i \ge 0} q^{2 (i + j)} (qt_c)^{2 r j - 2i j  + 4r i }\frac{(-a^2q^{-2}t_c;q^2t_c^2)_{j}}{(q^{2(r-i+1)}t_c^{2(r-i)};q^2t_c^2)_{i}}{r \brack j}_{q^2t_c^2}{j \brack i}_{q^2t_c^2}.\nonumber
\eea
As in the case of the Hopf link, the summand for each pair $(i,j)$ corresponds to a homological block.
Taking into account how each block slides, one can obtain the Poincar\'e polynomial of the unreduced $([r_1],[r_2])$-colored
HOMFLY homology of the $(2,4)$ torus link in the $(a,q,t_c)$-gradings
\bea\label{unreduced-T24+diff}
&&\overline\scP^{(+)}_{[r_1],[r_2]}({T_{2,4}};a,q,t_c) = \cr
&=&a^{r_1+r_2}q^{-r_1-r_2}\frac{(-a^2t_c;q^2 t_c^2)_{r_1}}{(q^2;q^2 t_c^2)_{r_1}}\\
&& \sum_{r_2 \ge j\ge i \ge 0} q^{2 (i + j)} (qt_c)^{2 r_1 j - 2i j  + 2 i (r_1 + r_2)}\frac{(-a^2q^{-2}t_c;q^2t_c^2)_{j}}{(q^{2(r_2-i+1)}t_c^{2(r_2-i)};q^2t_c^2)_{i}}{r_2\brack j }_{q^2t_c^2}{j \brack i}_{q^2t_c^2}.\nonumber
\eea
Furthermore, extrapolating these results, one finds
the Poincar\'e polynomial of the unreduced $([r_1],[r_2])$-colored HOMFLY homology of the $(2,2p)$ torus link in the $(a,q,t_c)$-gradings
\begin{align}\label{torus-link-diff-rk+}
&\overline\scP^{(+)}_{[r_1],[r_2]}(T_{2,2p};a,q,t_c) = \cr
&=(aq^{-1})^{(p-1)(r_1+r_2)}\frac{(-a^2t_c;q^2 t_c^2)_{r_1}}{(q^2;q^2 t_c^2)_{r_1}} \tag{Po+s}\\
&\sum_{{r_2}=s_{p+1}\ge s_p\ge\cdots\ge s_1\ge0}\frac{(-a^2q^{-2}t_c;q^2t_c^2)_{s_p}}{(q^{2({r_2}-s_1+1)}t_c^{2({r_2}-s_1)};q^2t_c^2)_{s_1}} \prod_{i=1}^{p}q^{2s_i}(qt_c)^{2({r_1}+{r_2})s_i-2s_is_{i+1}}{s_{i+1} \brack s_{i}}_{q^2t_c^2}. \nonumber
\end{align}
In Appendix \ref{sec:RR-id}, we prove using Bailey's Lemma that this is equal to the refined Chern-Simons invariants \eqref{refined-CS-sym+} of the $(2,2p)$ torus link $T_{2,2p}$. Interestingly, the equality can be regarded as a generalization of the celebrated Rogers-Ramanujan (Andrews-Gordon) identity. If we make use of the sliding property, we can indeed obtain a simplified version of the identity, which looks like a more familiar form of the Rogers-Ramanujan identity. Given the Poincar\'e polynomial $\overline\scP^{(+)}_{[r_1],[r_2]}(L)$ of the $(2,2p)$ torus link $T_{2,2p}$, the sliding property tells us that the superpolynomial of the unknot can be obtained by setting one of two colors to be the singlet
\bea\nonumber
\overline\scP^{(+)}_{[0],[r]}(T_{2,2p})=\overline\scP^{(+)}_{[r]}(\unknot)~.
\eea
For example, in the case of the Hopf link \eqref{unreduced-Hopf+}, we obtain the identity after making an appropriate change of variables
\bea\nonumber
\sum_{i=0}^{r}q^{i}t_c^{-2i} \frac{(-aq^{-1}t^3;q)_{i}}{(q^{({r}-i+1)}t^{-2};q)_{i}}{r \brack i}_{q} &=&\frac{(-at;q)_{r}}{(qt^{-2};q)_{r}}~,
\eea
which is a special case of the $q$-Chu-Vandermonde identity:
\bea\nonumber
{}_2\varphi_1(a,q^{-r};c;q,q)=\frac{(c/a;q)_r}{(c;q)_r}a^r.
\eea
We leave the verification of this as a good exercise to the reader.

Therefore, from the Poincar\'e polynomial  \eqref{torus-link-diff-rk+} of the $(2,2p)$ torus link $T_{2,2p}$, we obtain identities of Rogers-Ramanujan type for $p\in\bN$:
\be\label{QDI}
\sum_{{r}=s_{p+1}\ge s_p\ge\cdots\ge s_1\ge0}\frac{(-aq^{-1}t^3;q)_{s_p}}{(q^{{r}-s_1+1}t^{-2};q)_{s_1}} \prod_{i=1}^{p}q^{{(r+1)}s_i-s_is_{i+1}} t^{-2s_i}{s_{i+1}\brack s_{i} }_{q}=\frac{(-at;q)_{r}}{(qt^{-2};q)_{r}}.
\ee
The proof of this identity follows from a specialization of the equivalence between the Poincar\'e polynomial  \eqref{torus-link-diff-rk+} and the refined Chern-Simons invariants \eqref{refined-CS-sym+}, presented in Appendix \ref{sec:RR-id}.

\subsection{$(3,3)$ torus link}

The sliding property of homological blocks is ubiquitous in the HOMFLY homology of torus links colored by symmetric representations.
A beautiful illustration of this behavior comes from the uncolored HOMFLY homology of the $(3,3)$ torus link $T_{3,3}$:
\bea\label{uncolor-T33-redblue}
\overline\scP^{(+)}_{\yng(1),\yng(1),\yng(1)}(T_{3,3};a,q,t_c)&=&\overline\scP^{(+)}_{\yng(1)}(\unknot;a,q,t_c)\Big[\tfrac{a^4}{q^4}+{\color{red}a^4\left(t_c^{2}+\tfrac{(1+q^2)q^{2}t_c^4}{1-q^2} \right)+a^6\left(q^{-2}t_c^{3}+\tfrac{(1+q^{2})t_c^5}{1-q^2}
\right)}\cr
&&+{\color{blue}a^{4}\left(\tfrac{q^4t_c^4}{1-q^2} +\tfrac{q^8t_c^6}{(1-q^2)^2}\right)+a^6\left(\tfrac{q^2t_c^5}{1-q^2}+\tfrac{(1+q^{2})q^2t_c^7}{(1-q^2)^2}\right)+a^8\tfrac{q^2t_c^8}{(1-q^2)^2} }\Big]~.\qquad
\eea
Here, the terms colored in red and blue correspond, respectively, to the red and blue dots in Figure~\ref{fig:HOMFLY-T33-1-uncolor}.
Comparing this to the colored HOMFLY invariants $\overline P_{[1],[1],[r]}(T_{3,3};a,q)$,
one finds that the generators colored in red compose a homological block with the ``shift speed'' $(qt_c)^2$,
while the blue ones constitute a block with the ``shift speed'' $(qt_c)^4$, thereby resulting in
\bea\label{HOMFLY-T33-1r}
&&\overline\scP^{(+)}_{[1],[1],[r]}(T_{3,3};a,q,t_c)=\cr
&=&\overline\scP^{(+)}_{[r]}(\unknot;a,q,t_c)\tfrac{a^{r-1}}{q^{r-1}}\Big[\tfrac{a^4}{q^4}+{\color{red}(qt_c)^{2(r-1)}\left\{a^4\left(t_c^{2}+\tfrac{(1+q^2)q^{2}t_c^4}{1-q^2} \right)+a^6\left(q^{-2}t_c^{3}+\tfrac{(1+q^{2})t_c^5}{1-q^2}
\right)\right\}}\cr
&&+{\color{blue}(qt_c)^{4(r-1)}\left\{a^{4}\left(\tfrac{q^4t_c^4}{1-q^2} +\tfrac{q^8t_c^6}{(1-q^2)^2}\right)+a^6\left(\tfrac{q^2t_c^5}{1-q^2}+\tfrac{(1+q^{2})q^2t_c^7}{(1-q^2)^2}\right)+a^8\tfrac{q^2t_c^8}{(1-q^2)^2} \right\}}\Big]~.
\eea

\begin{figure}[h]
 \centering
    \includegraphics[width=13cm]{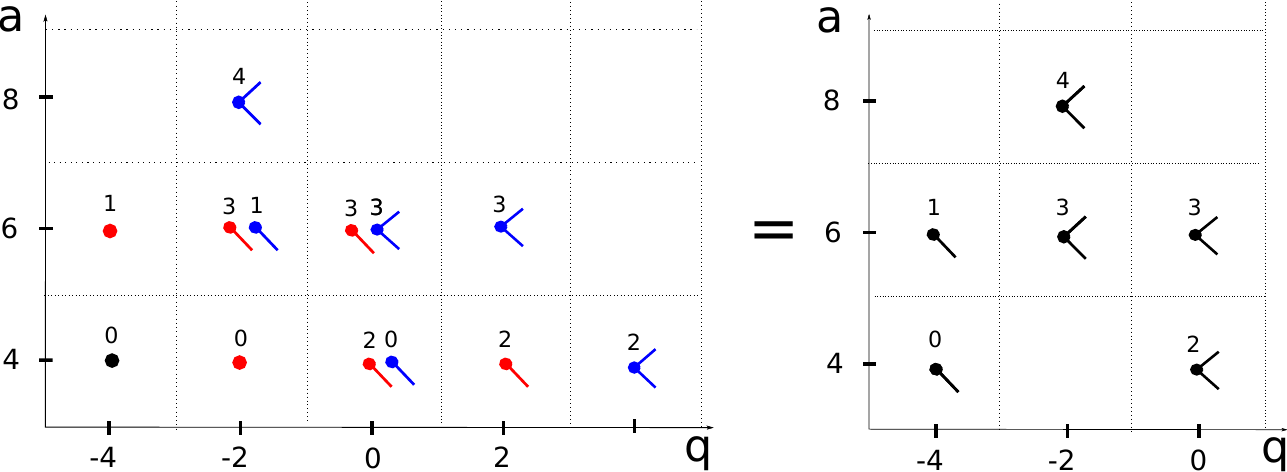}
    \caption{The left diagram is the $([1],[1],[0])$-colored HOMFLY homology of the $(3,3)$ torus link and the right diagram is the unreduced $([1],[1])$-colored HOMFLY homology of the Hopf link. The homological blocks colored by red and blue correspond to the terms colored by red and blue in \eqref{HOMFLY-T33-10}. The barbs represent semi-infinite tails associated with the factors $1/(1-q^2)$.}\label{fig:T33-T22-sliding}
\end{figure}

The shift in $(q,t_c)$-degree can be justified as follows.
Specializing \eqref{HOMFLY-T33-1r} to $r=0$, one finds
\bea\label{HOMFLY-T33-10}
\overline\scP^{(+)}_{[1],[1],[0]}(T_{3,3};a,q,t_c)
&=&\tfrac{q}{a}\Big[\tfrac{a^4}{q^4}+{\color{red}(qt_c)^{-2}\left\{a^4\left(t_c^{2}+\tfrac{(1+q^2)q^{2}t_c^4}{1-q^2} \right)+a^6\left(q^{-2}t_c^{3}+\tfrac{(1+q^{2})t_c^5}{1-q^2}
\right)\right\}}\cr
&&+{\color{blue}(qt_c)^{-4}\left\{a^{4}\left(\tfrac{q^4t_c^4}{1-q^2} +\tfrac{q^8t_c^6}{(1-q^2)^2}\right)+a^6\left(\tfrac{q^2t_c^5}{1-q^2}+\tfrac{(1+q^{2})q^2t_c^7}{(1-q^2)^2}\right)+a^8\tfrac{q^2t_c^8}{(1-q^2)^2}\right\}}\Big]\cr
&=&\tfrac{a^3}{q^3}\tfrac{q(1+a^2t_c)}{a(1-q^2)}\left[a\left(q^{-1}+\tfrac{q^3t_c^2}{1-q^2}\right)+ \tfrac{a^3 qt_c^3}{1-q^2} \right]\cr
&=&\tfrac{a^3}{q^3} \overline\scP^{(+)}_{[1],[1]}(T_{2,2};a,q,t_c)~.
\eea
Up to a suitable normalization, this is  the \emph{unreduced} uncolored HOMFLY homology of the Hopf link, {\it cf.} \eqref{Hopf-uncolor}.
Thus, what we find is consistent with the fact that assigning the trivial representation to one of the link components is equivalent to eliminating that component. In the present case, this turns the $(3,3)$ torus link into the Hopf link; see Figure~\ref{fig:T33-T22-sliding}.

\begin{figure}[h]
 \centering
    \includegraphics[width=13cm]{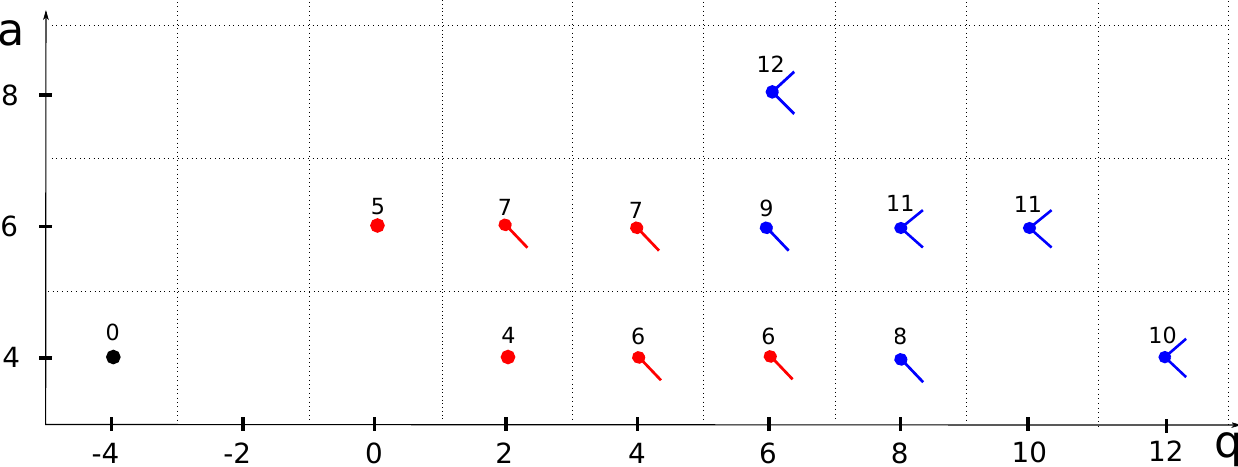}
    \caption{The diagram for $q\overline\scP^{(+)}_{[1],[1],[2]}(T_{3,3})/a\overline\scP^{(+)}_{[2]}(\bigcirc)$.  It is easy to see that the diagram is obtained from Figure~\ref{fig:HOMFLY-T33-1-uncolor} by shifting the red homological block by $(qt_c)^2$ and the blue homological block by $(qt_c)^4$. The barbs represent semi-infinite tails coming from the factors $1/(1-q^2)$.}\label{fig:HOMFLY-T33-12}
\end{figure}

We note that the color-sliding property enables us to obtain, via \eqref{HOMFLY-T33-1r}, the $([1],[1],[r])$-colored HOMFLY homology of the torus link $T_{3,3}$
from the knowledge of the uncolored HOMFLY homologies of $T_{3,3}$ and $T_{2,2}$. As an example, for $r=2$, Figure \ref{fig:HOMFLY-T33-12} illustrates how homological blocks shift when one of the ranks is increased.

\subsection{The sliding property in Chern-Simons theory}\label{sec:sliding-CS}

\begin{figure}[h]
\begin{LARGE}
\bea\nonumber
{\raisebox{-2.5cm}{\includegraphics[width=5cm]{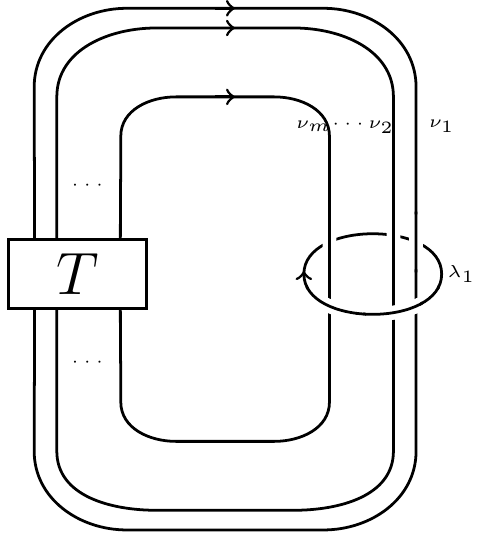}}}=\sum_{\mu_2,\cdots,\mu_m} N_{\nu_1,\nu_2,\cdots,\nu_m}^{\mu_2,\cdots,\mu_m}  ~~{\raisebox{-2.5cm}{\includegraphics[width=5cm]{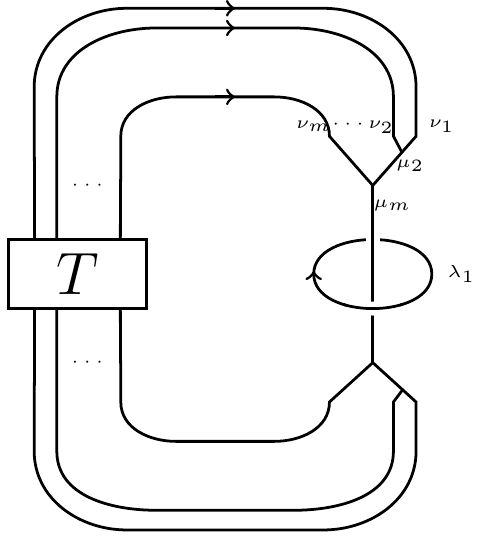}}}
\eea
\end{LARGE}
\caption{Tangle diagram.}\label{fig:tangle}
\end{figure}

At the decategorified level, the sliding property of link invariants can be explained in ordinary Chern-Simons theory.
Let us first consider $\fraksl(N)$ colored quantum invariants $\overline J^{\fraksl(N)}_{\lambda_1,\lambda_2,\cdots,\lambda_n}(L)$ of an $n$-component link $L$ with an unknot component. Without loss of generality, we can assume that the unknot is the first component colored by $\lambda$ linking the remaining $(n-1)$-component link $L'$ obtained as a closure of the $(m,m)$-tangle $T$, as shown in the left part of Figure \ref{fig:tangle}. Therefore, the color $\nu_i$ $(i=1,\cdots, m)$ is one of either $\lambda_2,\cdots,\lambda_n$ or their conjugate representations $\bar \lambda_2,\cdots,\bar\lambda_n$ depending on the orientation. A tangle diagram can be evaluated by using braid operations on multi-point conformal blocks in the $\wh\fraksl(N)_k$ WZNW model \cite{Nawata:2013qpa,Witten:1988hf}.

Since the tangle diagram at the right of Figure \ref{fig:tangle} can be regarded as a connected sum of the tangle $T$ and the Hopf link colored by $\lambda_1$ and $\mu_m$, the  $\fraksl(N)$ colored quantum invariants of $L$ can be expressed as
\bea\label{link-tangle}
\overline J^{\fraksl(N)}_{\lambda_1,\lambda_2,\cdots,\lambda_n}({L})&=&\sum_{\mu_2,\cdots,\mu_m} N_{\nu_1,\nu_2,\cdots,\nu_m}^{\mu_2,\cdots,\mu_m} \overline J^{\fraksl(N)}(T) \frac{S_{\lambda_1\mu_m}}{S_{0\mu_m}}\cr
&=&s_{\lambda_1}(q^{\varrho})  \sum_{\mu_2,\cdots,\mu_m} N_{\nu_1,\nu_2,\cdots,\nu_m}^{\mu_2,\cdots,\mu_m} \overline J^{\fraksl(N)}(T) \frac{s_{\mu_m}(q^{\varrho+\lambda_1})}{s_{\mu_m}(q^{\varrho})}
.
\eea
That is, $\overline J^{\fraksl(N)}(T)$ can be evaluated using braiding operations on WZNW conformal blocks. $N_{\nu_1,\nu_2,\cdots,\nu_m}^{\mu_2,\cdots,\mu_m} $ is the dimension of the following conformal block:
  \begin{figure}[H]
  \begin{minipage}[b]{8cm}\centering
\includegraphics[scale=.9]{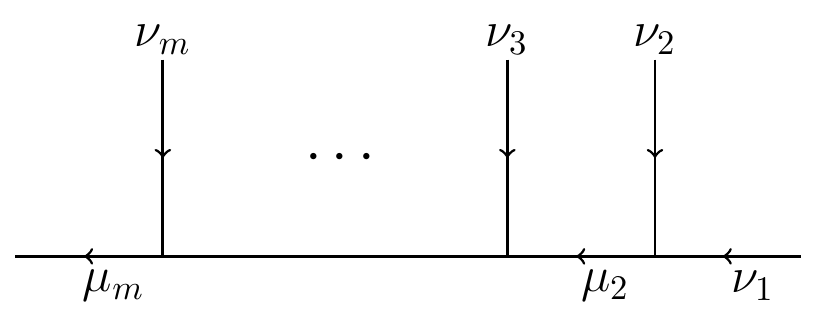}
\label{fig:newton52}
\end{minipage}
\hspace{.5cm}
  \begin{minipage}[b]{8cm}\centering
\includegraphics[scale=0.55]{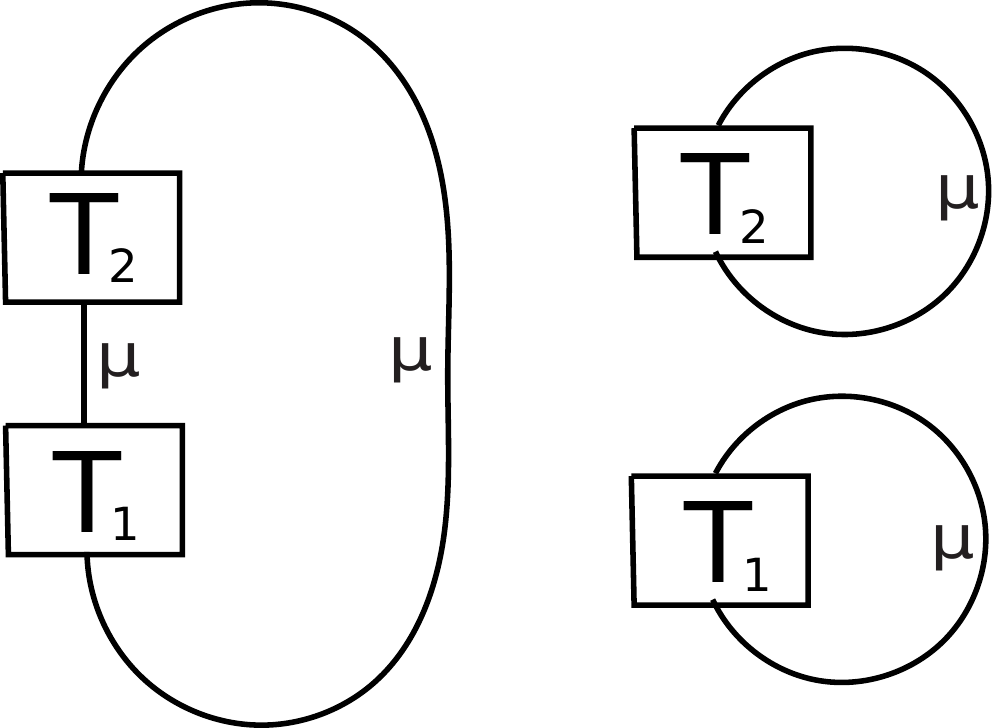}
\label{fig:newton61}
\end{minipage}
\end{figure}

\noindent Here we use the fact that the $\fraksl(N)$ quantum invariant of a connected sum (composition) of $(1,1)$-tangles is
\bea\nonumber
\overline J^{\fraksl(N)}(T_1\sharp T_2)=\sum_\mu \overline J^{\fraksl(N)}(T_1) \overline J^{\fraksl(N)}(T_2) \frac{S_{00}}{S_{0\mu}}~,
\eea
and   the $\fraksl(N)$ quantum invariant of  the Hopf link is equal to the modular $S$-matrix $S_{\lambda\mu}/S_{00}=s_\lambda(q^\varrho)s_\mu(q^{\varrho+\lambda})$ that can be expressed in terms of  the Schur function $s_\lambda(x)$ labeled by a representation $\lambda$. From the second line of \eqref{link-tangle}, one can convince oneself that the sliding property of $\fraksl(N)$ colored quantum invariants amounts to the fact that the dependence on the color $\lambda_1$ in the summand only appears in the factor $s_{\mu_m}(q^{\varrho+\lambda_1})$. In addition, when $\lambda_1$ is the trivial representation, the formula \eqref{link-tangle} reduces to the quantum invariants of the remaining $(n-1)$-component link $L'$. This dependence holds not only for symmetric representations but also for arbitrary representations.  As an example, we demonstrate an explicit evaluation of the colored quantum invariants of the Whitehead link:
\bea
&&\overline J^{\fraksl(N)}_{\lambda_1,\lambda_2}({\WL})=\left\langle {\raisebox{-1.2cm}{\includegraphics[width=3cm]{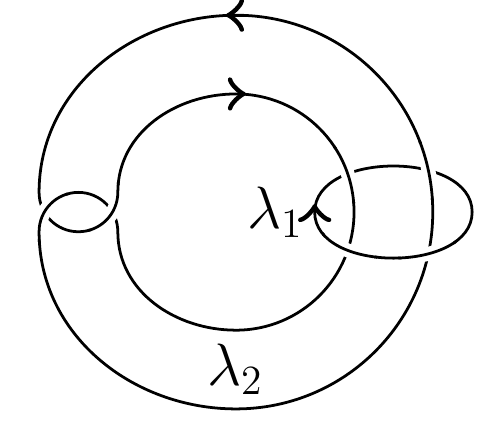}}} \right\rangle=\sum_{\mu_1\in \lambda_1\otimes \overline \lambda_1}\left\langle {\raisebox{-1.2cm}{\includegraphics[width=3cm]{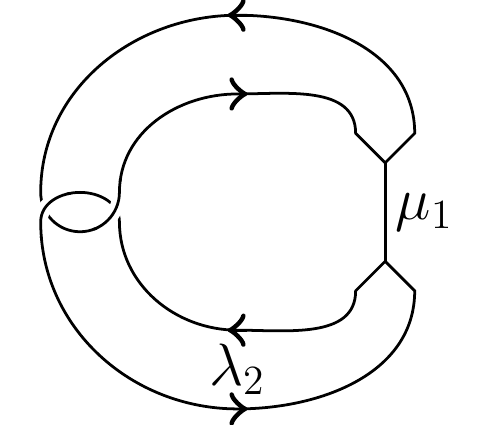}}} \right\rangle \frac{S_{\lambda_1\mu_1}}{S_{0\mu_1}} \\
&=&\left(\frac{S_{0\lambda_2}}{S_{00}} \right)^2\sum_{\mu_1,\mu_2} B^{(-)}_{\mu_2}(\lambda_2,\overline\lambda_2)^2F_{0\mu_1} \left[\small{\begin{array}{cc} \lambda_2 & \overline \lambda_2 \\ \lambda_2 & \overline \lambda_2 \end{array}}\right]F_{0\mu_2} \left[\small{\begin{array}{cc} \lambda_2 & \overline \lambda_2 \\ \lambda_2 & \overline \lambda_2 \end{array}}\right]F_{\mu_1\mu_2} \left[\small{\begin{array}{cc} \lambda_2 & \overline \lambda_2 \\ \lambda_2 & \overline \lambda_2 \end{array}}\right]\frac{S_{\lambda_1\mu_1}}{S_{0\mu_1}}\nonumber
\eea
where $F$ is a fusion matrix (quantum $6j$-symbol) of WZNW conformal blocks. (See \cite{Nawata:2013qpa} for more details.)

If the tangle $T$ can be represented by a braid $B_m$, one can provide a simpler explanation of the sliding property. For this class of link, it was shown in \cite{Mironov:2011ym,Mironov:2011aa,Itoyama:2012qt} using the Reshetikhin-Turaev construction \cite{Reshetikhin:1990} that the $\fraksl(N)$ quantum invariants of the remaining $(n-1)$-component link $L'$ admit character expansions
\bea\label{character-exp}
\overline J^{\fraksl(N)}_{\lambda_2,\ldots,\lambda_n} (L') =\sum_{\mu~ \vdash \sum |\nu_i|} h_{\nu_i}^\mu(B_m)  \frac{S_{0\mu}}{S_{00}}~,
\eea
where the choice $\nu_i\in (\lambda_2,\cdots,\lambda_n)$ for $i=1,\cdots,m$ depends on the braid $B_m$.
The Ross-Jones formula \cite{Rosso:1993vn} can be thought of as a special case of these character expansions for torus links. It is worth emphasizing that character expansions depend on the braid representations of a link so that they are not unique for a given link.

If we place the link $L'$ in the solid torus, the states on the boundary torus can be expressed in terms of the basis comprised of integrable representations of the affine Lie algebra  $\wh\fraksl(N)_k$ at level $k$, which are in one-to-one correspondence with Wilson loops along the longitude (non-contractible cycle) in the solid torus.
Since the link $L$ in $S^3$ can be obtained by gluing of the solid torus with $L'$ and  the other solid torus with the unknot by the modular $S$-transformation, its $\fraksl(N)$ quantum invariants can therefore be expanded in terms of the modular $S$-matrices as follows:
\bea\label{s-exp}
\overline  J^{\fraksl(N)}_{\lambda_1,\lambda_2,\ldots,\lambda_n} ( L)&=& {\raisebox{-.55cm}{\includegraphics[width=6.5cm]{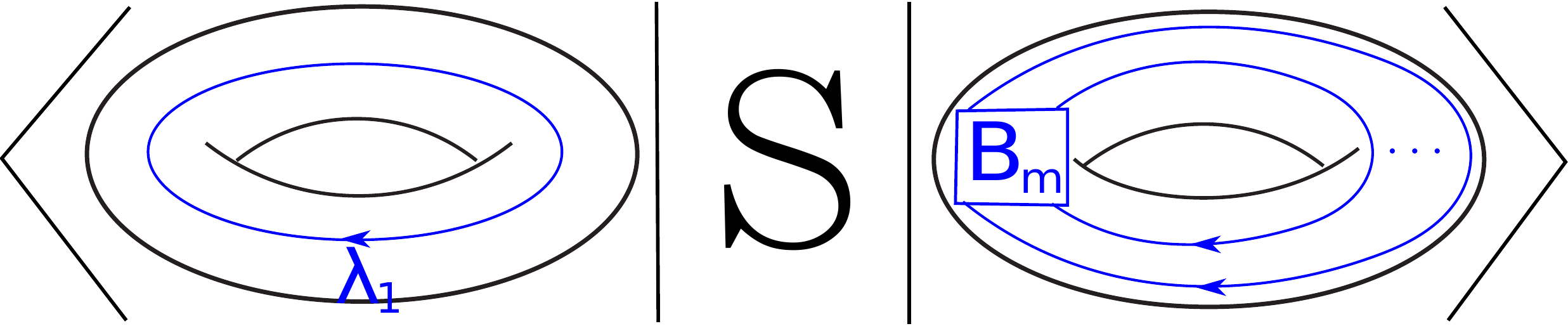}}}\cr
& =&\sum_{\mu~ \vdash \sum |\nu_i|} h_{\nu_i}^\mu(B_m)  \frac{S_{\lambda_1\mu}}{S_{00}}\cr
&=& s_{\lambda_1}(q^{\varrho}) \sum_{\mu~ \vdash \sum |\nu_i|} h_{\nu_i}^\mu(B_m)  s_{\mu}(q^{\varrho+\lambda_1})~.
\eea
Here, it is also easy to see that the dependence on the color $\lambda_1$ in the summand only appears through the factor $s_{\mu}(q^{\varrho+\lambda_1})$, and \eqref{s-exp} reduces to \eqref{character-exp} at $\lambda_1=0$. As examples, let us look at the Hopf link $T_{2,2}$ and the (2,4) torus link $T_{2,4}$. Even though each component is the unknot for these links, they are drawn by different braid representations. Hence, the quantum invariants are expanded in a different way:
\bea\label{S-surgery}
\overline J^{\fraksl(N)}_{\lambda_1,\lambda_2} (T_{2,2})&=&{\raisebox{-.55cm}{\includegraphics[width=6.5cm]{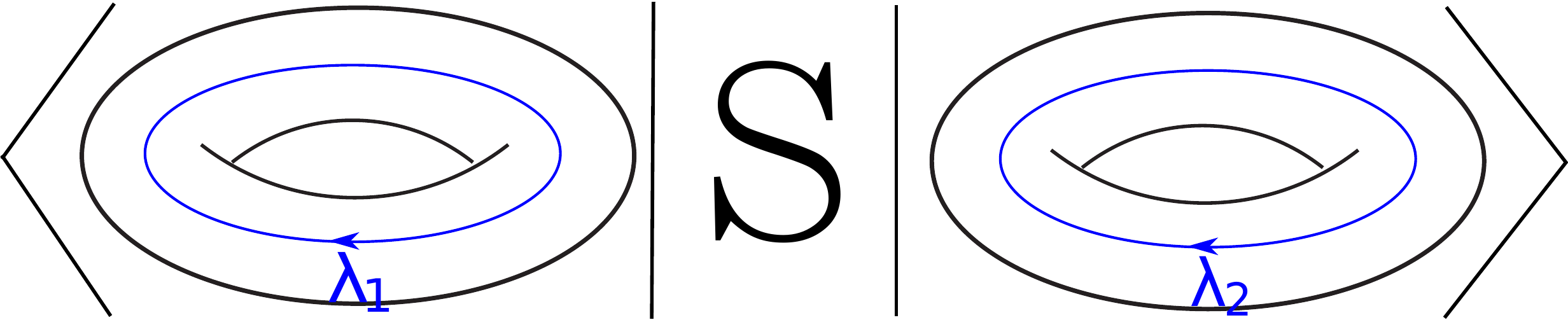}}}=\frac{S_{\lambda_1\lambda_2}}{S_{00}}\\
\overline J^{\fraksl(N)}_{\lambda_1,\lambda_2} (T_{2,4})&=&{\raisebox{-.55cm}{\includegraphics[width=6.5cm]{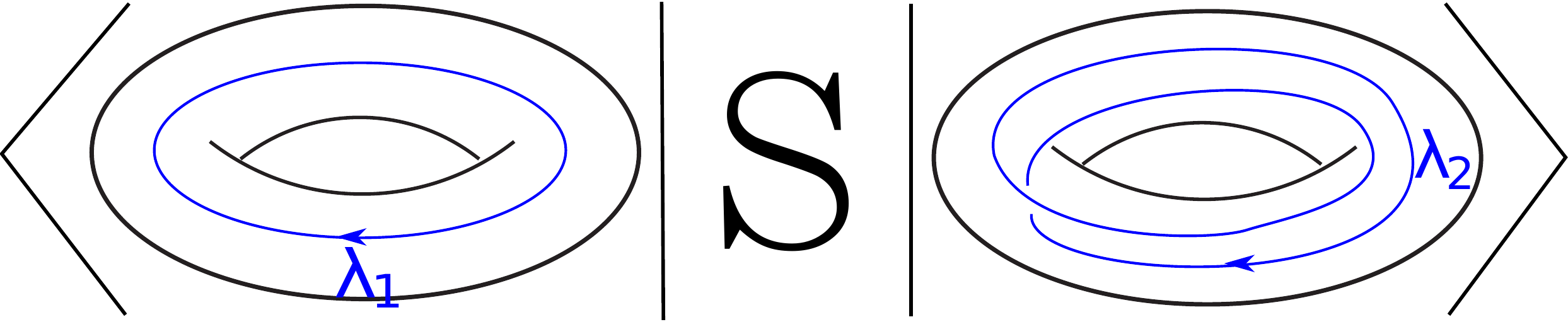}}}=\sum_{\mu\in \lambda_2\otimes\lambda_2}q^{C_2(\mu)}\frac{S_{\lambda_1\mu}}{S_{00}}~,\nonumber
\eea
where $C_2(\mu)$ is the quadratic Casimir of the representation $\mu$.

In fact, the expansions of quantum invariants in terms of the modular $S$-matrices can be generalized to links of the following type. Let us suppose that there are two links $L_1$ and $L_2$ that can be represented by braids $B_{m_1}$ and $\wt B_{m_2}$ so that their colored HOMFLY polynomials admit character expansions:
\be\nonumber
\overline J^{\fraksl(N)}_{\lambda_1,\ldots,\lambda_{n_1}} (L_1) =\sum_{\mu~ \vdash \sum |\nu_i|} h_{\nu_i}^\mu(B_{m_1})  \frac{S_{0\mu}}{S_{00}}~,\qquad
\overline J^{\fraksl(N)}_{\wt \lambda_1,\ldots,\wt \lambda_{n_2}} (L_2) =\sum_{\xi~ \vdash \sum |\rho_i|} \wt h_{\rho_i}^\xi(\wt  B_{m_2})  \frac{S_{0\xi}}{S_{00}}~.
\ee
If we glue these two braid representations by the modular  $S$-transformation, the $\fraksl(N)$ quantum invariants of the resulting link $L$ can be expanded in terms of the modular $S$-matrices:
\bea\label{general-S}
\overline J^{\fraksl(N)}_{\lambda_1,\ldots,\lambda_{n_1},\wt \lambda_1,\ldots,\wt \lambda_{n_2}} (L) &=& {\raisebox{-.55cm}{\includegraphics[width=6.5cm]{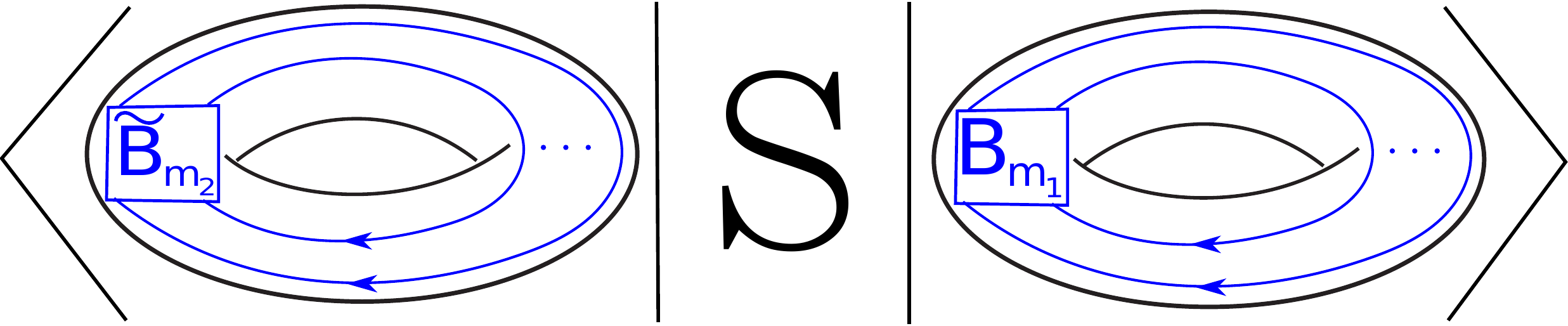}}}\cr
&=&\sum_{\mu~ \vdash \sum |\nu_i|, ~\xi~ \vdash \sum |\rho_i|}  h_{\nu_i}^\mu(B_{m_1}) \wt h_{\rho_i}^\xi(\wt  B_{m_2})  \frac{S_{\mu\xi}}{S_{00}}~.
\eea

\subsection*{Extension to refined Chern-Simons theory}

The refined Chern-Simons invariants of torus links can be evaluated by the refined version of the Rosso-Jones formula \cite{DuninBarkowski:2011yx}. In particular, the invariant of the torus link $T_{(n-1),k(n-1)}$ is
\bea\nonumber
\overline{\mathrm{rCS}}_{\lambda_2,\cdots,\lambda_n}(T_{(n-1),k(n-1)})=\sum_{\mu~ \vdash \sum |\lambda_i|} q_1^{-k\nu(\mu^T)}q_2^{k\nu(\mu)} \wh N^\mu_{\lambda_2,\cdots,\lambda_n} \frac{S^{\mathrm{ref}}_{0\mu}}{S^{\mathrm{ref}}_{00}}~,
\eea
where we define the function $\nu(\mu)$ for $\mu=\mu_1\ge\mu_2\ge\cdots\ge0$ and $ \wh N^\mu_{\lambda_2,\cdots,\lambda_n} $ as
\bea\nonumber
\nu(\mu)=\sum_i (i-1)\mu_i~,\qquad \prod_{i=2}^{n} \frac{S^{\mathrm{ref}}_{0\lambda_i}}{S^{\mathrm{ref}}_{00}}=\sum_{\mu~ \vdash \sum |\lambda_i|}\wh N^\mu_{\lambda_2,\cdots,\lambda_n} \frac{S^{\mathrm{ref}}_{0\mu}}{S^{\mathrm{ref}}_{00}}~.
\eea
Since the torus link $T_{n,kn}$ can be obtained by the $S$-gluing of the torus link $T_{(n-1),k(n-1)}$ and the unknot as above, the refined Chern-Simons invariants of the torus link $T_{n,kn}$ can be expanded in terms of the refined modular $S$-matrix \eqref{ST-refine}:
\bea\nonumber
\overline{\mathrm{rCS}}_{\lambda_1,\lambda_2,\cdots,\lambda_n}(T_{n,kn})&=&\sum_{\mu} q_1^{-k\nu(\mu^T)}q_2^{k\nu(\mu)} \wh N^\mu_{\lambda_2,\cdots,\lambda_n} \frac{S^{\mathrm{ref}}_{\lambda_1\mu}}{S^{\mathrm{ref}}_{00}}\cr
&=&M_{\lambda_1}(q_2^{\varrho})\sum_{\mu} q_1^{-k\nu(\mu^T)}q_2^{k\nu(\mu)} \wh N^\mu_{\lambda_2,\cdots,\lambda_n}M_{\mu}(q_2^{\varrho}q_1^{\lambda_1})~.
\eea
Here we can easily see that the simple dependence on the color $\lambda_1$ in the summand accounts for the sliding property of the torus link $T_{n,kn}$ at the refined level.

Let us conclude this subsection by mentioning some implications for mathematics. It is well-known that the Hilbert space of Chern-Simons theory on a torus is the space of characters of integrable representations of affine Lie algebra $\hat \frakg_k$ at level $k$, and that it admits an action of the modular $S$ and $T$ matrices. Apparently, the relation between refined Chern-Simons theory and homology of torus links predicts that there exists a homological generalization of these facts. Indeed, the $\fraksl(N)$ homology theory for a link in the thickened annulus (a solid torus) has been constructed in \cite{Queffelec:2015}. Hence, it is natural to expect that  there exists a HOMFLY homology theory of knots and links in the thickened annulus, which is the stabilization of the annular $\fraksl(N)$ homology theory \cite{Queffelec:2015} at the large $N$ limit. Furthermore, we conjecture that there should be a spectral sequence from the annular HOMFLY homology of a link to HOMFLY homology of the corresponding link  in $S^3$.
Moreover, we expect that there are the $S$ and $T$ functors, which satisfy the ordinary $\SL(2,\bZ)$ transformations, on chain complex of an annular link.
These functors should be consistent with the modular transformations in refined Chern-Simons theory after the spectral sequence.
This conjecture would provide a first-principles explanation of why refined Chern-Simons theory  \cite{Aganagic:2011sg} computes HOMFLY homology of torus knots/links. It would also lead to a proof of the sliding property for torus link homology.

  \begin{figure}[H]
\centering
\includegraphics[scale=.5]{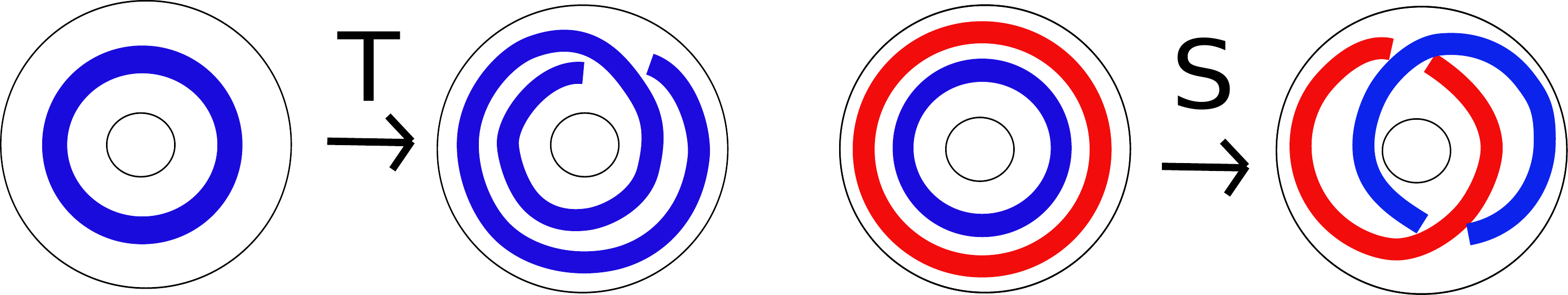}
 \caption{Modular transformations on links in  the thickened annulus,  where a link represented by a braid is placed in a colored strip.}
\end{figure}


\section{Associated varieties}
\label{sec:associated}

In this section we analyze the recursion relations and the asymptotic ``large color behavior'' of homological and polynomial link invariants.
As we will see, the asymptotic behavior is encoded in the geometry of higher-dimensional algebraic varieties, which arise as classical limits of recursion relations. We call these algebraic varieties the ``associated varieties'' of links and we initiate the analysis of their properties. However, we only scratch the surface of this very intricate subject, and associated varieties undoubtedly deserve further study.

Before moving on to the case of links, let us briefly recall what happens for knots. Apart from quantum group invariants discussed earlier, another kind of knot invariant is the $A$-polynomial.
Its zero locus, $A(K;x,y)=0$, defines a planar algebraic curve (in the space parametrized by complex variables $x$ and $y$)
which has a nice geometric interpretation.
Namely, it is the $\SL(2,\bC)$ character variety, which parametrizes inequivalent $\SL(2,\mathbb{C})$ representations of the knot group $\eta:\pi_1(S^3\backslash K)\to \SL(2,\mathbb{C})$.
Equivalently, $A(K;x,y)=0$ can be viewed as the moduli space of flat $\SL(2,\bC)$ connections on the knot complement, $S^3\backslash K$,
and this is how it enters the discussion of Chern-Simons gauge theory with complex gauge group.

The perspective of complex Chern-Simons theory also leads one to view $A(K;x,y)=0$ as a holomorphic Lagrangian subvariety
in the Hitchin moduli space $\frac{\bC^*\times \bC^*}{\bZ_2}$ associated to the boundary torus $T^2 = \partial (S^3\backslash K)$:
\bea\nonumber
\scM_{\rm flat}(S^3\backslash K,\SL(2,\mathbb{C})) = \left\{(x,y) \in \frac{\bC^*\times \bC^*}{\bZ_2} | A(K;x,y)=0  \right\}~.
\eea
Here $\bC^*\times \bC^*$ is parametrized by the holonomy eigenvalues of the meridian $\mu$ and the longitude $\lambda$ of the 2-torus,
\be\nonumber
\eta:\mu \mapsto \left(\begin{array}{cc} x&*\\ 0&x^{-1}\end{array}\right)~, \qquad \eta:\lambda \mapsto \left(\begin{array}{cc} y&*\\ 0&y^{-1}\end{array}\right)~,
\ee
and $\bZ_2: (x,y)\leftrightarrow (x^{-1},y^{-1})$ is the Weyl group of $\SL(2,\bC)$.
Note that the holomorphic symplectic form on the Hitchin moduli space in these coordinates is given by $\omega=d\log x\wedge d\log y$.
It was conjectured in \cite{Gukov:2003na} that the $A$-polynomial $A(K;x,y)$ can be
obtained by taking the large-color limit of the colored Jones polynomial $J^{\fraksl(2)}_{[r]}(K;q)$.
More precisely, the conjecture amounts to the statement that,
in the double scaling limit $q=e^\hbar\to1$ and $r\to \infty$ with $q^{2r}=x$ fixed, the colored Jones polynomial behaves as
\bea\label{c-apoly}
J^{\fraksl(2)}_{[r]}(K;q) \overset{{r \to \infty \atop \hbar \to 0}}{\sim}
\exp\left(\frac1{2\hbar} \wt\cW(K;x) + \cdots \right)
\eea
where the ellipsis denotes terms regular in $\hbar$, and  $\wt\cW(K;x)$ is given by an integral on the zero locus of the $A$-polynomial, $A(K;x,y)=0$:
\be\nonumber
\wt\cW(K;x) \; = \; \int \log y \frac{dx}{x}.
\ee
Put differently, the $A$-polynomial encodes the algebraic relation between $x=q^{2r}$ (that is kept fixed in
the large-color limit) and the variable $y$ defined through the asymptotic behavior \eqref{c-apoly}:
\bea
y:=\exp\left(x\frac{\partial \wt\cW(K;x)}{\partial x}\right)~.
\eea
In what follows we often refer to the function $\wt\cW(K;x)$ as the \emph{twisted superpotential}.
This name is motivated by applications to physics, which will be the subject of \S \ref{sec:phys}.

Moreover, the quantization of $A(K;x,y)$ turns out to be an operator that annihilates the colored Jones polynomials \cite{Gukov:2003na,Garoufalidis:2003,Garoufalidis:2004,Gelca}:
\bea
\label{q-apoly}
\widehat{A}(K;\hat x, \hat y;q) J_*(K;q)  = 0,
\eea
where the operators $\hat x$ and $\hat y$ act on the colored Jones polynomials as
\bea\label{xyhat}
\hat x J^{\fraksl(2)}_{[r]}=q^{2r}J^{\fraksl(2)}_{[r]} ~, \qquad \hat y J^{\fraksl(2)}_{[r]}=J^{\fraksl(2)}_{[r+1]} ~,\qquad\textrm{so that}\qquad \hat{y}\hat{x} = q^2 \hat{x}\hat{y}~.
\eea
In other words, the quantum $A$-polynomial $\widehat{A}(K;\hat x, \hat y;q)$ encodes the recurrence relation of minimal order with respect to the color. In the classical limit $q\to 1$, the operator $\widehat{A}(K;\hat x, \hat y;q) $ reduces to the $A$-polynomial $A(K;x,y)$.

Recently, this story was generalized in various directions.
First, the generalized volume conjecture \eqref{c-apoly} and its quantum version \eqref{q-apoly} have analogues
that control the color-dependence of the categorified colored Jones polynomial \cite{Fuji:2012pm}.
The resulting algebraic curve is the zero locus of the $t$-deformed $A$-polynomial, where $t$ is the same variable
we use to keep track of the homological grading in the Poincar\'e polynomials (and can be either $t_r$ or $t_c$.)
Second, a generalization of \eqref{c-apoly}-\eqref{q-apoly} in a different direction comes from replacing
the colored Jones polynomial by the colored HOMFLY polynomial \cite{Aganagic:2012jb,Garoufalidis:2012rt}.
In this case, the corresponding algebraic curve is defined by the zero locus of the $Q$-deformed $A$-polynomial.
This curve was further conjectured \cite{Aganagic:2012jb} to coincide with the augmentation polynomial
in knot contact homology \cite{Ng,NgFramed}, and to describe the geometry of the mirror B-model.
Recently, strong support for this conjecture has been found by using $Q$-deformed topological recursion \cite{Gu:2014yba}.

These two deformations of the $A$-polynomial are commutative, {\it i.e.}  they do not affect the commutation
relation $\hat{y}\hat{x} = q^2 \hat{x}\hat{y}$, and they can be combined in a two-variable deformation of the $A$-polynomial, called the super-$A$-polynomial, that was introduced in \cite{Fuji:2012nx} and further analyzed in \cite{Fuji:2012pm,Fuji:2012pi,3d3drevisited,Nawata:2012pg}.
In particular, it was shown in these works that the zero locus of the super-$A$-polynomial, $A^{\rm super}(K;x,y;a,t)=0$, defines
a holomorphic Lagrangian subvariety in the Hitchin moduli space $\frac{\bC^*\times \bC^*}{\bZ_2}$
where the $\bZ_2$ Weyl group action is also ``deformed'' by the two parameters, $a$ and $t$:
\bea
(x,y) \leftrightarrow  \left(\frac{-1}{a^2t^3x} ,  \frac{1}{t^{2S(K)}y} \right).  \label{z2-at}
\eea
Here, $S(K)$ is the $S$-invariant of the knot $K$, as in \cite{Rasmussen:2004}.
Indeed, in all examples that have been studied so far, the super-$A$-polynomials exhibit the symmetry
\bea\nonumber
A^{\rm super}(K;-(a^2t^3x)^{-1},(t^{2S(K)}y)^{-1};a,t)\propto A^{\rm super}(K;x,y;a,t)~.
\eea
This involution also  plays an important role in the $Q$-deformed topological recursion~\cite{Gu:2014yba}.
\\

Since the finite-dimensional HOMFLY homology of links inherits all the properties of the HOMFLY homology of knots, one can also investigate the large color asymptotics and recursion relations of Poincar\'e polynomials of finite-dimensional HOMFLY homologies of links, with all components colored by the same symmetric representation. As a result, analogues of super-$A$-polynomials of knots can be obtained that we call \emph{diagonal super-$A$-polynomials} of links.
The diagonal super-$A$-polynomials of links enjoy many properties \cite{Fuji:2012nx}
of the super-$A$-polynomials of knots.
In particular, at the classical level, diagonal super-$A$-polynomials define complex algebraic curves
which agree with the $q\to 1$ limits of the corresponding recursion relations.
Sometimes, we will refer to these recursion relations as ``quantum diagonal super-$A$-polynomials.''
Also, as in the case of knots, diagonal super-$A$-polynomials satisfy certain quantizability conditions \cite{Gukov:2011qp}.

For example, the quantum diagonal super-$A$-polynomial for the Hopf link takes
the form $\widehat{A}^{\textrm{super}} = a_0 + a_1 \hat{y} + a_2 \hat{y}^2$, with coefficients $a_0$, $a_1$, and $a_2$ given in  (\ref{T22-difference}). In the classical limit (or, equivalently, from the asymptotic analysis)
we obtain the classical diagonal super-$A$-polynomial for the Hopf link:
\bea
&&A^{\textrm{super}}(T_{2,2};x,y;a,t) =\cr
&=&(1 + a^2 t^3 x) y^2 \cr
 & &-a^2 (1 + t x - t^2 x + 2 t^2 x^2 + 2 a^2 t^3 x^2 + 2 a^2 t^4 x^2 -
   a^2 t^4 x^3 + a^2 t^5 x^3 + a^4 t^6 x^4) y\cr
 & & -a^4 t^3 (1 - x) x^2 (1 + t x) (1- a^2 t^2 x)   ~. \nonumber
\eea
More detailed analysis of this diagonal super-$A$-polynomial for the Hopf link, as well as examples for several other links, are presented in Appendix \ref{sec:diagonal}.
\\


Although the diagonal super-$A$-polynomials are interesting new invariants of links, they still belong to the realm of algebraic curves,
and are not conceptually different from ordinary super-$A$-polynomials.
One can obtain more interesting higher-dimensional objects by considering links with several components
colored by arbitrary symmetric representations.
Each of these representations can then be varied independently,
resulting in a more general system of recursion relations and a higher-dimensional limit shape.

In what follows we introduce and analyze such varieties for links with multiple components.
In order to carry out such analysis one needs to know homological knot invariants of links,
whose components are independently colored by different representations.
This is precisely where the notion of homological blocks introduced in \S \ref{sec:sliding} becomes especially useful.

To start with, we recall that the $\SL(2,\bC)$ character variety of an $n$-component link
is a Lagrangian subvariety of the Hitchin moduli space $\scM_H (\Sigma,G)$, as explained {\it e.g.} in \cite[\S 6]{Gadde:2013wq},
\be
\label{z2}
\scM_{\rm flat}(S^3\backslash L,\SL(2,\mathbb{C}))=\left\{(\vec{x},\vec{y}) \in \scM_H (\Sigma,G) | A_i(L;\vec{x},\vec{y})=0 \ \ (i=1,\cdots,n)  \right\}~
\ee
where $\Sigma := \partial (S^3\backslash L) = T^2 \times \ldots \times T^2$ and
\be
\label{MhitM}
\scM_H (\Sigma,G) \; \cong \; \scM_{\rm flat}(\Sigma,G_{\mathbb{C}}) \; = \; \left( \frac{\bC^*\times \bC^*}{\bZ_2} \right)^n \,.
\ee
This space comes equipped with a holomorphic symplectic form $\omega=\sum_id\log x_i\wedge  d\log y_i$
whose real and imaginary parts are two symplectic forms, usually denoted $\omega_I$ and $\omega_K$.
The $n$-dimensional subvariety \eqref{z2} is Lagrangian with respect to both of these symplectic structures and, therefore,
is a complex Lagrangian subvariety with respect to $\omega$.
In complex structure $J$, it is a holomorphic subvariety of the Hitchin moduli space $\scM_H (\Sigma,G)$
defined by $n$ complex equations $A_i(L;\vec{x},\vec{y})=0$, $i = 1, \ldots, n$.

As a result, the moduli space \eqref{z2} inherits a significant part of the rich structure of the hyper-K\"ahler space \eqref{MhitM}.
Namely, $\scM_{\rm flat}(S^3\backslash L,\SL(2,\mathbb{C}))$ is a holomorphic Lagrangian subvariety, or $(A,B,A)$-brane.
The latter name is supposed to indicate that it is a good $A$-brane for $A$-model of $\scM_H (\Sigma,\SU(2))$ with respect to $\omega_I$ and $\omega_K$
and a $B$-brane in complex structure $J$.
Note that of all complex and symplectic structures on $\scM_H (\Sigma,G)$, the triple $(\omega_I,J,\omega_K)$ is special in the sense that
these are the only structures whose definition does not require a complex structure on $\Sigma$.

In what follows, we extend both the generalized and the quantum volume conjecture to
hyperbolic links whose character variety \eqref{z2} is irreducible (has one component).
The twist links and the Borromean rings shown in Figure \ref{fig:surgery} belong to this class.
On the other hand, the Hopf link and, more generally, torus links turn out to be more delicate.
With this in mind, let us consider a hyperbolic link $L$ whose character variety consists of a single branch.
Then, in the large color limit $r_i\to \infty$ with $q=e^\hbar\to1$ and $q^{2r_i}=x_i$ fixed,
its unreduced colored Jones polynomial $\overline J^{\fraksl(2)}_{[r_1],\cdots,[r_n]}(L;q)$ takes the form \cite{Dimofte:2010ep}:
\bea
\label{cc-apoly}
\overline J^{\fraksl(2)}_{[r_1],\cdots,[r_n]}(L;q) \; \overset{{r_i \to \infty \atop \hbar \to 0}}{\sim} \;
e^{\tfrac{1}{2\hbar}\wt\cW(L;\vec{x})} := \exp\left(\frac1{2\hbar}\int \sum_i \log y_i \frac{dx_i}{x_i} + \cdots \right)
\eea
where the integral in the exponent is carried over a path on the higher-dimensional character variety \eqref{z2}
defined by the polynomial equations:
\be\label{character-variety}
A_j (L; \vec{x},\vec y)=0 ,\qquad j=1,\ldots ,n  ~.
\ee
Since, as we explained earlier, the hypersurface defined by these equations is complex Lagrangian with respect to
the holomorphic symplectic form $\omega=\sum_id\log x_i\wedge  d\log y_i$, the function $\wt\cW(L;\vec{x})$ defined in \eqref{cc-apoly}
exists (provided the quantizability conditions \cite{Gukov:2011qp} are satisfied).
Moreover, \eqref{cc-apoly} implies that the algebraic equations \eqref{character-variety}
are simply the relations
\bea
\label{yifi}
y_j=\exp\left(x_j\frac{\partial \wt \cW (L;\vec{x})}{\partial x_j}\right)~,   \qquad j=1,\ldots,n  ~.
\eea
As in the case of knots, the function $\wt \cW (L;\vec{x})$ is the twisted superpotential
of a certain 3d $\mathcal{N}=2$ theory that will be briefly discussed in \S \ref{sec:phys}.

The quantization of the higher-dimensional character variety also leads to recursion relations
(of minimal order) with respect to the color of each link component, such as
\be
b_k^{(j)}(\hat x_1,...,\hat x_n,q)\overline J^{\fraksl(2)}_{[r_1],\cdots,[r_j+k],\cdots,[r_n]}(L;q)+\cdots+b_0^{(j)}(\hat x_1,...,\hat x_n,q)\overline J^{\fraksl(2)}_{[r_1],\cdots,[r_j],\cdots,[r_n]}(L;q)=0~.\label{q-holo}
\ee
Indeed, quantization of the complex symplectic manifold \eqref{MhitM}
with the holomorphic symplectic form $\omega=\sum_id\log x_i\wedge  d\log y_i$
gives the so-called quantum torus algebra generated by the operators $\hat x_j ,\hat y_j $,  $j=1,\cdots,n$
that obey the $q$-commutation relations and act on the set of colored Jones polynomials as
\bea
\label{xyhat2}
\hat x_j \overline J^{\fraksl(2)}_{[r_1],\cdots,[r_n]}(L;q)&=&q^{2r_j} \overline J^{\fraksl(2)}_{[r_1],\cdots,[r_n]}(L;q) \cr
\hat y_j \overline J^{\fraksl(2)}_{[r_1],\cdots,[r_n]}(L;q)&=&\overline J^{\fraksl(2)}_{[r_1],\cdots,[r_j+1],\cdots,[r_n]}(L;q).
\eea
In this algebra, the recursion relations \eqref{q-holo} define the annihilator of an ideal,
\be
\label{quantum-cv}
{\widehat A}_j(L;\hat y_j,\hat x_1,\ldots,\hat x_n; q)\overline J^{\fraksl(2)}_{[r_1],\cdots,[r_n]}(L;q)=0~
\ee
where
\be
{\widehat A}_j(L;\hat y_j,\hat x_1,\ldots,\hat x_n ;q)=\sum_{\ell=0}^k b_\ell^{(j)}(\hat x_1,...,\hat x_n,q)\hat y_j^\ell~.\label{gamaaj}
\ee
Of course, in general these operators, as well as the corresponding $q$-difference equations \eqref{q-holo},
involve combinations of operators $\hat y_j$ with different values of $j$.
Here, to avoid clutter in writing \eqref{q-holo} and ${\widehat A}_j(L;\hat y, \hat x; q)$,
we assumed that each such operator can be ``diagonalized,'' \emph{i.e.} written in the form that involves only one kind of $\hat y_j$-operator.
(This will actually turn out to be the case in in simple examples.)

In general, the set of operators ${\widehat A}_j(L;\hat y, \hat x; q)$  ($i=1,...,n$)
provides a quantization of the character variety $\scM_{\rm flat}(S^3\backslash L,\SL(2,{\bC}))$
defined by \eqref{character-variety}, in the sense that
\be\label{classical-limit}
\lim_{q \to 1} \; {\widehat A}_j(L;\hat y, \hat x; q) \; = \; A_j(L; y, x)~.
\ee



As in the case of knots, one can consider $t$- and $a$-deformations of the character variety
of links,  corresponding to  homological and HOMFLY invariants respectively.
We collectively refer to all these varieties as \emph{associated varieties}.
In the context of the $a$-deformation, such analysis for links was performed in \cite{Arthamonov:2013rfa}.
For hyperbolic links, homological invariants colored by arbitrary symmetric representations
are not known at present and, therefore, we can only analyze their associated varieties at the level of $a$-deformations
using the cyclotomic expansions \eqref{cyclotomic-HOMFLY}.
In this class of links, one can generalize the above conjectures by replacing colored Jones polynomials by colored HOMFLY invariants.
Then, the corresponding associated variety is a Lagrangian subvariety of the Hitchin moduli space $(\bC^*\times\bC^*/\bZ_2)^n$
where the Weyl group symmetries are also ``deformed'' by the HOMFLY parameter $a$:
\be\label{z2-associate}
(x_i,y_i) \leftrightarrow \left( \frac{1}{a^2x_i} ,  \frac{1}{y_i} \right), \qquad  i=1,\ldots,n ~.
\ee
In subsection \ref{sec:AV-TL}, we demonstrate these symmetries for twist links $L_p$.
Furthermore, for a link $L$ in this family, the associated variety coincides with the augmentation variety $V_L(n)$ in \cite{Aganagic:2013jpa,Ngweb}.

In the case of torus links the situation is more subtle. The $a$-deformation preserves the Lagrangian condition but breaks the Weyl group symmetries. Therefore, the associated variety of a torus link can at best be a Lagrangian subvariety of the covering space $(\bC^*\times\bC^*)^n$.
Furthermore, the $a$-deformed associated variety of a torus link consists of several components,
only one of which coincides with the augmentation variety $V_L(n)$ in \cite{Aganagic:2013jpa,Ngweb}.

In the remainder of this section we illustrate the properties of associated varieties in explicit examples, for the Hopf link and twist links $L_p$. More examples of associated varieties for several other links are presented in Appendix \ref{sec:app-varieties}.

\subsection{Associated varieties for the Hopf link}

As in other parts of this paper, the first example which we consider is that of the Hopf link. We first present quantum associated varieties, \emph{i.e.} recursion relations satisfied by various invariants of the Hopf link. Then, we show that the classical limit of the recursion relations \emph{almost} agrees with the variety defined by the asymptotic behavior of these invariants. To be precise, the latter variety consists of two components, only one of which is singled out by comparison with the recursion relations.

\subsubsection*{Character variety} \label{sec:associated-Hopf}

\begin{figure}
 \centering
    \includegraphics[width=11cm]{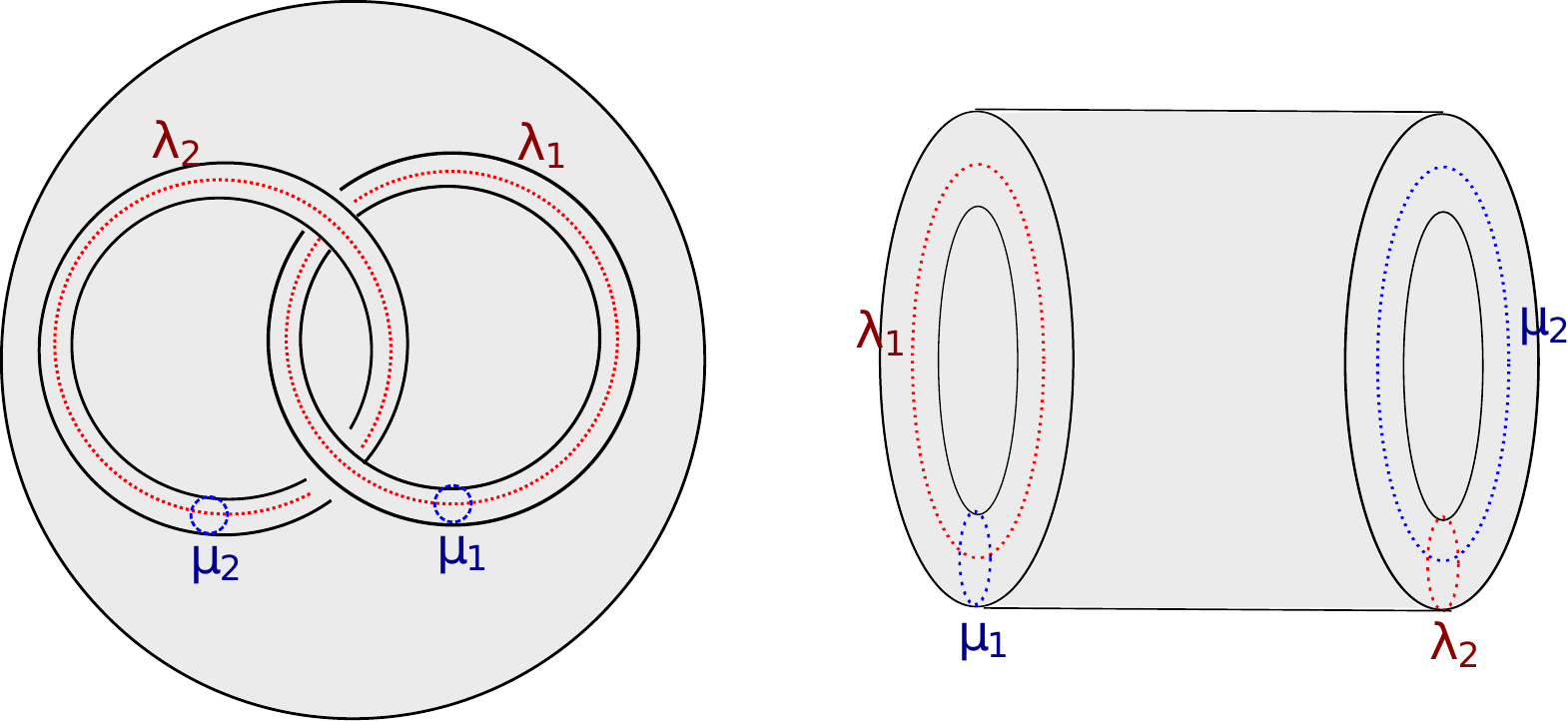}
    \caption{The complement of the Hopf link in $S^3$ is topologically $T^2\times I$. }\label{fig:Hopf-complement}
\end{figure}

We start our analysis with the recursion relations for the colored Jones polynomials. Recall, that for the Hopf link the colored Jones polynomials take a very simple form \cite{Witten:1988hf}:
\bea\label{Jones-Hopf}
\overline J^{\fraksl(2)}_{[r_1],[r_2]}(T_{2,2};q)=\frac{q^{(r_1+1)(r_2+1)}-q^{-(r_1+1)(r_2+1)}}{q-q^{-1}}~,
\eea
and have an elegant interpretation in terms of the modular $S$-matrix. Geometrically, the complement of the Hopf link is homeomorphic to the product $T^2\times I$ of a torus $T^2$ and an interval $I=[0,1]$, where the meridian of the torus at one end of the interval $I$ is identified with the longitude of the torus at the other end of this interval, and vice versa (see Figure \ref{fig:Hopf-complement}).
In order to see the geometry of the link complement, it is convenient to introduce operators $\hat m_1$ and $\hat m_2$ acting on (\ref{Jones-Hopf}) as
\bea\nonumber
\hat m_i \overline J^{\fraksl(2)}_{[r_1],[r_2]}=q^{r_i}\overline J^{\fraksl(2)}_{[r_1],[r_2]} ~,
\eea
relation to $\hat x_i$ in \eqref{xyhat2} via $\hat x_i=\hat m_i ^2$.
Then, one can find recursion relations of the form $\widehat{A}_i(T_{2,2})~\overline J^{\fraksl(2)}_{[r_1],[r_2]}(T_{2,2};q)=0$, with
\bea\label{recursion-Hopf-Jones}
\wh A_1(T_{2,2};\hat m_1,\hat m_2,\hat y_1;q)&=& \hat y_1^2 -(q \hat m_2+q^{-1} \hat m_2^{-1} )\hat y_1+1 ~,\cr
\wh A_2(T_{2,2};\hat m_1,\hat m_2,\hat y_2;q)&=&\hat y_2^2 -(q \hat m_1+q^{-1} \hat m_1^{-1} )\hat y_2+1 ~, \\
\wh A_3(T_{2,2};\hat m_1,\hat m_2,\hat y_1,\hat y_2;q)&=& (\hat m_1^2-1) q \hat m_2\hat y_1-( \hat m_2^2-1) q \hat m_1\hat y_2 -q^2(\hat m_1^2-\hat m_2^2)~. \nonumber
\eea
Instead of providing an independent derivation, we note that these recursion relations can be obtained as $a=-t=1$ limit of the $a$- and $t$-deformed recursions derived in appendix \ref{sec:app-B}.
The first and the second operator above represent the recursion relations with respect to one of the colors, while the third operator involves both colors. In the classical limit $q\to1$, the above recursion relations reduce to the classical variety
\bea
 A_1(T_{2,2};m_1,m_2,y_1)&=&  (y_1-m_2) (y_1-m_2^{-1})=0 ~,\cr
 A_2(T_{2,2};m_1,m_2,y_2)&=& (y_2-m_1) (y_2-m_1^{-1}) =0 ~, \label{recursion-Hopf-Jones-class}\\
 A_3(T_{2,2};m_1,m_2,y_1,y_2)&=& (  m_1^2-1) (  m_2 y_1-1)-(m_2^2-1) ( m_1 y_2 -1)=0~.\nonumber
\eea
The solutions to these equations are given by
\bea\label{Hopf-solution}
\left\{ \begin{array}{c} y_1=m_2\\ y_2=m_1  \end{array} \right. \qquad \textrm{and} \qquad \left\{ \begin{array}{c} y_1=m_2^{-1}\\ y_2=m_1^{-1}  \end{array} \right.~.
\eea
These solutions indicate that the meridian of one torus is indeed identified with the longitude of the other torus, and it is easy to see that they determine a Lagrangian subvariety of the Hitchin moduli space
\be\nonumber
\frac{\bC^*\times\bC^*}{\bZ_2^{(1)}}\times \frac{\bC^*\times\bC^*}{\bZ_2^{(2)}}
\ee
where the Weyl group symmetries are
\bea\nonumber
\bZ_2^{(1)}: ~ (m_1,y_1) \leftrightarrow (m_1^{-1},y_1^{-1})~,\qquad
\bZ_2^{(2)}: ~ (m_2,y_2) \leftrightarrow (m_2^{-1},y_2^{-1})~.
\eea

We note that in general the number of recursion relations for the link invariants can be greater than the dimension of the character variety. This occurs when the classical limit of recursion relations leads to equations that are not all independent. In our example, the system of equations (\ref{recursion-Hopf-Jones-class}) is indeed redundant, so that we do not all three equations to determine the set of solutions (\ref{Hopf-solution}). Indeed, these two solutions are determined by either the first and the third equation, or the second and the third equation. However, on the quantum level, all three operators (\ref{recursion-Hopf-Jones}) are independent.


\subsubsection*{$a$-deformation and augmentation variety}
Next, let us look at the recursion relations of colored HOMFLY invariants of the Hopf link:
\bea
\overline P_{[r_1],[r_2]}({T_{2,2}};a,{q})=q^{-r_1-r_2-r_1r_2}\frac{(a^2;{q}^2)_{r_1}}{({q}^2;{q}^2)_{r_1}}\sum_{i=0}^{r_2} {q}^{2(r_1+1)i}\frac{(a^2 {q}^{-2};{q}^2)_{i}}{({q}^{2};{q}^2)_{i}}~.~~\label{unreduced-Hopf-HOMFLY}
\eea
Note that this expression can be obtained by taking $t=-1$ specialization of \eqref{unreduced-Hopf+} and multiplying by the factor $q^{-r_1-r_2-r_1r_2}$, so that it agrees with the colored Jones polynomial \eqref{Jones-Hopf} at $a=q^2$.

The generalization of the three operators \eqref{recursion-Hopf-Jones} to the HOMFLY case can be written as
\bea\label{recursion-Hopf-HOMFLY}
&&\wh A_1(T_{2,2};\hat m_1,\hat m_2,\hat y_1;a,q)=\cr
&=& (1-\hat m_1^2)q^2\hat m_2^2\hat y_1^2 -(1-q^2\hat m_1^2 +q^2 \hat m_2^2-a^2 \hat m_1^2\hat m_2^2 )q\hat m_2\hat y_1+q^2 \hat m_2^2 (1-a^2\hat m_1^2) ~,\cr
&&\wh A_2(T_{2,2};\hat m_1,\hat m_2,\hat y_2;a,q)=\cr
&=& (1-\hat m_2^2)q^2\hat m_1^2\hat y_2^2 -(1-q^2\hat m_2^2 +q^2 \hat m_1^2-a^2\hat m_1^2\hat m_2^2 )q\hat m_1\hat y_2+q^2 \hat m_1^2 (1-a^2\hat m_2^2) ~, \cr
&&\wh A_3(T_{2,2};\hat m_1,\hat m_2,\hat y_1,\hat y_2;a,q)=\cr
&=& ( \hat m_1^2-1) q \hat m_2\hat y_1-(\hat m_2^2-1) q \hat m_1\hat y_2 -q^2(\hat m_1^2-\hat m_2^2)~.
\eea
Again, instead of providing an independent derivation, we can obtain the above operators from specializing $t=-1$ in the recursion relations derived in appendix \ref{sec:app-B}; we note that the corresponding recursion relations $\widehat{A}_i(T_{2,2}) \overline P_{[r_1],[r_2]}(T_{2,2};a,q)=0$ were also found in \cite{Aganagic:2013jpa}.
The classical limits of these recursion relations take form
\bea\label{classical-Hopf-HOMFLY}
 A_1(T_{2,2}; m_1, m_2, y_1;a)&=&m_2( (1- m_1^2)  y_1 - m_2(1-a^2 m_1^2) ) (m_1y_2-1)=0~,\cr
 A_2(T_{2,2}; m_1, m_2, y_2;a)&=&m_1( (1- m_2^2)  y_2 - m_1(1-a^2 m_2^2) ) (m_2y_1-1)=0~, \cr
 A_3(T_{2,2}; m_1, m_2, y_1, y_2;a)&=& (  m_1^2-1) (  m_2 y_1-1)-(  m_2^2-1) (  m_1 y_2 -1)=0~.
\eea
We find the following solutions to the equations above
\bea\label{Hopf-solution-HOMFLY}
\left\{ \begin{array}{c} (1- m_1^2) y_1=(1-a^2 m_1^2)m_2\\ (1- m_2^2) y_2=(1-a^2 m_2^2)m_1  \end{array} \right. \qquad \textrm{and} \qquad \left\{ \begin{array}{c} y_1=m_2^{-1}\\ y_2=m_1^{-1}  \end{array} \right.~.
\eea
It turns out that only one of the solutions in  \eqref{Hopf-solution} gets deformed by the complex structure $a$. Hence, it is clear that the solutions break the Weyl group symmetry and it is no longer a subvariety of the Hitchin moduli space. Rather, it is a Lagrangian subvariety in the covering space $\bC^*\times \bC^*$. Moreover, the second solution in \eqref{Hopf-solution-HOMFLY} coincides with the augmentation variety of the Hopf link given in \cite{Aganagic:2013jpa}.

\subsubsection*{Recursions for Poincar\'e polynomials}

We recall that the Poincar\'e polynomial of the $([r_1],[r_2])$-colored homology of the Hopf link is given in \eqref{unreduced-Hopf+}.
Rewriting it in terms of $\tilde{q}^2=q^2 t_c^2$ and $t=t_c$ we get
\bea
\overline\scP^{(+)}_{[r_1],[r_2]}({T_{2,2}};a,\wt{q},t)=\frac{(-a^2 t;\wt{q}^2)_{r_1}}{(\tilde{q}^2t^{-2};\wt{q}^2)_{r_1}}\sum_{i=0}^{r_2} \wt{q}^{2(r_1+1)i} t^{-2i}\frac{(-a^2 \wt{q}^{-2}t^3;\wt{q}^2)_{i}}{(\wt{q}^{2(r_2-i+1)}t^{-2};\wt{q}^2)_{i}}{r_2\brack i }_{\wt{q}^2}~.~~\label{unreduced-Hopf+bis}
\eea
A derivation of recursion relations for this Poincar\'e polynomial is quite technical. We therefore present it in Appendix \ref{sec:app-B}, where it is shown that these  recursion relations take the form $\wh A_i\, \overline\scP^{(+)}_{[r_1],[r_2]} = 0$, with
\bea
\wh A_1 
&=&\left(\wt{q}^2 \hat{x}_1-t^2\right) \left(\wt{q}^4 \hat{x}_1-t^2\right)\hat{y}_1^2\cr
&&+\left(\wt{q}^2 \hat{x}_1-t^2\right) \left(t^2+a^2 \wt{q}^2 t^3 \hat{x}_1 \hat{x}_2+\wt{q}^2 (-\wt{q}^2 \hat{x}_1+\hat{x}_2)\right)\hat{y}_1\cr
&&+\wt{q}^2 t^2 (1-\wt{q}^2 \hat{x}_1) (1+a^2 t \hat{x}_1) \hat{x}_2,   \label{A1-hat-HL} \\
\wh A_2
&=&\left(\wt{q}^2 \hat{x}_2-t^2\right) \left(\wt{q}^4 \hat{x}_2-t^2\right)\hat{y}_2^2\cr
&&+\left(\wt{q}^2 \hat{x}_2-t^2\right) \left(t^2+a^2 \wt{q}^2 t^3 \hat{x}_2 \hat{x}_1+\wt{q}^2 (-\wt{q}^2 \hat{x}_2+\hat{x}_1)\right)\hat{y}_2    \label{A2-hat-HL}  \\
&&+\wt{q}^2 t^2 (1-\wt{q}^2 \hat{x}_2) (1+a^2 t \hat{x}_2) \hat{x}_1~,\cr
\wh A_3 &=& (\wt{q}^2 \hat{x}_1 - t^2) \hat{y}_1 - (\wt{q}^2 \hat{x}_2 - t^2) \hat{y}_2 + \wt{q}^2(\hat{x}_2 - \hat{x}_1).    \label{A3-hat-HL}
\eea
Note that the first equation involves only the $\hat{y}_1$ operator and the second equation only the operator~$\hat{y}_2$. These first two equations are symmetric under the exchange of $(\hat{x}_1,\hat{y}_1)$ and $(\hat{x}_2,\hat{y}_2)$, and they are of second order in shift operators ($\hat{y}_1$ and $\hat{y}_2$ respectively). The third equation is of the first order and it involves both  $\hat{y}_1$ and $\hat{y}_2$ operators.

In the $\wt{q}\to 1$ limit, the above quantum equations reduce to
\bea
A_1 &=& (x_1-t^2)^2 y_1^2 + (x_1-t^2)(t^2 + x_2 - x_1 + a^2 t^3 x_1 x_2) y_1 + t^2 x_2 (1-x_1)(1+a^2 tx_1)  =  0 ,  \cr
A_2 &=& (x_2-t^2)^2 y_2^2 + (x_2-t^2)(t^2 + x_1 - x_2 + a^2 t^3 x_1 x_2) y_2 + t^2 x_1 (1-x_2)(1+a^2 tx_2)  =  0 ,  \cr
A_3 &=& (y_1 - 1)(x_1 - t^2) - (y_2-1)(x_2-t^2) = 0. \label{variety-hopf-class}
\eea
Note that the third equation is not independent from the first two equations; indeed, the difference of the first two equations includes the third equation as a factor
\be
 A_1 - A_2 = A_3\, \big(a^2 t^3 x_1 x_2 + x_1 y_1 + x_2 y_2 + t^2(1-y_1-y_2)   \big) .
\ee
One should therefore regard $A_1=A_{3}=0$, or $A_2=A_3=0$ (instead of $A_1=A_{2}=0$) as a set of two fundamental equations defining the variety.  This choice of a branch (encoded in the third equation $A_3=0$) is not fixed by the asymptotic analysis presented below, and only follows from the classical limit of the third recursion relation (\ref{A3-hat-HL}). This is a manifestation, on a classical level, of the quantum origin of the associated variety.

\subsubsection*{Large color asymptotics of Poincar\'e polynomial}

To see the large color asymptotics of the Poincar\'e polynomial (\ref{unreduced-Hopf+bis})
we take the limit $\wt{q}\to1$, keeping $\wt{q}^{2r_1}=x_1$, $\wt{q}^{2r_2}=x_2$ and $\wt{q}^{2i}=z$ fixed. This leads to the following twisted superpotential
\bea
\widetilde{\mathcal{W}}(T_{2,2};x_1,x_2,z;a,t )& = &  \log x_1\log z-2\log z\log t-\Li_2(x_2)+\Li_2(z)+\Li_2(x_2z^{-1})\cr
&&+\Li_2(-a^2 t^3)-\Li_2(-a^2 t^3 z)-\Li_2(x_2z^{-1}t^{-2})+\Li_2(x_2t^{-2})\cr
&&+\Li_2(-a^2 t)-\Li_2(-a^2 tx_1)-\Li_2(t^{-2})+\Li_2(t^{-2}x_1)~. \label{Wtilde-at-T22}
\eea
Solving the Neumann-Zagier equations (\ref{yifi}),
\bea\nonumber
y_1&=&e^{x_1\partial_{x_1}\widetilde{\mathcal{W}}(T_{2,2})}=\frac{(1 + a t x_1) z}{1 - x_1t^{-2}}~,\cr
y_2&=&e^{x_2\partial_{x_2}\widetilde{\mathcal{W}}(T_{2,2})}=\frac{(x_2-1 ) ( t^2 z-x_2)}{(t^2 - x_2) (x_2 - z)}~,\cr
1&=&e^{z\partial_{z}\widetilde{\mathcal{W}}(T_{2,2})}=\frac{x_1 (x_2 - z) (1 + a t^3 z)}{(z-1 ) (t^2 z-x_2)}~,
\eea
we subsequently obtain
\bea\label{cAv-Hopf}
A_1(T_{2,2};x_1,x_2,y_1;a,t)&=&(x_1-t^2 )^2 y_1^2+(x_1-t^2 ) ( a^2 t^3 x_1 x_2- x_1 + x_2+t^2  )y_1\cr
&&-t^2 ( x_1-1) (1 + a^2 t x_1) x_2~,\cr
A_2(T_{2,2};x_1,x_2,y_2;a,t)&=&
(x_2-t^2 )^2 y_2^2+(x_2-t^2 ) ( a^2 t^3 x_1 x_2- x_2 + x_1+t^2  )y_2\cr
&&-t^2 ( x_2-1) (1 + a^2 t x_2) x_1~.
\eea
Actually, these equations can be also obtained from the asymptotic analysis of the refined Chern-Simons invariants \eqref{refined-CS-sym+} of the Hopf link. The corresponding twisted superpotential is given in \eqref{twisted-rCS} and the saddle point equation is given in \eqref{saddle-torus-link}. This provides evidence that the Poincar\'e polynomial of colored HOMFLY homology \eqref{unreduced-Hopf+}  coincides with the refined Chern-Simons invariant \eqref{refined-CS-sym+} in the case of the Hopf link. Taking $t=-1$ we get
\bea
A_1(T_{2,2};x_1,x_2,y_1;a,t=-1)&=& (x_1-1) (y_1-1) (x_2-a^2 x_1 x_2-y_1+x_1 y_1)~,\cr
A_2(T_{2,2};x_1,x_2,y_2;a,t=-1)&=&(x_2-1) (y_2-1) (x_1-a^2 x_1 x_2-y_2+x_2 y_2)~.
\eea
By changing $y_1\to y_1m_2$ and $y_2\to y_2m_1$, this can be identified with the first and second equations \eqref{classical-Hopf-HOMFLY}, up to trivial factors of $(x_i-1)$.

We note the variety (\ref{cAv-Hopf}) consists of two branches, and only one of these branches should be identified as the Hopf link associated variety (\ref{variety-hopf-class}). It turns out that one of these branches is distinguished by the third equation in (\ref{variety-hopf-class}), which arises from the classical limit of refined recursion relations. The fact that one of the two branches is distinguished can be thought of as a remnant of the quantum origin of this system.


\subsection{Associated varieties for twist links}\label{sec:AV-TL}
For hyperbolic links, the HOMFLY homology colored by arbitrary symmetric representation is currently not known. We can however analyze associated varieties of hyperbolic links at the level of the $a$-deformation, taking advantage of the cyclotomic expansions of colored HOMFLY invariants of the twist links \eqref{cyclotomic-HOMFLY}. Using the expression \eqref{cyclotomic-HOMFLY} for the twist link $L_p$, we find the twisted superpotential
\be\nonumber
\widetilde{\mathcal{W}} (L_p;x_1,x_2,w,z,a)= \widetilde{\mathcal{W}}_F(x_1,z,a)+\widetilde{\mathcal{W}}_F(x_2,z,a)+\widetilde{\mathcal{W}}_G(z,a)+\widetilde{\mathcal{W}}_{\wt\omega_p}(z,w,a)~.
\ee
where
\bea\label{twisted-sp-blocks}
\widetilde{\mathcal{W}}_F(x,z,a)&=& i \pi \log z-\log a\log ( \tfrac{x}{z})+\tfrac34 (\log z)^2-\log z\log x-\Li_2 (z)+\Li_2 (a^2z^2)\cr
&&-\Li_2 (a^2xz)+\Li_2 (x/z)\cr
\widetilde{\mathcal{W}}_G(x,z,a)&=&-3 \log a \log z-\tfrac34 (\log z)^2+2\Li_2(a^2)-\Li_2(a^2z^2)-\Li_2(a^2z)+2\Li_2(z)~.\cr
&&\cr
\widetilde{\mathcal{W}}_{\wt\omega_p}(z,w,a)&=& \log a \log z + \tfrac14 (\log z)^2 + i\pi \log w +
 2 p \log a \log w + ( p + \tfrac 12) (\log w)^2\cr
 && - \Li_2( a^2 w) +
 \Li_2( a^2 w z )+ \Li_2( z )- \Li_2( w )- \Li_2( z/w)~.
\eea
Then, the saddle point equations can be read off:
\bea\label{saddle-twistlinks}
y_1 & = & \frac{a^2 x_1 z - 1}{a(x_1 - z)}~, \qquad \qquad   y_2  =  \frac{a^2 x_2 z - 1}{a(x_2 - z)}~, \cr
1&=&\frac{(x_1 - z) (x_2 - z) (z-w ) (a^2 z-1 ) (a^2 x_1 z-1) (a^2 x_2 z-1)}{w x_1 x_2 (z-1) (a^2 w z-1) (a^2 z^2-1)^2}~,\cr
1&=&\frac{a^{2 p}  w^{2 + 2 p}(w-1) (a^2 w-1)}{(w - z) (a^2 w z-1)   }~.
\eea
It is straightforward to see that the above set of equations is invariant under the $\bZ_2$  symmetry
\bea
(x_i,y_i) \leftrightarrow \left(\frac{1}{a^2x_i}, \frac{1}{y_i}\right) ~, \qquad\qquad i=1,2~.
\eea
Therefore, the associated variety of the twist link $L_p$ actually enjoys the $\bZ_2$ Weyl symmetry, unlike in the case of torus links. Furthermore, the holomorphic symplectic form $\omega=\sum_{i=1,2}d\log x_i\wedge d\log y_i$  vanishes on the above set of equations. Therefore, the associated variety of a twist link is a Lagrangian subvariety of the Hitchin moduli space $\frac{\bC^*\times\bC^*}{\bZ_2}\times \frac{\bC^*\times\bC^*}{\bZ_2}$. As an example, we present explicit equations for the associated variety of the Whitehead link in Appendix \ref{sec:AV-WL}, and we show that the associated variety of the Borromean rings also has the $\bZ_2$ symmetry in Appendix \ref{sec:AV-BR}.


\section{Mutant pairs}
\label{sec:mutants}
\begin{figure}[htb]
\centering
\includegraphics[width=1\textwidth]{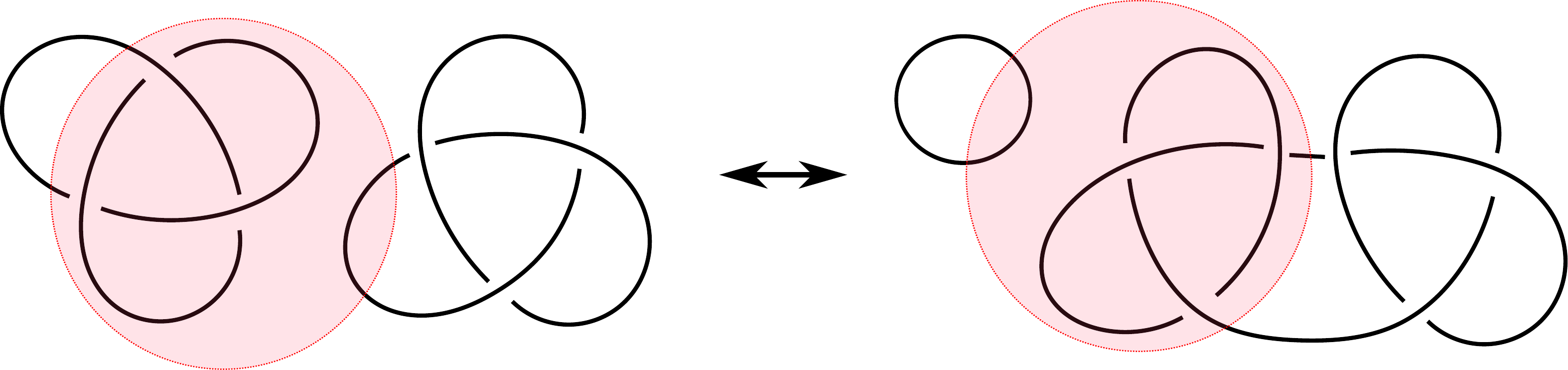}
\caption{A mutant pair: ${\bf 3_1 \sqcup \overline{3}_1}$ in the left and $\bigcirc \sqcup ({\bf 3_1 \sharp\, \overline{3}_1})$ in the right. This mutant pair is obtained by rotating the 3-ball represented in red by $180^\circ$ along the axis perpendicular to the projection plane. } \label{fig:mutant}
\end{figure}

In this section we show that the unreduced $\fraksl(2)$ homology distinguishes a certain family of mutant link pairs.
The mutation of links in $S^3$ is defined in the following way \cite{Conway:1970}. The 3-sphere $S^3$ is cut along an $S^2$ which intersects a link exactly four times (forming a 2-tangle), ending up with two 3-balls. Then, one of these 3-balls is rotated by $180^\circ$ along a certain axis of $S^2$ and glued back to the other 3-ball; see Figure \ref{fig:mutant}.
It has been proven that a mutant pair cannot be distinguished by polynomial invariants, including Jones, HOMFLY and Kauffman, colored by symmetric representations \cite{Morton:1996}. However, it is observed in \cite{Wehrli:2003} that Khovanov homology detects mutant pairs of the following class.

Let us start with  two torus knots or links $T_{2,n_1}$ and $T_{2,n_2}$ with $n_1,n_2 > 2$. Then, taking a disjoint union and a connected sum, we have a mutant pair
\be
L' = T_{2,n_1} \sqcup T_{2,n_2}~, \qquad \textrm{and} \qquad L = \unknot \sqcup \left( T_{2,n_1} \sharp\, T_{2,n_2} \right)
\label{mutantlinks}
\ee
where, as usual, $\unknot$ denotes the unknot. The \emph{unreduced} $\fraksl(2)$ homology for a disjoint union of two knots $K_1$ and $K_2$ is the tensor product of unreduced  $\fraksl(2)$ homology of both these knots
\bea\nonumber
\overline\scH^{\fraksl(2)}_{\yng(1)}(K_1\sqcup K_2)= \overline\scH^{\fraksl(2)}_{\yng(1)}(  K_1 ) \otimes\overline\scH^{\fraksl(2)}_{\yng(1)}( K_2 )~,
\eea
which relates their Poincar\'e polynomials by
\bea\label{disjoint}
\overline\scP^{\fraksl(2)}_{\yng(1)}(K_1\sqcup K_2)=\overline\scP^{\fraksl(2)}_{\yng(1)}(K_1)\cdot \overline\scP^{\fraksl(2)}_{\yng(1)}(K_2)~.
\eea
It is straightforward to deduce the Poincar\'e polynomial of the unreduced $\fraksl(2)$ homology of the torus links $T_{2,n}$ for $n>2$
\bea
\overline\scP^{\fraksl(2)}_{\yng(1)}(T_{2,2p+1};q,t)&=&q^{2p}\left[1+q^2+q^4t^2(1+q^4t)\frac{1-q^{4p}t^{2p}}{1-q^4t^2}\right]\\
\overline\scP^{\fraksl(2)}_{\yng(1)}(T_{2,2p};q,t)&=&q^{2p-1}\left[(1+q^2)(1+q^{4p}t^{2p})+q^4t^2(1+q^4t)\frac{1-q^{4(p-1)}t^{2(p-1)}}{1-q^4t^2}\right]~.\nonumber
\eea
Since neither of the two polynomials is  divisible by the unknot factor
\be\nonumber
\overline\scP^{\fraksl(2)}_{\yng(1)}(\unknot;q,t)=q+q^{-1},
\ee
while the Poincar\'e polynomial of $L$ obviously is, we conclude that the unreduced $\fraksl(2)$ homology distinguishes the mutant pair \eqref{mutantlinks}:
\bea\nonumber
\overline\scP^{\fraksl(2)}_{\yng(1)}(L;q,t)\neq\overline\scP^{\fraksl(2)}_{\yng(1)}(L';q,t)~.
\eea


As an example let us consider a disjoint union $\bf 3_1\sqcup\overline{3}_1$ of the trefoil knot and its mirror image, and the corresponding mutant configuration $\unknot \sqcup \big({\bf 3_1 \sharp\, \overline{3}_1}\big)$, see Figure \ref{fig:mutant}. For a connected sum $K_1\sharp\, K_2$, its \emph{reduced} HOMFLY polynomials colored by symmetric representations can be written as
\bea\label{connected}
P_{[r]}(K_1\sharp\, K_2;a,q)=P_{[r]}(K_1;a,q)\cdot P_{[r]}(K_2;a,q)~.
\eea
Since $\bf 3_1 \sharp\, \overline{3}_1$ is homologically thin,  its \emph{reduced} HOMFLY homology simply follows from \eqref{delta-grading}:
\bea\nonumber
\scP_{\yng(1)}({\bf 3_1 \sharp\, \overline{3}_1};a,q,t)&=&\scP_{\yng(1)}({\bf 3_1};a,q,t)\cdot \scP_{\yng(1)}({\bf \overline{3}_1};a,q,t)=\cr
&=&(a^2(q^{-2}+q^2t^2)+a^4t^3)(a^{-2}(q^{2}+q^{-2}t^{-2})+a^{-4}t^{-3})~.
\eea
Note that, for the mirror image, we have the relationship $\scP_{\yng(1)}({\bf \overline{3}_1};a,q,t)= \scP_{\yng(1)}({\bf 3_1};a^{-1},q^{-1},t^{-1})$.
Multiplying with the unknot factor $\scP_{\yng(1)}(\unknot)=a^{-1}q(1+a^2t)/(1-q^2)$, the action of the $d_2$ differential on $\overline\scH_{\yng(1)}({\bf 3_1 \sharp\, \overline{3}_1})$ becomes non-trivial, leaving
\bea\nonumber
\overline\scP^{\fraksl(2)}_{\yng(1)}({\bf 3_1 \sharp\, \overline{3}_1};q,t)=q^{-7} t^{-3} + (t^{-2} + t^{-1})q^{-3} +  2 q^{-1}+ 2 q  + q^3 (t^2 + t^3)+ q^7 t^3~.
\eea
On the other hand, the unreduced $\fraksl(2)$ homology of the trefoil is given by
\bea\nonumber
\overline\scP^{\fraksl(2)}_{\yng(1)}({\bf 3_1};q,t) =q + q^3 + q^5 t^2 + q^9 t^3~.
\eea
Therefore, we conclude from \eqref{disjoint} that
\begin{small}
\bea\label{sl2-mutant}
\overline\scP^{\fraksl(2)}_{\yng(1)}({\bf 3_1 \sqcup \overline{3}_1};q,t) &=&
(q + q^3 + q^5 t^2 + q^9 t^3)(q^{-1}+q^{-3}  + q^{-5}t^{-2} + q^{-9}t^{-3} ),\\
\overline\scP^{\fraksl(2)}_{\yng(1)}(\unknot \sqcup ({\bf 3_1 \sharp\, \overline{3}_1});q,t)  &=&(q^{-1}+q)(q^{-6} t^{-3} + (t^{-2} + t^{-1})q^{-2} +  2 + 2 q^2  + q^4 (t^2 + t^3)+ q^8 t^3)~,\nonumber
\eea\end{small}
which clearly shows that the unreduced $\fraksl(2)$ homology distinguishes this mutant pair
\bea\nonumber
\overline\scH^{\fraksl(2)}_{\yng(1)}(\unknot \sqcup ({\bf 3_1 \sharp\, \overline{3}_1}))\ncong\overline\scH^{\fraksl(2)}_{\yng(1)}({\bf 3_1 \sqcup \overline{3}_1})~.
\eea
In particular, $\dim \overline\scH^{\fraksl(2)}_{\yng(1)}({\bf 3_1 \sqcup \overline{3}_1})=16$ while  $\dim \overline\scH^{\fraksl(2)}_{\yng(1)}(\unknot \sqcup ( {\bf 3_1 \sharp\, \overline{3}_1}))=20$. However, setting $t=-1$ in \eqref{sl2-mutant}, we obtain the identity at the level of the Jones polynomials as we expected
\bea\nonumber
\overline J (\unknot \sqcup ({\bf 3_1 \sharp\, \overline{3}_1});q)=\overline J ({\bf 3_1 \sqcup \overline{3}_1};q)=(q+q^{-1})^2(1 +q^4 -q^6)(1 +q^{-4} -q^{-6})~.
\eea
\\

Having studied uncolored homology of the mutant pairs, another question we shall address is whether or not the identity \eqref{connected} can be uplifted to the level of the colored HOMFLY homology. When $K_1$ and $K_2$ are homologically thin, we claim that the equality
\be
\scP_{[r^\rho]}(K_1\sharp\, K_2)=  \scP_{[r^\rho]}(K_1)\cdot \scP_{[r^\rho]}(K_2)  \label{P-connected-sum}
\ee
holds for the Poincar\'e polynomials.
Now, if $K_1$ and $K_2$ are homologically thin, so is their connected sum $K_1\sharp\, K_2$ \cite{Khovanov:2003b}. Therefore, it is easy to verify the identity in the uncolored case by using the $\delta$-grading. In addition, when the color is specified by a rectangular Young diagram, the exponential growth property and the differential structure guarantee the identity. It would be interesting to study the case when either $K_1$ or $K_2$ is homologically thick.

For instance, let us explicitly compute the Poincar\'e polynomials of $[r]$-colored HOMFLY homology when $K_1={\bf 3_1}$ and $K_2={\bf \overline{3}_1}$.
The Poincar\'e polynomials of the $[r]$-colored HOMFLY homology for the trefoil in the $(a,q,t_r)$-grading can be expressed \cite{Fuji:2012pi}
\be\nonumber
\scP_{[r]} ({\bf 3_1};a,q,t) =a^{2 r} q^{-2 r} \sum_{k=0}^{r} q^{2 k (r+1)} t^{2 k} (-a^2 t q^{-2},q^2)_k{r \brack k}_{q^2} ~,
\ee
and for its mirror image we have $\scP_{[r]} ({\bf \overline{3}_1};a,q,t)=\scP_{[r]} ({\bf  3_1};a^{-1},q^{-1},t^{-1})$. On the other hand, for a connected sum, we find the $[r]$-colored HOMFLY homology in the $(a,q,t_r)$-gradings
\bea
\scP_{[r]}({\bf {\bf 3_1 \sharp\, \overline{3}_1}};a,q,t) &=& \sum_{r\ge j\ge i\ge0}{r \brack j}_{q^2}  {j \brack i}_{q^2} a^{-2j} t^{-j-2i} q^{-2j r -2 j i +j^2 +3j -2i}   \cr
& & \qquad \times (-a^2 t^3 q^{2r};q^2)_j (-a^2 t q^{-2};q^2)_i (-a^2 t q^{-2};q^2)_{j-i} ~. \label{P31connected31}
\eea

\begin{figure}[htb]
\begin{center}
\includegraphics[width=0.5\textwidth]{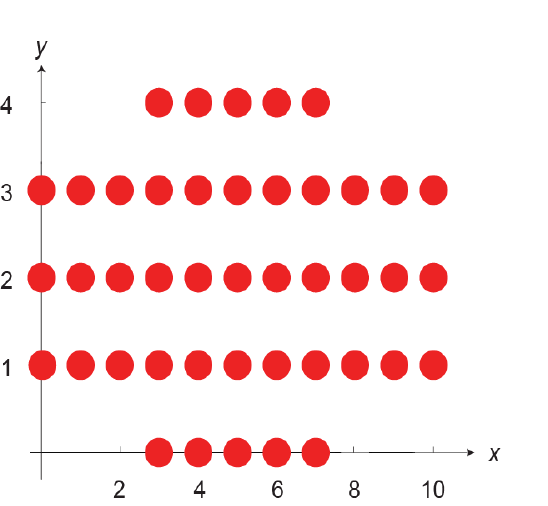}
\caption{\emph{Newton polygon of the super-$A$-polynomial for  $\scP_{[r]}({\bf 3_1 \sharp\, \overline{3}_1})$ and $ \scP_{[r]}({\bf 3_1})\cdot \scP_{[r]}({\bf \overline{3}_1})$.}   } \label{fig-mutants-Newton}
\end{center}
\end{figure}

The large color asymptotics for the above two links take different form. In particular, we find the twisted superpotential for $ \scP_{[r]}({\bf 3_1})\cdot \scP_{[r]}({\bf \overline{3}_1})$
\bea\nonumber
\widetilde{\mathcal{W}}({\bf 3_1 \cdot \overline{3}_1};x,z_1,z_2,a) & = & \log (z_1 z_2^{-1}) \,\log(x t^2)  - \textrm{Li}_2(-(a^2 t)^{-1}) + \textrm{Li}_2(-a^2 t)  + \textrm{Li}_2(x^{-1})\cr
& & - \textrm{Li}_2(x)
+ \textrm{Li}_2(x z_1^{-1}) + \textrm{Li}_2(z_1) -
  \textrm{Li}_2(-a^2 t z_1) - \textrm{Li}_2(z_2^{-1}) \cr
  &&+ \textrm{Li}_2(-(a^2 t z_2)^{-1}) -
  \textrm{Li}_2(z_2 x^{-1})~,
\eea
where $z_1=q^{2k}$ encodes an auxiliary summation variable $k$ in the expression for the superpolynomial for $\bf 3_1$, and similarly $z_2$ encodes such variable in the superpolynomial for $\bf  \overline{3}_1$. In a similar way, we can write the twisted superpotential
 for $\scP_{[r]}({\bf 3_1 \sharp\, \overline{3}_1})$
\bea\nonumber
\widetilde{\mathcal{W}}({\bf 3_1 \sharp\, \overline{3}_1};x,z_1,z_2,a) & = &
- \log(a^2 x z_2)\log z_1 - \log t \, \log(z_1 z_2^2) + \frac{1}{2} (\log z_1)^2
-\textrm{Li}_2(x)  \cr
&&+ \textrm{Li}_2(x z_1^{-1})+ \textrm{Li}_2(z_2) +
 \textrm{Li}_2(z_1 z_2^{-1})  +
 \textrm{Li}_2(-a^2 t^3 x) - \textrm{Li}_2(-a^2 t^3 x z_1) \cr
&&+ \textrm{Li}_2(-a^2 t) - \textrm{Li}_2(-a^2 t z_2) + \textrm{Li}_2(-a^2 t) - \textrm{Li}_2(-a^2 t z_1 z_2^{-1}) ~,
\eea
where $z_1=q^{2j}$ and $z_2=q^{2i}$ encode the dependence on $i$ and $j$ in (\ref{P31connected31}). While these two twisted superpotentials lead to different sets of equations, after eliminating $z_1$ and $z_2$ in both cases we find the same super-$A$-polynomial, which can be written as
\bea
&&A^{\textrm{super}}({\bf 3_1 \cdot \overline{3}_1};x,y,a,t)=A^{\textrm{super}}({\bf 3_1 \sharp\, \overline{3}_1};x,y,a,t) \cr
 & = & a^4 t^3 x^3 (1 + a^2 t^3 x)^4y^4 \nonumber\\[.1cm]
 &&-a^2 (1 + a^2 t^3 x)^2 (1 + a^2 t x + 2 a^2 t^3 x^2 +
   2 a^4 t^4 x^2 - a^4 t^4 x^3 + a^4 t^6 x^4)   \cr
   &&\quad(1 - t^2 x + 2 t^2 x^2 + 2 a^2 t^3 x^2 + a^2 t^5 x^3 + a^4 t^6 x^4)y^3\nonumber\\[.1cm]
   &&+t (1 - x) (1 + a^2 t^3 x) (-1 + a^6 t - 2 a^2 t x - 2 a^6 t^3 x - a^4 t^2 x^2 - 4 a^2 t^3 x^2 +
 4 a^6 t^3 x^2 - 4 a^4 t^4 x^2\cr
   &&\quad +4 a^8 t^4 x^2 + a^6 t^5 x^2 - 2 a^4 t^4 x^3 - 6 a^6 t^5 x^3 -
 2 a^8 t^6 x^3 + 4 a^6 t^5 x^4 - 6 a^4 t^6 x^4 + 8 a^8 t^6 x^4 \cr
   &&\quad-8 a^6 t^7 x^4 + 6 a^{10} t^7 x^4 - 4 a^8 t^8 x^4 + 2 a^6 t^7 x^5 +
 6 a^8 t^8 x^5 + 2 a^{10} t^9 x^5 - a^8 t^8 x^6 - 4 a^6 t^9 x^6\cr
   &&\quad+4 a^{10} t^9 x^6 - 4 a^8 t^{10} x^6 + 4 a^{12} t^{10} x^6 + a^{10} t^{11} x^6 +
 2 a^8 t^{10} x^7 + 2 a^{12} t^{12} x^7 - a^8 t^{12} x^8 \cr
 &&\quad+ a^{14} t^{13} x^8)y^2\nonumber\\[.1cm]
   &&+a^4 t^3 (1-x)^2 (1 + a^2 t x + 2 a^2 t^3 x^2 +
   2 a^4 t^4 x^2 - a^4 t^4 x^3 + a^4 t^6 x^4) \cr
   &&\quad(1 - t^2 x + 2 t^2 x^2 + 2 a^2 t^3 x^2 + a^2 t^5 x^3 + a^4 t^6 x^4)y\nonumber\\[.1cm]
   &&+a^8 t^9 (1 - x)^4 x^3~. \label{Asuper-3131}
\eea
Its Newton polygon is shown in Figure \ref{fig-mutants-Newton}, and all quantizability conditions are met. In the sequel, we conclude that
\be\nonumber
\scP_{[r]}({\bf 3_1 \sharp\, \overline{3}_1};a,q,t)=  \scP_{[r]}({\bf 3_1};a,q,t)\cdot \scP_{[r]}({\bf \overline{3}_1};a,q,t)~ ,
\ee
confirming the claim (\ref{P-connected-sum}).

\section{Thick torus knots and HFK-like differentials}\label{sec:HFK}

In this section we study the properties of colored HOMFLY homology of thick torus knots, specifically, the $(3,4)$ and~$(3,5)$ torus knots. For these knots we obtain the $[r]$-colored HOMFLY homology and the homology with respect to the HFK-like differentials. In addition, analysis of large color asymptotics of $[r]$-colored HOMFLY homology provides the super-$A$-polynomial of the $(3,4)$ torus knot.

All the torus knots except $(2,2p+1)$ torus knots are homologically thick. However, their colored HOMFLY homologies are subject to the refined exponential growth property, which enables us to study HOMFLY homology carrying symmetric representations of arbitrary rank.  Besides, since the colored HOMFLY homology of the $(3,3)$ torus link, one of the simplest three-component links, resembles that of the $(3,4)$ torus knot, it is important to understand the colored HOMFLY homology of the $(3,4)$ torus knot.

Another important aspect of thick homology is the relation to knot Floer homology. The knot Floer homology $\HFK(K)_{i,j}$ provides a categorification of the Alexander polynomial $\Delta (K;q)$, which can be obtained by the $a=-1$ specialization of the HOMFLY polynomial: $\Delta (K;q)=P(K;a=-1,q)$. Specifically, the Alexander polynomial is realized as the $q$-graded Euler characteristic of knot Floer homology,
\begin{equation}\nonumber
\Delta (q) = \sum_{i,j} (-1)^j q^i \dim \HFK(K)_{i,j}.
\end{equation}
It was conjectured in \cite{Dunfield:2005si} that the HOMFLY homology is endowed with the  HFK differential $d_0$ whose homology is isomorphic to the knot Floer homology.  Defining the Poincar\'e polynomial of the knot Floer homology as
\begin{equation}
\label{hfkdef}
\hfk (K;q,t) := \sum_{i,j} q^i t^j  \dim \HFK(K)_{i,j}~,
\end{equation}
the regrading for this isomorphism is expressed by
\bea\nonumber
\scP(\scH_{\yng(1)}(K),d_0)(a=t^{-1},q,t)=\hfk (K;q,t)~.
\eea
Furthermore, the authors of \cite{Gorsky:2013jxa} proposed that there exists the $d_{1-r}$ ($d_{1|r}$) differential in $[r]$-colored HOMFLY homology, in the $(a,Q,t_r,t_c)$-gradings, which is an extension of  the HFK differential  $d_0$  to the colored case, \textit{i.e.} $d_{1|1}=d_0$. This is a hybrid of ``positive'' colored differentials and ``negative'' colored differentials, where the $(a,Q,t_r,t_c)$-gradings of the $d_{1-r}$ differential are given by
\be\nonumber
\text{deg} ~d_{1-r}  = (-2,0,-3,-2r-1)~.
\ee
The $(a,Q,t_r,t_c)$-gradings of the $d_{1-r}$ differential relate two generators with different $\delta$-gradings. Hence, it acts trivially for homologically thin knots, so that  interesting cases can be found only in thick homologies.
The $(a,Q,t_r,t_c)$-gradings have the further advantage that the homology with respect to the $d_{1-r}$ differential  exhibits the refined exponential growth property:
\bea\label{REGP-HFK}
\wt \scP(\wt\scH_{[r]}(K),d_{1-r})(a,Q=q,t_r=t,t_c=1)=\Big[\scP(\scH_{\yng(1)}(K),d_{1|1})(a,q,t)\Big]^r~.
\eea

\begin{figure}[htb]
\begin{center}
\includegraphics[width=0.7\textwidth]{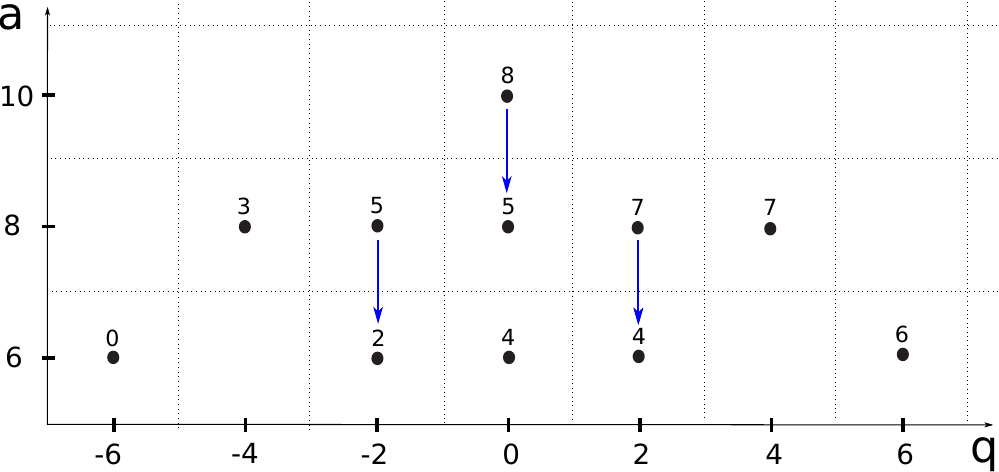}
\caption{The HOMFLY homology of the $(3,4)$ torus knot. Blue arrows represent the action of the HFK differential.} \label{fig:HOMFLY-T34}
\end{center}
\end{figure}

Now, let us look at the HOMFLY homology of the $(3,4)$ torus knot $T_{3,4}$. The Poincar\'e polynomial of the uncolored HOMFLY homology of $T_{3,4}$ is expressed by
\be\nonumber
\scP_{\yng(1)}(T_{3,4};a,q,t)= a^6 (q^{-6} + q^{-2}t^2 + t^4 + q^2 t^4 + q^6 t^6) +
 a^8 (q^{-4}t^3 + t^5 + q^{-2}t^5 + q^2 t^7 + q^4 t^7)+a^{10} t^8~,
\ee
and the $S$-invariant is $S(T_{3,4})=6$. Figure \ref{fig:HOMFLY-T34} depicts the action of the HFK differential $d_0$ (or $d_{1|1}$) of $(a,q,t)$-degree $(-2,0,-3)$ on $\scH_{\yng(1)}(T_{3,4})$, leaving
\be\nonumber
\scP(\scH_{\yng(1)}(T_{3,4}),d_{1|1})(a,q,t)=a^6q^{-6}+a^8q^{-4}t^3+a^6t^4+a^8q^{4}t^7+a^6q^{6}t^6~.
\ee
Setting $a=t^{-1}$ we get
\begin{equation}\nonumber
\scP(\scH_{\yng(1)}(T_{3,4}),d_{1|1})(a=t^{-1},q,t)= q^{-6}t^{-6}+q^{-4}t^{-5}+q^0t^{-2}+q^4t^{-1}+q^6~,
\end{equation}
which agrees with $\hfk(T_{3,4})$.  The $[2]$-colored HOMFLY homology of $T_{3,4}$ has been first obtained in \cite[Appendix B]{Gukov:2011ry} and the homology with respect to $d_{1|2}$ has been shown in \cite[\S4.3]{Gorsky:2013jxa}.

Although the $(3,4)$ torus knot is homologically thick, by taking advantage of the refined exponential growth property, one can find two different but equivalent expressions for the quadruply-graded $[r]$-colored HOMFLY  homology of $T_{3,4}$
\bea
&&\wt\scP_{[r]}(T_{3,4};a,Q,t_r,t_c) = \cr
&=&a^{6 r}t_r^{4r} t_c^{4 r^2}\sum _{j=0}^r  \sum _{j\ge k_1\ge k_2\ge k_3\ge 0}  Q^{-6 j + 4 (k_1 + k_2 + k_3)}  t_r^{2 (-2 j + k_1 + k_2 + k_3 )}\cr
 &&\times t_c^{2 ( -k_1 -k_2 -k_2 k_3  +(k_1+ k_2  + k_3) r + j (k_2 + k_3 - 2 r) )}  {r\brack j}_{t_c^2}{j\brack k_1}_{t_c^2}{k_1\brack k_2}_{t_c^2}{k_2\brack k_3}_{t_c^2}\cr
&&\times\left(-a^2 Q^{-2}t_r t_c;t_c^2\right)_{r-j}\left(-a^2 Q^2t_r^3 t_c^{1+2 r} ;t_c^2\right)_{r-j}\left(-a^2Q^{-2}t_r t_c^{1+2 (r-j)} ;t_c^2\right)_{k_1} \label{T34-HOMFLY-1} \\
& = & (a Q t_r)^{6r} t_c^{6r^2} \sum_{r\ge j \ge \gamma \ge \beta\ge\a\ge 0}  a^{2(j-\gamma)} Q^{-4\alpha-4\beta+4\gamma-8j}t_r^{-2(\alpha+\beta+\gamma)-(j-\gamma)} \cr
& &\times   t_c^{-2(\alpha^2+\beta^2+\gamma^2) -(j-\gamma)^2 -2(\alpha+\beta+\gamma)(j-\gamma)} {r \brack j}_{t_c^{-2}} {j \brack \gamma}_{t_c^{-2}} {\gamma \brack \beta}_{t_c^{-2}} {\beta \brack \alpha}_{t_c^{-2}}  \cr
& &\times (-a^2 Q^2 t_r^3 t_c^{2r+1}; t_c^2)_j (-a^{-2} Q^2 t_r^{-1} t_c^{-1}; t_c^{-2})_{j-\gamma} ~.\label{T34-HOMFLY-2}
\eea
We checked that this Poincar\'e polynomial, specialized as $\wt\scP_{[r]}(T_{3,4};a,q,q^{-1},qt)$, agrees with the corresponding refined Chern-Simons invariant computed in \cite{Shakirov:2013moa}  up to $r=3$. In addition, we find that the Poincar\'e polynomial of the homology with respect to the HFK-like differential $d_{1-r}$ takes the form
\bea
\wt\scP(\wt\scH_{[r]}(T_{3,4}),d_{1-r})(a,Q,t_r.t_c)&=&a^{6 r} \sum _{k=0}^r \sum _{i=0}^k Q^{6 (k-2 i)} t_r^{2 k+4 r-6 i} t_c^{-4 i^2+i (4 k-6 r)+2 r (k+2 r)} \cr
&&\times {r\brack k}_{t_c^2}{k\brack i}_{t_c^2}\left(-a^2Q^{-2} t_c t_r;t_c^2\right)_{k-i}\left(-a^2 Q^2t_r^3 t_c^{1+2 r} ;t_c^2\right)_i~.\nonumber
\eea
As a consistency check, this expression indeed  satisfies the refined exponential growth property \eqref{REGP-HFK}.

\begin{figure}[htb]
\begin{center}
\includegraphics[width=0.9\textwidth]{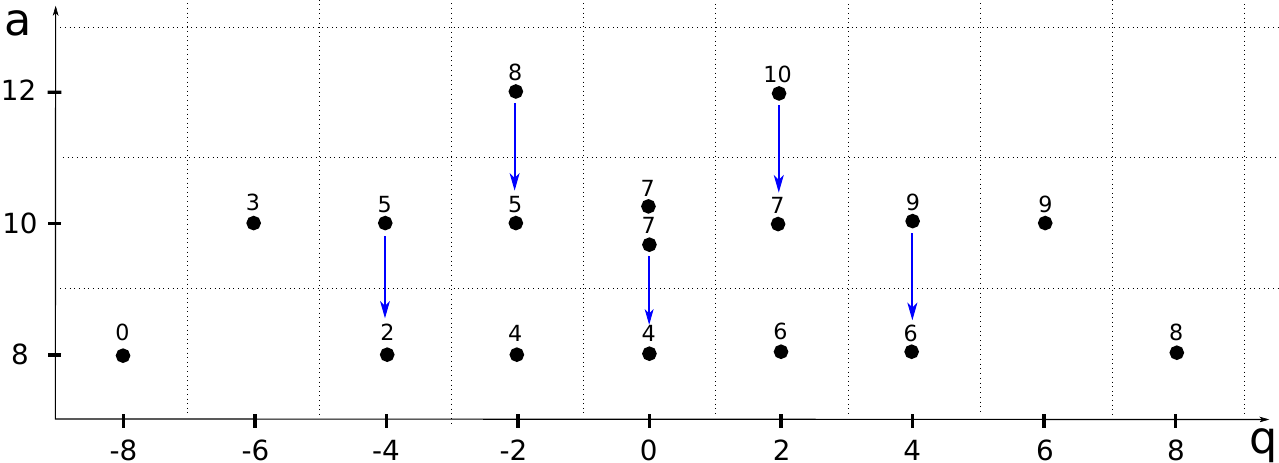}
\caption{The HOMFLY homology of the $(3,5)$ torus knot. Blue arrows represent the action of the HFK differential.} \label{fig:HOMFLY-T35}
\end{center}
\end{figure}

Next, we consider  the $(3,5)$ torus knot $T_{3,5}$.
The uncolored HOMFLY homology of $T_{3,5}$ is presented in Figure \ref{fig:HOMFLY-T35}, and its Poincar\'e polynomial is given by
\bea\nonumber
\scP_{\yng(1)}(T_{3,5};a,q,t)&=&a^8 (q^{-8} + q^{-4}t^2 + q^{-2} t^4+ t^4 + q^2 t^6 + q^4 t^6 + q^8 t^8) \cr
&&+  a^{10} (q^{-6}t^3 +q^{-4}t^5 + q^{-2}t^5 + 2 t^7 + q^2 t^7 + q^4 t^9 +
    q^6 t^9) \cr
    &&+ a^{12} (q^{-2}t^8 + q^2 t^{10})~.
\eea
As shown in Figure \ref{fig:HOMFLY-T35}, the HFK differential $d_{1|1}$ leaves 7 generators:
\be\label{HFK-T35}
\scP(\scH_{\yng(1)}(T_{3,5}),d_{1|1})(a,q,t)=a^8 (q^{-8} + q^{-2} t^4 + q^2 t^6 + q^8 t^8)+ a^{10} (q^{-6}t^3 + t^7 + q^6 t^9)~,
\ee
which, therefore, provides the HFK homology of $T_{3,5}$ upon setting $a=t^{-1}$
\bea\nonumber
\hfk(T_{3,5};q,t)= q^{-8} t^{-8} + q^{-6} t^{-7} + q^{-2} t^{-4} + t^{-3} + q^2t^{-2} + q^6t^{-1}+q^8 ~.
\eea
The dimension of the colored HOMFLY homology of $T_{3,5}$ is rather large, for instance, $\dim \scH_{[2]}(T_{3,5})=289$. In spite of this difficulty, we have succeeded in finding the $[r]$-colored HOMFLY homology of $T_{3,5}$ in the $(a,Q,t_r,t_c)$-gradings
\bea\label{HOMFLY-T35}
&&\wt\scP_{[r]}(T_{3,5};a,Q,t_r,t_c) = \cr
&=&a^{8 r}Q^{-2r}t_r^{4r} t_c^{4 r^2}\sum _{j=0}^r\sum _{i=0}^{r-j}  \sum _{j\ge k_1\ge k_2\ge k_3\ge k_4\ge 0}  Q^{4 i+10 j-4 ( k_1+ k_2+ k_3+ k_4)}  t_r^{2 (i+2 j-k_1-k_2-k_3-k_4)}\cr
&&\times t_c^{-2 \left(k_1 k_2+k_2 k_3+k_3 k_4+(k_1 +k_2 +k_3 +k_4) r-i (k_1+r)-j (k_2+k_3+k_4+2 r)\right)}\cr
&&\times {r\brack j}_{t_c^2}{r-j\brack i}_{t_c^2}{j\brack k_1}_{t_c^2}{k_1\brack k_2}_{t_c^2}{k_2\brack k_3}_{t_c^2}{k_3\brack k_4}_{t_c^2}\cr
&&\times\left(-a^2 Q^{-2}t_r t_c;t_c^2\right)_{r-j}\left(-a^2 Q^2t_r^3 t_c^{1+2 r} ;t_c^2\right)_{r-j}(-a^2Q^{-2}t_r t_c^{1+2 (r-j)} ;t_c^2)_{j-k_4}~.
\eea
In this case, the Poincar\'e polynomials $\wt\scP_{[r]}(T_{3,5};a,q,q^{-1},qt)$ also turn out to be equal to the corresponding refined Chern-Simons invariants, which were computed in \cite{Shakirov:2013moa} up to $r=3$.\footnote{The authors would like to thank S. Shakirov for sharing a Maple file.} Furthermore, the homology with respect to the HFK-like differential $d_{1-r}$ is given by
\bea\label{HFK-like-T35}
&&\wt\scP(\wt\scH_{[r]}(T_{3,5}),d_{1-r})(a,Q,t_r,t_c) = \\
&=&a^{8 r} \sum _{k=0}^r \sum _{j=0}^{r-k} \sum _{i=0}^k Q^{-2 (8 i - 2 j - 5 k + r)} t_r^{2 (-4 i + j + 2 (k + r))}t_c^{-2 (3 i^2 + k^2 + j (k - r) -  3 k r - 2 r^2 + i (-2 j - 4 k + 5 r))} \cr
&&\times {r \brack k}_{t_c^2}{k \brack i}_{t_c^2} {r-k \brack j}_{t_c^2}(-a^2 Q^{-2}t_rt_c;t_c^2)_{k-i} (-a^2 Q^{-2} t_r t_c^{1+2 (k-i)} ;t_c^2)_j(-a^2 Q^2 t_r^3 t_c^{1+2 r} ;t_c^2)_i~.\nonumber
\eea
It is easy to check, by comparing with \eqref{HFK-T35}, that \eqref{HFK-like-T35} obeys the refined exponential growth property \eqref{REGP-HFK}.

\begin{figure}[htb]
\begin{center}
\includegraphics[width=\textwidth]{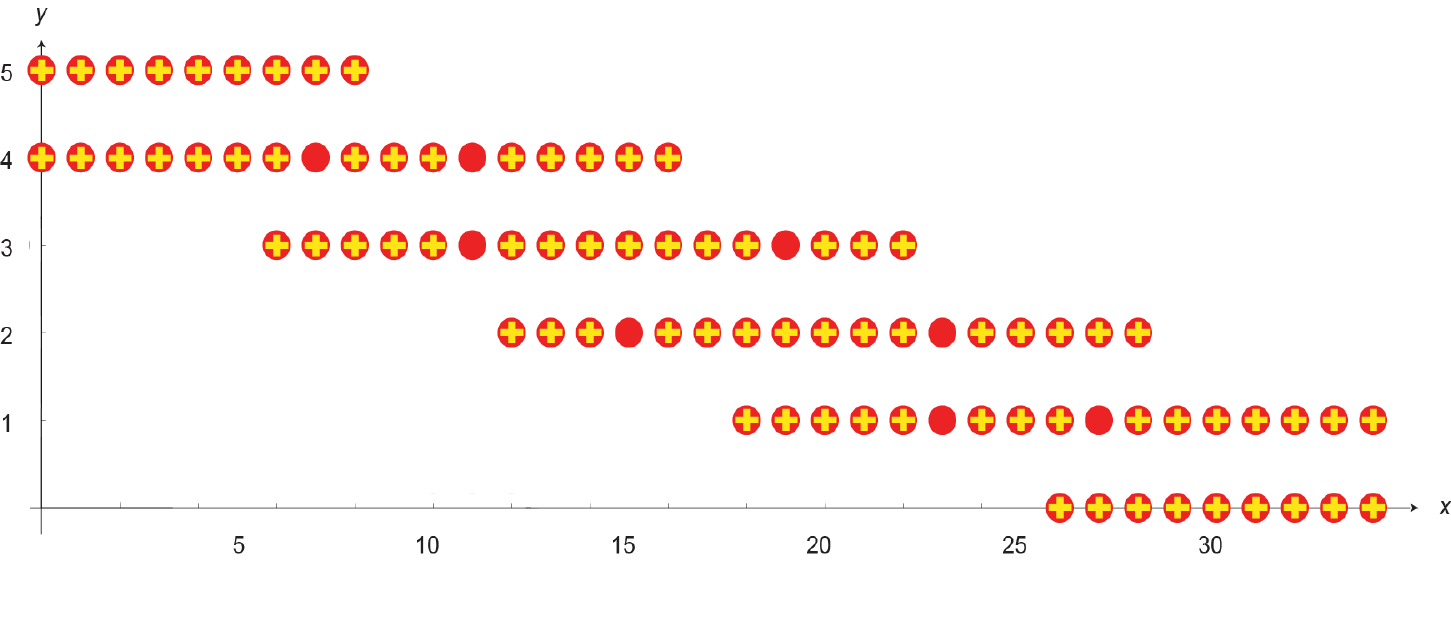}
\caption{\emph{Newton polygon of the super-$A$-polynomial for the $(3,4)$ torus knot, i.e. $8_{19}$.}   } \label{fig-819-Newton}
\end{center}
\end{figure}

Since we have found an expression for the $[r]$-colored HOMFLY homology of $T_{3,4}$, we can obtain its super-$A$-polynomial\footnote{In principle, the super-$A$-polynomial of $T_{3,5}$ can be obtained from \eqref{HOMFLY-T35}. However, an attempt has failed because  the expression \eqref{HOMFLY-T35} is too complicated.}.
Writing the Poincar\'e polynomials of the colored HOMFLY homology of the $(3,4)$ torus knot in the $(a,q,t_r)$-grading by using \eqref{T34-HOMFLY-2},
\bea
&&\scP_{[r]} (T_{3,4};a,q,t)  = \wt\scP_{[r]} (T_{3,4};a,q,t/q,q) =  \\
&=& (a t)^{6r} q^{6r^2} \sum_{r\ge j\ge \g \ge \b\ge\a\ge0} t^{-2(\alpha+\beta+j) }q^{2(-rj+j\gamma-\beta\gamma -\alpha\beta -\gamma^2-\alpha-\beta+2\gamma-3j-(\alpha+\beta+\gamma)(j-\gamma))}   \cr
&&\hspace{3.5cm}\times \frac{(q^2;q^2)_{r} (-a^2t^3q^{2r};q^2)_{j} (-a^2tq^{-2};q^2)_{j-\gamma}}{(q^2;q^2)_{r-j} (q^2;q^2)_{j-\gamma} (q^2;q^2)_{\gamma-\beta} (q^2;q^2)_{\beta-\alpha} (q^2;q^2)_{\alpha}}  ~,\nonumber
\eea
one can obtain the twisted superpotential
\bea
&&\widetilde{\mathcal{W}} (T_{3,4};x,z,z_\a,z_\b,z_\c,a,t)=\\
& = &
6 \log(a t) \log x + 3 (\log x)^2 - 2 \log(z z_{\alpha} z_{\beta}) \log t -
  \log z_{\beta} \log z - \log z_{\alpha} \log (z z_{\beta} z_{\gamma}^{-1}) \cr
& & - \log z \log x  - \textrm{Li}_2( x) + \textrm{Li}_2(x z^{-1}) + \textrm{Li}_2( z z_{\gamma}^{-1}) +
  \textrm{Li}_2( z_{\gamma} z_{\beta}^{-1}) + \textrm{Li}_2( z_{\beta} z_{\alpha}^{-1}) \cr
& & + \, \textrm{Li}_2(z_{\alpha} ) +
  \textrm{Li}_2( -a^2 t^3 x) - \textrm{Li}_2( -a^2 t^3 x z) + \textrm{Li}_2(-a^2 t) -
  \textrm{Li}_2(-a^2 t z z_{\gamma}^{-1})~,\nonumber
\eea
where we fix
\be
x = q^{2r},\quad z = q^{2j},\quad z_{\alpha} = q^{2{\alpha}},\quad z_{\beta} = q^{2{\beta}},\quad z_{\gamma}=q^{2{\gamma}}~,
\ee
in the double scaling limit $q\to 1$, $r\to \infty$. Then, one can read off the saddle point equations
\bea
y & = & e^{x\partial_x \widetilde{\mathcal{W}}(T_{3,4})} =  \tfrac{a^6 t^6 x^6(x-1)(1+a^2t^3 xz)}{(1+a^2t^3 x)(x-z)}    \cr
1 & = & e^{z \partial_{z} \widetilde{\mathcal{W}}(T_{3,4})} =  \tfrac{(x-z)(1+a^2t^3 xz)(a^2tz+z_{\gamma})}{t^2 x z z_{\alpha} z_{\beta}(z-z_{\gamma})}        \cr
1 & = & e^{z_{\alpha} \partial_{z_{\alpha}} \widetilde{\mathcal{W}}(T_{3,4})} = \tfrac{(z_{\beta}-z_{\alpha})z_{\gamma}}{t^2 z (z_{\alpha}-1)z_{\alpha} z_{\beta}}         \cr
1 & = & e^{z_{\beta} \partial_{z_{\beta}} \widetilde{\mathcal{W}}(T_{3,4})} =   \tfrac{z_{\beta}-z_{\gamma}}{t^2 z (z_{\alpha} - z_{\beta}) z_{\beta}}       \cr
1 & = & e^{z_{\gamma} \partial_{z_{\gamma}} \widetilde{\mathcal{W}}(T_{3,4})} =   \tfrac{z_{\alpha} z_{\beta}(z_{\gamma}-z)}{(z_{\beta}-z_{\gamma})(a t z + z_{\gamma})} ~.
\eea
Eliminating $z,z_{\alpha},z_{\beta},z_{\gamma}$ from these equations, we get the super-$A$-polynomial for the $(3,4)$ torus knot
\begin{footnotesize}
\bea
& & A^{\textrm{super}}(T_{3,4};x,y; a,t)  =
\left(1+a^2 t^3 x\right)^4 \left(1+a^2 t^5 x^2\right)^2y^5\nonumber\\[.1cm]
&&-a^6 \left(1+a^2 t^3 x\right)^3 (1-t^2 x+2 t^2 x^2+2 a^2 t^3 x^2-2 t^4 x^2+t^4 x^3+a^2 t^5 x^3+2 t^6 x^3+3 t^4 x^4+4 a^2 t^5 x^4\cr
&&-3 t^6 x^4+a^4 t^6 x^4-3 a^2 t^7 x^4+t^8 x^4-2 t^6 x^5+3 a^2 t^7 x^5-t^8 x^5+5 a^4 t^8 x^5-a^2 t^9 x^5-t^{10} x^5+4 t^6 x^6+6 a^2 t^7 x^6\cr
&&+2 a^4 t^8 x^6+4 a^2 t^9 x^6+t^{10} x^6+3 a^4 t^{10} x^6+a^2 t^{11} x^6+2 a^2 t^9 x^7+3 a^4 t^{10} x^7-9 a^2 t^{11} x^7+3 a^6 t^{11} x^7-6 a^4 t^{12} x^7\cr
&&+3 a^4 t^{10} x^8+10 a^2 t^{11} x^8+16 a^4 t^{12} x^8+7 a^6 t^{13} x^8+a^4 t^{14} x^8-2 a^4 t^{12} x^9+a^6 t^{13} x^9+3 a^4 t^{14} x^9-3 a^4 t^{16} x^9\cr
&&+a^4 t^{12} x^{10}+8 a^6 t^{15} x^{10}+4 a^4 t^{16} x^{10}+3 a^8 t^{16} x^{10}+3 a^6 t^{17} x^{10}-a^6 t^{15} x^{11}+a^6 t^{17} x^{11}+4 a^8 t^{18} x^{11}+a^6 t^{19} x^{11}\cr
&&-a^8 t^{18} x^{12}+2 a^8 t^{20} x^{12}+a^{10} t^{21} x^{13})y^4\nonumber\\[.1cm]
&&+a^{12} t^8  (1-x) x^6 \left(1+a^2 t^3 x\right)^2 (-1-3 x+4 t^2 x-3 t^2 x^2-8 a^2 t^3 x^2-2 t^4 x^2-6 t^2 x^3-8 a^2 t^3 x^3\cr
&&+7 t^4 x^3-3 a^4 t^4 x^3+5 a^2 t^5 x^3-4 t^6 x^3-9 a^2 t^5 x^4+3 t^6 x^4-9 a^4 t^6 x^4+9 a^2 t^7 x^4+3 t^8 x^4-6 t^4 x^5-12 a^2 t^5 x^5\cr
&&-12 a^4 t^6 x^5-16 a^2 t^7 x^5-6 a^6 t^7 x^5-4 t^8 x^5-14 a^4 t^8 x^5-5 a^2 t^9 x^5-6 a^2 t^7 x^6-7 a^4 t^8 x^6+15 a^2 t^9 x^6-a^6 t^9 x^6\cr
&&+14 a^4 t^{10} x^6-a^2 t^{11} x^6-9 a^4 t^8 x^7-20 a^2 t^9 x^7-12 a^6 t^9 x^7-48 a^4 t^{10} x^7-3 a^8 t^{10} x^7-31 a^6 t^{11} x^7-3 a^4 t^{12} x^7\cr
&&+6 a^4 t^{10} x^8+3 a^6 t^{11} x^8-11 a^4 t^{12} x^8-2 a^8 t^{12} x^8-8 a^6 t^{13} x^8+3 a^4 t^{14} x^8-4 a^4 t^{10} x^9-6 a^6 t^{11} x^9-5 a^8 t^{12} x^9\cr
&&-20 a^6 t^{13} x^9-6 a^4 t^{14} x^9-19 a^8 t^{14} x^9-7 a^6 t^{15} x^9+2 a^6 t^{13} x^{10}+6 a^8 t^{14} x^{10}-3 a^6 t^{15} x^{10}-3 a^{10} t^{15} x^{10}\cr
&&-12 a^8 t^{16} x^{10}-3 a^6 t^{17} x^{10}-3 a^8 t^{14} x^{11}+5 a^8 t^{16} x^{11}-2 a^{10} t^{17} x^{11}-5 a^8 t^{18} x^{11}+2 a^{10} t^{17} x^{12}-a^{10} t^{19} x^{12}\cr
&&+a^{12} t^{20} x^{13})y^3\nonumber\\[.1cm]
&&+a^{18} t^{16} (-1+x)^2 x^{12} \left(1+a^2 t^3 x\right) (1-2 x+t^2 x-3 x^2+5 t^2 x^2-2 a^2 t^3 x^2-5 t^4 x^2-2 t^2 x^3\cr
&&-6 a^2 t^3 x^3+3 t^4 x^3+3 a^4 t^4 x^3+12 a^2 t^5 x^3+3 t^6 x^3-4 t^2 x^4-6 a^2 t^3 x^4-5 a^4 t^4 x^4-20 a^2 t^5 x^4-6 t^6 x^4\cr
&&-19 a^4 t^6 x^4-7 a^2 t^7 x^4-6 a^2 t^5 x^5-3 a^4 t^6 x^5+11 a^2 t^7 x^5+2 a^6 t^7 x^5+8 a^4 t^8 x^5-3 a^2 t^9 x^5-9 a^4 t^6 x^6-20 a^2 t^7 x^6\cr
&&-12 a^6 t^7 x^6-48 a^4 t^8 x^6-3 a^8 t^8 x^6-31 a^6 t^9 x^6-3 a^4 t^{10} x^6+6 a^4 t^8 x^7+7 a^6 t^9 x^7-15 a^4 t^{10} x^7+a^8 t^{10} x^7\cr
&&-14 a^6 t^{11} x^7+a^4 t^{12} x^7-6 a^4 t^8 x^8-12 a^6 t^9 x^8-12 a^8 t^{10} x^8-16 a^6 t^{11} x^8-6 a^{10} t^{11} x^8-4 a^4 t^{12} x^8-14 a^8 t^{12} x^8\cr
&&-5 a^6 t^{13} x^8+9 a^8 t^{12} x^9-3 a^6 t^{13} x^9+9 a^{10} t^{13} x^9-9 a^8 t^{14} x^9-3 a^6 t^{15} x^9-6 a^8 t^{12} x^{10}-8 a^{10} t^{13} x^{10}+7 a^8 t^{14} x^{10}\cr
&&-3 a^{12} t^{14} x^{10}+5 a^{10} t^{15} x^{10}-4 a^8 t^{16} x^{10}+3 a^{10} t^{15} x^{11}+8 a^{12} t^{16} x^{11}+2 a^{10} t^{17} x^{11}-3 a^{12} t^{16} x^{12}+4 a^{12} t^{18} x^{12}\cr
&&+a^{14} t^{19} x^{13})y^2\nonumber\\[.1cm]
&&+a^{24} t^{24} (-1+x)^3 x^{18} (-1-x+2 t^2 x+x^2-t^2 x^2-4 a^2 t^3 x^2-t^4 x^2+x^3+8 a^2 t^3 x^3+4 t^4 x^3\cr
&&+3 a^4 t^4 x^3+3 a^2 t^5 x^3+2 a^2 t^3 x^4-a^4 t^4 x^4-3 a^2 t^5 x^4+3 a^2 t^7 x^4+3 a^4 t^4 x^5+10 a^2 t^5 x^5+16 a^4 t^6 x^5+7 a^6 t^7 x^5\cr
&&+a^4 t^8 x^5-2 a^4 t^6 x^6-3 a^6 t^7 x^6+9 a^4 t^8 x^6-3 a^8 t^8 x^6+6 a^6 t^9 x^6+4 a^4 t^6 x^7+6 a^6 t^7 x^7+2 a^8 t^8 x^7+4 a^6 t^9 x^7\cr
&&+a^4 t^{10} x^7+3 a^8 t^{10} x^7+a^6 t^{11} x^7+2 a^6 t^9 x^8-3 a^8 t^{10} x^8+a^6 t^{11} x^8-5 a^{10} t^{11} x^8+a^8 t^{12} x^8+a^6 t^{13} x^8+3 a^8 t^{10} x^9\cr
&&+4 a^{10} t^{11} x^9-3 a^8 t^{12} x^9+a^{12} t^{12} x^9-3 a^{10} t^{13} x^9+a^8 t^{14} x^9-a^{10} t^{13} x^{10}-a^{12} t^{14} x^{10}-2 a^{10} t^{15} x^{10}+2 a^{12} t^{14} x^{11}\cr
&&+2 a^{14} t^{15} x^{11}-2 a^{12} t^{16} x^{11}+a^{14} t^{17} x^{12}+a^{16} t^{18} x^{13})y\nonumber\\[.1cm]
&&-a^{30} t^{34} (1-x)^4  \left(1+a^2 t x^2\right)^2 x^{26}
\eea
\end{footnotesize}
The Newton polygon for this super-$A$-polynomial is shown in Figure \ref{fig-819-Newton}. This super-$A$-polynomial enjoys the involution symmetry \eqref{z2-at}. For $a=-t=1$, the above result reduces to
\be
A^{\textrm{super}}(T_{3,4};x,y;a= 1,t=-1) = (y-x^{12}) (1-y)(y+x^6)(y-x^8)(1+x)^2  (x-1)^6,
\ee
which includes the ordinary $A$-polynomial of the $(3,4)$ torus knot, $A(T_{3,4};x,y)=(1-y)(y+x^{6})$.
As conjectured in \cite{Fuji:2012pm,Gorsky:2013jxa}, we can also verify the relationship between the super-$A$-polynomial  and the Poincar\'e polynomial of the uncolored HOMFLY homology of $T_{3,4}$:
\bea
A^{\textrm{super}}(T_{3,4};x=1,y;a,t) &=&(1+a^2 t) \left(1+a^2 t^3\right)^4 \left(1+a^2 t^5\right)^2 y^4\\
&&\times\left[y-(a^6(1+ t^2+2  t^4+ t^6)+a^8( t^3+2  t^5+2  t^7)+a^{10} t^8)\right] \cr
&=&(1+a^2 t) \left(1+a^2 t^3\right)^4 \left(1+a^2 t^5\right)^2 y^4(y-\scP_{\yng(1)}(T_{3,4};a,q=1,t))~.\nonumber
\eea

As another confirmation that the above super-$A$-polynomial is correct, we verified that for $t=-1$ it reproduces the augmentation (or $Q$-deformed) polynomial. Such a polynomial for $T_{3,4}$ knot was derived, using the method of the topological vertex, in \cite{Jockers:2012pz}. More specifically, we should recall that super-$A$-polynomials in our convention arise from reduced superpolynomials, while in the literature (including \cite{Jockers:2012pz}) one typically finds $Q$-deformed polynomials corresponding to unreduced HOMFLY invariants. To translate one into another, we need to make a change of variables from $(x,y)$ into $(\alpha,\beta)$, as derived and explained in \cite{Fuji:2012nx}
\be
x=\beta,\qquad y=\alpha\frac{Q(1-\beta)}{1-\beta Q},\qquad Q=a^2~.
\ee
After an additional rescaling
$$
\widetilde{\alpha} = \frac{\alpha}{Q^2 \beta^{12}}~,
$$
we find that the above super-$A$-polynomial reduces to
\bea
&& A^{\rm Q-def}(T_{3,4};\tilde{ \alpha}, \beta;Q)  \cr
 &=&  \tfrac{ (1-Q\beta^2)^2 (\beta-1)^4 Q^{15} \beta^{26}} {1-Q\beta}\Big[
\big(1 - Q \beta\big)  -\beta^{4}\big(1-\beta+\beta^{2} - 2 Q \beta^2 - 5 \beta^3 + 9 Q \beta^3 - 3 Q^2 \beta^3  \cr
&&  + 4 Q \beta^4 - 4 Q^2 \beta^4 - Q^2 \beta^5 +
 Q^3 \beta^5 - 3 Q^3 \beta^6 + 3 Q^4 \beta^6+Q^5 \beta^8 - Q^6 \beta^9\big)\tilde{\alpha}\cr
 &&-\beta^{10}\big(1 - \beta - 3 \beta^2 + 4 Q \beta^2 + 4 \beta^3 - 8 Q \beta^3 + 3 Q^2 \beta^3
 - 10 \beta^4 + 27 Q \beta^4 - 17 Q^2 \beta^4 + 6 Q \beta^5 \cr
 &&- 10 Q^2 \beta^5 +
 4 Q^3 \beta^5  - 3 Q^2 \beta^6 + 5 Q^3 \beta^6 - 3 Q^4 \beta^6 - 5 Q^3 \beta^7 + 6 Q^4 \beta^7 + Q^4 \beta^8
-  Q^5 \beta^9\big)\tilde{\alpha}^{2} \cr
&&+\beta^{16}\big(1- \beta + 5 \beta^2 - 6 Q \beta^2 + 3 \beta^3 - 5 Q \beta^3 + 3 Q^2 \beta^3  - 6 \beta^4 + 10 Q \beta^4 - 4 Q^2 \beta^4 + 10 \beta^5 -\cr
&& 27 Q \beta^5 + 17 Q^2 \beta^5 - 4 Q \beta^6 + 8 Q^2 \beta^6 - 3 Q^3 \beta^6 + 3 Q^2 \beta^7 - 4 Q^3 \beta^7 + Q^3 \beta^8 - Q^4 \beta^9\big)\tilde{\alpha}^{3}\cr
&&+\beta^{22} \big(1- \beta + 3 \beta^3 - 3 Q \beta^3 + \beta^4 - Q \beta^4 - 4 \beta^5 + 4 Q \beta^5 + 5 \beta^6 - 9 Q \beta^6 + 3 Q^2 \beta^6 - Q \beta^7  \cr
&&+ 2 Q^2 \beta^7+ Q^2 \beta^8 - Q^3 \beta^9\big)\tilde{\alpha}^{4}-\beta^{34}\big(1-\beta\big)\tilde{\alpha}^{5}\Big] ~.
\eea
Apart from a rather simple overall rational factor, the terms in the square bracket precisely reproduces the result in~\cite[(4.42)]{Jockers:2012pz}.

\section{Physical interpretations}
\label{sec:phys}

In the bulk of this paper, we have developed and stressed mathematical aspects of link homology. However, the various results we have presented are also intimately related to gauge theories and physics. On one hand, the HOMFLY invariants (\emph{e.g.} obtained as $t=-1$ specialization of the Poincar\'e polynomials) of links which we found can be reformulated into Labastida-Mari{\~n}o-Ooguri-Vafa (LMOV) invariants \cite{Ooguri:1999bv,Labastida:2001ts,Labastida:2000zp,Labastida:2000yw}, which have an M-theory interpretation as counts of BPS states formed by M2-branes ending on M5'-branes. It was discussed how to generalize this picture to Poincar\'e polynomials in \cite{Garoufalidis:2015ewa}, and in what follows we explain how to reinterpret Poincar\'e polynomials of links (which we found earlier) in terms of such putative refined LMOV invariants; in particular this generalizes the results for the Hopf link (and the unknot) obtained in the refined topological vertex formalism \cite{Iqbal:2007ii,Iqbal:2011kq}.

From a little different perspective, in recent years it has been realized that various quantities related to Chern-Simons theories with non-compact gauge groups and associated to a given knot, can be reinterpreted in terms of a certain 3d $\mathcal{N}=2$ theory. This correspondence between knot invariants and 3d $\mathcal{N}=2$ theories is referred to as the 3d/3d correspondence. In particular, decompositions of knot complements into ideal tetrahedra encode matter contents of corresponding 3d theories and various dualities between them \cite{Dimofte:2010tz,Terashima:2011qi,Dimofte:2011ju,Dimofte:2011py}. The 3d/3d correspondence can be best understood from the perspective of M5-branes, and it has been shown in \cite{Yagi:2013fda,Lee:2013ida,Cordova:2013cea} that the localization of the 6d $(2,0)$ theory on $S^1\times S^2$ or $S^3_b$ gives rise to Chern-Simons theory with complex gauge group. Furthermore, the 3d/3d correspondence was generalized to the case of homological knot invariants in \cite{Fuji:2012pm,Fuji:2012nx,Fuji:2012pi,3d3drevisited}.

One important manifestation of the 3d/3d correspondence is the identification of the asymptotics of colored knot invariants (HOMFLY polynomials or superpolynomials) with the twisted superpotential $\widetilde{\cW}$, which encodes properties of the dual 3d $\mathcal{N}=2$  gauge theory. This twisted superpotential turns out to be a sum of dilogarithms and quadratic terms, which represent a matter content of the 3d $\mathcal{N}=2$ theory and Chern-Simons (or FI couplings) respectively. The arguments of dilogarithm terms that appear in the twisted superpotential are monomials in $a$, $q$ and $t$, whose powers encode charges of a given field under various $\U(1)$ symmetries. The twisted superpotential can also be reconstructed from an algebraic curve, which---as we reviewed in \S \ref{sec:associated}---is called the super-$A$-polynomial \cite{Fuji:2012nx} in the context of homological knot invariants, and whose specializations include the augmentation polynomial \cite{NgFramed} or Q-deformed $A$-polynomial \cite{Aganagic:2012jb}, as well as the $A$-polynomial \cite{Cooper:1994}. From the physics perspective, this algebraic curve represents the space of supersymmetric vacua of a dual 3d $\mathcal{N}=2$ theory.

There is no obstacle to generalize the 3d/3d correspondence to links. We postulate that to a given link one can associate some 3d $\mathcal{N}=2$ theory, and read off its properties from the twisted superpotential which is identified with the asymptotics, with respect to all colors, of colored link superpolynomials. As we will see in what follows, twisted superpotentials constructed in this way are expressed as sums of dilogarithms and quadratic terms, just as in the case of knots. We postulate that the dictionary presented in \cite{Fuji:2012nx,Fuji:2012pi,3d3drevisited} extends to the case of links, and one can read off the matter content of a dual $\mathcal{N}=2$ theory from the twisted superpotential; in particular, each dilogarithm term in the twisted superpotential represents one chiral field in $\mathcal{N}=2$ theory.

For a link, the colors associated to its components (for simplicity, assume that these colors correspond to symmetric representations $S^{r_i}$) are encoded in parameters $x_i=q^{2r_i}$. As reviewed in \S\ref{sec:setup}, from the physics perspective, to each component of a link one can associate a Lagrangian brane, so that this system of branes intersects ${S}^3$ precisely along the link. From this viewpoint, in the 3d $\cN=2$ gauge theory, the $x_i$ is interpreted as a twisted mass parameter for a global flavor symmetry arising from the gauge symmetry on the non-compact $i$-th Lagrangian brane, associated to the $i$-th link component. Furthermore, $y_i=\exp(x_i\frac{\partial \widetilde{\cW}}{\partial x_i})$ is interpreted as the corresponding effective FI parameters.
Finally, the associated varieties introduced in \S \ref{sec:associated}, which encode relations between $x_i$ and $y_i$, from the physics perspective play role of spaces of supersymmetric vacua of the effective 3d $\mathcal{N}=2$ gauge theories.

In case of links there are new phenomena which come into the game, such as the appearance of homological blocks and the sliding property, or higher dimensionality of associated varieties. In this section we summarize interesting features of these phenomena from the physics perspective, and list some corresponding open problems.

\subsection{Modular transformations in 3d/3d correspondence}
Using the 3d/3d correspondence, a knot invariant has been realized in \cite{Fuji:2012pi,3d3drevisited} as a 3d $\cN=2$ partition function of $\mathcal{T}_{\fraksl(2)}[M_K]:=\mathcal{T}[K]$ theory. It is straightforward to generalize this relation to a link invariant where 3d $\cN=2$ theory have $\U(1)^n$ flavor symmetry where $n$ is the number of the link component. Now, it is natural to ask how the sliding property of link invariants can be interpreted in 3d $\cN=2$ gauge theory. In particular, we have seen in \S\ref{sec:sliding-CS} that the sliding property can be understood as a result of the modular $S$-transformation. Therefore, in this subsection, we also study the meaning of modular transformations in the 3d/3d correspondence.

To start with, let us closely look at the colored HOMFLY polynomial of the Hopf link. From \eqref{unreduced-Hopf+}, it is expressed as
\bea
\overline P_{[r_1],[r_2]}({T_{2,2}};a,{q})=\frac{(a^2;{q}^2)_{r_1}}{({q}^2;{q}^2)_{r_1}}\sum_{i=0}^{r_2} {q}^{2(r_1+1)i}\frac{(a^2 {q}^{-2};{q}^2)_{i}}{({q}^{2};{q}^2)_{i}}~.~~\label{unreduced-Hopf-HOMFLY}
\eea
From this formula, one can see that the unknot invariant gets changed by the effect of the simple $S$-surgery \eqref{S-surgery} with the other unknot via
\bea\nonumber
S: \quad \frac{(a^2;q^2)_{r_2}}{(q^2;q^2)_{r_2}} \ \longrightarrow\ \sum_{i=0}^{r_2} {\color{red}{q}^{2(r_1+1)i}}  {\color{blue}  \frac{(a^2 {q}^{-2};{q}^2)_{i}}{({q}^{2};{q}^2)_{i}}}~.
\eea
To translate this change into  3d $\cN=2$ theories, let us take the large color asymptotics $r_1,r_2\to \infty$ as in \S\ref{sec:associated} and read off change of the effective twisted superpotentials:
\bea\label{S-W}
S:\quad  \wt\cW(\unknot;a,x_2)\ \longrightarrow\ {\color{red}\log x_1 \log z}+  {\color{blue}  \wt\cW(\unknot;a,z)}~,
\eea
where we fix $x_1=q^{2r_1}$, $x_2=q^{2r_2}$, $z=q^{2i}$, and
\bea
 \wt\cW(\unknot;a,x_2)=\Li_2(a^2)- \Li_2(a^2x_2)-\Li_2(1)+\Li_2(x_2)~.
\eea
Via the 3d/3d correspondence, the variables $x_1,x_2$ can be regarded as fugacities of flavor $\U(1)$ symmetries and the variable $z$ corresponds to $\U(1)$ gauge symmetry in 3d $\cN=2$ index on $S^1\times_q \bR^2$
\be
{\cal I}[\mathcal{T}]=\Tr(-1)^F q^{\tfrac R2+J_3} x^e a^f~.
\ee
Since each dilogarithm comes from one chiral field, the 3d $\cN=2$ theory $\cT[\unknot]$ for the unknot \cite{3d3drevisited}  has four chiral fields $\Phi_i$ with charges
\be
\cT[\unknot]:\qquad
\begin{array}{c|cccc}
 & \Phi_{1}   & \Phi_{2}   & \Phi_{3} & \Phi_{4} \\
    \hline
U(1)_{x_2}    &1 &0 &0    & -1   \\
U(1)_{a}    &0  &0 &2     & -2   \\ \hline
U(1)_{R}    &0  &2 &2   & 0
\end{array}\label{tab:3d-unknot}\quad.
\ee
In fact, the $S$-surgery \eqref{S-W} actually makes the $\U(1)_{x_2}$ flavor symmetry dynamical $\U(1)_z$ and couples the dynamical $\U(1)_z$ with a new background $\U(1)_{x_1}$ by
\be\nonumber S\,:\quad \cL(A_{x_2})\ {\longrightarrow} \  {\color{blue} \cL(A_{z})}+{\color{red} \frac{1}{2 \pi} A_{x_1} \wedge dA_z} \,,\ee
which actually leads to the change of  the 3d $\cN=2$ index
\begin{equation}
S: \quad {\cal I}[\unknot](x_2,a)\ {\longrightarrow} \ \int\frac{dz}{z}{\color{red} \frac{\theta(x_1;q^2)\theta(z;q^2)}{\theta(x_1z;q^2)}}  {\color{blue}  {\cal I}[\unknot](a,z)}~.
\label{S-transf-Z}
\end{equation}
Remarkably, this is exactly the $S$-transformation in 3d $\cN=2$ theories first described in \cite{Witten:2003ya}.

Now let us consider the $T$ transformation in knot theory. This is merely a framing correction to a knot invariant $\overline P_{\lambda}(K)$ which amounts to the multiplication $q^{C_2(\lambda)}$ ($C_2(\lambda)$ is the quadratic Casimir of the representation $\lambda$). In the case of symmetric representations, the transformation is
\be\nonumber
T: \quad \overline P_{[r]}(K) \ \longrightarrow \  q^{r(2r+1)} \overline P_{[r]}(K)~.
\ee
Though only this operation does not change knot invariants intrinsically, it is important because it does not commute with the $S$-transformation. Subsequently, the $T$-transformation changes the twisted superpotential by
\begin{equation}\label{T-transf-W}
T: \quad \wt \cW(K;x,a)\ {\longrightarrow} \ \wt\cW(K;x,a)+(\log x)^2,
\end{equation}
where $x=q^{2r}$.
In 3d $\cN=2$ theory, this is simply accomplished by adding a background Chern-Simons interaction at level $k=1$,
\be T\,:\quad \cL\ {\longrightarrow} \ \cL + \frac{1}{4\pi} A_x\wedge dA_x\,, \ee
which modifies the 3d $\cN=2$ index by
\begin{equation}
T:\quad {\cal I}[K](x,a)\ {\longrightarrow} \ x^{\frac{1}{2}}\theta(x;q^2)^{-1}{\cal I}[K](x,a)~.
\label{T-transf-Z}
\end{equation}
Again, this is exactly the $T$-transformation in 3d $\cN=2$ theories \cite{Witten:2003ya}.

\begin{center}
 \shabox{\parbox{.85\hsize}{Therefore, the 3d/3d correspondence identifies the modular transformations introduced by Witten in two physical theories: 3d pure Chern-Simons theory \cite{Witten:1988hf} and  3d $\cN=2$ gauge theory  \cite{Witten:2003ya}.}}
\end{center}

Having identified the modular transformations in the 3d/3d correspondence, we can consider the sliding property in 3d $\cN=2$ gauge theory. In terms of homological blocks, Poincar\'e polynomials of links can be written as in (\ref{colslide}), where the function $H_{[r_2],\ldots,[r_n]}(L;a,q,t_c,x_1)$ takes the form
\be\label{colslide2}
H_{[r_2],\ldots,[r_n]}(L;a,q,t,{\color{red}x_1})= \sum_{i_1,i_2\ldots} {\color{red}x_1^{i_1+\ldots+i_m} }  {\color{blue} \textrm{hb}(a,q,t,r_2,\cdots,r_n,\{i_k\}) }~,
\ee
where $x=q^{2r_1}$ depends on the color $r_1$ of the unknot component of a link that is responsible for the sliding property, and the homological block ${\color{blue} \textrm{hb}}$ is in general expressed by a product of $q$-Pochhammer symbols with a monomial of $a,q,t$.
Therefore,  the Poincar\'e polynomial of unreduced HOMFLY homology of a link $L$ can be generally written as
\be
\overline\scP_{[r_1],\ldots,[r_n]}(L;a,q,t)= \frac{(-a^2 t;q^2)_{r_1}}{({q}^2t^{-2} ;q^2)_{r_1}} \sum_{i_1,\ldots,i_m} {\color{red} q^{2r_1(i_1+\ldots+i_m)}}  {\color{blue}  \textrm{hb}(a,q,t,r_2,\cdots,r_n,\{i_k\}) }  \label{Pbar-slide}
\ee
where the sliding property is encoded in the factor given in red. For example, the Poincar\'e polynomial (\ref{torus-link-diff-rk+}) takes the above form after redefining $qt_c\to q$. At $r_1=0$, the left hand side becomes the Poincar\'e polynomial $\bar{\scP}_{[r_2],\ldots,[r_n]}(L';a,q,t)$ of the remaining $(n-1)$-component link $L'$
\be\nonumber
\overline\scP_{[r_2],\ldots,[r_n]}(L';a,q,t)=  \sum_{i_1,\ldots,i_m}    {\color{blue} \textrm{hb}(a,q,t,r_2,\cdots,r_n,\{i_k\})   }~.
\ee

\begin{figure}[H]
\includegraphics[width=\textwidth]{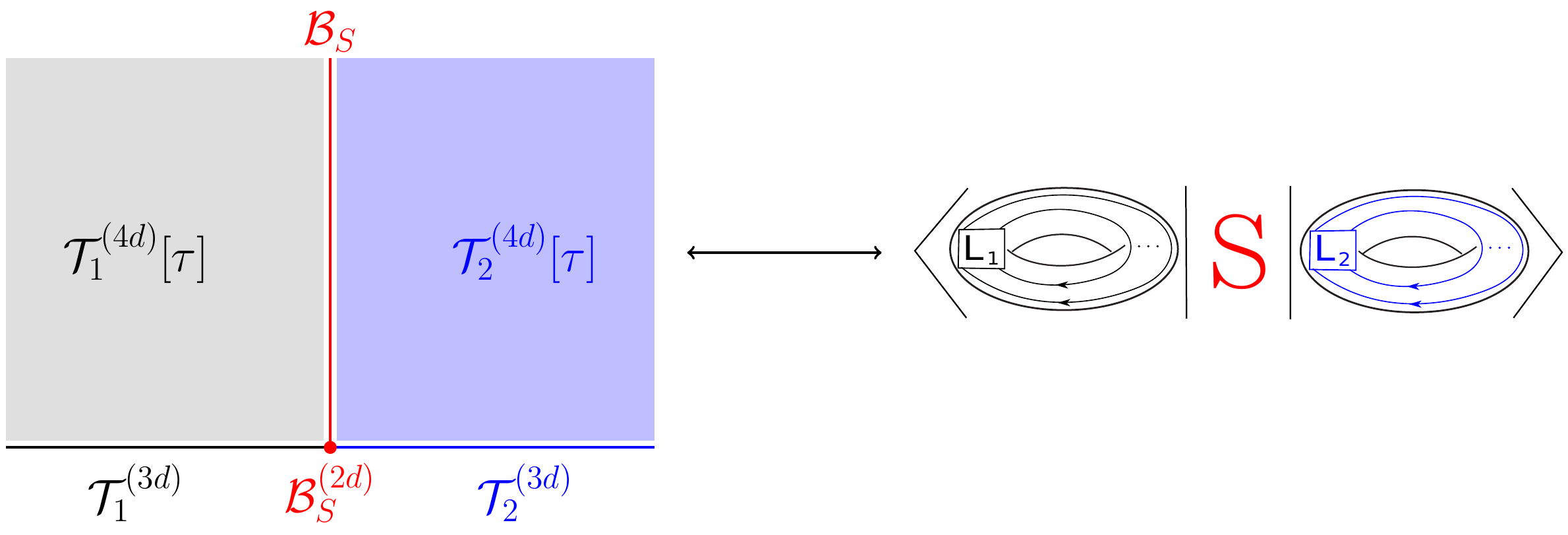}
\caption{The $S$-surgery of two links in pure Chern-Simons theory is equivalent to the joining of two 3d $\cN=2$ theories along the 2d $S$-transformation wall ${\cal B}_{\varphi=S}^{(2d)}$ in the 3d/3d correspondence.}\label{fig:duality wall}
\end{figure}

In the 3d/3d correspondence, the factor colored by red in \eqref{Pbar-slide} corresponds to the Chern-Simons term $A_x dA_z$, and the blue  homological blocks $\color{blue}\textrm{hb}$ are represented by vortex partition function of the 3d $\cN=2$ theory on $S^1 \times D^2$. Therefore, the 3d $\cN=2$ index for a link $L$ with one unknot component generally takes the form
\be\nonumber
{\cal I}[L](x_i,a,t)={\cal I}[\unknot](x_1,a,t)\int\prod_{k=1}^m\left[\frac{dz_k}{z_k}{\color{red} \frac{\theta(x_1;q^2)\theta(z_k;q^2)}{\theta(x_1z_k;q^2)}}\right]  {\color{blue}  \textrm{hb}[L'](x_2,\cdots,x_n,\{z_k\},a,t)}~,
\ee
where $\color{blue}\text{hb}[L']$ is a vortex partition function for the remaining link $L'$.
Here, the kernel of integration can be regarded as the index of a 2d $S$-transformation wall introduced in \cite{Gadde:2013wq}:
\begin{equation}\label{S-transf-W}
{\cal I}_{{\cal B}_{S}^{(2d)}}(x,z)=\frac{\theta(x;q^2)\theta(z;q^2)}{\theta(xz;q^2)}~,
\end{equation}
which is an elliptic genus of 2d (0,2) theory on the wall ${\cal B}_{S}^{(2d)}$. (See Figure \ref{fig:duality wall}.) On the $S$-transformation wall, the (0,2) theory  has $\U(1)_{x_1}\times \U(1)_{z}$ flavor symmetry where $\U(1)_z$ factor is gauged in the bulk on the right side of the wall. In addition, it consists of two Fermi multiplets and one chiral multiplet with $\U(1)_{x_1}\times \U(1)_{z}$ charge $(1,0),(0,1)$ and $(1,1)$, respectively. Thus, the sliding property of link invariants can be understood as the consequence of gluing ${\cal T}[L']$ with ${\cal T}[\unknot]$ via a 2d $S$-transformation wall.

More generally, one can consider a link $L$ that is obtained by the $S$-surgery of two non-trivial links $L_1$ and $L_2$ (right in Figure \ref{fig:duality wall}). As we have seen in \eqref{general-S}, the $\fraksl(N)$ link invariant can be represented as
\bea\nonumber
\overline J^{\fraksl(N)}_{\lambda_1^{(1)},\ldots,\lambda_{n_1}^{(1)}, \lambda_1^{(2)},\ldots, \lambda_{n_2}^{(2)}} (L)
&=&\sum_{\mu~ \vdash \sum |\nu_i|, ~\xi~ \vdash \sum |\rho_i|}  h_{\nu_i}^\mu(L_1) {\color{red}\frac{S_{\mu\xi}}{S_{00}}} {\color{blue}h_{\rho_i}^\xi(  L_2) } ~.
\eea
Correspondingly, the 3d $\cN=2$ index schematically takes form
\be\nonumber
{\cal I}^{3d}[L](\{x\})= \int \prod_{k,\ell} \frac{dw_\ell}{w_\ell}\frac{dz_k}{z_k}  ~\textrm{hb}[L_1](\{x^{(1)}\},\{w\}){\color{red} {\cal I}_{{\cal B}_{S}^{(2d)}} (\{w\},\{z\}) } {\color{blue}  \textrm{hb}[L_2](\{x^{(2)}\},\{z\})},
\ee
where $\U(1)_w$ ($\U(1)_z$) are dynamical, and $\U(1)_{x^{(1)}}$ ($\U(1)_{x^{(2)}}$) is a flavor symmetry in $\mathcal{T}[L_1]$ ($\mathcal{T}[L_2]$). Hence, one can convince oneself that a 2d $S$-transformation wall indeed \emph{links} two 3d $\cN=2$ theories  $\mathcal{T}[L_1]$  and  $\mathcal{T}[L_2]$.

\subsection{Homological blocks and the structure of (refined) BPS states}

Apart from the above consequences for 3d $\mathcal{N}=2$ theories, the sliding property also imposes strong constraints on the form of BPS invariants, or equivalently Labastida-Mari{\~n}o-Ooguri-Vafa (LMOV) invariants \cite{Ooguri:1999bv,Labastida:2001ts,Labastida:2000zp,Labastida:2000yw}. More precisely, in the context of homological link invariants we should consider refined LMOV invariants \cite{Garoufalidis:2015ewa}, which capture the dependence on the parameter $t$; we recall that in the brane picture, $t$ is a fugacity for rotations in the two-dimensional subspace of $\bR^4$, perpendicular to the Lagrangian brane $M_L$ in the brane setup (\ref{surfeng}).

In fact, to date a refined version of these invariants has not been formulated. Despite this fact we can discuss some of its putative general features, and also specialize the discussion below to the unrefined case $t=t_c=-1$ to get specific predictions for the original LMOV invariants. The LMOV invariants arise from a redefinition of HOMFLY polynomials (or superpolynomials, in the refined case). For a link with $n$ components colored by symmetric representations $S^{r_1},\ldots S^{r_n}$, it is useful to introduce first the generating function
\be
\overline\scP(z_1,\ldots,z_n) = \sum_{r_1,\ldots,r_n} \overline\scP^{(+)}_{[r_1],\ldots,[r_n]} z_1^{r_1}\cdots z_n^{r_n}.   \label{P-genfun}
\ee
We expect, at least in favorable circumstances (\emph{i.e.} in appropriate framing, or with a convenient choice of the preferred direction in the language of \cite{Iqbal:2007ii}), that the redefinition which leads to the refined BPS invariants amounts to the rewriting of this generating function in the product form (for $t=-1$ this reduces to the usual LMOV invariants \cite{Labastida:2001ts})
\be
\overline\scP(z_1,\ldots,z_n) = \prod_{r_i} \prod_{i,j,k} \prod_{l=1}^{\infty} (1 - z_1^{r_1}\cdots z_n^{r_n} \cdot a^i t^{j}(-t)^{\alpha l} q^{k}q^{l})^{N_{r_1,\ldots,r_n;i,j,k}},  \label{Pbar-prod}
\ee
where for fixed $r_1,\ldots,r_n$ the range of $i,j,k$ is finite, each product over $l$ represents the quantum dilogarithm function, and $\alpha$ is an appropriately adjusted parameter. Indeed, it was found in \cite{Iqbal:2007ii} that the refined topological string amplitude for a single brane attached to one of the two particular legs of $\mathbb{C}^3$ -- which engineers the unknot -- takes form of the quantum dilogarithm, with the quantum parameter corresponding either to $\alpha=1/2$ or $\alpha=-1/2$ in the notation above. Such a single quantum dilogarithm corresponds to a single non-vanishing exponent in (\ref{Pbar-prod}). More generally, the exponents in the above expression $N_{r_1,\ldots,r_n;i,j,k}$ are the putative BPS numbers, and they are expected to be integers counting the numbers of M2-branes with appropriate charges ending on M5'-branes. 
In particular, we can introduce the generating function for the unknot
\be
\overline\scP(\unknot;z) = \sum_{r} \frac{(-a^2 t_c;q^2t_c^2)_{r}}{({q}^2 ;q^2t_c^2)_{r}} z^r.
\ee
It is well known that in the unrefined case (for $t_c=-1$) this expression can be written as a product of two quantum dilogarithms, and only two BPS numbers are nonvanishing \cite{Ooguri:1999bv}.

Let us consider now the structure of the generating function $\overline\scP(z_1,\ldots,z_n)$ constructed from the Poincar\'e polynomial of the form (\ref{Pbar-slide}), whose form follows from the sliding property. We recall first that $i_1,i_2,\ldots$ in (\ref{Pbar-slide}) are non-negative numbers. It is therefore convenient to replace the summation over all $i_1,i_2,\ldots$ by the summation over $i=i_1+i_2+\ldots$, and then $i_2,i_3,\ldots$. In consequence, in the construction of the generating function (\ref{P-genfun}) the parameter $z_1$ gets combined with $(qt_c)^{2r_1(i_1+i_2+\ldots)} = \tilde{q}^{2i r_1}$. Ultimately we find that the generating function takes form
\be
\overline\scP(z_1,\ldots,z_n)  = \sum_i \overline\scP(\unknot;z_1 \tilde{q}^{2i}) p(z_2,z_3,\ldots;i),   \label{Pbar-factor}
\ee
where $p(z_2,z_3,\ldots;i)$ represents the remaining part of the expression (\ref{Pbar-slide}), resummed over $r_2,r_3,\ldots$, and with the remaining dependence on $i$.

At this stage, to get (refined) BPS numbers $N_{r_1,\ldots,r_n;i,j,k}$, one should find a product expansion of (\ref{Pbar-factor}). This is still a non-trivial task, and the result depends on details of the particular link under consideration. However, as a consequence of the sliding property, this final expression has a factorized form, such that dependence on $z_1$ enters only through the factor representing the unknot superpolynomial, which is coupled to the remaining part of the expression by a single summation over a positive integer $i$. This structure imposes particular constraints on the form of resulting invariants $N_{r_1,\ldots,r_n;i,j,k}$.

Reversing the logic, the form of (\ref{Pbar-factor}) could also be used to test if a given set of integers -- possibly predicted by other considerations -- has a chance to represent (refined) BPS invariants of a given link (with an unknot component). Namely, if for any such set of integers the corresponding product (\ref{Pbar-prod}) can be rewritten in the form (\ref{Pbar-factor}), it is then consistent with the sliding property, and has a chance to represent a Poincar\'e polynomial of such a link.

\subsection{Associated varieties}

Associated varieties for links, introduced in \S \ref{sec:associated}, play a prominent role in the physical interpretation -- they represent spaces of supersymmetric vacua of the dual 3d $\mathcal{N}=2$ theories. In the context of homological knot invariants, this interpretation was discussed at length in \cite{Fuji:2012pm,Fuji:2012nx,Fuji:2012pi,3d3drevisited}, and it easily generalizes to the case of links. However, various results of this paper illustrate some new and interesting featrues of 3d gauge theories associated to links.

We recall first that to a given knot more than one (UV) $\mathcal{N}=2$ theory can be associated, and all such theories are expected to flow to the same effective theory. For example, the families of theories called A and B, associated to $(2,2p+1)$ torus knots, were found in \cite{Fuji:2012pi}.  Twisted superpotentials of theories A and B were found to be different; in particular they encoded a different number of parameters $z_k$ corresponding to dynamical $\U(1)$ gauge symmetries. However, integrating out these dynamical symmetries led to the same super-$A$-polynomials, confirming that both theories A and B flow to the same effective theory. This confirms an expectation that there should be a unique effective 3d $\mathcal{N}=2$ theory associated to each knot (or link). Moreover, one might hope that this relation is one-to-one, and in particular two different knots (or links) cannot correspond to the same effective $\mathcal{N}=2$ theory.

From the results of this paper we can generalize the above observation to the case of torus links. Indeed, we have found two equivalent expressions for Poincar\'e polynomials of $(2,2p)$ torus links \eqref{refined-CS-sym+} and \eqref{torus-link-diff-rk+}, using the formalism of refined Chern-Simons theory and structural properties respectively. Just as for knots, the former expression is written in terms of a single summation, and the latter one in terms of multiple summations. These summations are reinterpreted as dynamical $\U(1)$ symmetries in the corresponding (UV) $\mathcal{N}=2$ theories, so that -- similarly as for knots -- we find two classes of ``dual'' theories, which we also call A and B respectively. It is straightforward to write down their matter content and quiver diagrams based on the form of the corresponding twisted superpotential (\ref{twisted-rCS}) and the twisted superpotential read off from (\ref{torus-link-diff-rk+}), exactly as in the case of knots \cite{Fuji:2012pi}.

However, in this paper we have also come across an example which is not consistent with the expectation  that there should be a one-to-one correspondence between knots (or links)  and $\mathcal{N}=2$ theories---or at least we have found an example of two different links which give rise to the same (diagonal) super-$A$-polynomial, and therefore the same effective supersymmetric theory. This example involves a mutant pair of ${\bf 3_1 \sqcup \overline{3}_1}$ and $\bigcirc \sqcup ({\bf 3_1 \sharp\, \overline{3}_1})$. These two links have the same (diagonal) super-$A$-polynomial (\ref{Asuper-3131}), and so the corresponding effective $\mathcal{N}=2$ theory (or at least the space of its supersymmetric vacua) does not distinguish between the two links in this mutant pair.

Of course, there is much more information encoded in knot homologies than in super-$A$-polynomials---in \S \ref{sec:mutants} we showed that the unreduced $\fraksl(2)$ homology does distinguish this mutant pair. Hence, it would be interesting to understand if there are some quantities in gauge theories which can distinguish this and other mutant pairs. It is also possible that the effective  3d $\mathcal{N}=2$ gauge theory, whose space of vacua is given by (\ref{Asuper-3131}), appears as the common subsector of two -- possibly different -- more general theories, that encode the two $\U(1)$ flavor symmetries associated to both components of ${\bf 3_1 \sqcup \overline{3}_1}$ or $\bigcirc \sqcup ({\bf 3_1 \sharp\, \overline{3}_1})$ (the diagonal super-$A$-polynomial (\ref{Asuper-3131}) encodes information only about the diagonal $\U(1)_{\textrm{diag}}\subset \U(1)\times \U(1)$). Understanding those more general theories and their behavior under the mutation operation is worth further study.

There is also one more immediate problem in trying to realize higher-dimensional associated varieties in physics. It has been argued several times \cite{Gukov:2003na,Fuji:2012pi} that, in the context of 3d $\mathcal{N}=2$ gauge theories, $A$-polynomials (and their $a$- or $t$-deformations) play very similar role to the Seiberg-Witten curves for 4d $\mathcal{N}=2$ theories. The latter curves have a direct physical interpretation via the so-called geometric engineering \cite{Katz:1996fh}, i.e. a given 4d $\mathcal{N}=2$ theory can be recovered from type II string theory compactified on a Calabi-Yau manifold, whose mirror includes the relevant Seiberg-Witten curve as its part. One might expect that similar ideas could work for 3d $\mathcal{N}=2$ theories, which then could be engineered in a similar spirit, based on appropriate Calabi-Yau manifold.  However, associated varieties for links with arbitrary number of components have an arbitrary dimension, so that they cannot fit into the structure of any 3 complex-dimensional Calabi-Yau geometry. Understanding if associated varieties play any direct role in string theory is an interesting problem.


\section*{Acknowledgements}
We would like to thank H. Fuji for collaboration in the initial stages of this project. We greatly appreciate discussions and correspondence with
\begin{center}
N.~Carqueville, H.-J.~Chung, B.~Cooper, T.~Dimofte, E. Gorsky, K. Habiro, Siqi He, K.~Kawagoe, B.~Knudsen, M.~Khovanov, M.~Manabe, A.~Mironov, A.~Morozov, L.~ Ng, R.~Osburn, P.~Putrov, P.~Ramadevi, J.~Rasmussen, D.~Roggenkamp,  A.~Ros~Camacho, S.~Shakirov, A.~Sleptsov, M.~Tierz, R.~van~der~Veen, J.~Walcher, P.~Wedrich, J.~Yagi, Kevin~Ye, and Zodinmawia.
\end{center}
We would particularly like to thank the American Institute of Mathematics for supporting this collaboration with a SQuaRE grant ``New connections between link homologies and physics''.  In addition, we are grateful to the following institutions, workshops and organizers for their warm hospitality:
\begin{itemize}\setlength{\parskip}{-0.15cm}
\item``Topological recursion and quantum algebraic geometry'' at QGM (S.G. S.N. P.S.) \item Max Planck Institute for Mathematics at Bonn (S.N. M.S. P.S.) \item``Physics and Mathematics of Link Homology'' in Montreal (S.G. S.N. I.S. M.S. P.S.) \item Simons Center for Geometry and Physics, for `` Simons Summer Workshop 2013, 2014, 2015'', and ``Program on knot homologies, BPS states, and SUSY gauge theories'' (S.G. S.N. I.S. M.S. P.S.)\item``Synthesis of integrabilities in the context of duality between the string theory and gauge theories" in Moscow (S.N.) \item``Quantum Curves and Quantum Knot Invariants'' at BIRS (S.N. M.S. P.S.) \item``Quantum Topology and Physics'' at Fukuoka (S.N.) \item ``Group Theory and Knots'' at Natal (S.N. P.S.) \item ``New Developments in TQFT'' at QGM (S.G. S.N)
\end{itemize}
S.N. would like to express deep gratitude to the previous institutions, NIKHEF Amsterdam, University of Warsaw and Max Planck Institute for Mathematics at Bonn where most of this work was carried out.

This work has been supported by the ERC Starting Grant no.~335739 \textit{``Quantum fields and knot homologies''}, funded by the European Research Council under the European Union's Seventh Framework Programme, and by Walter Burke Institute for Theoretical Physics, California Institute of Technology.

The work of S.G. is partially supported by the DOE Grant DE-SC0011632.
The work of S.N. is partially supported by the ERC Advanced Grant no.~246974, \textit{``Supersymmetry: a window to non-perturbative physics''}, and also partially supported by the center of excellence
grant ``Center for Quantum Geometry of Moduli Spaces (QGM)'' from the Danish National Research
Foundation. M.S. was also partially supported by the Ministry of Education of Serbia, grant no.~174012. P.S. acknowledges the support of the Foundation for Polish Science.

\appendix
\section{Conventions and notations} \label{sec:notation}
\subsection{Skein relations}
Here we just present the skein relations for uncolored HOMFLY and Kauffman invariants used throughout this paper.
The unreduced HOMFLY invariant
$\bar P(L;a,q)$ is the invariant of oriented links in $S^3$ defined
by the skein relations on oriented planar diagrams
\bea
&&a \bar P\left({\raisebox{-.2cm}{\includegraphics[width=.6cm]{overcrossing}}}\right)
- a^{-1}\bar P\left({\raisebox{-.2cm}{\includegraphics[width=.6cm]{undercrossing}}}\right)
=
(q-q^{-1}) \bar P\left({\raisebox{-.2cm}{\includegraphics[width=.6cm]{smoothing}}}\right)\,,
\eea
with the normalization
\be
\bar P ({\raisebox{-.1cm}{\includegraphics[width=.4cm]{unknot}}}) = {a - a^{-1} \over q -
q^{-1} }~. \ee

Similarly, the unreduced Kauffman invariant $\bar F (L;a,q)$ is the invariant
of unoriented links which is defined by another set of combinatorial rules.
We first define an invariant of planar diagrams $E(L;a,q)$
via the skein relations
\bea
&&E \left({\raisebox{-.1cm}{\includegraphics[width=1cm]{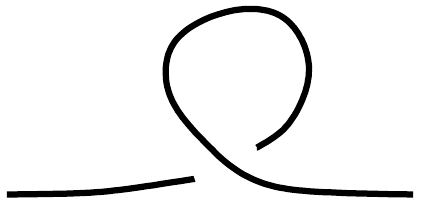}}}\right) = a E (-)
\qquad\qquad
E\left({\raisebox{-.1cm}{\includegraphics[width=1cm]{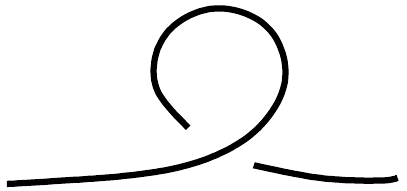}}}\right) = a^{-1}E (-)\cr
 &&E \left({\raisebox{-.2cm}{\includegraphics[width=.6cm]{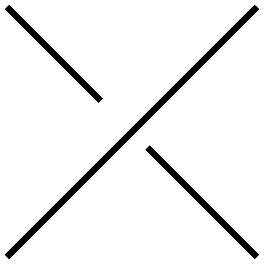}}}\right) -E
\left({\raisebox{-.2cm}{\includegraphics[width=.6cm]{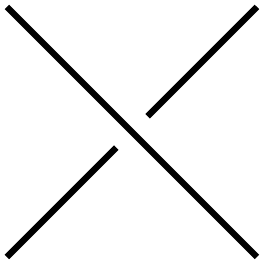}}}\right) =
(q-q^{-1}) \left[E \left({\raisebox{-.2cm}{\includegraphics[width=.6cm]{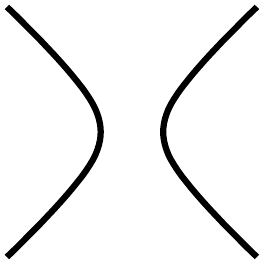}}}\right) -E
\left({\raisebox{-.2cm}{\includegraphics[width=.6cm]{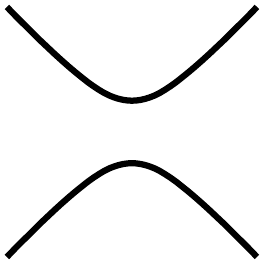}}}\right)\right]
\eea
and the normalization
\bea E ({\raisebox{-.1cm}{\includegraphics[width=.4cm]{unknot}}}) = {a - a^{-1} \over q -
q^{-1} } + 1. \eea
Then, if $w(L)=$(number of ``$+$'' crossings)$-$(number
of ``$-$'' crossings) is the writhe of $L$,
the Kauffman invariant is given by
\bea \bar F(L;a,q) = a^{-w(L)}E(L;a,q).\eea
The reduced versions of the HOMFLY and Kauffman invariants,
$P(L)$ and $F(L)$, can be defined by the same combinatorial rules,
with the normalization where the unknot evaluates to 1. In
particular, we have
\bea
\bar P (L) = \bar P ({\raisebox{-.1cm}{\includegraphics[width=.4cm]{unknot}}}) P(L)~, \quad\quad
\bar F (L) = \bar F ({\raisebox{-.1cm}{\includegraphics[width=.4cm]{unknot}}}) F(L) ~.
\eea

\newpage
\subsection{Notations}

\textbf{Link invariants}
\begin{table}[h]
\begin{tabular}{lp{11cm} l}
$ J^{\fraksl(N)}_{\lambda_1,\ldots,\lambda_n}(L;q)$& The reduced $\fraksl(N)$ quantum invariant of an $n$-component link $L$ colored by a representation $(\lambda_1,\ldots,\lambda_n)$ of $\fraksl(N)$ \cr
$\overline J^{\fraksl(N)}_{\lambda_1,\ldots,\lambda_n}(L;q)$& The unreduced $\fraksl(N)$ quantum invariant of an $n$-component link $L$ colored by a representation $(\lambda_1,\ldots,\lambda_n)$ of $\fraksl(N)$ \cr
$P_{\lambda_1,\ldots,\lambda_n}(L;a,q)$& The reduced HOMFLY invariant of an $n$-component link $L$ colored by a representation $(\lambda_1,\ldots,\lambda_n)$\cr
$\overline P_{\lambda_1,\ldots,\lambda_n}(L;a,q)$& The unreduced HOMFLY invariant of an $n$-component link $L$ colored by a representation $(\lambda_1,\ldots,\lambda_n)$\cr
$P^\fin_\lambda(L;a,q)$&  The finite-dimensional HOMFLY polynomial of an $n$-component link $L$ colored by  a representation $\lambda$. \cr
$F(L;a,q)$&  The reduced uncolored Kauffman invariant of an $n$-component link $L$\cr
$\overline{\textrm{rCS}}^{(\pm)}_{\lambda_1,\ldots,\lambda_n}(L;a,q,t)$&  The unreduced refined Chern-Simons invariant of an $n$-component link $L$ colored by a representation $(\lambda_1,\ldots,\lambda_n)$ in the range $|q|<1$ for the superscript $(+)$ or  in the range $|q|>1$ for the superscript $(-)$, respectively
\end{tabular}
\nonumber
\end{table}

\noindent\textbf{Knot homologies}
\begin{table}[H]
\begin{tabular}{lp{12cm} l}
${\scH}^\fin_{\lambda}(L)$& The quadruply-graded finite-dimensional HOMFLY homology of a link $L$ colored by a representation $\lambda$ with $(a,q,t_r,t_c)$-gradings \\
$\wt{\scH}^\fin_{\lambda}(L)$& The quadruply-graded finite-dimensional HOMFLY homology of a link $L$ colored by a representation $\lambda$ with $(a,Q,t_r,t_c)$-gradings \\
${\scH}^{(\pm)}_{\lambda,\ldots,\lambda}(L)$& The reduced quadruply-graded infinite-dimensional HOMFLY homology of a link $L$ colored by a representation $(\lambda,\ldots,\lambda)$ with $(a,q,t_r,t_c)$-gradings in the range $|q|<1$ for the superscript $(+)$ or  in the range $|q|>1$ for the superscript $(-)$, respectively \\
$\wt{\scH}^{(\pm)}_{\lambda,\ldots,\lambda}(L)$& The reduced quadruply-graded infinite-dimensional HOMFLY homology of a link $L$ colored by a representation $(\lambda,\ldots,\lambda)$ with $(a,Q,t_r,t_c)$-gradings in the range $|q|<1$ for the superscript $(+)$ or  in the range $|q|>1$ for the superscript $(-)$, respectively\\
$H_*(\scH_\lambda(L),d)$& The homology  with respect to a differential $d$ acting on $\scH_\lambda(L)$  with $(a,q,t_r,t_c)$-gradings \cr
$H_*(\wt\scH_\lambda(L),d)$& The homology  with respect to  a differential $d$ acting on $\wt\scH_\lambda(L)$  with $(a,Q,t_r,t_c)$-gradings \cr
$\scH^\Kauffman(L)$& The reduced triply-graded Kauffman homology  colored by a representation $\lambda$ with $(a,q,t)$-gradings \cr
\end{tabular}
\nonumber
\end{table}

\noindent \textbf{Poincar\'e polynomials}
\begin{eqnarray*}
\scP^{\fin}_{\lambda}(L;a,q, t_r,t_c)&:=& \sum_{i,j,k,\ell} a^i q^j t_r^k t_c^\ell ~\dim\; (\scH^{\fin}_{\lambda}(L))_{i,j,k,\ell}\cr
\wt\scP^{\fin}_{\lambda}(L;a,q, t_r,t_c)&:=& \sum_{i,j,k,\ell} a^i q^j t_r^k t_c^\ell ~\dim\; (\wt\scH^{\fin}_{\lambda}(L))_{i,j,k,\ell}\cr
\scP^{(\pm)}_{\lambda,\ldots,\lambda}(L;a,q, t_r,t_c)&:=& \sum_{i,j,k,\ell} a^i q^j t_r^k t_c^\ell ~\dim\; (\scH^{(\pm)}_{\lambda,\ldots,\lambda}(L))_{i,j,k,\ell}\cr
\wt\scP^{(\pm)}_{\lambda,\ldots,\lambda}(L;a,Q, t_r,t_c)&:=& \sum_{i,j,k,\ell} a^i Q^j t_r^k t_c^\ell ~\dim\;(\wt\scH^{(\pm)}_{\lambda,\ldots,\lambda}(L))_{i,j,k,\ell}\cr
\scP(\scH_\lambda(L),d)(a,q,t_r,t_c)&:=& \sum_{i,j,k,\ell} a^i q^j t_r^k t_c^\ell ~\dim H_*(\scH_\lambda(L),d)_{i,j,k,\ell}\cr
\wt\scP(\wt\scH_\lambda(L),d)(a,Q,t_r,t_c)&:=& \sum_{i,j,k,\ell} a^i Q^j t_r^k t_c^\ell ~\dim H_*(\wt\scH_\lambda(L),d)_{i,j,k,\ell}\cr
\scF(L;a,q, t)&:=& \sum_{i,j,k,\ell} a^i q^j t^k~\dim\;(\scH^\Kauffman_\lambda(L))_{i,j,k}\cr
\end{eqnarray*}

\section{Rogers-Ramanujan type identities}\label{sec:RR-id}
In this appendix, we shall prove the equivalence of two expressions for colored superpolynomials of the $(2,2p)$ torus links by using  the identities of multiple series Rogers-Ramanujan type \cite{Andrews:1984}. We have found the refined Chern-Simons invariants  (\ref{refined-CS-sym+}) and the Poincar\'e polynomial of unreduced HOMFLY homology (\ref{torus-link-diff-rk+}) for the $(2,2p)$ torus links. The equality of these two expressions can be proven based on the \emph{Bailey's Lemma} \cite{Andrews:1984,Paule}. In this appendix, we use the shorthand notation for $q$-Pochhammer symbols, $(x)_n=(x;q)_n$.

A pair of sequences $(\a_n,\b_n)_{n\ge0}$ is called a \emph{Bailey pair} if they are related by
\be
\b_n=\sum_{j=0}^n\frac{\a_j}{(q)_{n-j}(dq)_{n+j}}
\ee
or equivalently
\bea
\a_n = \frac{1-dq^{2n}}{1-d}\sum_{j=0}^n(-1)^{n-j}q^{(n-j)(n-j-1)/2}\frac{(d)_{n+j}}{(q)_{n-j}}\b_j \ .
\eea
Then, the Bailey's Lemma states that they satisfy the following identity
\be
\sum_{j\ge0}   \left( \frac{(\rho_1)_j(\rho_2)_j(dq/\rho_1\rho_2)^j\a_j}{(dq/\rho_1)_j(dq/\rho_2)_j} \right) \frac{1}{(q)_{n-j}(dq)_{n+j}}=   \sum_{j\ge0} \frac{(dq/\rho_1\rho_2)_{n-j}(\rho_1)_j(\rho_2)_j(dq/\rho_1\rho_2)^j\b_j}{(q)_{n-j}(dq/\rho_1)_n(dq/\rho_2)_n}~.
\ee
This identity can be considered that a Bailey pair $(\a_n,\b_n)_{n\ge0}$ gives rise to a new Bailey pair $(\a'_n,\b'_n)_{n\ge0}$
\be\label{new pair}
\a'_n=\frac{(\rho_1)_n(\rho_2)_n(dq/\rho_1\rho_2)^n\a_n}{(dq/\rho_1)_n(dq/\rho_2)_n} \ , \ \ \ \b'_n=\sum_{j\ge0} \frac{(dq/\rho_1\rho_2)_{n-j}(\rho_1)_j(\rho_2)_j(dq/\rho_1\rho_2)^j\b_j}{(q)_{n-j}(dq/\rho_1)_n(dq/\rho_2)_n} \ .  \ \  \
\ee
In this way, one can obtain a sequence of Bailey pairs $(\a,\b)\to (\a',\b')\to \cdots \to (\a^{(\ell)},\b^{(\ell)})\to \cdots $, called a \emph{Bailey chain}. In fact, Andrews has obtained the following identities \cite[Theorem 1]{Andrews:1984} by the iterative uses of the Bailey's Lemma:
\bea\label{Andrews}
&&\frac{(dq/b_p)_r(dq/c_p)_r}{(dq)_r(dq/b_pc_p)_r}\sum_{\ell\ge0}\Big(\frac{-d^{p+1}q^{p+1+r-(\ell-1)/2}}{b_0c_0\cdots b_pc_p}\Big)^\ell \frac{(q^{-r})_\ell~\alpha_\ell}{(dq^{r+1})_\ell}\prod_{i=0}^p\frac{(b_i)_\ell(c_i)_\ell }{(dq/b_i)_\ell(dq/c_i)_\ell}\cr
&=&  \sum_{{s_p}\ge  \cdots\ge s_{0}\ge0}\frac{q^{s_p}(b_p)_{s_p}(c_p)_{s_p}(q^{-r})_{s_p}\b_{s_0}}{(b_pc_pq^{-r}/d)_{s_p}} \prod_{i=0}^{p-1} \Big(\frac{dq}{b_ic_i}\Big)^{s_i}\frac{(b_i)_{s_i}(c_i)_{s_i}(dq/b_ic_i)_{s_{i+1}-s_i}}{(q)_{s_{i+1}-s_{i}}(dq/b_i)_{s_{i+1}}(dq/c_i)_{s_{i+1}}} \ .\cr
&&
\eea
The left hand side of the identity is expressed as a single summation and, therefore, it is suitable for the refined Chern-Simons invariants  (\ref{refined-CS-sym+}). On the other hand, the right hand side involves multiple-summations, which looks the Poincar\'e polynomial (\ref{torus-link-diff-rk+}). Thus, this identity can be used to show the equivalence of the two expressions. To prove the identity between (\ref{refined-CS-sym+}) and (\ref{torus-link-diff-rk+}),
we specialize the variables in \eqref{Andrews} as
\be\label{link-variable}
b_1=\cdots=b_{p}=c_0=\cdots=c_{p}=0  ,  \ b_0=q^{-r_1}  ,  \ b_{p}=-aq^{-1}t^3  ,  \ d=q^{-(r_1+r_2+1)}t^{2} , \ N=r_2\ .
\ee
In addition, we choose the following Bailey pair
\bea\label{link-bailey}
&&\a_n=(-1)^n q^{n(n-1)/2} \frac{1-q^{2n-r_1-r_2-1}t^2}{1-q^{-r_1-r_2-1}t^2}\frac{(q^{-r_1-r_2-1}t^2)_{n}}{(q)_{n}}  \ , \cr
&&\b_n=\delta_{n,0}  \ .
\eea
Then, the left hand side of \eqref{Andrews} can be arranged as
\bea
&&\frac{(-at)_{r_1}}{(qt^{-2})_{r_1}}\Big[\textrm{LHS of} \  \eqref{Andrews} \ \textrm{with} \ \eqref{link-variable} \ \textrm{and} \ \eqref{link-bailey}\Big] = \\
&=&\frac{
 (-at)_{r_1+r_2} }{(qt^{-2})_{r_1+r_2}  }\sum_{\ell\ge0}(-1)^\ell q^{ \ell (5 - \ell (1 + 2 p) + 2 r_1 + 2 r_2 + 2 p (1 + r_1 + r_2))/2} t^{-2\ell p}    \frac{\left(1-q^{r_1+r_2-2\ell+1}t^{-2}\right)}{\left(1-q^{r_1+r_2+1}t^{-2}\right)} \cr
   &&\hspace{4cm}\times\frac{\left(q^{-r_1}\right)_\ell
   \left(q^{-r_2}\right)_\ell \left(-a q^{-1} t^3\right)_\ell   \left(q^{-r_1-r_2-1} t^2\right)_\ell}{(q)_\ell
    \left(q^{-r_1} t^2\right)_\ell
   \left(q^{-r_2} t^2\right)_\ell \left(-a q^{r_1+r_2-\ell}
   t\right)_\ell}\cr
&=& q^{p r_1 r_2 }\sum_{\ell=0}^{\min(r_1,r_2)}  \frac{t^{-2 \ell (p+1)} q^{\ell (p+1)-  p(\ell -r_1) (\ell - r_2)}(q^{r_1-\ell+1})_\ell(q^{r_2-\ell+1})_\ell (-a q^{-1}t^3)_\ell(-at)_{r_1+r_2-\ell}}{(q)_\ell (qt^{-2})_{r_1+r_2-2\ell} (q^{r_1-\ell+1}t^{-2})_\ell(q^{r_2-\ell+1}t^{-2})_\ell(q^{r_1+r_2-2\ell+2}t^{-2})_\ell} ~.\nonumber
\eea
By multiplying $(aq^{-1})^{(p-1)(r_1+r_2)}$ after changing variables $q\to q^2t^2$ and  $a\to a^2$, this is exactly equal to  the refined Chern-Simons invariants  (\ref{refined-CS-sym+}). On the other hand, the right hand side can be manipulated as
\bea
&&\frac{(-at)_{r_1}}{(qt^{-2})_{r_1}}\Big[\textrm{RHS of} \  \eqref{Andrews}  \ \textrm{with} \ \eqref{link-variable} \ \textrm{and} \ \eqref{link-bailey} \Big] = \\
&=&\frac{(-at)_{r_1}}{(qt^{-2})_{r_1}}\sum_{{s_p}\ge  \cdots\ge s_{1}\ge s_0=0}q^{(r_1+1)s_p+(2 r_2 - s_p + 1) s_p/2-(2 r_2 - s_1 + 1) s_1/2}(-t^{2})^{s_1-s_p}\cr
&&\hspace{4cm}\times\frac{(-aq^{-1}t^3)_{s_p}(q^{-r_2})_{s_p}}{(q^{-r_2}t^2)_{s_1}} \prod_{i=0}^{p-1} \frac{t^{-2s_i}q^{({r_1}+{r_2}+1)s_i-s_is_{i+1}}}{(q)_{s_{i+1}-s_{i}}} \cr
&=&\frac{(-at)_{r_1}}{(qt^{-2})_{r_1}}\sum_{{r_2}=s_{p+1}\ge s_p\ge\cdots\ge s_1\ge0}\frac{(-aq^{-1}t^3)_{s_p}}{(q^{{r_2}-s_1+1}t^{-2})_{s_1}} \prod_{i=1}^{p}t^{-2s_i}q^{({r_1}+{r_2}+1)s_i-s_is_{i+1}}{s_{i+1} \brack s_{i}}_{q} ~.\nonumber
\eea
With the same modifications above (multiplying $(aq^{-1})^{(p-1)(r_1+r_2)}$ after changing variables $q\to q^2t^2$ and $a\to a^2$), this is precisely  the Poincar\'e polynomial (\ref{torus-link-diff-rk+}). Hence, we prove the identity.

In a similar fashion, one can prove the identity between two expressions for colored superpolynomials of the $(2,2p+1)$ torus knots $T_{2,2p+1}$.
The $[r]$-colored refined Chern-Simons invariants of $T_{2,2p+1}$ have been obtained in \cite{Fuji:2012pm} while the Poincar\'e polynomials of the $[r]$-colored HOMFLY homology have been presented in \cite{Fuji:2012pi}. To show they are identical,
we specialize the variables in \eqref{Andrews} as
\be\label{knot-variable}
b_0=\cdots=b_{p}=c_0=\cdots=c_{p}=0  ,  \ b_{p}=-aq^{-1}t^3 ,  \ d=q^{-(2r+1)}t^{2}, \ N=r.
\ee
In addition, we choose the following Bailey pair
\bea\label{knot-bailey}
&&\a_n=(-1)^n q^{n(n-1)/2} \frac{1-q^{2n-2r-1}t^2}{1-q^{-2r-1}t^2}\frac{(q^{-2r-1}t^2)_{n}}{(q)_{n}}  \ , \cr
&&\b_n=\delta_{n,0}  \ .
\eea
Then, the left hand side of the Andrews' identity \eqref{Andrews} can be rearranged as
\bea
&&\Big[\textrm{LHS of} \  \eqref{Andrews} \ \textrm{with} \ \eqref{knot-variable} \ \textrm{and} \ \eqref{knot-bailey}\Big] = \\
&=&\frac{
 (-aq^r t)_{r} }{(q^{r+1}t^{-2})_{r}  }\sum_{\ell\ge0} q^{ \ell (3 + p - \ell (1 + p) + 3 r + 2 p r)} t^{-2\ell (p+1)}    \frac{1-q^{2r-2\ell+1}t^{-2}}{1-q^{2r+1}t^{-2}} \frac{
   (q^{-r})_\ell (-a q^{-1} t^3)_\ell   (q^{-2r-1} t^2)_\ell}{(q)_\ell
   (q^{-r} t^2)_\ell (-a q^{2r-\ell}
   t)_\ell}\cr
&=& \sum_{\ell=0}^{r} (-1)^\ell  q^{\ell (2 r+3-\ell (2 p+1)+p (4 r+2))/2}(1-q^{2r-2\ell+1}t^{-2}) t^{-2 \ell p} \frac{
   (q^{-r})_\ell (-a q^{-1} t^3)_\ell  (-a q^{r}
   t)_{r-\ell}}{(q)_\ell
   (q^{-r} t^2)_\ell  (q^{r+1} t^{-2})_{r-\ell+1}} ~.\nonumber
\eea
After the change of variables $q\to q^2t^2$ and  $a\to a^2$, the last expression provides the \emph{reduced} refined Chern-Simons invariants of $T_{2,2p+1}$  with $t_c$-gradings. On the other hand, the right hand side can be manipulated into
\bea
&&\Big[\textrm{RHS of} \  \eqref{Andrews} \ \textrm{with} \ \eqref{knot-variable} \ \textrm{and} \ \eqref{knot-bailey}\Big] = \\
&=&\sum_{{s_p}\ge  \cdots\ge s_{1}\ge s_0=0}q^{(2r+1)s_p-( s_p - 1) s_p/2}(-t^{-2})^{s_p}(-aq^{-1}t^3)_{s_p}(q^{-r})_{s_p}\prod_{i=0}^{p-1} \frac{t^{-2s_i}q^{(2r+1)s_i-s_is_{i+1}}}{(q)_{s_{i+1}-s_{i}}} \cr
&=& \sum_{r=s_{p+1}\ge s_p\ge\cdots\ge s_1\ge0}(-aq^{-1}t^3)_{s_p}\prod_{i=1}^{p}t^{-2s_i}q^{(2r+1) s_i-s_is_{i+1}}{s_{i+1} \brack s_{i}}_{q}.\nonumber
\eea
The same change of variables gives the \emph{reduced} Poincar\'e polynomial with $t_c$-gradings (see \eqref{torusknot-HOMFLY}).

\section{Examples -- diagonal colored HOMFLY homology}
\label{sec:app-ExamplesDiag}

In this appendix we analyze structure of homologies and derive corresponding Poincar\'e polynomials for various links.

\subsection{The $(2,4)$ torus link $T_{2,4}$}

\noindent
\begin{minipage}{\linewidth}
      \centering
      \begin{minipage}{0.32\linewidth}
          \begin{figure}[H]\centering
              \includegraphics[scale=0.4]{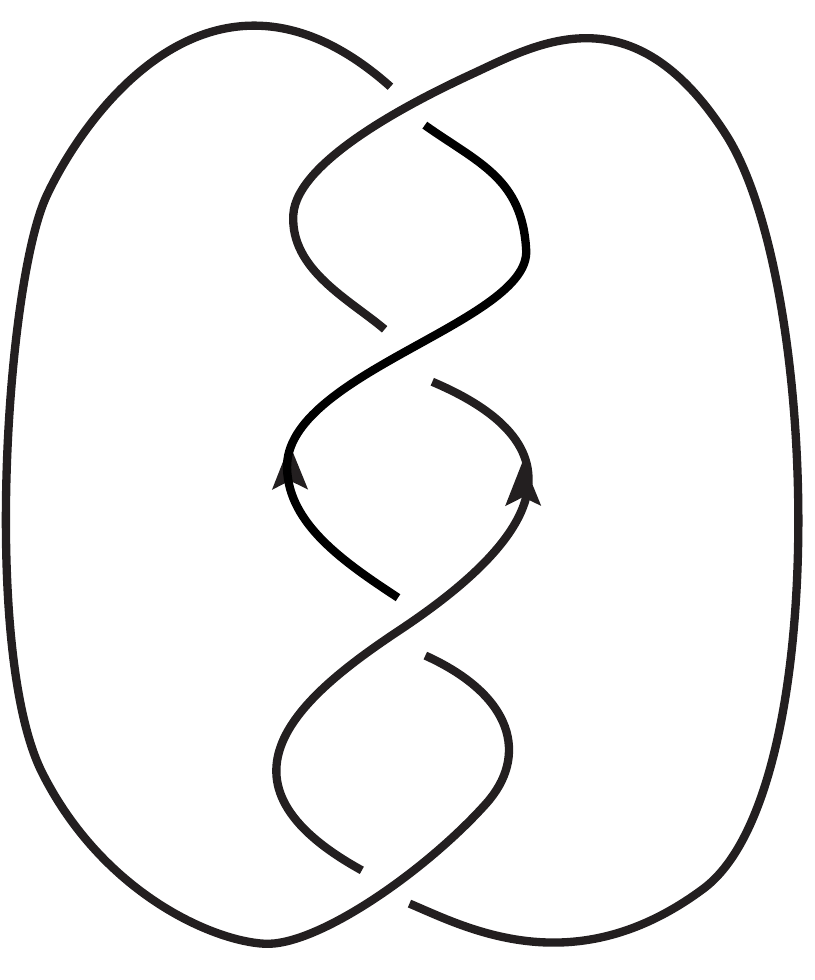}
              \caption{The torus link $T_{2,4}$.}\label{fig:T24}
          \end{figure}
      \end{minipage}
      \hspace{0.05\linewidth}
      \begin{minipage}{0.58\linewidth}
  \begin{figure}[H]
 \centering
    \includegraphics[width=9cm]{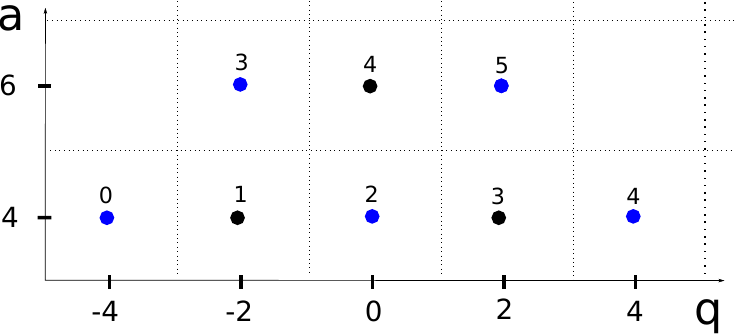}
    \caption{The finite-dimensional uncolored HOMFLY homology $\scH^{\rm fin}_{[1]}({T_{2,4}})$ of the $(2,4)$ torus link. The blue dots correspond to the generators of the  HOMFLY homology $\scH_{[1]}({T_{2,5}})$ of the $(2,5)$ torus knot. }\label{fig:HOMFLY-prime-T24-1}
\end{figure}
      \end{minipage}
  \end{minipage}

%
%
%
\subsubsection*{Finite-dimensional HOMFLY homology }

The uncolored finite-dimensional homology $\scH^{\rm fin}_{\yng(1)}(T_{2,4})$ of the $(2,4)$ torus link categorifies the 2-variable polynomial
\bea\nonumber
P^\fin_{\yng(1)}({T_{2,4}};a,q)&=&aq^{-1}(1-q^2)P_{\yng(1),\yng(1)}({T_{2,4}};a,q)\cr
&=&a^4(q^{-4}-q^{-2}+1-q^2 +q^4)-a^6(q^{-2}-1+q^2)~.
\eea
The homological grading can be simply read off from the $\delta$-grading \eqref{delta-grading} and yields
\bea\nonumber
\scP^{\rm fin}_{\yng(1)}({T_{2,4}};a,q,t) &=& a^4(q^{-4}+q^{-2} t+t^2+q^2t^3 +q^4 t^4)+a^6(q^{-2}t^3+t^4+q^2t^5)~.~~~~
\eea
As shown in Figure \ref{fig:HOMFLY-prime-T24-1}, the finite-dimensional HOMFLY homology of the $(2,4)$ torus link contains the HOMFLY homology of the $(2,5)$ torus knot.

\noindent
The Poincar\'e polynomial of the HOMFLY homology of the  $(2,4)$ torus link in the $(a,Q,t_r,t_c)$-gradings can be obtained with the help of the refined exponential growth \eqref{exp-growth-fin-1} and the differentials \eqref{regrading-finite-column}
\bea  \label{T24-prime-quad}
&&\wt\scP^{\fin}_{[r]}({T_{2,4}};a,Q,t_r,t_c) = \\
&=&a^{4r} Q^{-4r} \sum_{r \ge j\ge i \ge 0} Q^{4(j+i)}t_r^{2(j+i)} t_c^{4r(j+i)-2j(i+r)} (-a^2Q^{-2}t_rt_c;t_c^2)_{j}(-Q^2t_rt_c;t_c^2)_{r-i}{r \brack j}_{t_c^2}{j \brack i}_{t_c^2}.\nonumber
\eea
In particular, an appropriate specialization gives the Poincar\'e polynomial of the familiar triply-graded $[r]$-colored HOMFLY homology in the $t_r$-grading
\bea
&&\scP^{\rm fin}_{[r]}({T_{2,4}};a,q,t) = \label{T24-Prime} \\
&=&a^{4r} q^{-4r} \sum_{r \ge j\ge i \ge 0}q^{2(2r+1)(j+i)-2j(i+r)}t^{2(j+i)} (-a^2q^{-2}t;q^2)_{j}(-q^{2}t;q^2)_{r-i}{r \brack j}_{q^2}{j \brack i}_{q^2}.\nonumber
\eea
The explicit form of the finite-dimensional $[1^r]$-colored HOMFLY homology can be easily obtained from this expression by using the mirror symmetry \eqref{mirror-finite}.

\subsubsection*{Infinite-dimensional HOMFLY homology}

\begin{figure}[!htbp]
 \centering
    \includegraphics[width=9cm]{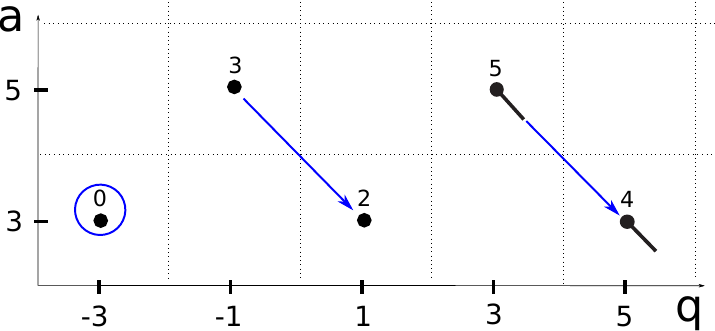}
    \caption{The uncolored positive HOMFLY homology $\scH^{(+)}_{[1],[1]}({T_{2,4}})$ of the (2,4) torus link where the straight tails represent the factors of $1/(1-q^2)$. The blue arrows represent the action of the canceling differential. The $S$-invariant of the (2,4) torus link is $S(T_{2,4})=3$.}\label{fig:HOMFLY-T24-1+}
\end{figure}

The HOMFLY invariant of the $(2,4)$ torus link has the form
\bea\label{HOMFLY-poly-T24}
P_{\yng(1),\yng(1)}({T_{2,4}};a,q) =\frac{a^3(1 - q^2 + q^4 - q^6 + q^8 - a^2 (q^2 - q^4 + q^6))}{q^3(1-q^2)}~.
\eea
Let us first consider a categorification of this expression -- understood as a power series in $a$ and $q$ -- in the range of $|q|<1$.
The resulting HOMFLY homology of the $(2,4)$ torus link is then uniquely determined by the HOMFLY invariant \eqref{HOMFLY-poly-T24} and the $\fraksl(2)$ homology, when combined with the structural properties described in \S \ref{sec:quadruply}. Its Poincar\'e polynomial looks like
\be\nonumber
\scP^{(+)}_{\yng(1),\yng(1)}({T_{2,4}};a,q,t) = a^3 \left[q^{-3} +q t^2 + \frac{q^5 t^4}{1-q^2}\right] + a^5 \left[q^{-1} t^3
+ \frac{q^3 t^5}{1-q^2}\right]~
\ee
and has two semi-infinite ``tails'' (associated with two $\frac{1}{1-q^2}$ terms) represented by the barbs in the dot diagram on Figure~\ref{fig:HOMFLY-T24-1+}.

\noindent Taking the homology with respect to the differential $d_2$ and specializing to $a=q^2$, we are left with the reduced $\fraksl(2)$ homology of the $(2,4)$ torus link
\be\nonumber
\scP^{\fraksl(2)}_{\yng(1),\yng(1)}({T_{2,4}};q,t)=q^3+t^2q^7+t^3q^{9}+t^4q^{11}~,
\ee
which is consistent with the result in \cite{Carqueville:2011zea}. (More generally, for the reduced Khovanov-Rozansky $\frak\fraksl(N)$ homology of $4^2_1$ (v2) see \cite{Carqueville:2011zea}.) The dot diagram of $\scH^{(+)}_{\yng(1),\yng(1)}({T_{2,4}})$ is presented in Figure \ref{fig:HOMFLY-T24-1+}. Apart from the geometric power series tails, it resembles that of the HOMFLY homology for the (2,5) torus knot, {\it cf.} Figure 6.5 in \cite{Dunfield:2005si}.

Moreover, the structure of the $([r],[r])$-colored positive HOMFLY homology $\wt\scH^{(+)}_{[r],[r]}({T_{2,4}})$ is controlled by the refined exponential growth property \eqref{REGP-infinite} and the differentials \eqref{color-diff-infinite}
\bea\label{T24-infinite+}
&&\wt\scP^{(+)}_{[r],[r]}({T_{2,4}};a,Q,t_r,t_c) = \\
&=&a^{3r} Q^{-3r} \sum_{r \ge j\ge i \ge 0}Q^{4(j+i)}t_r^{2(j+i)} t_c^{4r(j+i)-2j(i+r)} \frac{(-a^2Q^{-2}t_rt_c;t_c^2)_{j}}{(Q^{2}t_c^{2(r-i)};t_c^2)_{i}}{r \brack j}_{t_c^2}{j \brack i}_{t_c^2}.\nonumber
\eea
As in the case of the finite-dimensional homology, the $([1^r],[1^r])$-colored HOMFLY homology $\wt\scH^{(+)}_{[1^r],[1^r]}({T_{2,4}})$ follows directly from the mirror symmetry \eqref{mirror-infinite}, so we do not need to consider it separately.

\begin{figure}[!htbp]
 \centering
    \includegraphics[width=9cm]{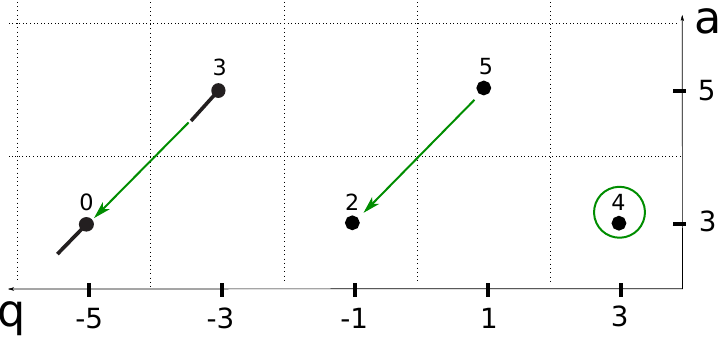}
    \caption{The uncolored negative HOMFLY homology $\scH^{(-)}_{[1],[1]}({T_{2,4}})$ of the (2,4) torus link, where the barbs represent the direction of semi-infinite ``tails'' resulting from the power series expansion of the factors $1/(1-q^{-2})$. The green arrows represent the action of the canceling differential.}\label{fig:HOMFLY-T24-1-}
\end{figure}

Next, we briefly discuss the infinite-dimensional homology categorifying the HOMFLY power series of the $(2,4)$ torus link in the regime $|q|>1$. We can write the Poincar\'e polynomial  of the uncolored HOMFLY homology $\scH^{(-)}_{\yng(1),\yng(1)}({T_{2,4}})$ as
\be\nonumber
\scP^{(-)}_{\yng(1),\yng(1)}({T_{2,4}};a,q,t)=a^3\left(q^3t^4+\frac{t^2}{q}+\frac{1}{q^5(1-q^{-2})}\right)+a^5\left(qt^5+\frac{t^3}{q^3(1-q^{-2})}\right)~,
\ee
where the diagram showing the generators is presented in Figure~\ref{fig:HOMFLY-T24-1-}. As in the case of the Hopf link, the uncolored homology diagram in Figure \ref{fig:HOMFLY-T24-1-} for $|q|>1$  is a mirror image of the diagram in Figure \ref{fig:HOMFLY-T24-1+} for $|q|<1$. Moreover, the corresponding Poincar\'e polynomials are related by
\bea\nonumber
\scP^{(-)}_{\yng(1),\yng(1)}(T_{2,4};a,q,t)=t^{-2} \scP^{(+)}_{\yng(1),\yng(1)}(T_{2,4};at^2,q^{-1},t^{-1})~.
\eea
Similarly, using the structural properties described in \S \ref{sec:quadruply}, one finds the Poincar\'e polynomial of the $([r],[r])$-colored HOMFLY homology $\wt\scH^{(-)}_{[r],[r]}({T_{2,4}})$:
\bea\label{T24-infinite-}
&&\wt\scP^{(-)}_{[r],[r]}({T_{2,4}};a,Q,t_r,t_c)\\
&=&a^{3r} Q^{3r}t_r^{4r}t_c^{r(3r+1)} \sum_{r \ge j\ge i \ge 0}Q^{-4(j+i)}t_r^{-2(j+i)} t_c^{-2j(i+r)} \frac{(-a^2Q^{2}t_r^3t_c^{2r+1};t_c^2)_{j}}{(Q^{-2};t_c^{-2})_{i}}{r \brack j}_{t_c^2}{j \brack i}_{t_c^2}.\nonumber
\eea
The explicit expressions for the anti-symmetric representations can again be obtained by applying the mirror symmetry \eqref{mirror-infinite}.

\subsubsection*{Kauffman homology }

The Kauffman invariant of the (2,4) torus link is
\bea\nonumber
&&F(T_{2,4};a,q) = \cr&=&\frac{a^4}{q^4(1-q^2)}\Big[1 - q^2 + q^4 - q^6 + q^8  +
 a (q - 2 q^3 + 2 q^5 - q^7) +
 a^2 (-1 + 2 q^2 - 3 q^4 + 2 q^6 - q^8)\cr
 &&+ a^3 (-q + 3 q^3 - 3 q^5 + q^7) + a^4 (-q^2 + 2 q^4 - q^6)\Big]~.
 \eea
The Poincar\'e polynomial of the Kauffman homology of the $(2,4)$ torus link follows from the structural properties and the differentials:
\bea\nonumber
\scF(T_{2,4};a,q,t)&=&a^3q^{-3} + a^4q^{-2}t^2 (1 +a^{-1}  q^{3})  (1 + a q^{-1} t) + a^5 q^{-1} t^4 (1+ a q^{-1} t) (1 + q^2 t)\cr
&&+\frac{a^4q^{2}t^4 (1 + a^{-1} q^{3})  (1 + a q^{-1} t) (1 + a^2 q^{-2} t^2 + a^2  t^3)}{ (1 - q^2)}\cr
&&+a^7qt^8\frac{ (1 + a q^{-1} t) (1 + q^2 t)}{(1 - a q^{-1})  (1 - q^2)}~.
\eea

\subsection{The $(2,2p)$ torus link $T_{2,2p}$}

\subsubsection*{Finite-dimensional HOMFLY homology}

The finite-dimensional $[r]$-colored HOMFLY homology of the $(2,2p)$ torus link contains the $[r]$-colored HOMFLY homology of the $(2,2p+1)$ torus knot. Note that the Poincar\'e polynomial of the $[r]$-colored HOMFLY homology of the $(2,2p+1)$ torus knot in the $(a,Q,t_r,t_c)$-gradings is given by
\bea\label{torusknot-HOMFLY}
&&\wt\scP_{[r]}(T_{2,2p+1};a,Q,t_r,t_c) = \\
&=&a^{2pr}Q^{-2pr} \sum_{r=s_{p+1}\ge s_p\ge\cdots\ge s_1\ge0}(-a^2Q^{-2}t_rt_c;t_c^2)_{s_p}\prod_{i=1}^{p}Q^{4s_i}t_r^{2s_i}t_c^{4r s_i-2s_is_{i+1}}{s_{i+1} \brack s_{i}}_{t_c^2}.\nonumber
\eea
Starting with this expression, the results \eqref{Hopf-prime-quad} and \eqref{T24-prime-quad} can be easily extrapolated to the $(2,2p)$ torus link:
\bea\nonumber
\wt\scP^{\fin}_{[r]}(T_{2,2p};a,Q,t_r,t_c)&=&a^{2pr}Q^{-2pr} \sum_{r=s_{p+1}\ge s_p\ge\cdots\ge s_1\ge0}(-a^2Q^{-2}t_rt_c;t_c^2)_{s_p}(-Q^2t_rt_c;t_c^2)_{r-s_1}\cr
&&\hspace{3cm}\times\prod_{i=1}^{p}Q^{4s_i}t_r^{2s_i}t_c^{4r s_i-2s_is_{i+1}}{s_{i+1} \brack s_{i}}_{t_c^2}.
\eea

\subsubsection*{Infinite-dimensional HOMFLY homology }

In the range $|q|<1$, the uncolored HOMFLY power series of $T_{2,2p}$ has a categorification with the Poincar\'e polynomial
\bea\nonumber
\scP^{(+)}_{\yng(1),\yng(1)}({T_{2,2p}};a,q,t)
&=& a^{2p-1} \left[ \frac{q^{-2p+1}(1-q^{4p}t^{2p})}{1-q^4t^2}+ \frac{q^{2p+1}t^{2p}}{1-q^2} \right]\cr
&&+a^{2p+1} \left[ \frac{q^{-2p+3}t^3(1-q^{4(p-1)}t^{2(p-1)})}{1-q^4t^2}+ \frac{q^{2p-1}t^{2p+1}}{1-q^2} \right]~~~~~
\eea
where the $S$-invariant of the $(2,2p)$ torus link is given by $S(2,2p)=2p-1$.
The $([r],[r])$-colored  HOMFLY homology of the $(2,2p)$ torus link is similar to the $[r]$-colored HOMFLY homology of the $(2,2p+1)$ torus knot, except for the semi-infinite tails associated with the geometric power series in the Poincar\'e polynomial.

Referring to the expression \eqref{torusknot-HOMFLY}, one can deduce the Poincar\'e polynomial of the infinite-dimensional $([r],[r])$-colored positive HOMFLY homology $\scH^{(+)}_{[r],[r]}(T_{2,2p})$
\bea\label{quad-T22p+}
&&\wt\scP^{(+)}_{[r],[r]}(T_{2,2p};a,Q,t_r,t_c) = \\
&=&a^{(2p-1)r}Q^{(1-2p)r}\sum_{r=s_{p+1}\ge s_p\ge\cdots\ge s_1\ge0}\frac{(-a^2Q^{-2}t_rt_c;t_c^2)_{s_p}}{(Q^{2}t_c^{2(r-s_1)};t_c^2)_{s_1}}\prod_{i=1}^{p}Q^{4s_i}t_r^{2s_i}t_c^{4r s_i-2s_is_{i+1}}{s_{i+1} \brack s_{i}}_{t_c^2}\nonumber
\eea
which satisfies all the required properties.
Similarly, in the regime $|q|>1$ we write the expression for $\scH^{(-)}_{[r],[r]}(T_{2,2p})$
\bea\label{quad-T22p-}
&&\wt\scP^{(-)}_{[r],[r]}(T_{2,2p};a,Q,t_r,t_c) = \cr
&=&(aQ)^{(2p-1)r} t_r^{2pr}t_c^{(2p-1)r^2+r}\\
&&\sum_{r=s_{p+1}\ge s_p\ge\cdots\ge s_1\ge0}\frac{(-a^2Q^{2}t_r^3t_c^{2r+1};t_c^2)_{s_p}}{(Q^{-2};t_c^{-2})_{s_1}}\prod_{i=1}^{p}Q^{-4s_i}t_r^{-2s_i}t_c^{-2s_is_{i+1}}{s_{i+1} \brack s_{i}}_{t_c^2} .\nonumber
\eea

Let us mention the relation to the refined Chern-Simons invariants of $T_{2,2p}$. From \eqref{quad-T22p+} and \eqref{quad-T22p-} it is straightforward to obtain the Poincar\'e polynomials of unreduced HOMFLY homology colored by $([r],[r])$ and $([1^r],[1^r])$ representations. When $|q|<1$, the Poincar\'e polynomial of unreduced HOMFLY homology with $t_c$-grading agrees with the refined Chern-Simons invariants \eqref{refined-CS-sym+} and \eqref{refined-CS-anti+}:
\bea\nonumber
\overline\scP^{(+)}_{[r],[r]}(T_{2,2p};a,q,t_r=1,t_c=t)&=&\overline{\rm rCS}^{(+)}_{[r],[r]}(T_{2,2p};a,q,t)~,\cr
\overline\scP^{(+)}_{[1^r],[1^r]}(T_{2,2p};a,q,t_r=1,t_c=t)&=&\overline{\rm rCS}^{(+)}_{[1^r],[1^r]}(T_{2,2p};a,q,t)~.
\eea
for both symmetric and anti-symmetric representations.
On the other hand, when $|q|<1$, they agree only in the case of symmetric representations
\eqref{refined-CS-sym-}
\bea\nonumber
\overline\scP^{(-)}_{[r],[r]}(T_{2,2p};a,q,t_r=t,t_c=1)&=&\overline{\rm rCS}^{(-)}_{[r],[r]}(T_{2,2p};a,q,t)~,\cr
\overline\scP^{(-)}_{[1^r],[1^r]}(T_{2,2p};a,q,t_r=t,t_c=1)&\neq&\overline{\rm rCS}^{(-)}_{[1^r],[1^r]}(T_{2,2p};a,q,t)~.
\eea

\subsubsection*{Kauffman homology}

The Poincar\'e polynomial of the Kauffman homology of $T_{2,2p}$ link that has all the expected structural properties
and carries the action of the differentials is easily evaluated to be
\bea\nonumber
&&\scF(T_{2,2p};a,q,t) = \cr
&=&\wt\scP^{(+)}_{\yng(2),\yng(2)}(T_{2,2p};a=a q^{-1},Q=a^{1/2} q^{-1/2},t_r=a^{1/2}q^{-3/2}t,t_c=a^{-1/2}q^{3/2})\cr
&=&a^{2p-1}q^{1-2p}\sum_{2=s_{p+1}\ge s_p\ge\cdots\ge s_1\ge0}\frac{(-a q^{-1}t;a^{-1} q^{3})_{s_p}}{(a^{-1+s_1} q^{5-3s_1};a^{-1} q^{3})_{s_1}}\\
&&\hspace{6cm}\times\prod_{i=1}^{p}a^{s_i ( s_{i+1}-1)} q^{s_i(7 - 3  s_{i+1})} t^{2 s_i}{s_{i+1} \brack s_{i}}_{a^{-1} q^{3}}.\nonumber
\eea

\subsection{The $(2,4)$ torus link $\wt T_{2,4}$ with opposite orientation}
\label{sec:4a1}

Now we come to an important point: the colored HOMFLY invariants of a link depends on the relative orientation of the link.
The Hopf link is exceptional in this sense: changing the relative orientation of the planar diagram in Figure~\ref{fig:Hopf} gives the same oriented link. However, this is no longer the case for any other torus link. The simplest example is the $(2,4)$ torus link. Reversing the relative orientation of its two components we obtain the oriented link denoted ${\bf 4a1}$ in the Knot Atlas \cite{knotatlas}. In this paper we denote it as $\wt T_{2,4}$ to emphasize its relation to the $(2,4)$ torus link and the role of orientation change.

\noindent
\begin{minipage}{\linewidth}
      \centering
      \begin{minipage}{0.32\linewidth}
          \begin{figure}[H]\centering
\includegraphics[width=3cm]{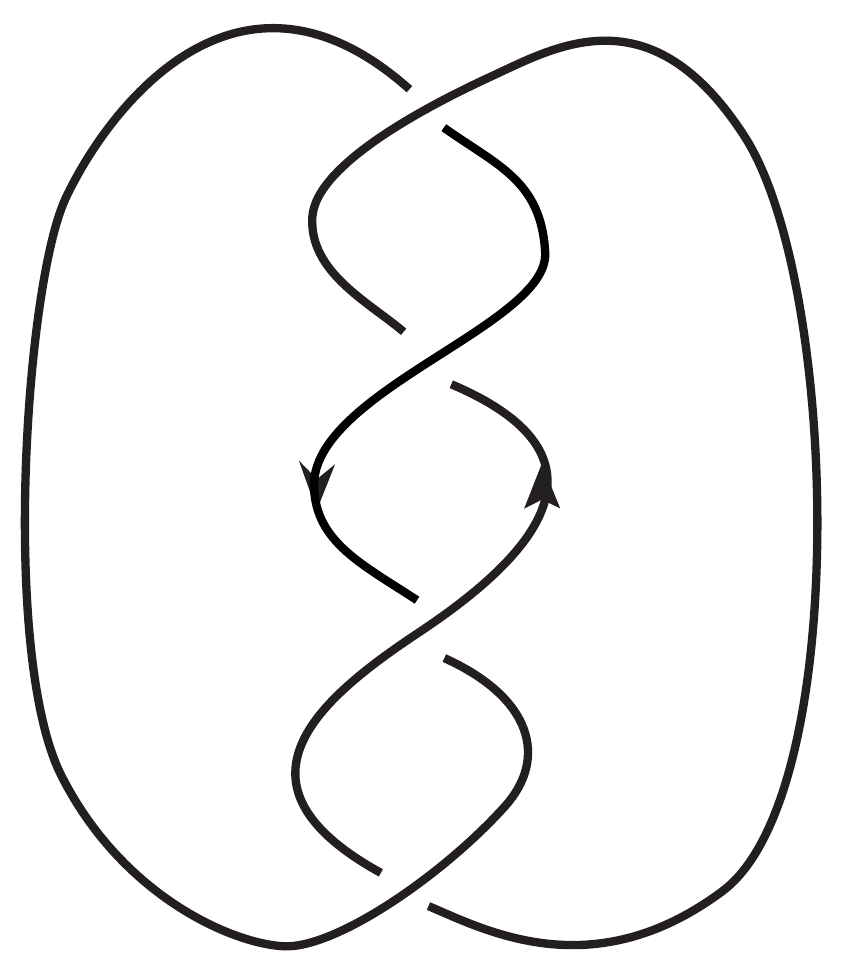}\caption{The link $\wt T_{2,4}$.}
\label{fig:4a1}
          \end{figure}
      \end{minipage}
      \hspace{0.05\linewidth}
      \begin{minipage}{0.58\linewidth}
  \begin{figure}[H]
 \centering
     \includegraphics[width=5.5cm]{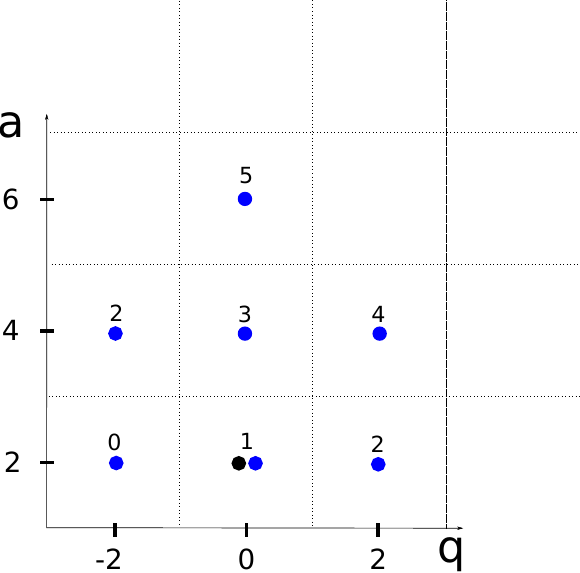}
    \caption{The finite-dimensional HOMFLY homology $\scH^{\rm fin}_{[1]}(\wt T_{2,4})$ of the link $\wt T_{2,4}$. The blue dots are also the generators of the HOMFLY homology of the $\bf 5_2$ knot.}\label{fig:HOMFLY-prime-4a1-1}\end{figure}
      \end{minipage}
  \end{minipage}


\subsubsection*{Finite-dimensional HOMFLY homology}

At the level of quantum group invariants, changing the orientation of one of the components colored by a representation $\lambda$ is equivalent to changing the color to the conjugate one, $\bar \lambda$, while keeping the orientation. This (not entirely trivial statement) can be shown {\it e.g.} by using axioms and properties of Chern-Simons TQFT. Therefore, as illustrated in Figure~\ref{fig:4a1-2}, the uncolored HOMFLY invariant of the $\wt T_{2,4}$ link is equal to the $([1],[1^{N-1}])$-colored HOMFLY invariant of the $(2,4)$ torus link $T_{2,4}$,
\bea\label{HOMFLY-poly-4a1}
P_{\yng(1),\yng(1)}(\wt T_{2,4};a,q)&=&\frac{a(1 - 2 q^2 + q^4 + a^2 (1 - q^2 + q^4) - a^4 q^2) }{q(1-q^2)}~.
\eea
Hence, in order to obtain a finite-dimensional HOMFLY homology of the $\wt T_{2,4}$ link, one needs to categorify the following Laurent polynomial:
\be\nonumber
P^\fin_{\yng(1)}({\wt T_{2,4}};a,q)=aq^{-1} (1-q^2)P_{\yng(1)}(\wt T_{2,4};a,q)=a^2(q^{-2}-2+q^2)+a^4(q^{-2}-1+q^2)-a^6~.
\ee
Since the link $\wt T_{2,4}$ is homologically thin, all generators of its finite-dimensional HOMFLY homology have the same the $\delta$-grading \eqref{delta-grading}, which immediately gives
\be\nonumber
\scP^{\rm fin}_{\yng(1)}(\wt T_{2,4};a,q,t)=a^2(q^{-2}+2t+q^2t^2)+a^4(q^{-2}t^2+t^3+q^2t^4)+a^6t^5~.
\ee
Remarkably, the finite-dimensional HOMFLY homology of the $\wt T_{2,4}$ link contains the HOMFLY homology of the ${\bf 5_2}$ knot (the twist knot $K_{s=1,t=2}$ in \S \ref{sec:cyclotomic}) as shown in Figure \ref{fig:HOMFLY-prime-4a1-1}.
This statement holds even for the colored HOMFLY homology. Hence,  with the help of the expression \eqref{superpoly-pretzel}, we find the Poincar\'e polynomial of the finite-dimensional $[r]$-colored HOMFLY homology of the $\wt T_{2,4}$ link
\bea\label{4a1-prime-quad}
&&\wt\scP^{\fin}_{[r]} (\wt T_{2,4}; a,Q,t_r,t_c)\\
 &=&\sum_{r \ge k \ge j\ge i \ge 0}   (-1)^{k+j} a^{2(r- k)} t_r^{r-2 k }t_c^{
 j (1 + j - 2 k) - k + (k - r)^2} {r \brack k}_{t_c^2}\cr
&&(-a^2Q^{-2} t_rt_c;t_c^2)_{j} (-a^2Q^2 t_r^3 t_c^{2r+1};t_c^2)_{j}{k \brack j}_{t_c^2}   (-1)^i (at_rt_c)^{4i} t_c^{5i(i-1)} \frac{1-a^2t_r^2t_c^{4i}}{(a^2t_r^2t_c^{2i};t_c^2)_{j+1}} {j \brack i}_{t_c^2}. \nonumber
\eea
Using this result, the Poincar\'e polynomial of the triply-graded $[r]$-colored HOMFLY homology in the $t_r$-grading can be written as
\bea\label{4a1-HOMFLYprime}
&&\scP^{\rm fin}_{[r]} (\wt T_{2,4}; a,q,t)=\wt\scP^{\fin}_{[r]} (\wt T_{2,4}; a,q,tq^{-1},q) = \\
 &=&\sum_{r \ge k \ge j\ge i \ge 0}  (-1)^{k+j} a^{2(r- k)} t^{r-2 k }q^{
 j (1 + j - 2 k) +k-r + (k - r)^2} {r \brack k}_{q^2}\cr
&&(-a^2q^{-2} t;q^2)_{j} (-a^2 q^{2 r} t^3;q^2)_{j}{k \brack j}_{q^2}   (-1)^i (at)^{4i} q^{5i(i-1)} \frac{1-a^2q^{4i-2}t^2}{(a^2q^{2i-2}t^2;q^2)_{j+1}} {j \brack i}_{q^2} .\nonumber
\eea

\subsubsection*{Infinite-dimensional HOMFLY homology}

The Poincar\'e polynomial of the $\fraksl(2)$ homology of the link $\wt T_{2,4}$ has the following form
({\it cf.} the reduced Khovanov-Rozansky $\frak\fraksl(N)$ invariant of $4^2_1$ (v1) in \cite{Carqueville:2011zea}):
\bea\nonumber
\scP^{\fraksl(2)}_{\yng(1),\yng(1)}(\wt T_{2,4};q,t)&=&q+q^3t+q^5t^2+q^9t^4~.
\eea
Starting with the $\fraksl(2)$ homology and the HOMFLY invariant \eqref{HOMFLY-poly-4a1} when $|q|<1$,
the differentials and other structural properties uniquely determine
the uncolored (positive) HOMFLY homology $\scH^{(+)}_{\yng(1),\yng(1)}(\wt T_{2,4})$:
\bea\label{HOMFLY-4a1+}
\scP^{(+)}_{\yng(1),\yng(1)}(\wt T_{2,4};a,q,t)&=&a(q^{-1}+qt)+a^3\left(q^{-1}t^2+\frac{q^3t^4}{1-q^2}\right)+\frac{a^5q t^5}{1-q^2}~.
\eea
Comparing with \eqref{rCS-4a1}, one can see that the Poincar\'e polynomial of $\scH^{(+)}_{\yng(1),\yng(1)}(\wt T_{2,4})$ is different from the refined Chern-Simons invariant of the $\wt T_{2,4}$ link.
\begin{figure}[h]
 \centering
    \includegraphics[width=6cm]{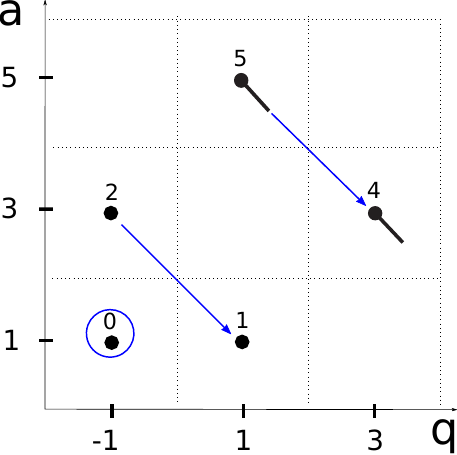}
    \caption{The uncolored positive HOMFLY homology $\scH^{(+)}_{[1],[1]}(\wt T_{2,4})$  of the $\wt T_{2,4}$ link where the barbs represent semi-infinite tails associated with the power series expansion of the factors $1/(1-q^2)$. The blue arrows show the action of the canceling differential $d^+_{[1]\to[0]}$. The $S$-invariant of the link $\wt T_{2,4}$ is $S(\wt T_{2,4})=1$.}\label{fig:HOMFLY-4a1-1}
\end{figure}

Although it is tedious, using differentials one can determine the $(\yng(2),\yng(2))$-colored HOMFLY homology $\scH^{(+)}_{\yng(2),\yng(2)}({\wt T_{2,4}})$:
\bea
&&\scP^{(+)}_{\yng(2),\yng(2)} (\wt T_{2,4}; a,q,t_r,t_c) = \\
&=&\frac{a^2}{q^2} (1 + q^2 t_c t_r) (1 + q^4 t_c^3 t_r) +
   \frac{a^4 t_r^2 t_c^2}{q^2}  (1 + q^2 t_c^2) (1 + q^4 t_rt_c^3 ) \left[1 + \frac{
      q^6 t_r^2 t_c^4  (1 + a^2q^{-2} t_r t_c)}{1 - q^4 t_c^2}\right]\cr
      && +
   a^6 q^2 t_r^4t_c^8  \left[1 + \frac{
      q^6 t_r^2  t_c^4 (1 + q^2 t_c^2)(1 + a^2q^{-2} t_r t_c)}{
      1 - q^4 t_c^2}+ \frac{
      q^{12} t_r^4t_c^8  (1 + a^2q^{-2} t_r t_c) (1 + a^2t_r t_c^3 )}{(1 -
         q^2) (1 - q^4 t_c^2)}\right]~.\nonumber
\eea
\begin{figure}[h]
 \centering
    \includegraphics[width=14cm]{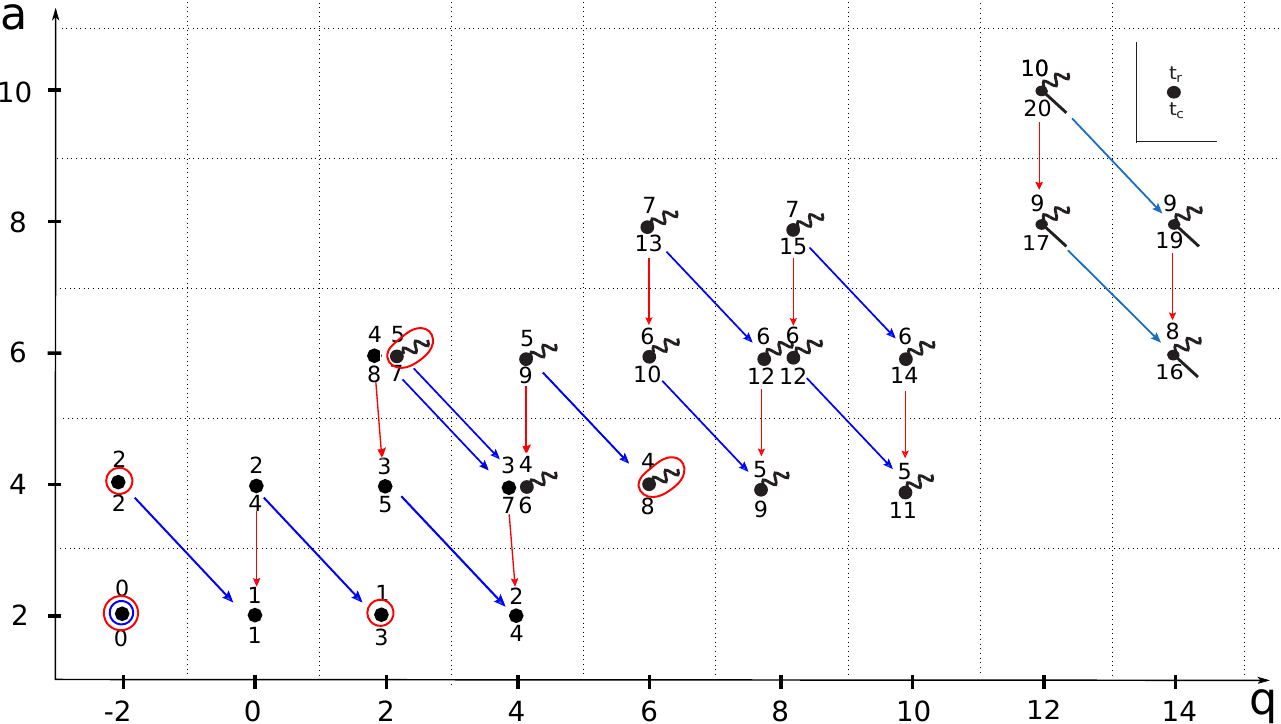}
    \caption{The $([2],[2])$-colored positive HOMFLY homology $\scH^{(+)}_{[2],[2]}({\wt T_{2,4}})$ of the  link $\wt T_{2,4}$ where the straight and wavy tails represent the power series expansions of the factors $1/(1-q^2)$ and  $1/(1- q^{4}t_c^2)$, respectively. The blue and red arrows show, respectively, the action of the canceling differential and the colored differential $d^+_{[2]\to [1]}$. The dots encircled in red represent the generators of $H_*(\scH^{(+)}_{[2],[2]}(\wt T_{2,4}),d^+_{[2]\to [1]})$, which is isomorphic to the uncolored homology $\scH^{(+)}_{[1],[1]}(\wt T_{2,4})$ shown in Figure~\ref{fig:HOMFLY-4a1-1}.}\label{fig:HOMFLY-4a1-2}
\end{figure}

With a little more effort, using the refined exponential growth property \eqref{REGP-infinite} and the differentials \eqref{color-diff-infinite}, we find the Poincar\'e polynomial of the $([r],[r])$-colored HOMFLY homology $\wt\scH^{(+)}_{[r],[r]}({\wt T_{2,4}})$ in the $(a,Q,t_r,t_c)$-gradings
\bea\label{T24-quad+}
\wt\scP^{(+)}_{[r],[r]} (\wt T_{2,4}; a,Q,t_r,t_c)&=&\left(aQ^{-1}\right)^{r}\sum_{r \ge j\ge i \ge 0}  a^{2j} t_r^{2j} t_c^{2j^2} (-Q^2 t_r t_c^{2 j + 1};t_c^2)_{r-j} {r \brack j}_{t_c^2} \cr
 &&\hspace{2.3cm}\times Q^{4i} t_r^{2i} t_c^{2ri}  \frac{ (-a^2 Q^{-2} t_r t_c;t_c^2)_{i}}{(Q^{2} t_c^{2 (r -i)};t_c^2)_i}  {j \brack i}_{t_c^2}.
\eea

Without repeating all the details, in the range $|q|>1$ we simply write the Poincar\'e polynomial of the $([r],[r])$-colored HOMFLY homology  $\wt\scH^{(-)}_{[r],[r]}({\wt T_{2,4}})$
\bea\label{T24-quad-}
\wt\scP^{(-)}_{[r],[r]} (\wt T_{2,4}; a,Q,t_r,t_c)&=&\left(aQt_r^2t_c^{r+1}\right)^{r}\sum_{r \ge j\ge i \ge 0}  a^{2j} t_r^{2j} t_c^{2j^2} (-Q^{-2} t_r^{-1} t_c^{2 (j -r)+ 1};t_c^2)_{r-j} {r \brack j}_{t_c^2} \cr
 &&\hspace{2cm}\times Q^{-4i} t_r^{-2i} t_c^{-2ri}  \frac{ (-a^2 Q^{2} t_r^3 t_c^{2r+1};t_c^2)_{i}}{(Q^{-2};t_c^{-2})_i}  {j \brack i}_{t_c^2}.\qquad
\eea

\subsubsection*{Kauffman homology}

The Kauffman invariant of the $\wt T_{2,4}$ link is
\bea\nonumber
&&F(\wt T_{2,4};a,q)=\cr
&=&\frac{1}{q^3(1-q^2)}\Big[q^2 - 2 q^4 + q^6 + a(-q + 3 q^3 - 3 q^5 + q^7)+a^2 (1 -
    2 q^2 + 3 q^4 - 2 q^6 + q^8) \cr
&&   +a^3 (q - 2 q^3 + 2 q^5 -
    q^7)  +a^4 (-1 + q^2 - q^4 + q^6 - q^8) \Big]~.
\eea
The Poincar\'e polynomial of the Kauffman homology of the link $\wt T_{2,4}$ that enjoys all the required properties is easily found to be
\bea\nonumber
\scF(\wt T_{2,4};a,q,t)&=&a(q^{-1}+qt)+a^2(q^{-2}t+2t^2+q^2 t^3)\cr
&&+a^3\left(q^{-3}t^2+q^{-1}t^3+qt^4\left(1+\frac1{1-q^2}\right)+\frac{q^3 t^5}{1-q^2}+\frac{q^5 t^6}{1-q^2}\right)\cr
&&+\frac{a^4}{1-q^2}\left(q^{-2}t^4+2t^5+2q^2t^6+q^4t^7 \right)\cr
&&+\frac{a^5}{1-q^2}\left(q^{-3}t^5+q^{-1}t^6 +qt^7 +\frac{q^3t^8}{1-a q^{-1}}+\frac{q^5t^9}{1-a q^{-1}} \right)\cr
&&+\frac{a^6}{(1-q^2)(1-a q^{-1})}(q^2t^9+q^4t^{10})~.
\eea

\begin{figure}[h]
 \centering
    \includegraphics[width=13cm]{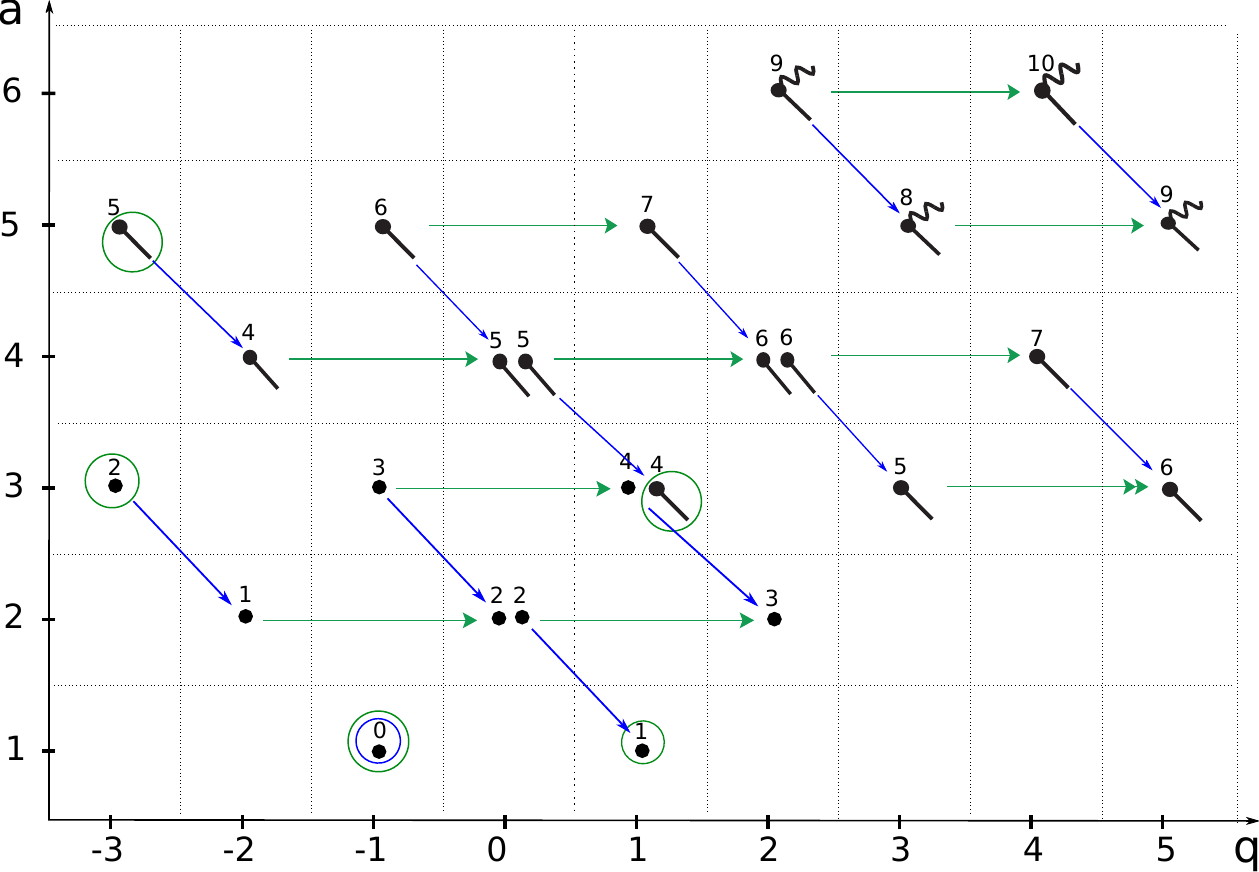}
    \caption{The Kauffman homology of the $\wt T_{2,4}$ link where the straight and wavy tails represent the factors $1/(1-q^2)$ and $1/(1-a q^{-1})$, respectively. The blue and green arrows show the action of the canceling differential and the universal differential, respectively.}\label{fig:Kauffman-4a1}
\end{figure}

\subsection{The $(2,2p)$ torus link $\wt T_{2,2p}$ with opposite orientation}

\subsubsection*{Finite-dimensional HOMFLY homology }

The finite-dimensional  $[r]$-colored homology of the $\wt T_{2,2p}$ link contains the $[r]$-colored HOMFLY homology \eqref{superpoly-pretzel} of the twist knot $K_{s=1,p}$. Thus, the Poincar\'e polynomial of the finite-dimensional  $[r]$-colored homology of the $\wt T_{2,2p}$ link can be easily obtained from \eqref{4a1-prime-quad} by incorporating the twist element $\varpi_{p,i}(a,t_r,t_c)$, {\it cf.} \eqref{twisting3},
\bea
&&\wt\scP^{\fin}_{[r]} (\wt T_{2,2p}; a,Q,t_r,t_c) = \cr
 &=&\sum_{r \ge k \ge j\ge i \ge 0}  (-1)^{k+j} a^{2(r- k)} t_r^{r-2 k }t_c^{
 j (1 + j - 2 k) - k + (k - r)^2} {r \brack k}_{t_c^2} \\
&&(-a^2Q^{-2} t_rt_c;t_c^2)_{j} (-a^2Q^2 t_r^3 t_c^{2r+1};t_c^2)_{j}{k \brack j}_{t_c^2} (-1)^i (at_rt_c)^{2pi} t_c^{(2p+1)i(i-1)} \frac{1-a^2t_r^2t_c^{4i}}{(a^2t_r^2t_c^{2i};t_c^2)_{j+1}} {j \brack i}_{t_c^2} .\nonumber
\eea

\subsubsection*{Infinite-dimensional HOMFLY homology}

One can extend the formulae \eqref{Hopf-quad+} and \eqref{T24-quad+} to obtain the Poincar\'e polynomial of the $([r],[r])$-colored positive HOMFLY homology $\scH^{(+)}_{[r],[r]} (\wt T_{2,2p})$ (that corresponds to $|q|<1$):
\bea
&&\wt\scP^{(+)}_{[r],[r]} (\wt T_{2,2p}; a,Q,t_r,t_c) = \cr
 &=&\left(aQ^{-1}\right)^{r} \sum_{r=s_{p}\ge \cdots\ge s_1\ge0} \prod_{i=2}^{p-1}  a^{2s_i} t_r^{2s_i} t_c^{2s_i^2} (-Q^2 t_r t_c^{2 s_i + 1};t_c^2)_{s_{i+1}-s_{i}} {s_{i+1} \brack s_{i}}_{t_c^2} \cr
 &&\hspace{5cm}\times Q^{4s_1} t_r^{2s_1} t_c^{2rs_1}  \frac{ (-a^2 Q^{-2} t_r t_c;t_c^2)_{s_1}}{(Q^{2} t_c^{2 (r -s_1)};t_c^2)_{s_1}}  {s_2 \brack s_1}_{t_c^2}.
\eea
In a similar manner, one can extend the formulae \eqref{Hopf-quad-} and \eqref{T24-quad-} to obtain the Poincar\'e polynomial of the $([r],[r])$-colored negative HOMFLY homology $\scH^{(-)}_{[r],[r]} (\wt T_{2,2p})$ (that corresponds to $|q|>1$):
\bea
&&\wt\scP^{(-)}_{[r],[r]} (\wt T_{2,2p}; a,Q,t_r,t_c) = \cr
 &=&\left(aQt_rt_c^r\right)^{r} \sum_{r=s_{p}\ge \cdots\ge s_1\ge0} \prod_{i=2}^{p-1}  a^{2s_i} t_r^{2s_i} t_c^{2s_i^2} (-Q^{-2} t_r^{-1} t_c^{2 (s_i -r)+ 1};t_c^2)_{s_{i+1}-s_{i}} {s_{i+1} \brack s_{i}}_{t_c^2} \cr
 &&\hspace{4cm}\times Q^{-4s_1} t_r^{-2s_1} t_c^{-2rs_1}  \frac{ (-a^2 Q^{-2} t_r t_c;t_c^2)_{s_1}}{(Q^{-2} ;t_c^{-2})_{s_1}}  {s_2 \brack s_1}_{t_c^2}.
\eea

\subsubsection*{Kauffman homology}

Since the link $\wt T_{2,2p}$ is homologically thin, its Kauffman homology can be obtained from the $(\yng(2),\yng(2))$-colored HOMFLY homology via \eqref{Kauffman-thin}:
\bea
&&\scF(\wt T_{2,2p}; a,q,t) = \cr
&=&\wt\scP^{(+)}_{\yng(2),\yng(2)}(\wt T_{2,2p};a=a q^{-1},Q=a^{1/2}q^{-1/2},t_r=a^{1/2}q^{-3/2},t_c=a^{-1/2}q^{3/2})\cr
 &=&a q^{-1} \sum_{2=s_{p}\ge \cdots\ge s_1\ge0} \prod_{i=2}^{p-1} a^{-( s_i-3) s_i} q^{s_i (3 s_i-5)} t^{2 s_i}  (-a^{1 - s_i}q^{3 s_i-1} t ;a^{-1}q^3)_{s_{i+1}-s_{i}} {s_{i+1} \brack s_i }_{a^{-1}q^3} \cr
 &&\hspace{5cm}\times a^{s_1}q^{s_1}t^{2s_1} \frac{ (-a q^{-1}t;a^{-1}q^3)_{s_1}}{(a^{ s_1 - 1}q^{ 5-3s_1};a^{-1}q^3)_{s_1}}  {s_2 \brack s_1}_{a^{-1}q^3}.
\eea

%

\subsection{The Whitehead link}

The Whitehead link $\textbf{WL}$ is the simplest two-component hyperbolic link that can be identified with the twist link $L_{p}$ for $p=1$, as explained in \S \ref{sec:cyclotomic}.

\noindent
\begin{minipage}{\linewidth}
      \centering
      \begin{minipage}{0.32\linewidth}
          \begin{figure}[H]\centering
\includegraphics[scale=0.34]{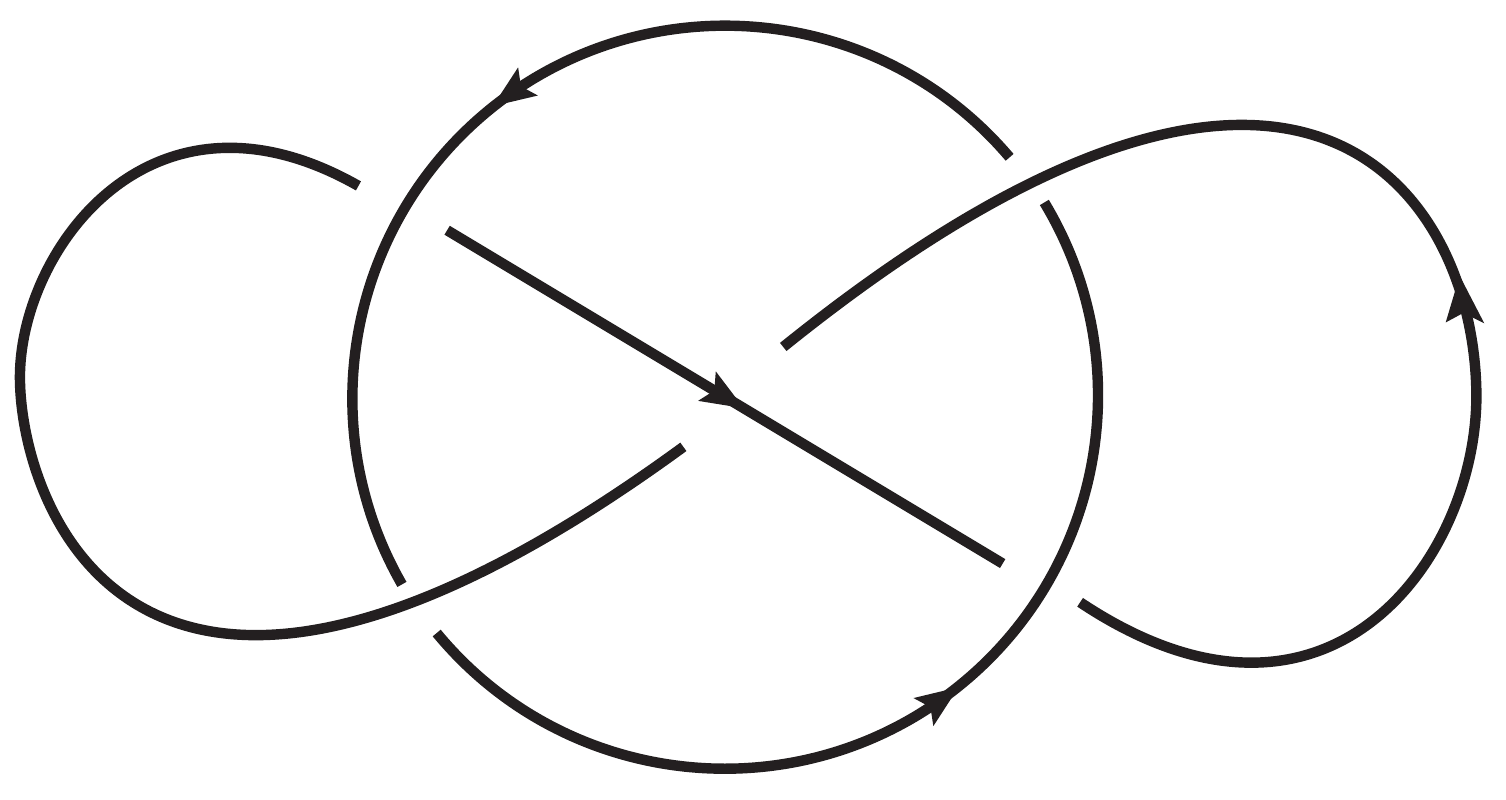}\caption{The Whitehead link $\WL$.}
\label{fig:WL}
          \end{figure}
      \end{minipage}
      \hspace{0.05\linewidth}
      \begin{minipage}{0.58\linewidth}
  \begin{figure}[H]
 \centering
     \includegraphics[width=9cm]{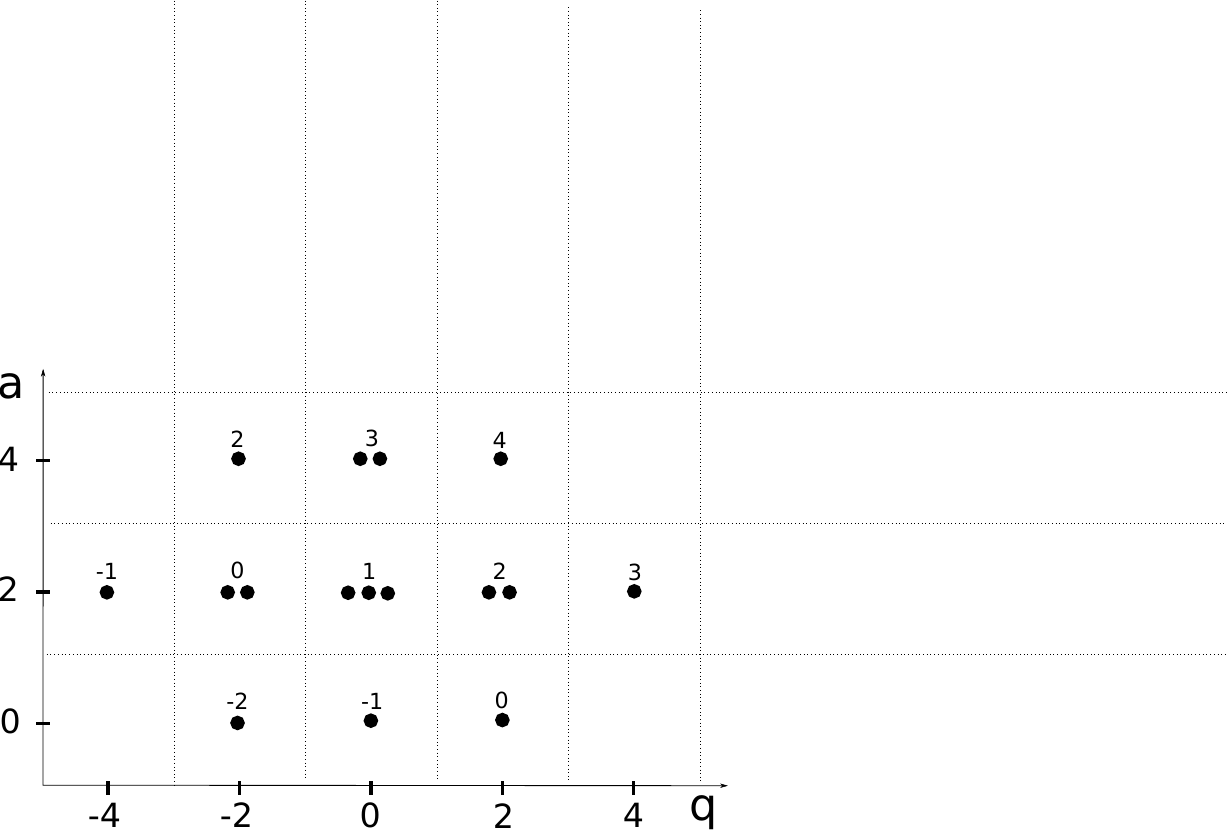}
    \caption{The finite-dimensional HOMFLY homology $\scH^{\rm fin}_{[1]}(\WL) $ of the Whitehead link. }\label{fig:HOMFLY-prime-WL}\end{figure}
      \end{minipage}
  \end{minipage}

\subsubsection*{Finite-dimensional HOMFLY homology}

In order to obtain the finite-dimensional HOMFLY homology, one needs to categorify the following polynomial
\bea
P^\fin_{\yng(1)}(\WL;a,q)&=&aq^{-1}(1-q^2)P_{\yng(1),\yng(1)}(\WL;a,q)  = \\
&=&\frac{1}{q^{2} }-1+q^2 +
 a^2 \left(-\frac{1}{q^4 } +\frac{2}{q^2} - 3+ 2 q^2  - q^4 \right) +
 a^4 \left(\frac{1}{q^2} - 2 + q^2 \right)~.\nonumber
\eea
Since the Whitehead link is homologically thin, one can easily obtain the finite-dimensional HOMFLY homology via \eqref{delta-grading}:
\bea
&&\scP^{\rm fin}_{\yng(1)}(\WL;a,q,t) = \\
&=&\frac{1}{q^{2} t^{2}} +\frac{1}{t} +q^2 +
 a^2 \left(\frac{1}{q^4 t} +\frac{2}{q^2} +  3 t + 2 q^2 t^2 + q^4 t^3\right) +
 a^4 \left(\frac{t^2}{q^2} + 2 t^3 + q^2 t^4\right)~.\nonumber
\eea

\noindent
The refined exponential growth \eqref{exp-growth-fin-1} and the differentials \eqref{regrading-finite-column} then determine the Poincar\'e polynomial of the finite-dimensional $[r]$-colored HOMFLY homology of the Whitehead link
\bea
&&\widetilde\scP^{\fin}_{[r]}(\WL;a,Q,t_r,t_c) = \cr
 &=&(- t_c t_r)^{-r} \sum _{i=0}^r (-t_r)^{-i}Q^{-2 i} t_c^{i (1 + 2 i - 4 r)}
 (-a^2Q^{-2} t_c t_r;t_c^2)_i (-a^2 Q^2 t_c^{2 r+1} t_r^3;t_c^2)_i   {r \brack i}_{t_c^2}\cr
&&\hspace{3cm}\times(-Q^2 t_c t_r;t_c^2)_i  (-Q^2 t_c^{2 (r-i)+1} t_r;t_c^2)_i (a^2 t_c^{4 i+2} t_r^2;t_c^2)_{r-i} \cr
&=&(- t_c t_r)^{-r} \sum _{i=0}^r a^{i} t_c^{i(i-1)/2}\scF^\prime_{r,i}(a,Q,t_r,t_c)\cr
&&\hspace{2cm} (-a^{2}t_c^{2(r-1)})^{-i} (-a ^2Q^{-2}  t_rt_c;t_c^2)_{i} (- a ^2Q^2t_r^3 t_c^{1+2r};t_c^2)_{i } {r \brack i}_{t_c^2},
\eea
where the factor $\scF^\prime_{r,i}(a,Q,t_r,t_c)$ is defined in \eqref{block}. Subsequently, the Poincar\'e polynomial of the triply-graded $[r]$-colored HOMFLY homology in the $t_r$-grading can be written as
\bea\label{Whitehead-HOMFLYprime}
\scP^{\rm fin}_{[r]}(\WL;a,q,t)&=&(-t)^{-r} \sum_{i=0}^r (-t)^{-i} q^{2 i (i - 2 r)} (-a^2 t q^{-2};q^2)_{i} (-a^2 t^3 q^{2 r};q^2)_{i}{r \brack i}_{q^2}\cr
&&\hspace{1.7cm}\times(-q^2t;q^2)_i(-q^{2(r-i+1)}t;q^2)_i(a^2q^{4i}t^2;q^2)_{r-i}~.
\eea

\subsubsection*{Infinite-dimensional HOMFLY homology}

The formula \eqref{cyclotomic-HOMFLY} gives the reduced HOMFLY invariant of the Whitehead link
\be\nonumber
P_{\yng(1),\yng(1)}(\WL;a,q) =\frac{q^2 - q^4 + q^6-
 a^2 (1 - 2 q^2 + 3 q^4 - 2 q^6 + q^8) + a^4 (q^2 - 2 q^4 + q^6) }{aq^3(1-q^2)}~.
\ee
In addition, the reduced $\fraksl(2)$ homology of the Whitehead link has the form \cite{Carqueville:2011zea}
\bea\nonumber
{\scP}^{\fraksl(2)}_{\yng(1),\yng(1)}(\WL;q,t)&=&  \frac{1}{q^3 t^2} + \frac{1}{q t} +2 q+ q^3 t + 2 q^5 t^2 + q^7 t^3~.
\eea
Using these data, one can determine the uncolored positive HOMFLY homology $\scH^{(+)}_{\yng(1),\yng(1)}(\WL)$
that enjoys all the required properties and has the Poincar\'e polynomial
\bea\nonumber
\scP^{(+)}_{\yng(1),\yng(1)}(\WL;a,q,t) &=&a^3 ( q^{-1} t^2 + q t^3 )
+a \left( q^{-3} t^{-1}+q^{-1}  + q t + q^3 t^2   +\frac{q t}{1-q^2} \right)\cr
&&
+ a^{-1} \left(q^{-1} t^{-2} + \frac{q^3}{ 1-q^2} \right)~.
\eea
\begin{figure}[h]
 \centering
    \includegraphics[width=7cm]{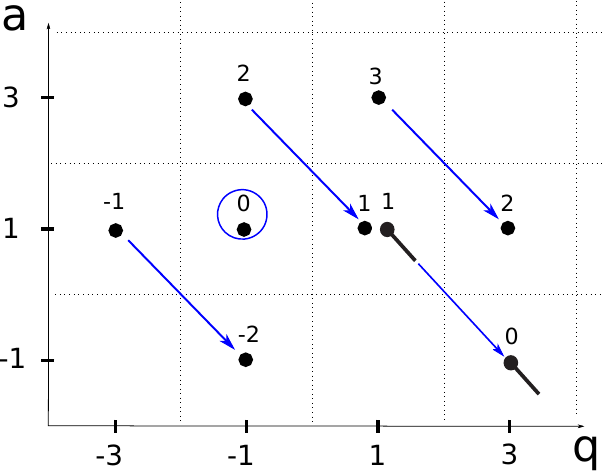}
    \caption{The ``$+$'' version of the uncolored HOMFLY homology of the Whitehead link, $\scH^{(+)}_{[1],[1]}(\WL)$, where the straight tails represent the factors $1/(1-q^2)$. The blue arrows show the action of the canceling differential, and the $S$-invariant of the Whitehead link is $S(\WL)=1$.}\label{fig:HOMFLY-WL-1}
\end{figure}

\noindent
Furthermore, the Poincar\'e polynomial of the $([r],[r])$-colored HOMFLY homology of the Whitehead link in the $(a,Q,t_r,t_c)$-gradings can be written as
\bea\label{WL-quad+}
&&\wt\scP^{(+)}_{[r],[r]}(\WL;a,Q,t_r,t_c) = \cr
&=&\left(-\frac{Q}{a t_r t_c} \right)^{r}\sum_{r \ge j\ge i \ge 0}  (-1)^{j } Q^{2j-4i}t_r^{j-2i} t_c^{i ( 2 j - 4 r)+j}  {r \brack j}_{t_c^2} {j \brack i}_{t_c^2} \\
&&\times \frac{(-a^2Q^{-2}t_rt_c;t_c^2)_i  (-a^2Q^2 t_r^3 t_c^{2r+1};t_c^2)_i(-Q^2 t_r t_c^{2 ( r- i )+1}  ;t_c^2)_{i}(-a^2Q^{-2} t_r t_c^{2i+1 } ;t_c^2)_{j-i}}{(Q^{2} t_c^{2(r-j+i)};t_c^2)_{j-i}}~ .  \nonumber
\eea

The analysis for $|q|>1$ can be performed in the same way, so that we just write the resulting answer for $\scH^{(-)}_{[r],[r]}(\WL)$
\bea\label{WL-quad-}
&&\wt\scP^{(-)}_{[r],[r]}(\WL;a,Q,t_r,t_c) = \cr
&=&\left(-aQ t_r t_c^r \right)^{-r}\sum_{r \ge j\ge i \ge 0}  (-1)^{j } Q^{4i-2j}t_r^{2i-j} t_c^{j ( 1+2 i- 2 r)}  {r \brack j}_{t_c^2} {j \brack i}_{t_c^2} \\
&&\times \frac{(-a^2Q^{2}t_r^3t_c^{2r+1};t_c^2)_i  (-a^2Q^{-2} t_r t_c;t_c^2)_i(-Q^{-2} t_r^{-1} t_c^{1- 2i }  ;t_c^2)_{i}(-a^2Q^{2} t_r^3 t_c^{2(r+i)+1 } ;t_c^2)_{j-i}}{(Q^{-2};t_c^{-2})_{j-i}}~ .  \nonumber
\eea

\subsubsection*{Kauffman homology}

The Kauffman invariant of the Whitehead link can be written as
\bea\nonumber
&&F(\WL;a,q) = \cr
&=&\frac{1}{a q^2(1-q^2)}\Big[-q + 2 q^3 - 3 q^5 +
 2 q^7 - q^9 + (1 - 4 q^2 + 6 q^4 - 6 q^6 + 4 q^8 -
    q^{10}) a \cr
    &&+ (3 q - 8 q^3 + 11 q^5 - 8 q^7 +
    3 q^9) a^2 + (-1 + 5 q^2 - 10 q^4 + 10 q^6 - 5 q^8 +
    q^{10}) a^3\cr
    && + (-2 q + 6 q^3 - 8 q^5 + 6 q^7 -
    2 q^9) a^4 + (-q^2 + 3 q^4 - 3 q^6 + q^8) a^5\Big]~.
\eea
The Poincar\'e polynomial of the Kauffman homology
\bea\nonumber
&&{\scF}(\WL;a ,q,t) = \cr
&=&a^{-1}\left[\tfrac{1}{q^3 t^4}+\tfrac{1}{q t^3}+\tfrac{q}{t^2} \left(1+\tfrac{1}{1-q^2}\right)+\tfrac{q^3}{\left(1-q^2\right) t}+\tfrac{q^5}{1-q^2}\right]\cr
&&+\tfrac{1}{q^4 t^3}+\tfrac1{q^2 t^2}\left(2+\tfrac{1}{1-q^2}\right)+\tfrac1t\left(2+\tfrac{2}{1-q^2}\right)+q^2 \left(2+\tfrac{2}{1-q^2}\right)+q^4 t\left(1+\tfrac{1}{1-q^2}\right) \cr
&&+a  \Big[\tfrac1{q^3 t}\left(2+\tfrac{1}{1-q^2}\right)+\tfrac1{q}\left(4+\tfrac{1}{1-q^2}+\tfrac{1}{\left(1-q^2\right) \left(1-a q^{-1}\right)}\right)\cr
&&\hspace{4cm} +q t \left(4+\tfrac{2}{1-q^2}+\tfrac{1}{\left(1-q^2\right) \left(1-a q^{-1}\right)}\right)+q^3 t^2\left(2+\tfrac{2}{1-q^2}\right) +\tfrac{ q^5 t^3}{1-q^2}\Big]\cr
&&+a ^2 \Big[\tfrac{1}{q^4}+\tfrac{t}{q^2} \left(3+\tfrac{1}{\left(1-q^2\right)}+\tfrac{1}{\left(1-q^2\right) \left(1-a q^{-1}\right)}\right)\cr
&&\hspace{2cm}+t^2 \left(4+\tfrac{1}{1-q^2}+\tfrac{1}{\left(1-q^2\right) \left(1-a q^{-1}\right)}\right)+q^2 \left(3+\tfrac{3}{1-q^2}\right) t^3+q^4 \left(1+\tfrac{1}{1-q^2}\right) t^4\Big]\cr
&&+a ^3\left[\tfrac{t^2}{q^3}\left(1+\tfrac{1}{1-q^2} \right)+\tfrac{ t^3}{q}\left(3+\tfrac{2}{1-q^2} \right)+ q t^4\left(4+\tfrac{1}{1-q^2} \right)+2 q^3 t^5\right] \cr
&&+a^4\left[\tfrac{t^4}{q^2}+2 t^5+q^2 t^6\right]~.
\eea

\subsection{Twist link $L_p$}

Now, let us consider the homological invariants of the twist link $L_p$ with a planar diagram shown in Figure~\ref{fig:surgery}.

\subsubsection*{Finite-dimensional HOMFLY homology}
For the class of links in Figure \ref{fig:surgery}, the Poincar\'e polynomials of the finite-dimensional $[r]$-colored HOMFLY homology have a structure similar to the cyclotomic expansions of the polynomial invariants \eqref{Habiro} and \eqref{cyclotomic-HOMFLY}. To see that, let us introduce the factor
\be\label{block}
\cF_{r,i}(a,Q,t_r,t_c)\equiv a^i Q^{-2 i} t_r^{-i} t_c^{ -i (1 - 3 i + 4 r)/2}(-Q^2 t_c t_r;t_c^2)_i  (-Q^2 t_c^{2 (r-i)+1} t_r;t_c^2)_i (a^2 t_c^{4 i+2} t_r^2;t_c^2)_{r-i} ~.
 \ee
Then, the Poincar\'e polynomial can be obtained by replacing the twist elements
$\varpi_{t,i}(a,t_r,t_c)$ ({\it cf.} \eqref{superpoly-pretzel}) by the factors $\cF_{r,i}(a,Q,t_r,t_c)$:
\bea\label{TL-prime-quad}
\wt\scP^\fin_{[r]}(L_p;a,Q,t_)r,t_c)
&=&(-t_rt_c)^{-r}\sum_{i=0}^r \varpi_{p,i}(a,t_r,t_c)\cF_{r,i}(a,Q,t_r,t_c) \\
&&\hspace{.5cm}\times(-a^{2}t_c^{2(r-1)})^{-i} (-a ^2Q^{-2}  t_rt_c;t_c^2)_{i} (- a ^2Q^2t_r^3 t_c^{1+2r};t_c^2)_{i } {r \brack i}_{t_c^2}.\nonumber
\eea

\subsubsection*{Infinite-dimensional HOMFLY homology}
The Poincar\'e polynomial of the $([r],[r])$-colored homology $\scH^{(\pm)}_{[r],[r]}(L_p)$ of the twist link $L_p$ can be easily obtained from \eqref{WL-quad+} and \eqref{WL-quad-} by incorporating the twist element $\varpi_{p,i}(a,t_r,t_c)$ ({\it cf.} \eqref{twisting3}):
\bea\nonumber
&&\wt\scP^{(+)}_{[r],[r]}(L_p;a,Q,t_r,t_c) = \cr
&=&\left(-\frac{Q}{a t_r t_c} \right)^{r}\sum_{r \ge k \ge j\ge i \ge 0} (-1)^{k } Q^{2k-4j}t_r^{k-2j} t_c^{j ( 2 k - 4 r)+k}  {r \brack k}_{t_c^2} {k \brack j}_{t_c^2} \cr
&&\times \frac{ (-a^2Q^{-2}t_rt_c;t_c^2)_j (-a^2Q^2 t_r^3 t_c^{2r+1};t_c^2)_j (-Q^2 t_r t_c^{2 ( r- j )+1}  ;t_c^2)_{j}(-a^2Q^{-2} t_r t_c^{2j+1 } ;t_c^2)_{k-j}}{ (Q^{2} t_c^{2(r-k+j)};t_c^2)_{k-j}}\cr
&&\times(-1)^i (at_rt_c)^{2pi} t_c^{(2p+1)i(i-1)} \frac{1-a^2t_r^2t_c^{4i}}{(a^2t_r^2t_c^{2i};t_c^2)_{j+1}} {j \brack i}_{t_c^2} ,\cr
&&\wt\scP^{(-)}_{[r],[r]}(L_p;a,Q,t_r,t_c) = \cr
&=&\left(-aQ t_r t_c^r \right)^{-r}\sum_{r \ge k\ge j \ge i\ge 0}  (-1)^{k } Q^{4j-2k}t_r^{2j-k} t_c^{k ( 1+2 j- 2 r)}  {r \brack k}_{t_c^2} {k \brack j}_{t_c^2} \cr
&&\times \frac{(-a^2Q^{2}t_r^3t_c^{2r+1};t_c^2)_j  (-a^2Q^{-2} t_r t_c;t_c^2)_j(-Q^{-2} t_r^{-1} t_c^{1- 2j }  ;t_c^2)_{j}(-a^2Q^{2} t_r^3 t_c^{2(r+j)+1 } ;t_c^2)_{k-j}}{(Q^{-2};t_c^{-2})_{k-j}}\cr
&&\times(-1)^i (at_rt_c)^{2pi} t_c^{(2p+1)i(i-1)} \frac{1-a^2t_r^2t_c^{4i}}{(a^2t_r^2t_c^{2i};t_c^2)_{j+1}} {j \brack i}_{t_c^2} .
\eea

\subsubsection*{Kauffman homology}
Since the twist link $L_p$ is homologically thin, the Kauffman homology can be obtained from the $(\yng(2),\yng(2))$-colored HOMFLY homology via \eqref{Kauffman-thin}
\bea\nonumber
&&\scF(L_p;a,q,t)=\wt\scP^{(+)}_{\yng(2),\yng(2)}(L_p;a=a q^{-1},Q=a^{1/2}q^{-1/2},t_r=a^{1/2}q^{-3/2},t_c=a^{-1/2}q^{3/2}) \cr
&=&\frac{q}{t^2 a} \sum_{k=0}^2 \sum_{j=0}^k \sum_{i=0}^j (-1)^{k} q^{-k+j(3k-7)} t^{k- 2 j} a^{j+k-jk }{r \brack k}_{a^{-1}q^3} {k \brack j}_{a^{-1}q^3} \cr
&&\times \frac{ (-a q^{-1}t;a^{-1}q^3)_j (-a^2t^3;a^{-1}q^3)_j(-a^{j-1}q^{5 - 3 j} t  ;a^{-1}q^3)_{j}(-a^{1 - j}q^{3 j-1} t ;a^{-1}q^3)_{k-j}}{ (q^{5 + 3 j - 3 k} a^{k-j-1};a^{-1}q^3)_{k-j}}\cr
&&\times(-1)^i a^{ 2 i p -  (2p+1)i(i-1)/2}   q^{ 3 (2p+1)i(i-1)/2-2 i p}t^{2 i p}\frac{1-a^{3 - 2 i}q^{ 6 i-5} t^2 }{(a^{3 - i} q^{3 i-5} t^2 ;a^{-1}q^3)_{j+1}} {j \brack i}_{a^{-1}q^3} .\nonumber\eea

\subsection{The Borromean rings}

The Borromean rings $\textbf{BR}$ shown in Figure~\ref{fig:surgery} provide the simplest example of a three-component hyperbolic link.

\subsubsection*{Finite-dimensional HOMFLY homology}

As in the case of the twist link \eqref{TL-prime-quad}, the Poincar\'e polynomial of the finite-dimensional $[r]$-colored HOMFLY homology of the Borromean rings can be obtained by substituting the factor $\cF_{r,i}(a,Q,t_r,t_c)$ for the twist elements in \eqref{superpoly-pretzel}:
\bea\label{BR-HOMFLYprime}
\wt\scP^{\fin}_{[r]}(\BR;a,Q,t_r,t_c)
&=&(-t_r^2t_c^2)^{-r}\sum_{i=0}^r\cF_{r,i}(a,Q,t_r,t_c)\cF_{r,i}(a,Q,t_r,t_c) \\
&&\hspace{.5cm}\times(-a^{2}t_c^{2(r-1)})^{-i} (-a ^2Q^{-2}  t_rt_c;t_c^2)_{i} (- a ^2Q^2t_r^3 t_c^{1+2r};t_c^2)_{i } {r \brack i}_{t_c^2}.\nonumber
\eea

\subsubsection*{Infinite-dimensional HOMFLY homology}

The formula \eqref{cyclotomic-HOMFLY} gives the reduced HOMFLY invariant of the Borromean rings:
\bea\nonumber
&&P_{\yng(1),\yng(1),\yng(1)}(\BR;a,q) = \cr
&=&\frac{1}{a^2q^4(1-q^2)}\Big[-q^2 + 4 q^4 - 5 q^6 + 4 q^8 - q^{10}+
 a^2 (1 - 4 q^2 + 7 q^4 - 10 q^6 + 7 q^8 - 4 q^{10} + q^{12}) \cr
 && +a^4 (-q^2 + 4 q^4 - 5 q^6 + 4 q^8 - q^{10})\Big]~.
\eea
In addition, the reduced $\fraksl(2)$ homology of the Borromean rings has the form \cite{Carqueville:2011zea}:
\be\nonumber
 \scP^{\fraksl(2)}_{\yng(1),\yng(1),\yng(1)}(\BR;q,t)=\frac{1}{q^6t^3}+\frac{3}{q^4t^2}+\frac{2}{q^2t}+4+2q^2t+3q^4t^2+q^6t^3~.
\ee
From these data, the structural properties and differentials uniquely determine the Poincar\'e polynomial of the positive HOMFLY homology of the Borromean rings:
\bea\label{uncolor-BR+}
&&\scP^{(+)}_{\yng(1),\yng(1),\yng(1)}(\BR;a,q,t) = \cr
&=&a^{-2}\left[q^{-2}t^{-3}+2t^{-2}+q^4 \left(\tfrac{1}{1-q^2}+\tfrac1{(1-q^2)^2} \right) \right]\cr
&&+q^{-4}t^{-2}+2q^{-2}t^{-1}+2+q^2t\left(1+\tfrac{1}{1-q^2}+\tfrac{2}{(1-q^2)^2}\right)+q^4t^2\left(1+\tfrac1{(1-q^2)^2}\right)\cr
&&+a^2\left[q^{-2}t+t^2\left(1+\tfrac1{(1-q^2)^2} \right)+q^{2}t^3\left(1+\tfrac1{1-q^2} \right)\right]~.
\eea
Similarly, the negative homology that categorifies the HOMFLY power series expansion in $|q|>1$
can be easily obtained from \eqref{uncolor-BR+}:
\bea\nonumber
\scP^{(-)}_{\yng(1),\yng(1),\yng(1)}(\BR;a,q,t) =t^2\scP^{(+)}_{\yng(1),\yng(1),\yng(1)}(\BR;at^2,q^{-1},t^{-1}) ~.
\eea
\begin{figure}[h]
 \centering
    \includegraphics[width=9cm]{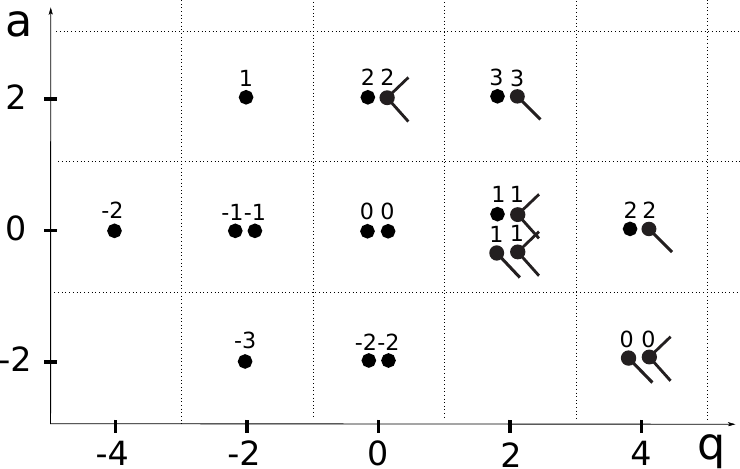}
    \caption{The uncolored positive HOMFLY homology $\scH^{(+)}_{[1],[1],[1]}(\BR)$ of the Borromean rings. The barbs represent semi-infinite tails associated with the factors $1/(1-q^2)$. The $S$-invariant of the  Borromean rings is $S(\BR)=0$.}\label{fig:HOMFLY-BR}
\end{figure}

It turns out that the \emph{unreduced} $\fraksl(2)$ homology of the Borromean rings is rather subtle. Applying the differential $d_2$ to the unreduced HOMFLY homology and substituting $a=q^2$, one finds
\bea\nonumber
&&\scP(\overline \scH^{(+)}_{\yng(1),\yng(1),\yng(1)}(\BR),d_2)(a=q^2,q,t) = \\
&=& \frac{1}{q^6t^3}+\frac{2}{q^4t^2}+\frac{1}{q^2t^2}+2t^{-1}+4+4q^2+2q^2t+2q^4t+3q^4t^2+2q^6t^2+q^8t^3~.\nonumber
\eea
However, the Poincar\'e polynomial of the unreduced $\fraksl(2)$ homology of the Borromean rings is \cite{Carqueville:2011zea}:
\be\nonumber
\overline \scP^{\fraksl(2)}_{\yng(1),\yng(1),\yng(1)}(\BR;q,t)=\frac{1}{q}\Big[\frac{1}{q^6t^3}+\frac{2}{q^4t^2}+\frac{1}{q^2t^2}+2t^{-1}+4+4q^2+2q^2t+q^4t^2+2q^6t^2+q^8t^3\Big]~.
\ee
This implies that the spectral sequence with respect to differentials \cite{Rasmussen:2004} in the unreduced HOMFLY homology of the Borromean rings is non-trivial even though the Borromean rings are homologically thin.

We leave the analysis of the colored HOMFLY homology of the Borromean rings for future work.

\begin{figure}[h]\centering
\includegraphics[scale=1]{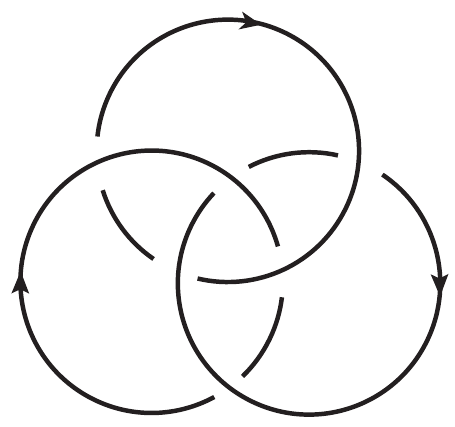}\caption{The $(3,3)$ torus link.}
\label{fig:T33}
\end{figure}

\subsection{The $(3,3)$ torus link $T_{3,3}$}

\subsubsection*{Finite-dimensional HOMFLY homology}

The finite-dimensional HOMFLY homology of the $(3,3)$ torus link $T_{3,3}$ can be obtained by categorifying the following polynomial
\bea\nonumber
P^\fin_{\yng(1)}(T_{3,3};a,q)&=& aq^{-1}(1-q^2)P_{\yng(1),\yng(1),\yng(1)}(T_{3,3};a,q) = \cr
&=&
 a^6 ( q^{-6} - 2q^{-4} + 2q^{-2}-1 + 2 q^2 - 2 q^4 + q^6)\cr
 &&+ a^8 ( - q^{-4} + q^{-2} -2+ q^2 - q^4) +a^{10}~.
\eea
The $(3,3)$ torus link is homologically thick, {\it i.e.} the $\delta$-gradings of the generators are not all equal. However, the differentials uniquely determine the finite-dimensional HOMFLY homology of the $(3,3)$ torus link to be
\bea\nonumber
\scP^{\rm fin}_{\yng(1)}(T_{3,3};a,q,t)&=&
 a^6 (q^{-6} + 2 tq^{-4} + 2 t^2q^{-2} + 2 t^3 + t^4 + 2 q^2 t^4 +
    2 q^4 t^5 + q^6 t^6) \\
    &&+ a^8 (t^3q^{-4} + 2 t^4q^{-2} + 2 t^5 + t^5q^{-2} + 2 q^2 t^6 + q^2 t^7 +
    q^4 t^7)+a^{10} t^8~.\nonumber
\eea
Curiously, the finite-dimensional HOMFLY homology of the $(3,3)$ torus link contains the HOMFLY homology of the $(3,4)$ torus knot. Therefore, making use of the expression \eqref{T34-HOMFLY-1} for the $(3,4)$ torus knot, one finds the Poincar\'e polynomial of the finite-dimensional  $[r]$-colored HOMFLY homology of $T_{3,3}$:
\bea\nonumber
&&\wt\scP^{\fin}_{[r]}(T_{3,3};a,Q,t_r,t_c) = \cr
&=&a^{6 r} \sum _{j=0}^r  \sum _{i=0}^{r-j}  \sum _{j \ge k_1\ge k_2\ge k_3\ge k_4=0}  Q^{4 (r+k_2+ k_3+ k_4- j)-6 k_1}  t_r^{2( k_2+ k_3+ k_4)-j-4 k_1+5 r} \cr
&&\times  t_c^{2 k_2 (j-k_3)+j^2+2 k_4 k_1-4 j k_1+2 k_3 (j+k_1-k_4)+2 k_4 r-2 j r+5 r^2}\cr
&&\times  { r \brack j }_{t_c^2} { r-j \brack i}_{t_c^{-2}}{ j \brack k_1}_{t_c^2} { k_1 \brack k_2}_{t_c^2}{ k_2\brack k_3}_{t_c^2} {k_3 \brack k_4}_{t_c^2} \cr
&&\times\left(-a^2 Q^{-2} t_r t_c ;t_c^2\right)_{r-j}(-a^2 Q^{-2} t_r t_c^{1+2 (r-j)} ;t_c^2)_{j-k_1} (-a^2 Q^2 t_r^3 t_c^{1+2 r} ;t_c^2)_{j-k_1} \cr
&&\times (-a^2 Q^{-2} t_r t_c^{1+2 (r-k_1)};t_c^2)_{k_2}   \left[ (-Q^2 t_r t_c;t_c^2)_{k_1-k_3}\right]^2 .
\eea

  \begin{figure}[h]\centering
\includegraphics[width=13cm]{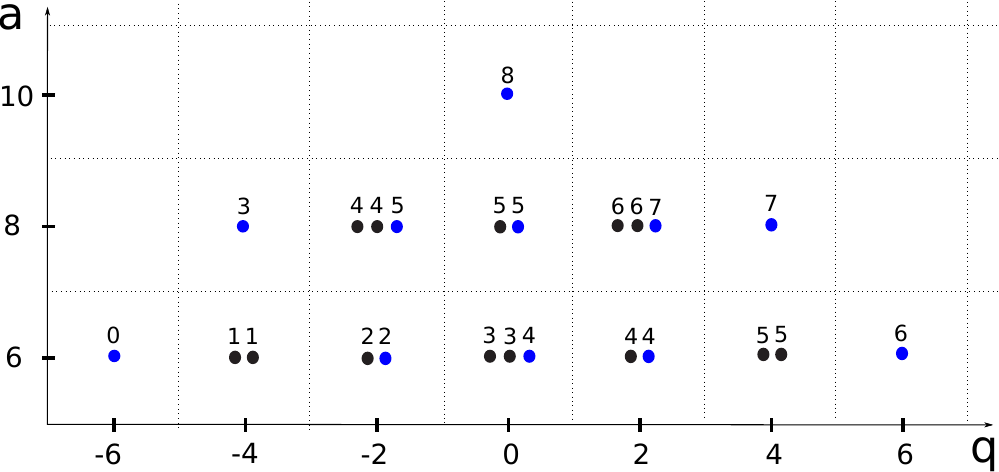}
 \caption{The finite-dimensional HOMFLY homology $\scH^{\rm fin}_{[1]}(T_{3,3})$ of the $(3,3)$ torus link. The $\delta$-gradings of the generators are not all equal. The blue dots are also the generators of the HOMFLY homology $\scH_{[1]}(T_{3,4})$ of the $(3,4)$ torus knot.}
\label{fig:HOMFLY-prime-T33}
\end{figure}

\begin{figure}[h]
 \centering
    \includegraphics[width=12cm]{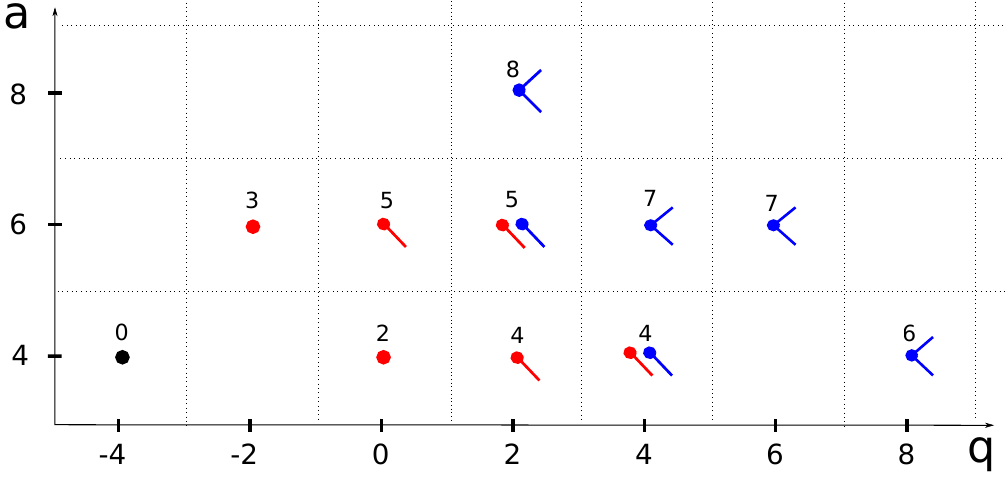}
    \caption{The uncolored positive HOMFLY homology $\scH^{(+)}_{[1],[1],[1]}(T_{3,3})$  of the $(3,3)$ torus link where the barbs represent the semi-infinite tails associated with the geometric series expansion of $1/(1-q^2)$. The $S$-invariant of the  $(3,3)$ torus link is $S(T_{3,3})=4$.}\label{fig:HOMFLY-T33-1-uncolor}
\end{figure}

\subsubsection*{Infinite-dimensional HOMFLY homology}

The HOMFLY invariant of the $(3,3)$ torus link is equal to
\bea\nonumber
&&P_{\yng(1),\yng(1),\yng(1)}(T_{3,3};a,q) = \\
&=&\frac{a^4(1 - 2 q^2 + 2 q^4 - q^6 + 2 q^8 - 2 q^{10} + q^{12} +
 a^2 (-q^2 + q^4 - 2 q^6 + q^8 - q^{10})+ a^4 q^6)}{q^4 (1 - q^2)^2} ~.\nonumber
\eea
On the other hand, the unreduced $\fraksl(2)$ homology of the $(3,3)$ torus link is \cite{Stosic:2006}:
\be\nonumber
\overline \scP^{\fraksl(2)}_{\yng(1),\yng(1)}(T_{3,3};q,t)=q^4+q^6+q^8t^2+q^{10}t^4+q^{12}(t^3+3t^4)+2q^{14}t^4~.
\ee
Combined with the properties of the differentials,
these data uniquely determine the Poincar\'e polynomial of the HOMFLY homology of the $(3,3)$ torus link:
\bea\nonumber
\scP^{(+)}_{\yng(1),\yng(1),\yng(1)}(T_{3,3};a,q,t) &=&a^{4}\left[q^{-4}+t^2+ \frac{(q^2+2q^{4})t^4}{1-q^2}+\frac{q^{8}t^6}{(1-q^2)^2} \right]\cr
&&+a^{6}\left[q^{-2}t^3+\frac{(1+2q^{2})t^{5}}{1-q^2}+\frac{(q^4+q^6)t^7}{(1-q^2)^2}\right]+\frac{a^{8}q^2t^8}{(1-q^2)^2}~.\quad
\eea
In fact, this expression agrees with the refined Chern-Simons invariant of the $(3,3)$ torus link computed in \cite{DuninBarkowski:2011yx}.
Similarly, the Poincar\'e polynomial of the negative HOMFLY homology categorifying the power series expansion with $|q|>1$ is
\bea\nonumber
\scP^{(-)}_{\yng(1),\yng(1),\yng(1)}(T_{3,3};a,q,t)=t^{-2}\scP^{(+)}_{\yng(1),\yng(1),\yng(1)}(T_{3,3};at^2,q^{-1},t^{-1})~.
\eea

\section{Derivation of recursion relations for the Hopf link} \label{sec:app-B}

In this Appendix, we derive three  recursion relations (\ref{A1-hat-HL})-(\ref{A3-hat-HL}) of the colored superpolynomial of the Hopf link. We recall that the colored superpolynomial for the Hopf link can be written as (see (\ref{unreduced-Hopf+bis}))
\be\nonumber
 \overline{\scP}^{(+)}_{[r_1],[r_2]}(T_{2,2};a,\wt{q},t) = \sum_{j=0}^{r_2} p(r_1,r_2,j),
\ee
where
\be
p(r_1,r_2,j) = \wt{q}^{2(r_1+1)j}t^{-2j} \frac{(-a^2 t,\wt{q}^2)_{r_1}}{(\wt{q}^2 t^{-2},\wt{q}^2)_{r_1}}
\frac{(-a^2 t^3 \wt{q}^{-2},\wt{q}^2)_j}{(\wt{q}^{2(r_2-j+1)}t^{-2},\wt{q}^2)_j}
\frac{(\wt{q}^2,\wt{q}^2)_{r_2}}{(\wt{q}^2,\wt{q}^2)_j (\wt{q}^2,\wt{q}^2)_{r_2-j}}.    \label{p-HL-r1r2j}
\ee
The recursion relations can be found from the analysis of various ratios involving summands $p(r_1,r_2,j)$ in (\ref{p-HL-r1r2j}); these ratios are
\bea
\frac{p(r_1+1,r_2,j)}{p(r_1,r_2,j)} & = & \wt{q}^{2j} \frac{1+a^2 t \wt{q}^{2r_1}}{1-t^{-2}\wt{q}^{2(r_1+1)}}~, \label{pp1} \\
\frac{p(r_1,r_2+1,j)}{p(r_1,r_2,j)} & = & \frac{(1-\wt{q}^{2(r_2+1)})(1-t^{-2}\wt{q}^{2(r_2-j+1)})}{(1-\wt{q}^{2(r_2-j+1)})(1-t^{-2}\wt{q}^{2(r_2+1)})} , \label{pp2}\\
\frac{p(r_1,r_2,j+1)}{p(r_1,r_2,j)} & = & t^{-2}\wt{q}^{2(r_1+1)} \frac{(1-\wt{q}^{2(r_2-j)})(1+a^2t^{3}\wt{q}^{2(j-1)})}{(1-\wt{q}^{2(j+1)})(1-t^{-2}\wt{q}^{2(r_2-j)})}~, \label{pp3} \cr
\frac{p(r_1+1,r_2+1,j)}{p(r_1,r_2,j)} & = & \wt{q}^{2j} \frac{(1-\wt{q}^{2(r_2+1)}) (1+a^2 t \wt{q}^{2r_1}) (1-t^{-2}\wt{q}^{2(r_2-j+1)})}{(1-\wt{q}^{2(r_2-j+1)}) (1-t^{-2}\wt{q}^{2(r_1+1)}) (1-t^{-2}\wt{q}^{2(r_2+1)})}~, \label{pp4}\cr
\frac{p(r_1+1,r_2,j+1)}{p(r_1,r_2,j)} & = & t^{-2} \wt{q}^{2(r_1+j+2)} \frac{(1-\wt{q}^{2(r_2-j)}) (1+a^2 t \wt{q}^{2r_1}) (1+a^2 t^{3}\wt{q}^{2(j-1)})}{(1-\wt{q}^{2(j+1)}) (1-t^{-2}\wt{q}^{2(r_1+1)}) (1-t^{-2}\wt{q}^{2(r_2-j)})}~, \label{pp5}\cr
\frac{p(r_1,r_2+1,j+1)}{p(r_1,r_2,j)} & = & t^{-2}\wt{q}^{2(r_1+1)} \frac{(1-\wt{q}^{2(r_2+1)})(1+a^2 t^{3}\wt{q}^{2(j-1)})}{(1-\wt{q}^{2(j+1)})(1-t^{-2}\wt{q}^{2(r_2+1)})}~, \label{pp6}\\
\frac{p(r_1+1,r_2+1,j+1)}{p(r_1,r_2,j)} & = & t^{-2} \wt{q}^{2(r_1+j+2)} \frac{(1-\wt{q}^{2(r_2+1)}) (1+a^2 t \wt{q}^{2r_1}) (1+a^2 t^{3} \wt{q}^{2(j-1)})}{(1-\wt{q}^{2(j+1)}) (1-t^{-2}\wt{q}^{2(r_1+1)}) (1-t^{-2}\wt{q}^{2(r_2+1)})}~.  \nonumber
\eea
Now we can write the above expressions in the linear form in $p_{r_1,r_2,_j}$ (i.e. cancel denominators), and resum over $j=0,\ldots,r_2$. It is also convenient to introduce auxiliary functions
$$
R_{r_1,r_2} = \sum_{j=0}^{r_2} \wt{q}^{2j} p(r_1,r_2,j), \
S_{r_1,r_2} = \sum_{j=0}^{r_2} \wt{q}^{4j} p(r_1,r_2,j),\
T_{r_1,r_2} = \sum_{j=0}^{r_2} \wt{q}^{6j} p(r_1,r_2,j).
$$
For example, writing the first equation (\ref{pp1}) in the form
$$
 (1-t^{-2}\wt{q}^{2(r_1+1)})\, p(r_1+1,r_2,j) = \wt{q}^{2j}  (1+a^2 t \wt{q}^{2r_1})\, p(r_1,r_2,j),
$$
and summing over $j=0,\ldots,r_2$, we get
$$
(1-t^{-2} \wt{q}^{2(r_1+1)}) \overline{\scP}^{(+)}_{[r_1+1],[r_2]}= (1+a^2 t \wt{q}^{2r_1}) R_{r_1,r_2}.
$$


\subsection{Recursions involving only $\widehat{y}_1$ or $\widehat{y}_2$}

To start with, we consider equations (\ref{pp2}) and (\ref{pp6}), which lead respectively to
\begin{small}
\bea\nonumber
(1 - t^{-2} \wt{q}^{2(r_2+1)}) \big(  R_{r_1,r_2+1} -\wt{q}^{2(r_2+1)}\overline{\scP}^{(+)}_{[r_1],[r_2+1]} \big) &=&  (1 - \wt{q}^{2(r_2+1)})\big(  R_{r_1,r_2}-t^{-2}\wt{q}^{2(r_2+1)}\overline{\scP}^{(+)}_{[r_1],[r_2]} \big)~,\cr
(1 - t^{-2} \wt{q}^{2(r_2+1)}) \big( \overline{\scP}^{(+)}_{[r_1],[r_2+1]} - R_{r_1,r_2+1} \big) &=& \wt{q}^{2(r_1+1)}t^{-2} (1 - \wt{q}^{2(r_2+1)})\big( \overline{\scP}^{(+)}_{[r_1],[r_2]} - a^2 t^3 \wt{q}^{-2}R_{r_1,r_2} \big)~.\nonumber\eea
\end{small}
We can solve these equations for $R_{r_1,r_2}$ and $R_{r_1,r_2+1}$, and express them in terms of $\overline{\scP}^{(+)}_{[r_1],[r_2]}$ and $\overline{\scP}^{(+)}_{[r_1],[r_2+1]}$
\begin{small}
\bea\nonumber
R_{r_1,r_2} & = & \frac{\wt{q}^2(\wt{q}^{2r_2}-\wt{q}^{2r_1}) \overline{\scP}^{(+)}_{[r_1],[r_2]} + (t^2-\wt{q}^{2(r_2+1)}) \overline{\scP}^{(+)}_{[r_1],[r_2+1]}   }{t^2(1+a^2 t \wt{q}^{2r_1})}  \\
R_{r_1,r_2+1} & = & \frac{\wt{q}^{2(r_1+1)}(\wt{q}^{2(r_2+1)}-1)(1+a^2 t \wt{q}^{2r_2})\overline{\scP}^{(+)}_{[r_1],[r_2]}+  (t^2-\wt{q}^{2(r_2+1)})(1+a^2 t \wt{q}^{2(r_1+r_2+1)}) \overline{\scP}^{(+)}_{[r_1],[r_2+1]}   }{(1+a^2 t \wt{q}^{2r_1})(t^2 - \wt{q}^{2(r_2+1)})} \nonumber
\eea
\end{small}
Now we can shift the index $r_2\to r_2+1$ in the first equation above, and take advantage of the fact that the left hand sides of both equations will be equal, so that their right hand sides must also be equal. This implies
\bea\nonumber
&&(\wt{q}^{2(r_2+1)} - t^2) (\wt{q}^{ 2(r_2+2)} - t^2) \overline{\scP}^{(+)}_{[r_1],[r_2+2]} + \cr
&+&   (\wt{q}^{2(r_2+1)} - t^2) (t^2 + \wt{q}^{2(r_1+1)}- \wt{q}^{2(r_2+2)}  +     a^2 t^3 \wt{q}^{2(r_1 + r_2 +1)} ) \overline{\scP}^{(+)}_{[r_1],[r_2+1]} + \cr
&+&t^2 \wt{q}^{2(r_1+1)}(1 - \wt{q}^{2(r_2+1)})  (1 + a^2  t \wt{q}^{2r_2}) \overline{\scP}^{(+)}_{[r_1],[r_2]} =0.
\eea
By symmetry of $\overline{\scP}_{[r_1],[r_2]}$, the same equation holds with respect to shifts in $r_1$
\bea\nonumber
&&(\wt{q}^{2(r1+1)} - t^2) (\wt{q}^{2( r1+2)} - t^2) \overline{\scP}^{(+)}_{[r_1+2],[r_2]}  + \cr
&+&  (\wt{q}^{2(r_1+1)} - t^2) (t^2 + \wt{q}^{2(r_2+1)}- \wt{q}^{2(r_1+2)}  +     a^2 t^3 \wt{q}^{2(r_1 + r_2 +1)} ) \overline{\scP}^{(+)}_{[r_1+1],[r_2]} + \cr
&+& t^2 \wt{q}^{2(r_2+1)}(1 - \wt{q}^{2(r_1+1)})  (1 + a^2  t \wt{q}^{2r_1}) \overline{\scP}^{(+)}_{[r_1],[r_2]} =0.
\eea
The above two equations are equivalent to (\ref{A1-hat-HL}) and (\ref{A2-hat-HL}).


\subsection{Recursions involving $\widehat{y}_1$ and $\widehat{y}_2$}

It is possible to find a few recursion relations which involve both $\widehat{y}_1$ and $\widehat{y}_2$.
Let us look first at equations for $\frac{p(r_1+1,r_2,j)}{p(r_1,r_2,j)}$ and $\frac{p(r_1,r_2+1,j+1)}{p(r_1,r_2,j)}$, i.e. (\ref{pp1}) and (\ref{pp6}), which lead respectively to
\begin{small}
\bea\nonumber
(1 - t^{-2} \wt{q}^{2(r_1+1)}) \overline{\scP}^{(+)}_{[r_1+1],[r_2]} &=& (1 + a^2 t \wt{q}^{2r_1})R_{r_1,r_2}~,  \\
(1 - t^{-2} \wt{q}^{2(r_2+1)}) \big( \overline{\scP}^{(+)}_{[r_1],[r_2+1]} - R_{r_1,r_2+1} \big) &=& \wt{q}^{2(r_1+1)}t^{-2} (1 - \wt{q}^{2(r_2+1)})\big( \overline{\scP}^{(+)}_{[r_1],[r_2]} - a^2 t^3 \wt{q}^{-2}R_{r_1,r_2} \big) ~.\nonumber
\eea
\end{small}
From the first equation above we now determine $R_{r_1,r_2}$ in terms of $\overline{\scP}^{(+)}_{[r_1+1],[r_2]}$ and substitute to the second equation. This gives the recursion of the form
\be
\alpha \overline{\scP}^{(+)}_{[r_1+1],[r_2+1]}+ \beta \overline{\scP}^{(+)}_{[r_1+1],[r_2]} + \gamma \overline{\scP}^{(+)}_{[r_1],[r_2+1]} + \delta  \overline{\scP}^{(+)}_{[r_1],[r_2]} = 0,  \label{y1y2-recursion}
\ee
where
\bea\nonumber
\alpha &=& (1 - t^{-2} \wt{q}^{2(r_1 + 1)}) (1 - t^{-2} \wt{q}^{2(r_2 + 1)})~,\cr
\beta &=& a^2 t \wt{q}^{2r_1} (1 - t^{-2} \wt{q}^{2(r_1 + 1)}) (1 - \wt{q}^{2(r_2 + 1)})~,\cr
\gamma &=& -(1 + a^2 t \wt{q}^{2r_1}) (1 - t^{-2} \wt{q}^{2(r_2 + 1)})~,\cr
\delta &=& t^{-2} \wt{q}^{2(r_1 + 1)} (1 + a^2 t \wt{q}^{2r_1}) (1 - \wt{q}^{2(r_2 + 1)})~.
\eea
It is not hard to rewrite this recursion in a symmetric form, i.e. invariant under the exchange of $r_1$ and $r_2$.

Another recursion relation follows from (\ref{pp1}) and (\ref{pp2}), which lead respectively to
\begin{small}
\bea\nonumber
(1 - t^{-2} \wt{q}^{2(r_1+1)}) \overline{\scP}^{(+)}_{[r_1+1],[r_2]}  &=& (1 + a^2 t \wt{q}^{2r_1})R_{r_1,r_2}~,\\
(1 - t^{-2} \wt{q}^{2(r_2+1)}) \big(  R_{r_1,r_2+1} -\wt{q}^{2(r_2+1)}\overline{\scP}^{(+)}_{[r_1],[r_2+1]}  \big) &=&  (1 - \wt{q}^{2(r_2+1)})\big(  R_{r_1,r_2}-t^{-2}\wt{q}^{2(r_2+1)}\overline{\scP}^{(+)}_{[r_1],[r_2]}  \big) ~.\nonumber
\eea
\end{small}
Again, from the first equation above we determine $R_{r_1,r_2}$ in terms of $\overline{\scP}_{[r_1+1],[r_2]}$ and substitute to the second equation. This also gives a recursion of the form
\be
\alpha' \overline{\scP}^{(+)}_{[r_1+1],[r_2+1]}  + \beta' \overline{\scP}^{(+)}_{[r_1+1],[r_2]}  + \gamma' \overline{\scP}^{(+)}_{[r_1],[r_2+1]}  + \delta' \overline{\scP}^{(+)}_{[r_1],[r_2]}   = 0,  \label{y1y2prim-recursion}
\ee
where
\bea\nonumber
\alpha' &=& (1 - t^{-2} \wt{q}^{2(r_1 + 1)}) (1 - t^{-2} \wt{q}^{2(r_2 + 1)})~,\cr
\beta' &=& -(1 - t^{-2} \wt{q}^{2(r_1 + 1)}) (1 - \wt{q}^{2(r_2 + 1)})~,\cr
\gamma' &=& -\wt{q}^{2(r_2+1)} (1 + a^2 t \wt{q}^{2r_1}) (1 - t^{-2} \wt{q}^{2(r_2 + 1)})~,\cr
\delta' &=& t^{-2} \wt{q}^{2(r_2 + 1)} (1 + a^2 t \wt{q}^{2r_1}) (1 - \wt{q}^{2(r_2 + 1)}).
\eea
It is also not hard to find a symmetric (w.r.t. $r_1\leftrightarrow r_2$) form of this recursion.


Two recursion relations for $\overline{\scP}^{(+)}_{[r_1],[r_2]}$, given in equations (\ref{y1y2-recursion}) and (\ref{y1y2prim-recursion}), have identical coefficients $\alpha$ and $\alpha'$. Therefore, subtracting one equation form another, and canceling some overall factor, we get the first order ``mixed'' recursion
\be\nonumber
(\wt{q}^{2(r_1+1)} - t^2) \overline{\scP}^{(+)}_{[r_1+1],[r_2]}- (\wt{q}^{2(r_2+1)} - t^2) \overline{\scP}^{(+)}_{[r_1],[r_2+1]} +\wt{q}^2(\wt{q}^{2r_2}-\wt{q}^{2r_1})\overline{\scP}^{(+)}_{[r_1],[r_2]} = 0.
\ee
In this way we reproduce (\ref{A3-hat-HL}).


\section{Diagonal super-$A$-polynomials}\label{sec:diagonal}


As we mentioned in \S \ref{sec:associated}, the finite-dimensional HOMFLY homology of links with components colored by the same representation inherits all the properties of the HOMFLY homology of knots, and it is interesting to analyze its large color behavior and recursion relations for corresponding Poincar\'e polynomials (which we also refer to as diagonal superpolynomials). This leads to algebraic curves which are the analogs of super-$A$-polynomials of knots, and which we call \emph{diagonal super-A-polynomials} of links.

In this Appendix we determine such diagonal super-$A$-polynomials for several links and discuss their properties. Once relevant superpolynomials are determined, we follow similar strategy as in \cite{Fuji:2012nx,Fuji:2012pi,Nawata:2012pg,Fuji:2013rra}. Namely, on one hand, we determine difference equations satisfied by superpolynomials; these equations are also referred to as quantum (diagonal) super-$A$-polynomials, and their classical ($q\to 1$) limits lead to diagonal super-$A$-polynomials. On the other hand, we show that the same algebraic curves are encoded in the asymptotics of (diagonal) superpolynomials.

One important property of (classical) super-$A$-polynomials discussed in \cite{Fuji:2012nx,Fuji:2012pi,Nawata:2012pg,Fuji:2013rra} in their ``quantizability'', which is the statement that the leading term $S_0=\int \log y\frac{dx}{x}$ in the large color expansion of colored HOMFLY polynomials is well defined (i.e. it does not depend on the integration path). There is a simple necessary condition for an algebraic curve to be quantizable: its face polynomials (i.e. polynomials $p_{\rm face}(z)=\sum_k a_k z^k$, where $k$ runs along a fixed face of the Newton polygon, and $a_k$ denote corresponding coefficients) must be tempered (i.e. their roots must be roots of unity), for details see \cite{Gukov:2011qp,Fuji:2012nx}. Similarly as for ordinary super-$A$-polynomials, we find that these conditions are satisfied for quite generic values of $a$ and $t$, and they imply that $a$ and $t$ parameters must be roots of unity.


\subsection{Hopf link}

\begin{figure}[H]
  \begin{minipage}[b]{7cm}\centering
\includegraphics[scale=.7]{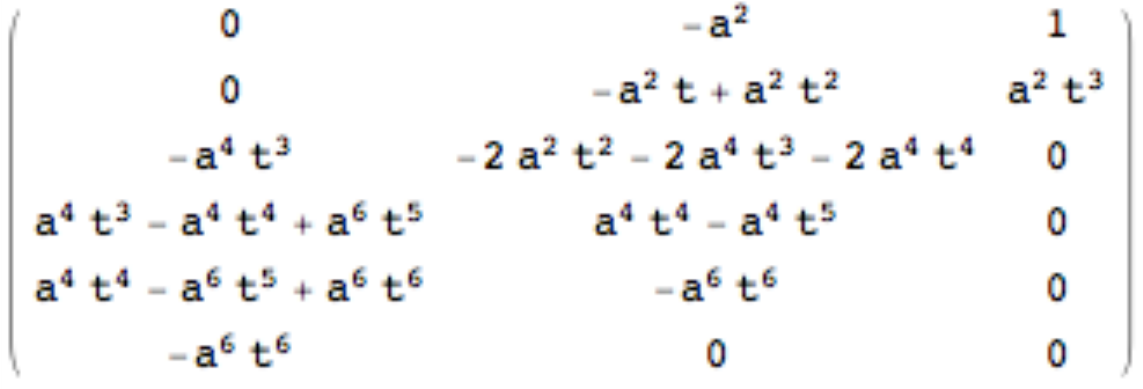}
\caption{Matrix representation of the ``diagonal'' super-$A$-polynomial for the Hopf link.}\label{fig-T22-matrix}
\end{minipage}
\hspace{.5cm}
  \begin{minipage}[b]{7.5cm}\centering
\includegraphics[scale=1.1]{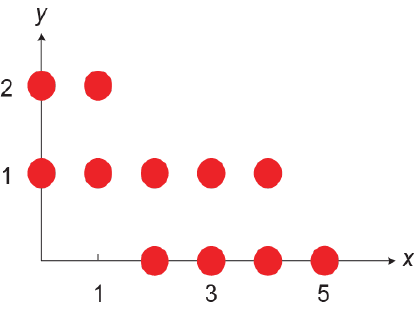}
\caption{Newton polygon of the ``diagonal'' super-$A$-polynomial for the Hopf link.}\label{fig-T22-Newton}
\end{minipage}
\end{figure}

The Poincar\'e polynomial of the finite-dimensional HOMFLY homology for the Hopf link is given in (\ref{T22-HOMFLYprime}). In this case the quantum super-$A$-polynomial (i.e. corresponding difference equation) takes form:
\be
\widehat{A}^{\textrm{super}} = a_0 + a_1 \hat{y} + a_2 \hat{y}^2,   \label{T22-difference}
\ee
where
\bea\nonumber
a_0 & = & a^4 q^2 t^3 \hat{x}^2 (1 - q^2 \hat{x})(1 + q^2 t \hat{x}) (1 - a^2 t^2 \hat{x}) (1 + a^2 q^6 t^3 \hat{x}^2)   \cr
a_1 & = & \frac{a^{2}}{q^{10}} (1 + a^2 q^4 t^3 \hat{x}^2) (q^8 + q^{12} t \hat{x} - q^{12} t^2 \hat{x} +
   q^{14} t^2 \hat{x}^2 + q^{16} t^2 \hat{x}^2 + a^2 q^{10} t^3 \hat{x}^2 + a^2 q^{14} t^3 \hat{x}^2 + \cr
   & &   + a^2 q^{12} t^4 \hat{x}^2 + a^2 q^{14} t^4 \hat{x}^2 - a^2 q^{16} t^4 \hat{x}^3 + a^2 q^{16} t^5 \hat{x}^3 + a^4 q^{16} t^6 \hat{x}^4)  \cr
a_2 & = & -(1 + a^2 q^2  t^3 \hat{x}) (1 + a^2 q^2 t^3 \hat{x}^2) ~. \nonumber
\eea
On the other hand, from the asymptotics of (\ref{T22-HOMFLYprime}) we obtain the following ``diagonal'' twisted superpotential:
\bea\nonumber
\widetilde{\mathcal{W}}^\fin (T_{2,2};x,z,a,t) & = &2 \log x \log a+\log x \log z + 2\log t \log z + \textrm{Li}_2(-a^2 t) - \textrm{Li}_2(-a^2 t z)   \cr
& & + \textrm{Li}_2(-t) - \textrm{Li}_2(-t xz^{-1}) - \textrm{Li}_2(x) + \textrm{Li}_2(z) + \textrm{Li}_2(x z^{-1}) ~.
\eea
This twisted superpotential gives rise to the following saddle equations:
\bea\nonumber
y & = & e^{x\partial_x \widetilde{\mathcal{W}}^\fin(T_{2,2})} = a^{2} z \frac{(x-1)(t x + z)}{x - z} \cr
1 & = & e^{z \partial_{z} \widetilde{\mathcal{W}}^\fin(T_{2,2})}  = t^2 x \frac{(x-z)(1+a^2 t z)}{(z - 1)(tx + z)}  ~. \nonumber
\eea
Eliminating $z_1$ from the above two equations we get the following diagonal super-$A$-polynomial
\bea
&&A^{\textrm{super}}(T_{2,2};x,y;a,t) = \cr
&=&(1 + a^2 t^3 x) y^2 \cr
 & &-a^2 (1 + t x - t^2 x + 2 t^2 x^2 + 2 a^2 t^3 x^2 + 2 a^2 t^4 x^2 -
   a^2 t^4 x^3 + a^2 t^5 x^3 + a^4 t^6 x^4) y\cr
 & & -a^4 t^3 (1 - x) x^2 (1 + t x) (1- a^2 t^2 x)   ~. \label{diag-superA-hopf}
\eea
The corresponding matrix of coefficients is shown in Figure \ref{fig-T22-matrix} and the Newton polygon is shown in Figure \ref{fig-T22-Newton}. This super-$A$-polynomial agrees with $q\to 1$ limit of the difference equation (\ref{T22-difference}).

The quantizability conditions in this case are satisfied as long as $a$ and $t$ are roots of unity, similarly as was the case for super-$A$-polynomials for knots. This condition follows as long as all the face polynomials factorize into binomials, which is indeed the case. For example, the face polynomial $f(z)$ corresponding to the bottom horizontal face in Figure \ref{fig-T22-Newton} factorizes as
$$
f(z) = a^4 t^3 (1-a^2 t^2 z) (z-1) (1+t z)~.
$$


\subsection{$(2,4)$ torus link $T_{2,4}$}

\begin{figure}[htb]
\begin{center}
\includegraphics[width=0.5\textwidth]{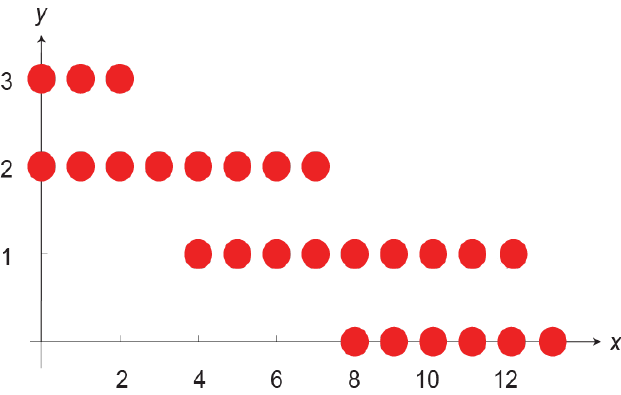}
\caption{\emph{Newton polygon of the ``diagonal'' super-$A$-polynomial for the $(2,4)$ torus link.}   } \label{fig-T24-Newton}
\end{center}
\end{figure}

For more complicated links we determine diagonal super-$A$-polynomials just from the asymptotics of corresponding diagonal Poincar\'e polynomials -- interested reader may check that the results agree with $q\to 1$ limit of difference equations for those Poincar\'e polynomials. The Poincar\'e polynomial of the finite-dimensional HOMFLY homology for $T_{2,4}$ link is given in (\ref{T24-Prime}) and its asymptotics encodes the following twisted superpotential:
\bea\nonumber
&&\widetilde{\mathcal{W}}^\fin (T_{2,4};x,z_1,z_2,a,t) = \cr
& = &8 \log x \log a + 2\log (tx )\log (z_1 z_2) - \log( x z_2)\log z_1 - \textrm{Li}_2(x) + \textrm{Li}_2(x z_1^{-1}) +  \cr
& &  - \textrm{Li}_2(t x z_2^{-1}) + \textrm{Li}_2(-a^2 t ) - \textrm{Li}_2(-a^2 t z_1) + \textrm{Li}_2(-t) + \textrm{Li}_2(z_1 z_2^{-1})
+ \textrm{Li}_2(z_2)~.
\eea
This twisted superpotential leads to the saddle equations:
\bea\nonumber
y & = & e^{x\partial_x \widetilde{\mathcal{W}}^\fin(T_{2,4})} = a^{8} z_1^2 z_2 \frac{(x-1)(t x + z_2)}{x - z_1} \cr
1 & = & e^{z_1 \partial_{z_1} \widetilde{\mathcal{W}}^\fin(T_{2,4})} = t^2 x \frac{(x-z_1)(1+a^2 t z_1)}{z_1 (z_1 - z_2)}   \cr
1 & = & e^{z_2 \partial_{z_2} \widetilde{\mathcal{W}}^\fin(T_{2,4})} = t^2 x^2 \frac{z_1-z_2}{z_1(z_2 - 1)(tx + z_2)}   ~.\nonumber
\eea
Eliminating $z_1$ and $z_2$ from the above two equations we get the following super-$A$-polynomial
\bea
& & A^{\textrm{super}}(T_{2,4};x,y;a,t) = \cr
&=& (1 + a^2 t^3 x)^2y^3\nonumber\\[.1cm]
&&-a^8 (1 + a^2 t^3 x) (1 + t x - t^2 x + 2 t^2 x^2 - t^3 x^2 +
   2 a^2 t^3 x^2 + 2 t^3 x^3 - 2 t^4 x^3 + 2 a^2 t^4 x^3 \cr
   && \quad-
   2 a^2 t^5 x^3  + 3 t^4 x^4 + 4 a^2 t^5 x^4 + 2 a^2 t^6 x^4 +
   a^4 t^6 x^4 - a^2 t^6 x^5 + a^2 t^7 x^5 + a^4 t^7 x^5 \cr
   && \quad-
   a^4 t^8 x^5 + 2 a^4 t^8 x^6)y^2\nonumber\\[.1cm]
   &&+a^{16} t^5 (-1 + x) x^4 (1 + t x) (2 + t x - t^2 x - a^2 t^2 x +
   a^2 t^3 x + 3 t^2 x^2 + 4 a^2 t^3 x^2 + 2 a^2 t^4 x^2 \cr
   && \quad +
   a^4 t^4 x^2 - 2 a^2 t^4 x^3 + 2 a^2 t^5 x^3 - 2 a^4 t^5 x^3 +
   2 a^4 t^6 x^3 + 2 a^4 t^6 x^4 - a^4 t^7 x^4 + 2 a^6 t^7 x^4\cr
   && \quad-
   a^6 t^8 x^5   + a^6 t^9 x^5 + a^8 t^{10} x^6)y\nonumber\\[.1cm]
   &&+a^{24} t^{10} (-1 + x)^2 x^8 (1 + t x)^2 (-1 + a^2 t^2 x)~.   \label{Asuper-diagT24}
\eea
The corresponding Newton polygon is shown in Figure \ref{fig-T24-Newton}. One can easily check that quantizability conditions in this case are satisfied. For example, face polynomial corresponding to the top and bottom faces in figure \ref{fig-T24-Newton} are given respectively (as polynomials in $x$) by the first and last lines in formula (\ref{Asuper-diagT24}) --  these expressions nicely factorize and, as usual, impose that $a$ and $t$ should be roots of unity.

\subsection{$\wt T_{2,4}$ link}

\begin{figure}[H]
  \begin{minipage}[b]{7cm}\centering
\includegraphics[scale=.75]{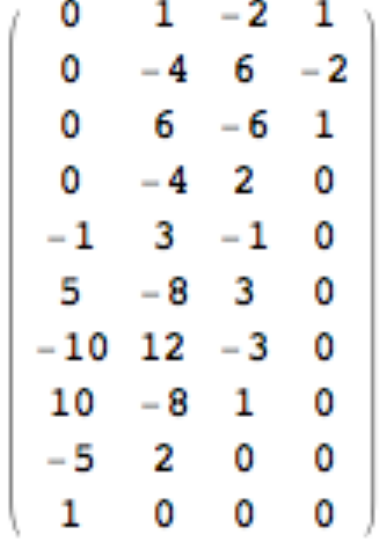}
\caption{Matrix representation of the $a=-t=1$ limit of the ``diagonal'' super-$A$-polynomial for the link $\wt T_{2,4}$.}\label{fig-4a1-matrix}
\end{minipage}
\hspace{.5cm}
  \begin{minipage}[b]{7.5cm}\centering
\includegraphics[scale=1.1]{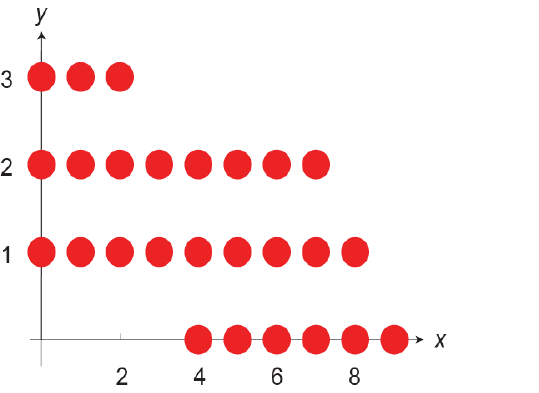}
\caption{Newton polygon of the ``diagonal'' super-$A$-polynomial for the link $\wt T_{2,4}$.}\label{fig-4a1-Newton}
\end{minipage}
\end{figure}

For $(2,4)$ torus link with opposite orientation of its components, denoted $\wt T_{2,4}$, the Poincar\'e polynomial of the finite-dimensional HOMFLY homology is given in (\ref{4a1-HOMFLYprime}). This gives rise to the twisted superpotential:
\bea\nonumber
&&\widetilde{\mathcal{W}}^\fin(\wt T_{2,4};x,z_1,z_2,z_3,a,t) = \cr
 & = &i\pi \log (z_1z_2z_3)  +2 \log a \log\Big(\frac{xz_3^2}{z_1}\Big)   + \log t \log\Big(\frac{xz_3^4}{z_1^2}\Big)  +\log z_2 (\tfrac12\log z_2-\log z_1)+ \tfrac{5}{2}(\log z_3)^2  \cr
 &&+\tfrac12 (\log z_1-\log x)^2  - \textrm{Li}_2(x) + \textrm{Li}_2(x z_1^{-1}) + \textrm{Li}_2(z_1 z_2^{-1}) + \textrm{Li}_2(z_2 z_3^{-1}) + \textrm{Li}_2(z_3)  \cr
& & + \textrm{Li}_2(-a^2t) - \textrm{Li}_2(-a^2 tz_2) + \textrm{Li}_2(-a^2 t^3 x) - \textrm{Li}_2(-a^2 t^3 x z_2)+\textrm{Li}_2(a^2 t^2 z_3) - \textrm{Li}_2(a^2 t^2 z_2z_3)~,\nonumber\eea
where
\be\nonumber
x = q^{2r},\quad z_1 = q^{2k}, \quad z_2 = q^{2j},\quad z_3 = q^{2i}~.
\ee
From this we obtain a system that determines the saddle solution:
\bea
y & = & e^{x\partial_x \widetilde{\mathcal{W}}^\fin(\wt T_{2,4})} = a^2 t x \frac{(1-x)(1 + a^2 t^3 x z_2)}{(z_1-x)(1 + a^2 r^3 x)} \cr
1 & = & e^{z_1 \partial_{z_1} \widetilde{\mathcal{W}}^\fin(\wt T_{2,4})} = \frac{z_1 - x}{a^2 t^2 x(z_1 - z_2)}  \cr
1 & = & e^{z_2 \partial_{z_2} \widetilde{\mathcal{W}}^\fin(\wt T_{2,4})} = \frac{z_3(z_2 - z_1)(1 + a^2 t z_2)(1 + a^2 t^3 x z_2)}{z_1(z_2 - z_3)}  \cr
1 & = & e^{z_3 \partial_{z_3} \widetilde{\mathcal{W}}^\fin(\wt T_{2,4})} = a^2 t^2 z_3 \frac{z_3 - z_2}{1 - z_3}~.  \nonumber
\eea
Eliminating $z_1,\, z_2$ and $z_3$ gives the diagonal super-$A$-polynomial
\begin{small}
\bea
&&A^{\textrm{super}} (\wt T_{2,4};x,y,a,t) = \cr
& = & (1 + a^2 t^3 x)^2y^3\nonumber\\[.1cm]
&&-a^2 (1 + a^2 t^3 x) (2 - x + 2 t x - t^2 x + 2 t^2 x^2 +
   a^2 t^2 x^2 + 4 a^2 t^3 x^2 + 3 a^2 t^4 x^2 - 2 a^2 t^3 x^3 +
   2 a^2 t^5 x^3\cr
   && - a^2 t^5 x^4 + 2 a^4 t^5 x^4 + 2 a^4 t^6 x^4 -
   a^4 t^6 x^5 + a^4 t^7 x^5 + a^6 t^8 x^6)y^2\nonumber\\[.1cm]
&&-a^4 (-1 + x) (1 + t x) (1 + t x - t^2 x - t^3 x^2 + 2 a^2 t^3 x^2 +
   2 a^2 t^4 x^2 + 2 a^2 t^4 x^3 - 2 a^2 t^6 x^3 + 2 a^2 t^6 x^4 \cr
   &&+
   a^4 t^6 x^4 + 4 a^4 t^7 x^4 + 3 a^4 t^8 x^4 + a^4 t^7 x^5 -
   2 a^4 t^8 x^5 + a^4 t^9 x^5 + 2 a^6 t^{10} x^6)y\nonumber\\[.1cm]
   &&-a^8 t^7 (-1 + x)^2 x^4 (1 + t x)^2 (-1 + a^2 t^2 x)~.    \label{Asuper-diag-wtT24}
\eea
\end{small}
The corresponding matrix representation in the limit $a=-t=1$ is shown in Figure \ref{fig-4a1-matrix}, and its Newton polygon  is shown in Figure \ref{fig-4a1-Newton}.

In this case the quantizability conditions are also satisfied for $a$ and $t$ roots of unity. This follows from the fact that all face polynomials factorize into binomials. For example, the face polynomials for the top and bottom horizontal faces in Figure \ref{fig-4a1-Newton} are given respectively (as polynomials in $x$) by the first and last lines in (\ref{Asuper-diag-wtT24}). As usual they nicely factorize into binomials.


\subsection{Whitehead link}

\begin{figure}[h]
  \begin{minipage}[b]{7cm}\centering
\includegraphics[scale=.65]{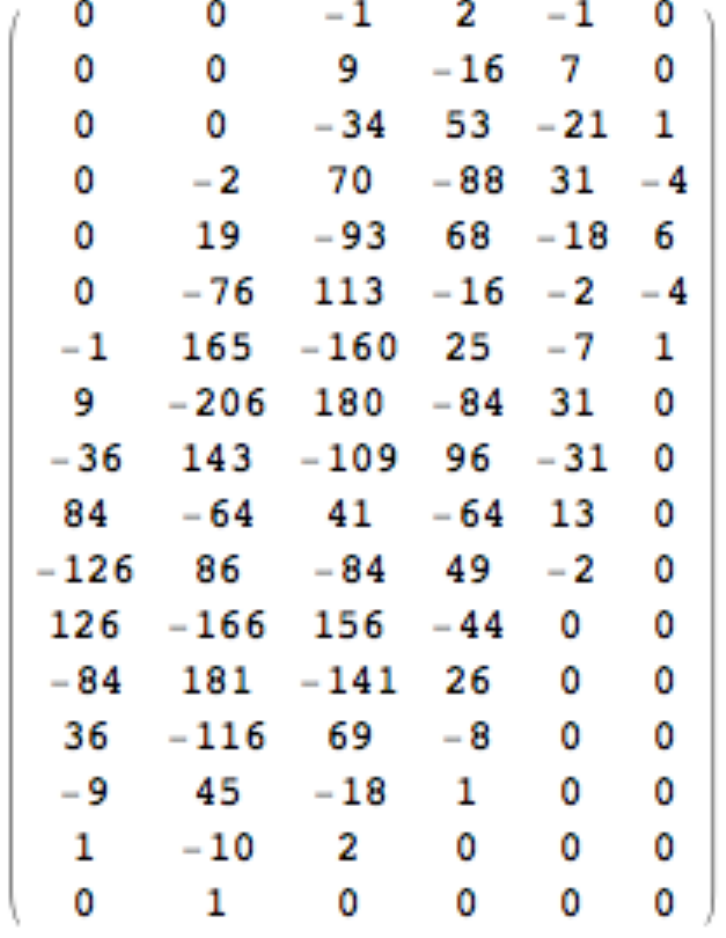}
\caption{Matrix representation of the $a=-t=1$ limit of the ``diagonal'' super-$A$-polynomial for the Whitehead link.}\label{fig-Whitehead-matrix}
\end{minipage}
\hspace{.5cm}
  \begin{minipage}[b]{7.5cm}\centering
\includegraphics[scale=1.1]{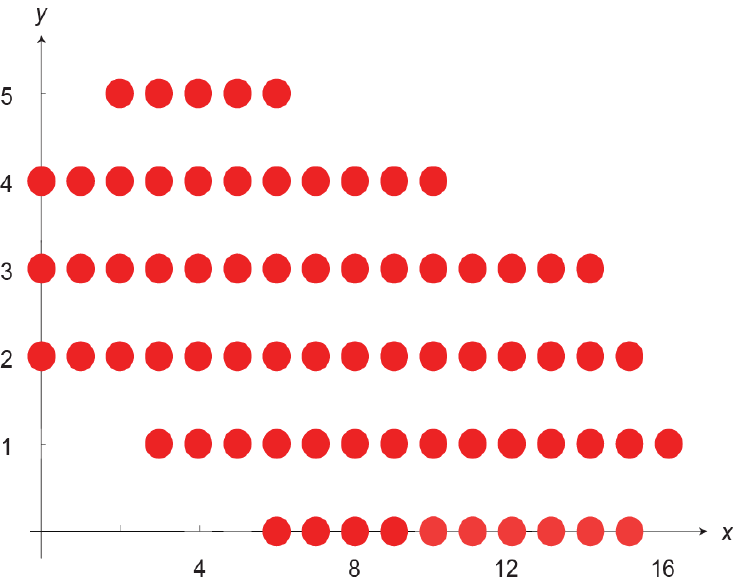}
\caption{Newton polygon of the ``diagonal'' super-$A$-polynomial for the Whitehead link.}\label{fig-Whitehead-Newton}
\end{minipage}
\end{figure}

The Poincar\'e polynomial of the finite-dimensional HOMFLY homology for the Whitehead link is given in (\ref{Whitehead-HOMFLYprime}). This gives rise to the twisted superpotential:
\bea
\widetilde{\mathcal{W}}^\fin (\WL;x,a,t)& = &(\log z)^2 - \log(-t)\log (xz) - 2\log x \log z +    \textrm{Li}_2(-a^2t)-\textrm{Li}_2(-a^2tz) \cr
&&+\textrm{Li}_2(-a^2t^3x)-\textrm{Li}_2(-a^2t^3xz)+ \textrm{Li}_2(-t)- \textrm{Li}_2(-tz)+ \textrm{Li}_2(-txz^{-1})\cr
&&+\textrm{Li}_2(-tx)+\textrm{Li}_2(a^2t^2z^2)-\textrm{Li}_2(a^2t^2xz)-\textrm{Li}_2(x)+\textrm{Li}_2(z)+\textrm{Li}_2(xz^{-1})~,\nonumber\eea
where
\be\nonumber
x = q^{2r},\quad z = q^{2i}.
\ee
This twisted superpotential gives rise to the following saddle equations:
\bea\nonumber
y & = & e^{x\partial_x \widetilde{\mathcal{W}}^\fin(\WL)} =  \frac{(1-x)(1 + t x)(1 - a^2 t^2 x z)(1 + a^2 t^3 x z)}{t (x-z)(t x + z)(1 + a^2 t^3 x)} \cr
1 & = & e^{z \partial_{z} \widetilde{\mathcal{W}}^\fin(\WL)} = \frac{(x-z)(t x + z)(1 + t z)(1+a^2 t z)(1 - a^2 t^2 x z)(1 + a^2 t^3 x z)}{t x^2(1-z)(1-a^2t^2z^2)^2} ~.\nonumber
\eea
After eliminating $z$, one can write the super-$A$-polynomial of the Whitehead link
\begin{footnotesize}
\bea
&&A^{\textrm{super}}(\WL;x,y;a,t) = \cr
&=&t^2 x^2 (1 + a^2 t^3 x)^4y^5\nonumber\\[.1cm]
&&+(1 + a^2 t^3 x)^3 (-1 - t x - a^2 t x + t^2 x + a^2 t^2 x -
   2 t^2 x^2 - 3 a^2 t^2 x^2 - a^2 t^4 x^2 - 4 a^2 t^3 x^3 +
   2 a^2 t^4 x^3 - 2 a^4 t^4 x^3 \cr
   &&\quad+ 2 a^2 t^5 x^3 + 2 a^4 t^5 x^3 -
   4 a^2 t^4 x^4 - 4 a^2 t^5 x^4 - 6 a^4 t^5 x^4 - 4 a^4 t^6 x^4 -
   2 a^4 t^7 x^4 - a^4 t^7 x^5 - a^6 t^7 x^5 \cr
   &&\quad+ a^4 t^8 x^5 +
   a^6 t^8 x^5 + a^4 t^7 x^6 - 2 a^4 t^8 x^6 - 3 a^6 t^8 x^6 -
   a^6 t^{10} x^6 - a^6 t^9 x^7 + a^6 t^{10} x^7)y^4\nonumber\\[.1cm]
   &&-t (-1 + x) (1 + t x) (1 + a^2 t^3 x)^2 (-1 - a^4 - 2 a^2 t x -
   2 a^4 t x + 2 a^2 t^2 x + 2 a^4 t^2 x - 4 a^2 t^2 x^2 -
   6 a^4 t^2 x^2 \cr
   &&\quad- 4 a^2 t^3 x^2 + 5 a^4 t^3 x^2 - 4 a^6 t^3 x^2 -
   2 a^4 t^4 x^2 - 6 a^4 t^3 x^3 + 2 a^4 t^4 x^3 - 4 a^6 t^4 x^3 +
   4 a^4 t^5 x^3 + 4 a^6 t^5 x^3 - 6 a^4 t^4 x^4\cr
   &&\quad - 6 a^4 t^5 x^4 -
   9 a^6 t^5 x^4 - 6 a^4 t^6 x^4 + 4 a^6 t^6 x^4 - 6 a^8 t^6 x^4 -
   3 a^6 t^7 x^4 - 3 a^6 t^7 x^5 - 3 a^8 t^7 x^5 + 3 a^6 t^8 x^5 +
   3 a^8 t^8 x^5\cr
   &&\quad + 4 a^6 t^7 x^6 - 4 a^6 t^9 x^6 + 12 a^8 t^9 x^6 -
   4 a^{10} t^9 x^6 - 4 a^8 t^9 x^7 + 2 a^8 t^{10} x^7 -
   2 a^{10} t^{10} x^7 + 2 a^8 t^{11} x^7 + 2 a^{10} t^{11} x^7\cr
   &&\quad +
   2 a^8 t^{11} x^8 + 3 a^{10} t^{11} x^8 - a^8 t^{12} x^8 + 4 a^{10} t^{12} x^8 -
    a^{12} t^{12} x^8 + a^{10} t^{13} x^8 - a^{10} t^{13} x^9 - a^{12} t^{13} x^9 +
   a^{10} t^{14} x^9 \cr
   &&\quad+ a^{12} t^{14} x^9 + a^{12} t^{15} x^{10})y^3\nonumber\\[.1cm]
   &&+a^4 t^2 (-1 + x)^2 (1 + t x)^2 (1 + a^2 t^3 x) (-1 - t x - a^2 t x +
   t^2 x + a^2 t^2 x - 2 t^2 x^2 - 3 a^2 t^2 x^2 + t^3 x^2 -
   4 a^2 t^3 x^2 \cr
   &&\quad+ a^4 t^3 x^2 - a^2 t^4 x^2 - 4 a^2 t^3 x^3 +
   2 a^2 t^4 x^3 - 2 a^4 t^4 x^3 + 2 a^2 t^5 x^3 + 2 a^4 t^5 x^3 -
   4 a^2 t^4 x^4 + 4 a^2 t^6 x^4 - 12 a^4 t^6 x^4 \cr
   &&\quad+ 4 a^6 t^6 x^4 -
   3 a^4 t^7 x^5 - 3 a^6 t^7 x^5 + 3 a^4 t^8 x^5 + 3 a^6 t^8 x^5 +
   6 a^4 t^7 x^6 + 6 a^4 t^8 x^6 + 9 a^6 t^8 x^6 + 6 a^4 t^9 x^6 -
   4 a^6 t^9 x^6 \cr
   &&\quad+ 6 a^8 t^9 x^6 + 3 a^6 t^{10} x^6 - 6 a^6 t^9 x^7 +
   2 a^6 t^{10} x^7 - 4 a^8 t^{10} x^7 + 4 a^6 t^{11} x^7 +
   4 a^8 t^{11} x^7 + 4 a^6 t^{11} x^8 + 6 a^8 t^{11} x^8\cr
   &&\quad +
   4 a^6 t^{12} x^8 - 5 a^8 t^{12} x^8 + 4 a^{10} t^{12} x^8 +
   2 a^8 t^{13} x^8 - 2 a^8 t^{13} x^9 - 2 a^{10} t^{13} x^9 +
   2 a^8 t^{14} x^9 + 2 a^{10} t^{14} x^9 \cr
   &&\quad+ a^8 t^{15} x^{10} + a^{12} t^{15} x^{10})y^2\nonumber\\[.1cm]
   &&-a^8 t^6 (-1 + x)^3 x^3 (1 + t x)^3 (-1 + t - t x + 2 t^2 x +
   3 a^2 t^2 x + a^2 t^4 x - a^2 t^4 x^2 - a^4 t^4 x^2 + a^2 t^5 x^2 +
    a^4 t^5 x^2 \cr
   &&\quad+ 4 a^2 t^4 x^3 + 4 a^2 t^5 x^3 + 6 a^4 t^5 x^3 +
   4 a^4 t^6 x^3 + 2 a^4 t^7 x^3 - 4 a^4 t^6 x^4 + 2 a^4 t^7 x^4 -
   2 a^6 t^7 x^4 + 2 a^4 t^8 x^4 \cr
   &&\quad+ 2 a^6 t^8 x^4 + 2 a^4 t^8 x^5 +
   3 a^6 t^8 x^5 + a^6 t^{10} x^5 - a^6 t^{10} x^6 - a^8 t^{10} x^6 +
   a^6 t^{11} x^6 + a^8 t^{11} x^6 + a^8 t^{12} x^7)y\nonumber\\[.1cm]
   &&-a^{12} t^{11} (-1 + x)^4 x^6 (1 + t x)^4 (-1 + a^2 t^2 x)~.  \label{Asuper-diag-WL}
\eea
\end{footnotesize}
Matrix representation in $a=-t=1$ limit of this super-$A$-polynomial (disregarding $(x-1)(1+t x)$ factors) is shown in Figure \ref{fig-Whitehead-matrix}, and the corresponding Newton polygon is shown in Figure \ref{fig-Whitehead-Newton}.

In this case the face polynomials again factorize into products of binomials, so that one concludes that quantizability conditions are met as long as $a$ and $t$ are roots of unity. For example, the face polynomials corresponding respectively to the top bottom horizontal faces in the Newton polygon in Figure \ref{fig-Whitehead-Newton} are given respectively by the first and the last line in expression (\ref{Asuper-diag-WL}).


\subsection{Borromean rings}

\begin{figure}[h]
\begin{center}
\includegraphics[width=0.7\textwidth]{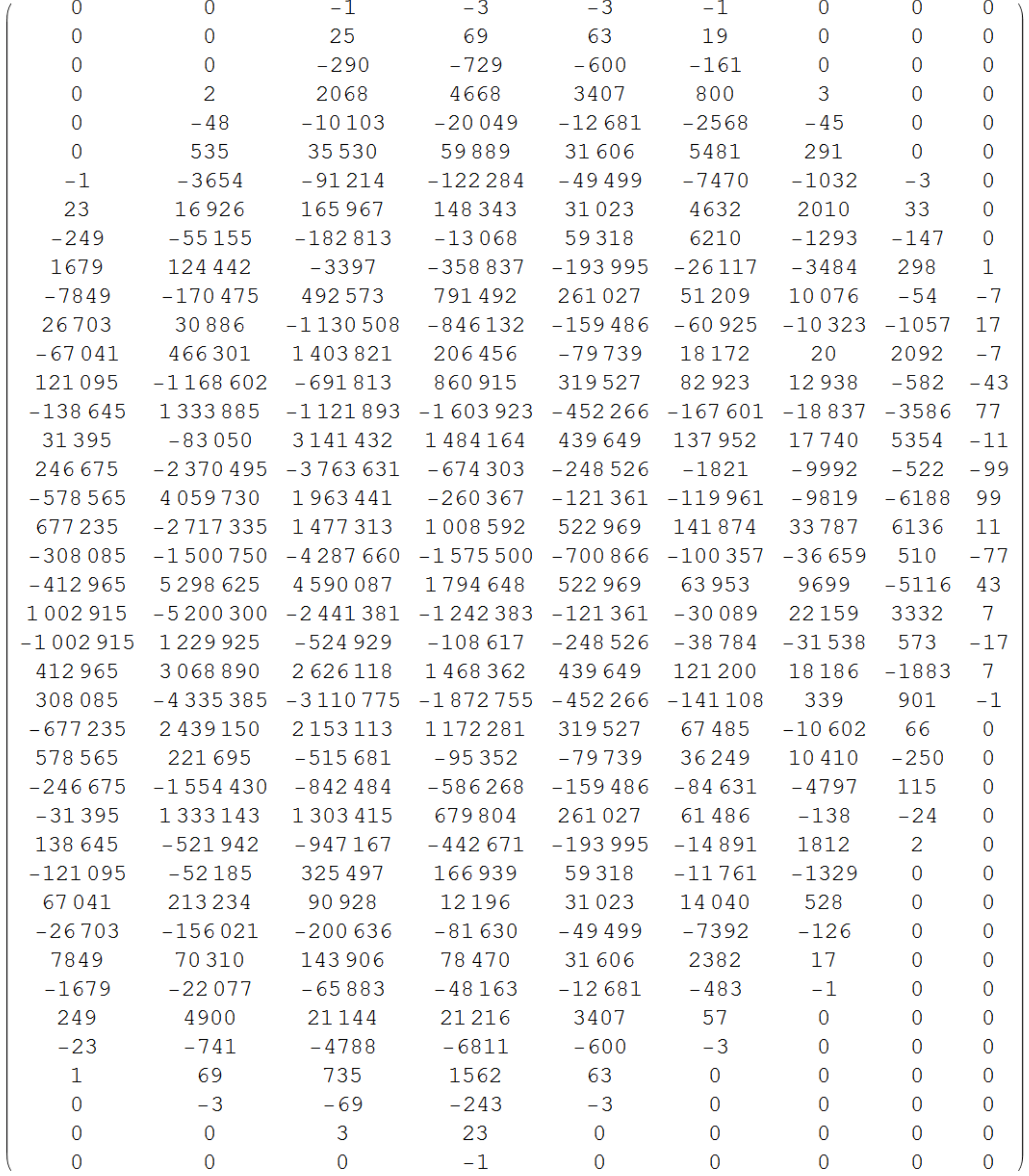}
\caption{{Matrix representation of the $a=-t=1$ limit of the ``diagonal'' super-$A$-polynomial for the Borromean rings.}   } \label{fig-BR-matrix}
\end{center}
\end{figure}

\begin{figure}[h]
\begin{center}
\includegraphics[width=0.8\textwidth]{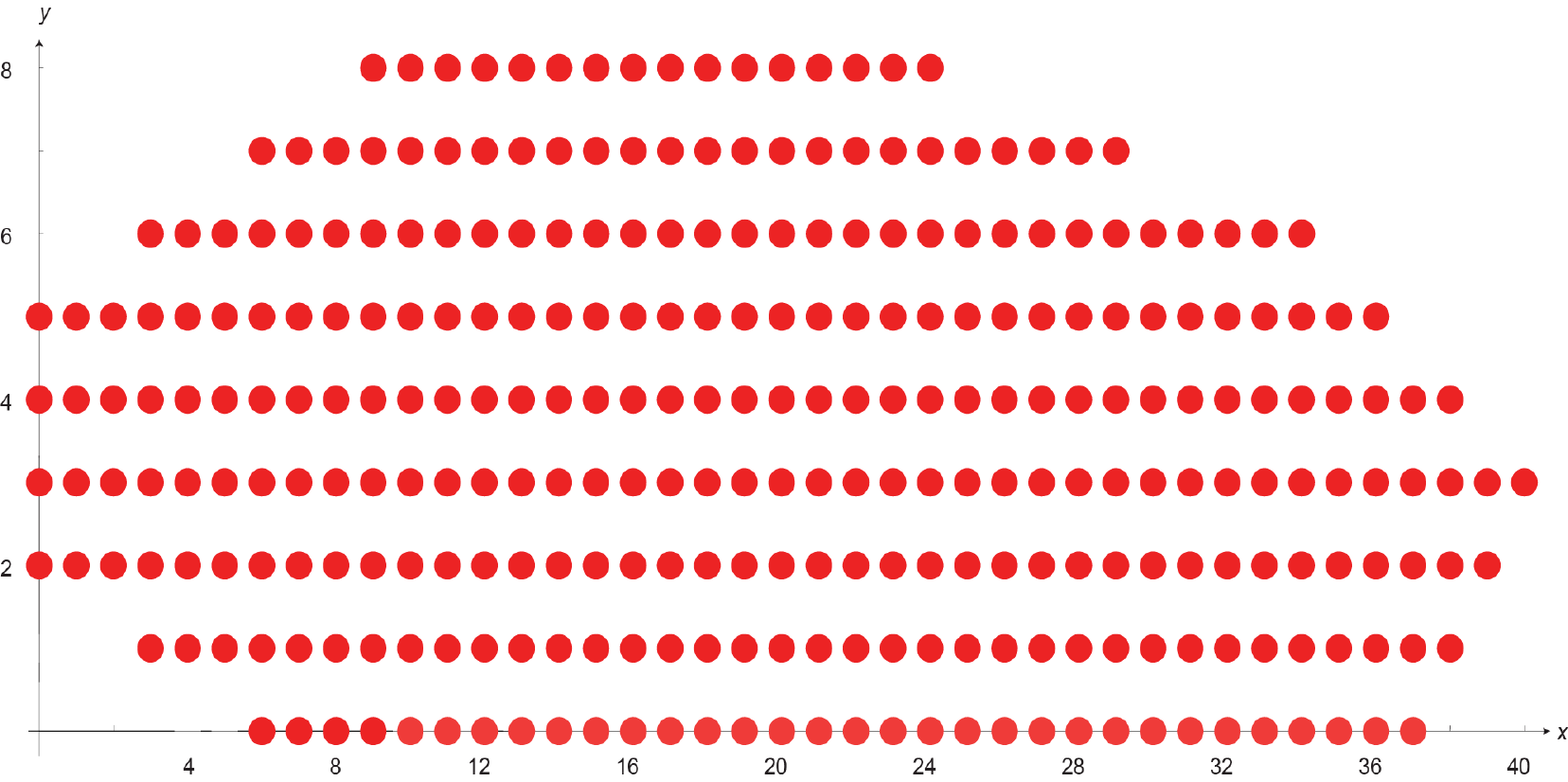}
\caption{{Newton polygon of the ``diagonal'' super-$A$-polynomial for the Borromean rings.}   } \label{fig-BR-Newton}
\end{center}
\end{figure}

The Poincar\'e polynomial of the finite-dimensional HOMFLY homology for the Borromean rings follows from specialization of (\ref{BR-HOMFLYprime}) to $(a,q,t)$ variables, and it gives rise to the twisted superpotential:
\bea\nonumber
\widetilde{\mathcal{W}}^\fin (\BR;x,a,t)&=& -2\log x \log(-t) +i\pi \log z-2 \log z \log a -\log x\log z + \textrm{Li}_2(-a^2t)  \cr
&&   -  \textrm{Li}_2(-a^2tz)  + \textrm{Li}_2(-a^2 t^3 x) - \textrm{Li}_2(-a^2 t^3 x z)  - \textrm{Li}_2(x) + \textrm{Li}_2(z) + \textrm{Li}_2(xz^{-1})  \cr
&&+ 2\Big[  \log z( \log a-\log t ) -\log z\log x + \frac{3}{4}(\log z)^2  + \textrm{Li}_2(- t) - \textrm{Li}_2(-t z)  \cr
&& + \textrm{Li}_2(- t x z^{-1}) - \textrm{Li}_2(- t x) + \textrm{Li}_2(a^2 t^2 z^2)   - \textrm{Li}_2(a^2 t^2 x z) \Big]  ~,
\eea
where
\be\nonumber
x = q^{2r},\quad z = q^{2i}~.
\ee
This twisted superpotential gives rise to the following saddle equations:
\bea\nonumber
y & = & e^{x\partial_x \widetilde{\mathcal{W}}^\fin(\BR)} =  \frac{(1-x)(1 + t x)^2(1 - a^2 t^2 x z)^2(1 + a^2 t^3 x z)}{ t^2 (x-z)(t x + z)^2(1 + a^2 t^3 x)} ~,\cr
1 & = & e^{z \partial_{z} \widetilde{\mathcal{W}}^\fin(\BR)} = \frac{(x-z)(t x + z)^2(1 + t z)^2(1+a^2 t z)(1 - a^2 t^2 x z)^2(1 + a^2 t^3 x z)}{t^2 x^3(1-z)(1-a^2t^2z^2)^4} ~.\nonumber
\eea
After eliminating $z$, the resulting super-$A$-polynomial is too long to type here -- it takes form
$$
A^{\textrm{super}} = (x-1)(1 + t x)(1 + a t^3 x)(1 - t^4 x^2 a) (1 - t^2 x^2 a) A^{\textrm{super}}_{\textrm{non-factorizable}}~,
$$
with $A^{\textrm{super}}_{\textrm{non-factorizable}}$ being a non-factorizable polynomial, which consists of 16249 monomial terms (with respect to $x,y,a$ and $t$)! Matrix representation in $a=-t=1$ limit of this $A^{\textrm{super}}_{\textrm{non-factorizable}}$ is shown in Figure \ref{fig-BR-matrix} and the corresponding Newton polygon is shown in Figure \ref{fig-BR-Newton}.

For Borromean rings, the face polynomials again factorize into products of binomials and quantizability conditions are met as long as $a$ and $t$ are roots of unity. For example, the face polynomials $f_1(z)$ and $f_2(z)$, corresponding respectively to the bottom and top horizontal faces in the Newton polygon in Figure \ref{fig-BR-Newton}, factorize as:
\bea
f_1(z) & = & a^{12} t^{13} (1 - z)^7 (1 + t z)^{14} (1 - a^2 t^2 z)^2 (1 - a^2 t^2 z^2)^4~, \cr
f_2(z) & = & a^{10} t^{12} (1 + a^2 t^3 z)^7 (1 - a^2 t^4 z^2)^4~.    \nonumber
\eea

\section{Explicit computations of associated varieties}
\label{sec:app-varieties}

In this appendix we analyze asymptotic expansions of various link invariants and determine their higher-dimensional associated varieties.

\subsection{$(2,4)$ torus link}

The Poincar\'e polynomial of the $([r_1],[r_2])$-colored homology of the $(2,4)$ torus link is given in \eqref{unreduced-T24+diff}.
To determine its large color asymptotics we define $\tilde{q}^2=q^2 t^2$, and rearrange this formula as
\bea
&&\overline\scP^{(+)}_{[r_1],[r_2]}({T_{2,4}};a,\wt{q},t) = \cr
&=&a^{(r_1+r_2)}\wt{q}^{-(r_1+r_2)}t^{(r_1+r_2)} \frac{(-a^2 t;\wt{q}^2)_{r_1}}{(\tilde{q}^2 t^{-2};\wt{q}^2)_{r_1}}\cr
&& \sum_{r_2\ge j\ge i\ge0} \wt{q}^{ 2(r_1 j  +  i (r_1 + r_2)+ (i + j-ij))} t^{-2(i+j)}\frac{(-a^2 \wt{q}^{-2}t^3;\wt{q}^2)_{j}}{(\wt{q}^{2(r_2-i+1)}t^{-2};\wt{q}^2)_{i}}{r_2\brack j }_{\wt{q}^2}{j \brack i}_{\wt{q}^2}~.
\eea
Taking the limit $\wt{q}\to1$ and keeping $\wt{q}^{2r_1}=x_1$, $\wt{q}^{2r_2}=x_2$, $\wt{q}^{2i}=z_1$ and $\wt{q}^{2j}=z_2$ fixed, we obtain the corresponding twisted superpotential
\bea
\widetilde{\mathcal{W}}(T_{2,4};x_1,x_2,z_1,z_2;a,t )& = &(\log x_1+\log x_2)(\log a+ \log t)+ \log x_1\log z_2     \label{Wtilde-at-T24} \\
&&+(\log x_1+\log x_2)\log z_1-\log z_1\log z_2-2(\log z_1+\log z_2)\log t\cr
&&-\Li_2(x_2)+\Li_2(x_2z_2^{-1})+\Li_2(z_2z_1^{-1})+\Li_2(z_1)\cr
&&+\Li_2(-a^2 t^3)-\Li_2(-a^2 t^3 z_2)-\Li_2(x_2z_1^{-1}t^{-2})+\Li_2(x_2t^{-2})\cr
&&+\Li_2(-a^2 t)-\Li_2(-a^2 t x_1)-\Li_2(t^{-2})+\Li_2(t^{-2}x_1)~. \nonumber
\eea
Eliminating $z_1$ and $z_2$ from the Neumann-Zagier equations \eqref{yifi} we determine the classical associated variety for the $(2,4)$ torus link
\bea\nonumber
&&A_1(T_{2,4};x_1,x_2,y_1;a,t) = \cr
&=&t \left(x_1-t^2\right)^3 y_1^3\cr
&&+a \left(x_1-t^2\right)^2 \left(t^4-t^2 x_1-x_1 x_2+t^2 x_1 x_2+a^2 t^5 x_1 x_2-a^2 t^3 x_1^2 x_2+x_2^2+a^2 t^3 x_1 x_2^2\right)y_1^2\cr
&&-a^2 tx_2 \left(x_1-t^2\right) (x_1-1)  (-t^2 x_1+x_1^2+t^2 x_2-x_1 x_2+t^2 x_1 x_2+a^2 t^5 x_1 x_2\cr
&&\quad+a^2 t^3 x_1 x_2^2+a^4 t^6 x_1^2 x_2^2)y_1\cr
&&-a^{3} t^4 x_1 x_2^3(x_1-1)^2  (1+a^2 t x_1) = 0 ~,\nonumber\\[.1cm]
&&A_2(T_{2,4};x_1,x_2,y_2;a,t)\cr
&=&t \left(x_2-t^2\right)^3 y_2^3\cr
&&+a \left(x_2-t^2\right)^2 \left(t^4-t^2 x_2-x_1 x_2+t^2 x_1 x_2+a^2 t^5 x_1 x_2-a^2 t^3 x_2^2 x_1+x_1^2+a t^3 x_2x_1^2\right)y_2^2\cr
&&-a^2 tx_1 \left(x_2-t^2\right) (x_2-1)  (-t^2 x_2+x_2^2+t^2 x_1-x_1 x_2+t^2 x_1 x_2+a^2 t^5 x_1 x_2\cr
&&\quad+a^2 t^3 x_2 x_1^2+a^4 t^6 x_1^2 x_2^2)y_2\cr
&&-a^{3} t^4 x_2 x_1^3(x_2-1)^2  (1+a^2 t x_2) = 0~.\nonumber
\eea
In this example, these equations can be also obtained from the refined Chern-Simons invariant \eqref{refined-CS-sym+} of  the $(2,4)$ torus link.


\subsection{$(2,2p)$ torus links}

Here we show that the associated varieties of the $(2,2p)$ torus links are indeed Lagrangian. We found the refined Chern-Simons invariants  (\ref{refined-CS-sym+}) and the Poincar\'e polynomial of unreduced HOMFLY homology (\ref{torus-link-diff-rk+}) for the $(2,2p)$ torus links. As we expected from the proof presented in Appendix \ref{sec:RR-id}, we verified that (for several values of $p$)  that they lead to the same classical varieties.  While both of these representations have certain advantageous features, the refined Chern-Simons representation (\ref{refined-CS-sym+}) is more convenient to check the Lagrangian condition since it involves  only a single summation.  Writing the formula in terms of $\tilde{q}^2=q^2 t^2$, the twisted superpotential for every $p$ is expressed by  $x_1=\tilde q^{2r_1}$, $x_2= \tilde q^{2r_2}$, and $z=\tilde q^{2l}$:
\bea\label{twisted-rCS}
\widetilde{\cW}_{\textrm{rCS}}^{(p)} & = & (p-1)(\log x_1 x_2)(\log a t)+p\log x_1 \log x_2 -p\log\frac{z}{x_1} \log\frac{z}{x_2} - 2(p+1)\log z \log t   \cr
& & + \Li_2\Big(\frac{x_1}{z}\Big) + \Li_2\Big(\frac{x_2}{z}\Big)- \Li_2(x_1)- \Li_2(x_2) + \Li_2\big(-a^2 t^3\big) \Li_2\big(-a^2 t^3 z\big) \cr
& & + \Li_2\big(-a^2 t\big)  - \Li_2\big(-a^2 t x_1 x_2 z^{-1} \big) + \Li_2(z) - \Li_2(t^{-2}) - \Li_2\Big( \frac{x_1}{t^2 z} \Big) - \Li_2\Big( \frac{x_2}{t^2 z} \Big)  \cr
& & + \Li_2\big( \frac{x_1}{t^2} \big) + \Li_2\big( \frac{x_2}{t^2} \big) + \Li_2\Big( \frac{x_1 x_2}{t^2 z} \Big).
\eea
Therefore, the associated variety is encoded in a system of three equations
\bea\label{saddle-torus-link}
y_1 & = & e^{x_1 \partial_{x_1}\widetilde{\cW}_{rCS}^{(p)} }  = a^{p-1} t^{p+1} z^p \frac{(x_1 - 1)(x_1-t^2 z)(a^2 t x_1 x_2 + z)}{(x_1-t^2)(x_1-z)(t^2z-x_1 x_2)} \cr
y_2 & = & e^{x_2 \partial_{x_2}\widetilde{\cW}_{rCS}^{(p)} }  = a^{p-1} t^{p+1} z^p \frac{(x_2 - 1)(x_2-t^2 z)(a^2 t x_1 x_2 + z)}{(x_2-t^2)(x_2-z)(t^2z-x_1 x_2)}  \cr
1 & = &  e^{z \partial_{z}\widetilde{\cW}_{rCS}^{(p)} }  =  \frac{x_1^p x_2^p}{z^{2p} t^{2p}} \frac{(x_1 - z)(x_2-z)(x_1 x_2 - t^2 z)(1+a^2t^3 z)}{(z-1)(x_1-t^2z)(x_2-t^2z)(a^2tx_1 x_2 +z)} ~.
\eea
Imposing these relations on the holomorphic symplectic form $\omega=\sum_{i=1,2}d\log x_i\wedge d\log y_i$, we verify that it vanishes.


\subsection{Whitehead link}\label{sec:AV-WL}
For twist links with small values of $p$, one can explicitly determine the associated varieties by eliminating the variable $z$ and $w$ from the set of equations \eqref{saddle-twistlinks} . In particular, for the Whitehead link ($p=1$), the associated variety of the Whitehead link is expressed as the zero locus of the following equations:
\bea\label{WH-associated}
&&A_1(\WL;x_1,x_2,y_1;a)\cr
 &=& a^3 x_1^2 x_2\cr
&&+\left(a^2 x_1-a^2 x_1^2-a^2 x_2+2 a^2 x_1 x_2-a^4 x_1^2 x_2+a^4 x_1 x_2^2-a^4 x_1^2 x_2^2\right) y_1\cr
&&+\left(a-2 a x_1+a^3 x_1^2+a x_2-4 a^3 x_1 x_2+a^3 x_1^2 x_2+a^3 x_2^2-2 a^3 x_1 x_2^2+a^5 x_1^2 x_2^2\right) y_1^2\cr
&&+\left(-1+a^2 x_1-a^2 x_2+2 a^2 x_1 x_2-a^4 x_1^2 x_2-a^2 x_2^2+a^4 x_1 x_2^2\right) y_1^3\cr
&&+a x_2 y_1^4 =0\cr
&&A_2(\WL;x_1,x_2,y_2;a)\cr
&=& a^3 x_1 x_2^2\cr
&&+\left(-a^2 x_1+a^2 x_2+2 a^2 x_1 x_2+a^4 x_1^2 x_2-a^2 x_2^2-a^4 x_1 x_2^2-a^4 x_1^2 x_2^2\right) y_2\cr
&&+\left(a+a x_1+a^3 x_1^2-2 a x_2-4 a^3 x_1 x_2-2 a^3 x_1^2 x_2+a^3 x_2^2+a^3 x_1 x_2^2+a^5 x_1^2 x_2^2\right) y_2^2\cr
&&+\left(-1-a^2 x_1-a^2 x_1^2+a^2 x_2+2 a^2 x_1 x_2+a^4 x_1^2 x_2-a^4 x_1 x_2^2\right) y_2^3\cr
&&+a x_1 y_2^4 = 0,
\eea
where $A_2(\WL;x_1,x_2,y_2;a)=A_1(\WL;x_2,x_1,y_2;a)$. The associated variety determined by the above set of equations actually agrees with the augmentation variety of the Whitehead link obtained in \cite{Aganagic:2013jpa} with a suitable change of variables.
It is straightforward to verify that the variety determined by \eqref{WH-associated} is the Lagrangian subvariety in the Hitchin moduli space
\be\nonumber
\frac{\bC^*\times\bC^*}{\bZ_2^{(1)}}\times \frac{\bC^*\times\bC^*}{\bZ_2^{(2)}}
\ee
where the Weyl group symmetries are given by
\bea\nonumber
\bZ_2^{(1)}: ~ (x_1,y_1) \leftrightarrow \left(\frac{1}{a^2x_1}, \frac{1}{y_1}\right)~,\qquad
\bZ_2^{(2)}: ~ (x_2,y_2) \leftrightarrow \left(\frac{1}{a^2x_2}, \frac{1}{y_2}\right)~.
\eea
In fact, one can see that
\bea\nonumber
A_1(\WL;1/(a^2x_1),x_2,1/y_1;a)&=&a^{-2}x_1^{-2}y_1^{-4}A_1(\WL;x_1,x_2,y_1;a),\cr
A_2(\WL;1/(a^2x_1),x_2,y_2;a)&=&a^{-2}x_2^{-2}A_2(\WL;x_1,x_2,y_2;a).
\eea


\subsection{Borromean rings}\label{sec:AV-BR}
One can also determine the associated variety of the Borromean rings at the level of $a$-deformation using the cyclotomic expression of the colored HOMFLY invariant \eqref{cyclotomic-HOMFLY}. Taking large color asymptotics, we find the corresponding twisted superpotential
\be\nonumber
\widetilde{\mathcal{W}} (\BR;x_1,x_2,x_3,z,a)= \widetilde{\mathcal{W}}_F(x_1,z,a)+\widetilde{\mathcal{W}}_F(x_2,z,a)+\widetilde{\mathcal{W}}_F(x_3,z,a)+\widetilde{\mathcal{W}}_G(z,a)~,
\ee
where $\widetilde{\mathcal{W}}_F$ and $\widetilde{\mathcal{W}}_G$ are given in \eqref{twisted-sp-blocks}.
Then, the saddle point equations can be read off
\bea\nonumber
y_i & = & \frac{a^2 x_i z - 1}{a(x_i - z)}~, \qquad\qquad i=1,2,3  \cr
1& = &- \frac{(x_1 -z)(x_2-z)(x_3-z)(z-1)(a^2z-1)(a^2x_1 z -1 ) (a^2 x_2 z-1)(a^2 x_3 z - 1) } {x_1 x_2 x_3(1 - a^2z^2)^4 } ~.  \nonumber
\eea
Eliminating $z$ in the above equations, we find that the $a$-deformed associated variety is determined by equations:
\begin{footnotesize}
\bea
&&A_1(\BR;x_1,x_2,x_3,y_1;a)\cr
&=&a^3 x_2 (1-a^2x_1^2)^2 y_1^6\nonumber\\[.1cm]
&&+(a^2 x_1-a^2 x_1^2-a^4 x_1^2+a^4 x_1^3-a^2 x_2+a^2 x_1 x_2+a^4 x_1 x_2-2 a^4 x_1^2 x_2+a^4 x_1^3 x_2+a^6 x_1^3 x_2-a^6 x_1^4 x_2\cr
&&+a^4 x_1 x_2^2-a^4 x_1^2 x_2^2-a^6 x_1^2 x_2^2+a^6 x_1^3 x_2^2-a^2 x_3+a^2 x_1 x_3+a^4 x_1 x_3-2 a^4 x_1^2 x_3+a^4 x_1^3 x_3+a^6 x_1^3 x_3\cr
&&-a^6 x_1^4 x_3-a^2 x_2 x_3-a^4 x_2 x_3+4 a^4 x_1 x_2 x_3-2 a^4 x_1^2 x_2 x_3-2 a^6 x_1^2 x_2 x_3+4 a^6 x_1^3 x_2 x_3-a^6 x_1^4 x_2 x_3\cr
&&-a^8 x_1^4 x_2 x_3-a^4 x_2^2 x_3+a^4 x_1 x_2^2 x_3+a^6 x_1 x_2^2 x_3-2 a^6 x_1^2 x_2^2 x_3+a^6 x_1^3 x_2^2 x_3+a^8 x_1^3 x_2^2 x_3-a^8 x_1^4 x_2^2 x_3\cr
&&+a^4 x_1 x_3^2-a^4 x_1^2 x_3^2-a^6 x_1^2 x_3^2+a^6 x_1^3 x_3^2-a^4 x_2 x_3^2+a^4 x_1 x_2 x_3^2+a^6 x_1 x_2 x_3^2-2 a^6 x_1^2 x_2 x_3^2+a^6 x_1^3 x_2 x_3^2\cr
&&+a^8 x_1^3 x_2 x_3^2-a^8 x_1^4 x_2 x_3^2+a^6 x_1 x_2^2 x_3^2-a^6 x_1^2 x_2^2 x_3^2-a^8 x_1^2 x_2^2 x_3^2+a^8 x_1^3 x_2^2 x_3^2)  y_1^5\nonumber\\[.1cm]
&&+(a-2 a x_1-2 a^3 x_1+6 a^3 x_1^2-2 a^3 x_1^3-2 a^5 x_1^3+a^5 x_1^4+a x_2+a^3 x_2-8 a^3 x_1 x_2+6 a^3 x_1^2 x_2+6 a^5 x_1^2 x_2\cr
&&-8 a^5 x_1^3 x_2+a^5 x_1^4 x_2+a^7 x_1^4 x_2+a^3 x_2^2-2 a^3 x_1 x_2^2-2 a^5 x_1 x_2^2+6 a^5 x_1^2 x_2^2-2 a^5 x_1^3 x_2^2-2 a^7 x_1^3 x_2^2+a^7 x_1^4 x_2^2\cr
&&+a x_3+a^3 x_3-8 a^3 x_1 x_3+6 a^3 x_1^2 x_3+6 a^5 x_1^2 x_3-8 a^5 x_1^3 x_3+a^5 x_1^4 x_3+a^7 x_1^4 x_3+3 a^3 x_2 x_3-8 a^3 x_1 x_2 x_3\cr
&&-8 a^5 x_1 x_2 x_3+26 a^5 x_1^2 x_2 x_3-8 a^5 x_1^3 x_2 x_3-8 a^7 x_1^3 x_2 x_3+3 a^7 x_1^4 x_2 x_3+a^3 x_2^2 x_3+a^5 x_2^2 x_3-8 a^5 x_1 x_2^2 x_3\cr
&&+6 a^5 x_1^2 x_2^2 x_3+6 a^7 x_1^2 x_2^2 x_3-8 a^7 x_1^3 x_2^2 x_3+a^7 x_1^4 x_2^2 x_3+a^9 x_1^4 x_2^2 x_3+a^3 x_3^2-2 a^3 x_1 x_3^2-2 a^5 x_1 x_3^2\cr
&&+6 a^5 x_1^2 x_3^2-2 a^5 x_1^3 x_3^2-2 a^7 x_1^3 x_3^2+a^7 x_1^4 x_3^2+a^3 x_2 x_3^2+a^5 x_2 x_3^2-8 a^5 x_1 x_2 x_3^2+6 a^5 x_1^2 x_2 x_3^2+6 a^7 x_1^2 x_2 x_3^2\cr
&&-8 a^7 x_1^3 x_2 x_3^2+a^7 x_1^4 x_2 x_3^2+a^9 x_1^4 x_2 x_3^2+a^5 x_2^2 x_3^2-2 a^5 x_1 x_2^2 x_3^2-2 a^7 x_1 x_2^2 x_3^2+6 a^7 x_1^2 x_2^2 x_3^2-2 a^7 x_1^3 x_2^2 x_3^2\cr
&&-2 a^9 x_1^3 x_2^2 x_3^2+a^9 x_1^4 x_2^2 x_3^2) y_1^4\nonumber\\[.1cm]
&&+(-1-a^2+6 a^2 x_1-4 a^2 x_1^2-4 a^4 x_1^2+6 a^4 x_1^3-a^4 x_1^4-a^6 x_1^4-2 a^2 x_2+6 a^2 x_1 x_2+6 a^4 x_1 x_2\cr
&&-20 a^4 x_1^2 x_2+6 a^4 x_1^3 x_2+6 a^6 x_1^3 x_2-2 a^6 x_1^4 x_2-a^2 x_2^2-a^4 x_2^2+6 a^4 x_1 x_2^2-4 a^4 x_1^2 x_2^2-4 a^6 x_1^2 x_2^2\cr
&&+6 a^6 x_1^3 x_2^2-a^6 x_1^4 x_2^2-a^8 x_1^4 x_2^2-2 a^2 x_3+6 a^2 x_1 x_3+6 a^4 x_1 x_3-20 a^4 x_1^2 x_3+6 a^4 x_1^3 x_3+6 a^6 x_1^3 x_3\cr
&&-2 a^6 x_1^4 x_3-2 a^2 x_2 x_3-2 a^4 x_2 x_3+24 a^4 x_1 x_2 x_3-20 a^4 x_1^2 x_2 x_3-20 a^6 x_1^2 x_2 x_3+24 a^6 x_1^3 x_2 x_3\cr
&&-2 a^6 x_1^4 x_2 x_3-2 a^8 x_1^4 x_2 x_3-2 a^4 x_2^2 x_3+6 a^4 x_1 x_2^2 x_3+6 a^6 x_1 x_2^2 x_3-20 a^6 x_1^2 x_2^2 x_3+6 a^6 x_1^3 x_2^2 x_3\cr
&&+6 a^8 x_1^3 x_2^2 x_3-2 a^8 x_1^4 x_2^2 x_3-a^2 x_3^2-a^4 x_3^2+6 a^4 x_1 x_3^2-4 a^4 x_1^2 x_3^2-4 a^6 x_1^2 x_3^2+6 a^6 x_1^3 x_3^2-a^6 x_1^4 x_3^2\cr
&&-a^8 x_1^4 x_3^2-2 a^4 x_2 x_3^2+6 a^4 x_1 x_2 x_3^2+6 a^6 x_1 x_2 x_3^2-20 a^6 x_1^2 x_2 x_3^2+6 a^6 x_1^3 x_2 x_3^2+6 a^8 x_1^3 x_2 x_3^2-2 a^8 x_1^4 x_2 x_3^2\cr
&&-a^4 x_2^2 x_3^2-a^6 x_2^2 x_3^2+6 a^6 x_1 x_2^2 x_3^2-4 a^6 x_1^2 x_2^2 x_3^2-4 a^8 x_1^2 x_2^2 x_3^2+6 a^8 x_1^3 x_2^2 x_3^2-a^8 x_1^4 x_2^2 x_3^2-a^{10} x_1^4 x_2^2 x_3^2)y_1^3\nonumber\\[.1cm]
&&+(a-2 a x_1-2 a^3 x_1+6 a^3 x_1^2-2 a^3 x_1^3-2 a^5 x_1^3+a^5 x_1^4+a x_2+a^3 x_2-8 a^3 x_1 x_2+6 a^3 x_1^2 x_2+6 a^5 x_1^2 x_2\cr
&&-8 a^5 x_1^3 x_2+a^5 x_1^4 x_2+a^7 x_1^4 x_2+a^3 x_2^2-2 a^3 x_1 x_2^2-2 a^5 x_1 x_2^2+6 a^5 x_1^2 x_2^2-2 a^5 x_1^3 x_2^2-2 a^7 x_1^3 x_2^2\cr
&&+a^7 x_1^4 x_2^2+a x_3+a^3 x_3-8 a^3 x_1 x_3+6 a^3 x_1^2 x_3+6 a^5 x_1^2 x_3-8 a^5 x_1^3 x_3+a^5 x_1^4 x_3+a^7 x_1^4 x_3+3 a^3 x_2 x_3\cr
&&-8 a^3 x_1 x_2 x_3-8 a^5 x_1 x_2 x_3+26 a^5 x_1^2 x_2 x_3-8 a^5 x_1^3 x_2 x_3-8 a^7 x_1^3 x_2 x_3+3 a^7 x_1^4 x_2 x_3+a^3 x_2^2 x_3+a^5 x_2^2 x_3\cr
&&-8 a^5 x_1 x_2^2 x_3+6 a^5 x_1^2 x_2^2 x_3+6 a^7 x_1^2 x_2^2 x_3-8 a^7 x_1^3 x_2^2 x_3+a^7 x_1^4 x_2^2 x_3+a^9 x_1^4 x_2^2 x_3+a^3 x_3^2-2 a^3 x_1 x_3^2\cr
&&-2 a^5 x_1 x_3^2+6 a^5 x_1^2 x_3^2-2 a^5 x_1^3 x_3^2-2 a^7 x_1^3 x_3^2+a^7 x_1^4 x_3^2+a^3 x_2 x_3^2+a^5 x_2 x_3^2-8 a^5 x_1 x_2 x_3^2+6 a^5 x_1^2 x_2 x_3^2\cr
&&+6 a^7 x_1^2 x_2 x_3^2-8 a^7 x_1^3 x_2 x_3^2+a^7 x_1^4 x_2 x_3^2+a^9 x_1^4 x_2 x_3^2+a^5 x_2^2 x_3^2-2 a^5 x_1 x_2^2 x_3^2-2 a^7 x_1 x_2^2 x_3^2+6 a^7 x_1^2 x_2^2 x_3^2\cr
&&-2 a^7 x_1^3 x_2^2 x_3^2-2 a^9 x_1^3 x_2^2 x_3^2+a^9 x_1^4 x_2^2 x_3^2) y_1^2\nonumber\\[.1cm]
&&+(a^2 x_1-a^2 x_1^2-a^4 x_1^2+a^4 x_1^3-a^2 x_2+a^2 x_1 x_2+a^4 x_1 x_2-2 a^4 x_1^2 x_2+a^4 x_1^3 x_2+a^6 x_1^3 x_2-a^6 x_1^4 x_2\cr
&&+a^4 x_1 x_2^2-a^4 x_1^2 x_2^2-a^6 x_1^2 x_2^2+a^6 x_1^3 x_2^2-a^2 x_3+a^2 x_1 x_3+a^4 x_1 x_3-2 a^4 x_1^2 x_3+a^4 x_1^3 x_3+a^6 x_1^3 x_3\cr
&&-a^6 x_1^4 x_3-a^2 x_2 x_3-a^4 x_2 x_3+4 a^4 x_1 x_2 x_3-2 a^4 x_1^2 x_2 x_3-2 a^6 x_1^2 x_2 x_3+4 a^6 x_1^3 x_2 x_3-a^6 x_1^4 x_2 x_3\cr
&&-a^8 x_1^4 x_2 x_3-a^4 x_2^2 x_3+a^4 x_1 x_2^2 x_3+a^6 x_1 x_2^2 x_3-2 a^6 x_1^2 x_2^2 x_3+a^6 x_1^3 x_2^2 x_3+a^8 x_1^3 x_2^2 x_3-a^8 x_1^4 x_2^2 x_3\cr
&&+a^4 x_1 x_3^2-a^4 x_1^2 x_3^2-a^6 x_1^2 x_3^2+a^6 x_1^3 x_3^2-a^4 x_2 x_3^2+a^4 x_1 x_2 x_3^2+a^6 x_1 x_2 x_3^2-2 a^6 x_1^2 x_2 x_3^2\cr
&&+a^6 x_1^3 x_2 x_3^2+a^8 x_1^3 x_2 x_3^2-a^8 x_1^4 x_2 x_3^2+a^6 x_1 x_2^2 x_3^2-a^6 x_1^2 x_2^2 x_3^2-a^8 x_1^2 x_2^2 x_3^2+a^8 x_1^3 x_2^2 x_3^2)  y_1\nonumber\\[.1cm]
&&+a^3 x_2 (1-a^2x_1^2)^2 = 0\nonumber\\[.1cm]
&&A_2(\BR;x_1,x_2,x_3,y_2;a)=A_1(\BR;x_2,x_1,x_3,y_2;a) = 0\nonumber\\[.1cm]
&&A_3(\BR;x_1,x_2,x_3,y_3;a)=A_1(\BR;x_3,x_2,x_1,y_3;a)=0~,\nonumber
\eea
\end{footnotesize}
which is subject to the Lagrangian condition with respect to the holomorphic symplectic form. The associated variety turns out to coincide with the augmentation variety computed in \cite{Ngweb} with a suitable change of variables. Also, it enjoys the involution symmetry \eqref{z2-associate}, for instance,
\bea\nonumber
A_1(\BR; x_1,x_2,x_3,y_1;a)&=&A_1(\BR;1/(a^2 x_1),x_2,x_3,1/y_1;a)~,\cr
 A_2(\BR; x_1,x_2,x_3,y_2;a)&=&A_2(\BR;1/(a^2 x_1),x_2,x_3,y_2;a)~,\cr
  A_3(\BR; x_1,x_2,x_3,y_3;a)&=&A_3(\BR;1/(a^2 x_1),x_2,x_3,y_3;a)~.\nonumber
\eea

%
\pagebreak

\bibliography{CS}{}
\bibliographystyle{JHEP}

\end{document}